\documentclass{scrartcl}
\usepackage[utf8]{inputenc}
\usepackage{hyperref}
\usepackage{authblk}

\title{Science Case for the new High-Intensity Muon Beams HIMB at PSI}
\subtitle{Edited by A.~Knecht, F.~Meier~Aeschbacher, T.~Prokscha,  S.~Ritt, A.~Signer}

\author[1]{M.~Aiba}
\author[1]{A.~Amato}
\author[1,2]{A.~Antognini}
\author[3]{S.~Ban}
\author[4]{N.~Berger}
\author[1,5]{L.~Caminada}
\author[6]{R.~Chislett}
\author[2]{P.~Crivelli}
\author[1,5]{A.~Crivellin}
\author[1,2]{G.~Dal~Maso}
\author[7]{S.~Davidson}
\author[8]{M.~Hoferichter}
\author[2]{R.~Iwai}
\author[3]{T.~Iwamoto}
\author[1,2]{K.~Kirch}
\author[1]{A.~Knecht}
\author[1]{U.~Langenegger}
\author[9]{A.~M.~Lombardi}
\author[1]{H.~Luetkens}
\author[1]{F.~Meier~Aeschbacher}
\author[3]{T.~Mori}
\author[1,2]{J.~Nuber}
\author[3]{W.~Ootani}
\author[1,10]{A.~Papa}
\author[1]{T.~Prokscha}
\author[11]{F.~Renga}
\author[1]{S.~Ritt}
\author[2]{M.~Sakurai}
\author[1]{Z.~Salman}
\author[1]{P.~Schmidt-Wellenburg}
\author[12]{A.~Sch\"oning}
\author[1,5,*]{A.~Signer}
\author[2]{A.~Soter}
\author[1]{L.~Stingelin}
\author[3]{Y.~Uchiyama}
\author[4]{F.~Wauters}

\date{\vspace{-5ex}}
  
\affil[1]{Paul~Scherrer~Institut, 5232~Villigen~PSI, Switzerland}
\affil[2]{Institute~for~Particle~Physics~and~Astrophysics, ETH~Zurich, 8093~Z\"urich, Switzerland~}
\affil[3]{ICEPP,~The~University~of~Tokyo, 7-3-1~Hongo,~Bunkyo-ku,~Tokyo,~113-0033,~Japan}
\affil[4]{Institute~for~Nuclear~Physics~and~PRISMA$^+$~Cluster~of~Excellence, Johannes~Gutenberg~University~Mainz, Germany}
\affil[5]{Physik-Institut, Universit\"at~Z\"urich, 8057~Z\"urich, Switzerland}
\affil[6]{Department of Physics and
Astronomy, University College London, WC1E 6BT, United Kingdom}
\affil[7]{LUPM, Universit\'e~de~Montpellier, Place~Eug\`ene~Bataillon, 34095~Montpellier, France}
\affil[8]{Albert Einstein Center for Fundamental Physics, Institute for Theoretical Physics, University of Bern, Sidlerstrasse 5, 3012 Bern, Switzerland}
\affil[9]{CERN, 1211~Geneva~23, Switzerland~}
\affil[10]{Dipartimento~di~Fisica~E.~Fermi~\&~INFN~Sezione~di~Pisa,~Largo~Bruno~Pontecorvo, Edificio~C,~208,~56127~Pisa,~Italy~}
\affil[11]{INFN~Sezione~di~Roma, Piazzale~A.~Moro~2, 00185~Roma, Italy~}
\affil[12]{Physikalisches~Institut,~Universit\"at~Heidelberg,~Im~Neuenheimer~Feld~226, 69120~Heidelberg,~Germany}
\affil[*]{Corresponding author: \href{mailto:adrian.signer@psi.ch}{adrian.signer@psi.ch}}

\usepackage{hyperref}

\addtokomafont{caption}{\small}
\setkomafont{captionlabel}{\sffamily\bfseries}
\setcapindent{1em}

\usepackage[style=numeric-comp,sorting=none,backend=biber,maxcitenames=1,maxbibnames=1]{biblatex}
\usepackage{graphicx}

\usepackage{subfig}

\usepackage{amsmath}
\usepackage{amsfonts}
\usepackage{amssymb}
\usepackage{amssymb}
\usepackage{url}
\usepackage{color}

\usepackage{lineno}
\usepackage{tabularx}

\usepackage[per-mode=symbol-or-fraction,separate-uncertainty=true,binary-units]{siunitx}    
\DeclareSIUnit\micron{\micro\metre}
\DeclareSIUnit\ecm{e\cm}

\usepackage{booktabs}

\usepackage{hepnames}

\usepackage{acro}
\acsetup{single}
\DeclareAcronym{PSI}{ short = PSI, long  = Paul Scherrer Institut, long-indefinite = the}
\DeclareAcronym{MAPS}{ short = MAPS, long  = monolithic active pixel sensor}
\DeclareAcronym{DMAPS}{ short = D-MAPS, long  = depleted monolithic active pixel sensor}
\DeclareAcronym{HVMAPS}{ short = HVMAPS, long  = high-voltage monolithic active pixel sensor}
\DeclareAcronym{HVCMOS}{ short = HV-CMOS, long  = high-voltage complementary metal-oxide-semiconductor}
\DeclareAcronym{HDI}{ short = HDI, long  = high-density interconnect, long-indefinite = a, short-indefinite = an}
\DeclareAcronym{EDM}{ short = EDM, long  = electric dipole moment}
\DeclareAcronym{MDM}{ short = MDM, long  = magnetic dipole moment}
\DeclareAcronym{AMM}{ short = AMM, long  = anomalous magnetic moment}
\DeclareAcronym{HVP}{ short = HVP, long  = hadronic vacuum polarisation}
\DeclareAcronym{HLbL}{ short = HLbL, long  = hadronic light-by-light scattering}
\DeclareAcronym{BR}{ short = BR, long  = branching ratio}
\DeclareAcronym{EFT}{ short = EFT, long  = effective field theory}
\DeclareAcronym{MC}{ short = MC, long  = Monte-Carlo}
\DeclareAcronym{LFV}{ short = LFV, long  = lepton flavour violation}
\DeclareAcronym{MFV}{ short = MFV, long  = minimal flavour violation}
\DeclareAcronym{cLFV}{ short = cLFV, long  = charged lepton flavour violation}
\DeclareAcronym{HEP}{ short = HEP, long  = high energy physics}
\DeclareAcronym{PCB}{ short = PCB, long  = printed circuit board}
\DeclareAcronym{SM}{ short = SM, long  = Standard Model of particle physics}
\DeclareAcronym{BSM}{ short = BSM, long  = beyond the Standard Model}
\DeclareAcronym{MSSM}{ short = MSSM, long  = minimal supersymmetric Standard Model}
\DeclareAcronym{LHC}{ short = LHC, long  = Large Hadron Collider}
\DeclareAcronym{DAQ}{ short = DAQ, long  = data acquisition}
\DeclareAcronym{muSR}{ short = $\mu$SR, long  = muon spin rotation}
\DeclareAcronym{LEmuSR}{ short = LE-$\mu$SR, long  = low-energy muon spin rotation}
\DeclareAcronym{LEmu}{ short = LE-$\mu^+$, long  = low-energy muon}
\DeclareAcronym{SmuS}{ short = S$\mu$S, long  = Swiss Muon Source}
\DeclareAcronym{LMU}{ short = LMU, long  = Laboratory for Muon Spin Spectroscopy}
\DeclareAcronym{HIMB}{ short = HIMB, long  = High-Intensity Muon Beams}
\DeclareAcronym{LEM}{ short = LEM, long  = Low Energy Muons}
\DeclareAcronym{HPGe}{ short = HPGe, long  = high-purity germanium}
\DeclareAcronym{APV}{ short = APV, long  = atomic parity violation}
\DeclareAcronym{PV}{ short = PV, long  = parity violation}
\DeclareAcronym{SFHe}{ short = SFHe, long  = superfluid helium}
\DeclareAcronym{NMR}{ short = NMR, long  = nuclear magnetic resonance}
\DeclareAcronym{MRI}{ short = MRI, long  = magnetic resonance imaging}
\DeclareAcronym{ESR}{ short = ESR, long  = electron spin resonance}
\DeclareAcronym{FMR}{ short = FMR, long  = ferromagnetic resonance}
\DeclareAcronym{ZF}{ short = ZF, long  = zero-field}
\DeclareAcronym{HIPA}{ short = HIPA, long  = high-intensity proton accelerator}
\DeclareAcronym{DC}{ short = DC, long  = direct current}
\DeclareAcronym{RF}{ short = RF, long  = radio frequency}
\DeclareAcronym{RFQ}{ short = RFQ, long  = radio frequency quadrupole}
\DeclareAcronym{CDW}{ short = CDW, long  = charge density wave}
\DeclareAcronym{TRSB}{ short = TRSB, long  = time-reversal symmetry breaking}
\DeclareAcronym{CIGS}{ short = CIGS, long  = copper indium gallium selenide}
\DeclareAcronym{RMS}{ short = RMS, long  = root mean square}
\DeclareAcronym{QED}{ short = QED, long  = quantum electrodynamics}
\DeclareAcronym{HFS}{ short = HFS, long = hyperfine splitting}
\DeclareAcronym{FNAL}{ short = FNAL, long  = Fermi National Accelerator Laboratory}
\DeclareAcronym{SES}{ short = SES, long  = single event sensititivity}
\DeclareAcronym{GPU}{ short = GPU, long  = graphics processing unit}
\DeclareAcronym{RMD}{ short = RMD, long  = radiative muon decay}
\DeclareAcronym{GasPM}{ short = Gas~PM, long  = gaseous photo multiplier tube}
\DeclareAcronym{SiPM}{ short = SiPM, long  = silicon photomultiplier}
\DeclareAcronym{MPPC}{ short = MPPC, long  = multi-pixel photon counter}
\DeclareAcronym{LYSO}{ short = LYSO, long  = Lutetium–Yttrium oxyorthosilicate}
\DeclareAcronym{TPC}{ short = TPC, long  = time projection chamber}
\DeclareAcronym{RPC}{ short = RPC, long  = resistive plate chamber}
\DeclareAcronym{mRPC}{ short = mRPC, long  = multi-layer resistive plate chamber}
\DeclareAcronym{DR}{ short = DR, long  = dilution refrigerator}
\DeclareAcronym{BG}{ short = BG, long  = background}
\DeclareAcronym{TDC}{ short = TDC, long = time-to-digital converter}
\DeclareAcronym{LXe}{ short = LXe, long = liquid xenon}
\DeclareAcronym{pTC}{ short = LXe, long = pixelated timing counter}
\DeclareAcronym{CDCH}{ short = CDCH, long = cylindrical drift chamber}
\DeclareAcronym{SLS}{ short = SLS, long  = Swiss Light Source}
\DeclareAcronym{ALP}{ short = ALP, long  = axion-like particle}
\DeclareAcronym{MPGD}{ short = MPGD, long = micro-pattern gaseous detector}
\DeclareAcronym{VUV}{ short = VUV, long  = vacuum ultraviolet}

\numberwithin{equation}{section}

\begin{document}

\maketitle

\begin{abstract}
In April 2021, scientists active in muon physics met to discuss and work out the physics case for the new High-Intensity Muon Beams (HIMB) project at PSI that could deliver of order $10^{10}$\,s$^{-1}$ surface muons to experiments. Ideas and concrete proposals were further substantiated over the following months and assembled in the present document. The high intensities will allow for completely new experiments with considerable discovery potential and unique sensitivities. The physics case is outstanding and extremely rich, ranging from fundamental particle physics via chemistry to condensed matter research and applications in energy research and elemental analysis.  In all these fields, HIMB will ensure that the facilities S$\mu$S and CHRISP on PSI's High Intensity Proton Accelerator complex HIPA remain world-leading, despite the competition of muon facilities elsewhere.  
\end{abstract}

\newpage

\tableofcontents

\newpage


\section{Introduction}

Muon physics covers research fields from fundamental particle physics to materials science. Muons are leptons with either negative or positive electric charge, such as electrons and positrons, but 207 times more massive and unstable. Muons are efficiently produced in weak decays of pions which are usually produced by proton beams hitting nuclei in some target material. Muons themselves decay again weakly, with a lifetime of about 2.2\,$\mu$s. They have spin 1/2 and a magnetic moment.

In particle physics, since their first discovery, muons played an important role to help develop the theory and establish the present \ac{SM}. Muons are today used to perform some of the most sensitive tests to probe the limits of our theoretical understanding. In almost all presently existing tensions of experimental measurements with precision predictions of the \ac{SM} muons are involved, suggesting they might be playing a key role in finding and establishing the breaking of this best theory to date.   
With a similarly long history, a broad range of research topics in solid-state physics, chemistry and materials science is being addressed by muon spin spectroscopy, usually using positive muons as highly sensitive local magnetic probes.

In all these fields, the availability of low-momentum, high-intensity muon beams is a prerequisite, and technological progress from muon production to detector development and sample environments boosted reach and capabilities of the research over decades.

World-wide, several large-scale facilities provide muons to experiments and user instrumentation. They are located at RAL (UK), at J-PARC (Japan), at TRIUMF (Canada) and at PSI in Switzerland. They are all active in condensed matter research, while broader programs in particle physics are pursued mostly at J-PARC and at PSI. A dedicated program in particle physics also exists at the muon campus of FNAL (USA). Several other accelerator facilities are developing or considering a future muon physics program, among them the CSNS (China), RAON (Korea), SNS (USA), and the ESS (Sweden). Some facilities provide pulsed, others continuous muon beams, complementing each other and allowing for different kinds of experiments.

The High Intensity Proton Accelerator facility HIPA at PSI~\cite{Grillenberger:2021kyv} provides one of the most powerful proton beams to target stations, with 1.4\,MW average beam power presently only matched by the pulsed SNS. HIPA has a 50\,MHz time structure leading to quasi-continuous beams of slow muons. Many experiments today use so-called ‘surface muons’, positive muons generated in the decay of positive pions stopped close to the surface of a production target. These muons have well-defined momenta, are fully polarised and can be efficiently transported into secondary beam areas. The muon beams at PSI are presently leading the high-intensity frontier, with surface muon rates of order $10^8$\,s$^{-1}$, and PSI is home to world-leading research in particle physics (see~\cite{SciPost:ppPSI} for a recent review) and condensed matter research with muons (see~\cite{RevModPhys.69.1119, doi:10.7566/JPSCP.2.010201, AMATO2009606, doi:10.1080/08957959.2016.1173690, condmat5020042,10.3389/fphy.2021.651163}  for reviews).

Installation of two new \ac{HIMB}, proposed within the IMPACT project (Isotope and Muon Production using Advanced Cyclotron and Target technology) at PSI, will constitute a leap forward for muon physics. Surface muon intensities will be boosted to $10^{10}$\,s$^{-1}$, serving particle physics in one and condensed matter research in a second beamline. These unprecedented muon intensities will allow for completely new experiments.

The interested national and international muon science community met for a \ac{HIMB} physics case workshop, April 6-9, 2021 at PSI, as a kick-off event for the work presented in this paper. While we report on ongoing work and many aspects will be worked out over the coming years and many new, additional applications and ideas for \ac{HIMB} will continue to appear, we have assembled here a compelling physics case documenting unprecedented opportunities for muon science. \ac{HIMB} will secure a world-wide leading position of fundamental and applied muon science at PSI and attract many national and international user groups. Specific findings of future research cannot be predicted, however, it is clear that \ac{HIMB} will have great impact. Projects to be conducted on this new facility will push the limit of the known far into presently unknown territory with plenty of opportunities for ground-breaking discovery, furthering basic knowledge and understanding of nature, development of novel technologies and fertilising spin-offs to other fields.

This paper is structured according to the two main use cases of the two beamline branches of \ac{HIMB} as follows:
Opportunities for particle physics are discussed in~\autoref{sec:particlephysics}. Muon spin spectroscopy and materials science follow in~\autoref{sec:musr}. In~\autoref{sec:facilities} we deal with facility aspects, a short description of the lay-out and the properties of HIMB as well as further add-ons and technological aspects.


\section{Particle physics with HIMB} \label{sec:particlephysics}

\subsection{Muon flavour physics}\label{sec:CLFV}
Flavour is arguably the least understood sector of the SM. There is
still no compelling explanation as to why three copies (or flavours)
of matter field families exist. Put specifically for charged leptons,
why is there a muon and a tau in addition to the electron? According
to the SM, the only difference between the various flavours is the
coupling to the Higgs field, resulting in (widely)
different masses.

Given the need to go beyond the SM it is natural to explore the sector
that is least understood. Does the muon indeed behave precisely as the
electron? There have been several recent measurements that cast doubt
on this statement. The so-called $B$-anomalies seem to indicate that
the decay of some $B$ mesons involving a $\mu^+\,\mu^-$ pair in the
final state do not precisely follow the pattern expected from
corresponding decays with an electron-positron or tau pair in the
final state ~\cite{Graverini:2018riw,HFLAV:2019otj,Fischer:2021sqw}. Also,
measurements of the anomalous magnetic moment (AMM) of the muon~\cite{Abi:2021gix,Bennett:2006fi} are in
tension with the prediction of the SM~\cite{Aoyama:2020ynm}. Thus, it is not
inconceivable
that the role of the muon in particle physics will evolve from ``who
ordered that'' to being the key to unlock the door to physics beyond
the SM. Ever more precise investigations of processes involving muons
are required to address these questions.

In the \ac{SM} without right-handed (and therefore with massless)
neutrinos, lepton flavour is conserved. Hence, a muon cannot decay
into an electron without a muon neutrino and an electron antineutrino
in the final state. The neutrinos are required to balance muon and
electron flavour. However, this symmetry is accidental. Writing down
all operators with the fields of the \ac{SM}, compatible with gauge
invariance, Lorentz invariance, and renormalisability (i.e. of
dimension 4 or less) it just so happens that there is no operator that
violates lepton flavour symmetry.  It is by no means a fundamental
ingredient in the construction of the \ac{SM}. And, more importantly, we
know that this symmetry is broken in nature. Indeed, neutrinos have
tiny but non-vanishing masses and as a result they oscillate,
i.e. they change their flavour. This can be seen for example by letting
muons decay and observe at a distant detector the emergence of electrons. This
is precisely the observation of lepton flavour violation. Accordingly, the
\ac{SM} has to be modified by the introduction of right-handed neutrinos. This
also leads \ac{cLFV} in muon decays investigated at PSI such as 
$\mu\to e \gamma$ or $\mu\to e e e$. In the  \ac{SM} with massive neutrinos,
these decays can happen, albeit with a \ac{BR} smaller than $10^{-54}$. Thus, while
there is nothing sacred about
lepton flavour, for practical purposes \ac{cLFV} processes are forbidden in
the SM and any measurement of such a process is a clear signal of \ac{BSM}
physics.  This offers a unique opportunity to search for \ac{BSM} physics and
has triggered a wealth of experimental and theoretical activities~\cite{Calibbi:2017uvl}.

Similar to the introduction of right-handed neutrinos, a generic \ac{BSM}
model induces \ac{cLFV}. As one typical example we mention supersymmetry.
Even in the \ac{MSSM} the soft breaking terms include numerous \ac{cLFV} operators
and currently there is no
satisfactory explanation as to why they are absent or suppressed. The
situation is similar for virtually all extensions of the \ac{SM}.  Thus,
from a \ac{BSM} point of view there is absolutely nothing exotic about
\ac{cLFV}. To the contrary, the absence of cLFV is exotic.  

This can also be understood from an \ac{EFT} point
of view. If we assume \ac{BSM} physics is at a high scale well beyond the
energy scale of the process, we can parameterise BSM effects through
operators of dimension larger than 4. Using the SM fields, there is a
single dimension 5 operator~\cite{Weinberg:1979sa}, closely linked
to neutrino masses. At dimension 6, there are numerous operators
including contact interactions (four-fermion operators) and dipole
interactions among others \cite{Grzadkowski:2010es,Jenkins:2017jig}. The flavour-diagonal dipole operators contribute to the \ac{EDM} and \ac{AMM} of the
corresponding particles. Their off-diagonal variants directly induce for example
the \ac{cLFV} decay $\mu\to e \gamma$. Similarly, off-diagonal four-fermion
operators contribute to decays like $\mu\to eee$.  Including a flavour-diagonal
dimension~6 operator to parameterise \ac{BSM} effects, but avoiding the
corresponding \ac{cLFV} operators requires either the introduction of an additional
symmetry as explanation, or a fine-tuning for which there is
currently no theoretical justification at all.
In this connection it should also be mentioned that the
conventional split into contributions from dipole and contact
interactions is applicable at the high scale only. Through
renormalisation-group evolution these operators mix and contact
interactions directly impact, e.g., the $\mu\to e \gamma$ process
\cite{Pruna:2014asa,Crivellin:2017rmk}.

The mixing of different effects also implies that a diverse program is
required to investigate cLFV. In case a signal is measured in a single
process, it is impossible to pinpoint what precisely causes this cLFV
decay. As for the current tension for the AMM of the muon, there will
be many possible explanations. In order to narrow down the nature of
BSM that causes a potential cLFV decay it is imperative to measure or
constrain as many such processes as possible. While this also concerns
cLFV tau decays, the most stringent limits come from cLFV muon decays.

\begin{figure}
  \centering
   \includegraphics[trim=0 0 0 0, clip,width=0.9\textwidth]{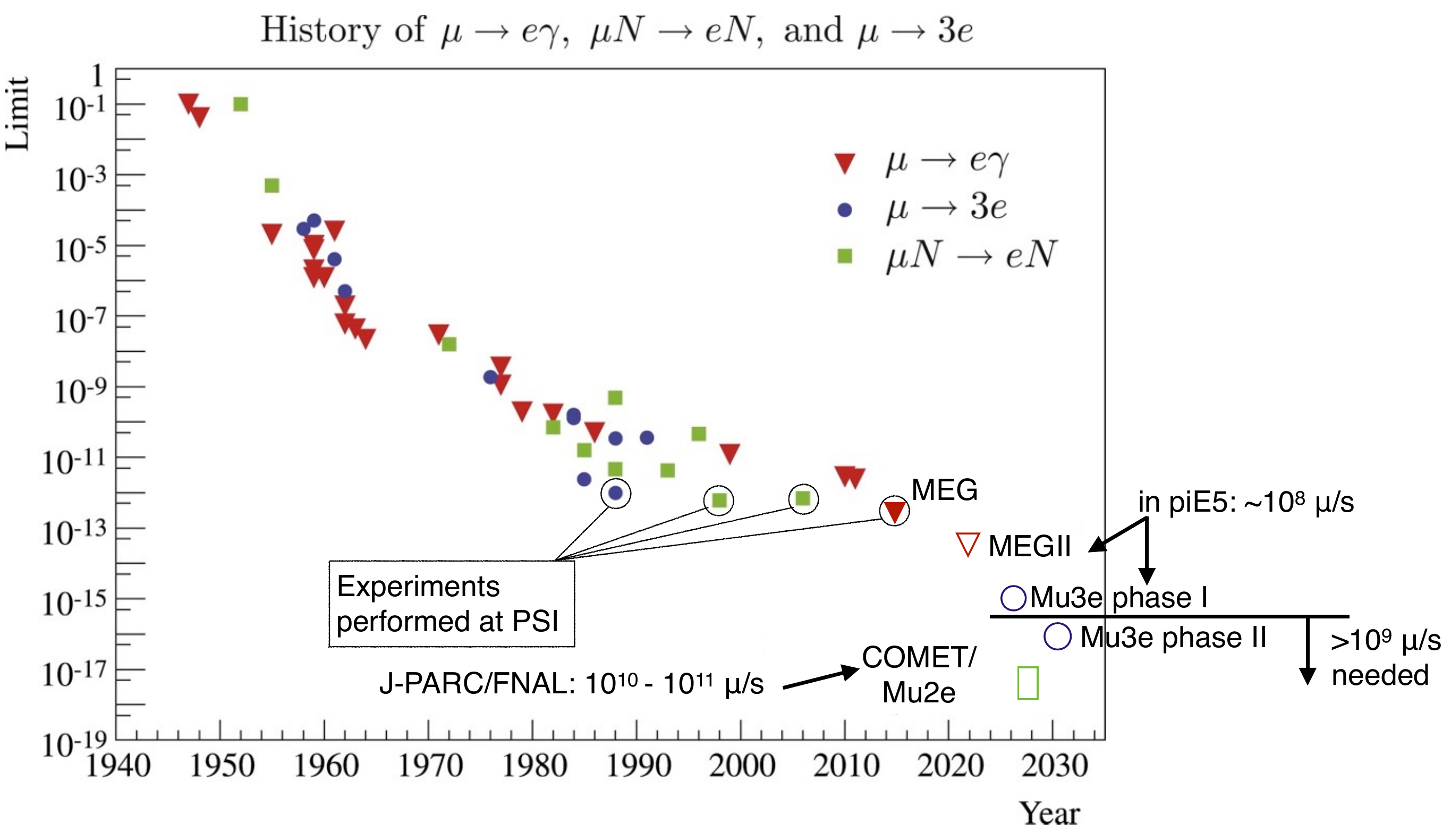}
   \caption{The history of limits for branching ratios of the three golden cLFV channels. Symbols that are not filled in indicate projected measurements. Plot modified from \cite{Bernstein:2013hba}.}
    \label{fig:cLFV_history}
\end{figure}

There are three golden cLFV muon-decay channels that have dramatic
impact. In addition to the already mentioned $\mu\to e\,\gamma$, these are the
decay $\mu\to e\, (e^+e^-)$ (or $\mu\to 3e$ for short) as well as the conversion
of a muon to an electron in the field of a nucleus, $\mu\,N\to e\,N$. As
illustrated in \autoref{fig:cLFV_history}, 
there is a long history of searches for such decays. The current best
limits on the \ac{BR} for all three decays have been obtained at PSI with
$\mathrm{BR}(\mu\to 3e) <  1\cdot 10^{-12}$~\cite{SINDRUM:1987nra},
$\mathrm{BR}(\mu\to e\gamma) < 4.2\cdot 10^{-13}$~\cite{MEG:2016leq}
and $\mathrm{BR}(\mu\,N\to e\,N) < 7 \cdot 10^{-13}$~\cite{SINDRUMII:2006dvw}.

The sensitivity to $\mu\to e\,\gamma$ and $\mu\to 3e$ will
be improved further in the coming years. Using the currently existing
beamline $\pi$E5 at PSI with $10^8\,\mu\,\mathrm{s}^{-1}$,
MEG~II~\cite{Baldini:2018nnn} and phase~I of Mu3e~\cite{Arndt:2020obb}
aim to reach a sensitivity of $\mathrm{BR}(\mu\to e\gamma)=
6\cdot10^{-14}$ and $\mathrm{BR}(\mu\to 3e) \sim 2\cdot 10^{-15}$,
respectively.

To improve the sensitivity for the third golden channel, $\mu\,N\to
e\,N$, it is advantageous to use a pulsed beam.  At J-PARC and
Fermilab a dedicated effort is ongoing by COMET~\cite{COMET:2018auw}
and Mu2e~\cite{Mu2e:2014fns} with the long-term goal to increase the
sensitivity for muon conversion by several orders of magnitude. For
the other two channels a continuous muon beam is better suited. As
illustrated in \autoref{fig:cLFV_history}, improving upon MEG~II and
phase~I of Mu3e requires an increased muon intensity, as provided by the
\ac{HIMB} project. In fact, from the beginning it was envisaged to have a phase~II
for Mu3e and in \autoref{sec:mu3e} the decisive impact of \ac{HIMB} for
Mu3e will be described in detail. Prospects for measurements of
$\mathrm{BR}(\mu\to e\gamma)$ beyond MEG~II will be discussed in
\autoref{sec:mueg}.

There are several observables linked to cLFV that can also be
investigated more precisely with HIMB. As already mentioned, the \ac{AMM} and
the \ac{EDM} of the muon are related to
the same dipole operator, albeit with diagonal flavour indices. Plans
for future PSI activities in this direction are discussed in
\autoref{sec:moments}.  Muonium, a bound state of a positive muon with
an electron $M=(\mu^+\,e^-)$ also serves as a clean probe for testing
cLFV through oscillation into antimuonium $\overline{M}=(\mu^-\,e^+)$. More
generally, muonium serves as a laboratory for \ac{QED} and gravity tests. This
will be elaborated upon in \autoref{sec:muonium}. In \autoref{sec:pool} we
will touch upon further possible applications of HIMB for particle
physics with muons. Taken together, HIMB is an essential facility to
fully explore the opportunities provided by high-intensity, low-energy
experiments. It will ensure PSI remains at the forefront of this
branch of particle physics, lead to further insights into its theory
landscape~\cite{Colangelo:2021xix} and trigger progress in
experimental techniques that often has applications well beyond
particle physics as well.


\newcommand{\mte}{$\mu^+\rightarrow e^+e^-e^+$\xspace}
\newcommand{\mtenn}{$\mu^+\rightarrow e^+e^-e^+\nu\bar{\nu}$\xspace}

\subsection{The Mu3e experiment} \label{sec:mu3e}

The Mu3e experiment at PSI is aiming to search for the \ac{LFV}
decay of the muon in the reaction $\mu^+ \rightarrow e^+ e^+ e^-$ with
unprecedented sensitivity.

The ultimate experimental sensitivity depends mainly on the achievable energy
(momentum) resolution and has been estimated~\cite{Blondel:2013ia} to be
$\mathcal{O}(\num{E-16}$) \ac{SES} for the new and innovative conceptual design
of the Mu3e experiment, and assuming state-of-the-art detector technologies.
Another limitation comes from the available muon rate which, for a given number of muon stops, not only
determines  the total running time of the experiment but also the accidental \ac{BG} rate, and thus the maximum achievable sensitivity.

The validity of the experimental concept and the performance of the applied high-rate detector technologies will be tested during Mu3e phase I at the $\pi$E5 compact muon beam line, with an anticipated start of data taking in the year 2023. Experience collected during phase~I will provide important input for the design of the phase~II detector.

\begin{figure}[tb!]
  \centering
     \includegraphics[width=0.69\textwidth]{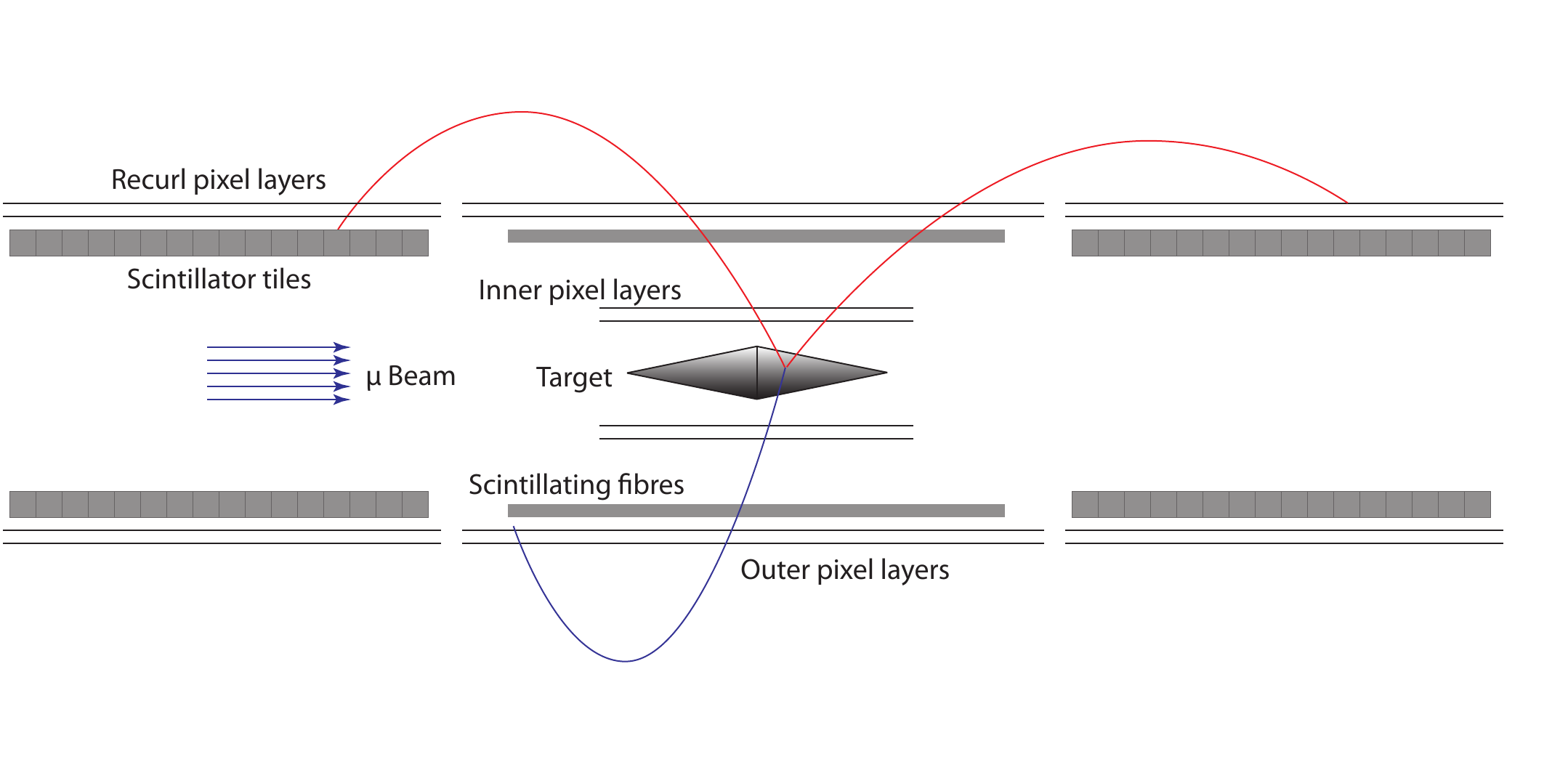}
    \caption{Schematic view of the Mu3e phase I experiment. Taken from
      \cite{Arndt:2020obb}.
    }
    \label{fig:Mu3e_PhaseI_design}
\end{figure}
The design of the Mu3e phase~I experiment is sketched in \autoref{fig:Mu3e_PhaseI_design} and shortly described here. A full description of the experimental setup is given in the Mu3e phase~I technical design report~\cite{Arndt:2020obb}.
Positively charged muons with a momentum of about $p=\SI{28}{MeV}/c$ (``surface muons'') pass a moderator and are then stopped on a hollow double cone shaped target.
The \SI{10}{cm} long muon stopping target is surrounded by two layers of silicon pixel sensors which have a pixel size of $\SI{80}{\mu m} \times \SI{80}{\mu m}$.
Tracking is completed by the two outer pixel layers in the central region, and the up- and downstream recurl stations which register hits from back-curling tracks in the strong solenoidal magnetic fields of $B=\SI{1}{T}$.
All silicon pixel layers use the monolithic MuPix chip produced in \ac{HVMAPS} technology~\cite{Peric:2007zz,Peric:2013cka}.
The ultra-thin pixel modules have a radiation length of only about $\SI{1e-3}{X_0}$ and are cooled by an innovative gaseous helium system.
The timing system consists of scintillating fibres in the central region and scintillating tiles placed up- and downstream 
from the central region.
Both timing detector systems provide sub-nanosecond resolution to measure time coincidences.
Vertex and timing information combined are crucial to suppress accidental \ac{BG}, mainly originating from Bhabha scattered positrons in combination with ordinary Michel decays, see \autoref{sec:Mu3e_bg}.
The data are continuously readout and processed by a \ac{GPU} based filter farm
which searches for three tracks pointing to a common vertex (3-prong) and
reconstructs the charge and momentum of the particles.

In the following the proposed detector design for Mu3e phase~II is
presented, taking into account the \ac{HIMB} constraints as well as the expected \ac{BG}.

\subsubsection{Mu3e final focus}
\label{sec:Mu3eFF}
The transverse beam emittance $\epsilon_\textrm{trans}=\epsilon_x \cdot
\epsilon_y$ of the \ac{HIMB} will be about one order of magnitude larger compared
to the compact muon beam line used for Mu3e phase I.
The so-called matched beam size depends on the magnetic field $B$ of the
experimental solenoid
\begin{eqnarray}
\sigma_{x,y}^\textrm{matched} & = & \sqrt{\frac{2 \epsilon_{x,y} \, p}{e \,
B}}
\label{eq:mu3e_matched_beam}
\end{eqnarray}
with $p$ being the muon momentum at the stopping target. To achieve an overall muon stopping efficiency similar to Mu3e phase I the
 diameter of the stopping target  (\SI{38}{mm} in phase~I) would have to be significantly increased.

However, a wider muon stopping target is detrimental to the requirement of an excellent vertex resolution which is mainly given by the track extrapolation uncertainties
due to multiple scattering.
As a consequence the accidental \acs{BG} rejection scales with the muon stopping target diameter as $\approx 1/D^2$.
Thus a small target diameter, and correspondingly small beam size, is favoured to achieve highest sensitivity in an accidental \acs{BG} limited environment,
see also discussion in \autoref{sec:Mu3e_bg}

\begin{figure}[b!]
  \centering
     \includegraphics[trim=0 0 0 0,clip,width=0.499\textwidth]{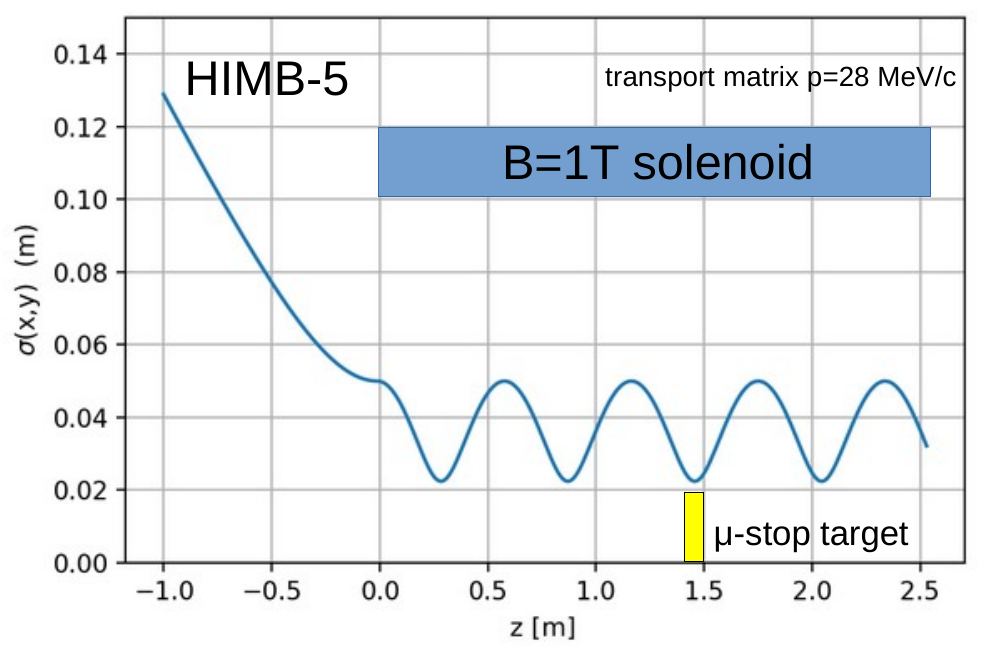}
     \includegraphics[trim=0 0 0 0,clip,width=0.49\textwidth]{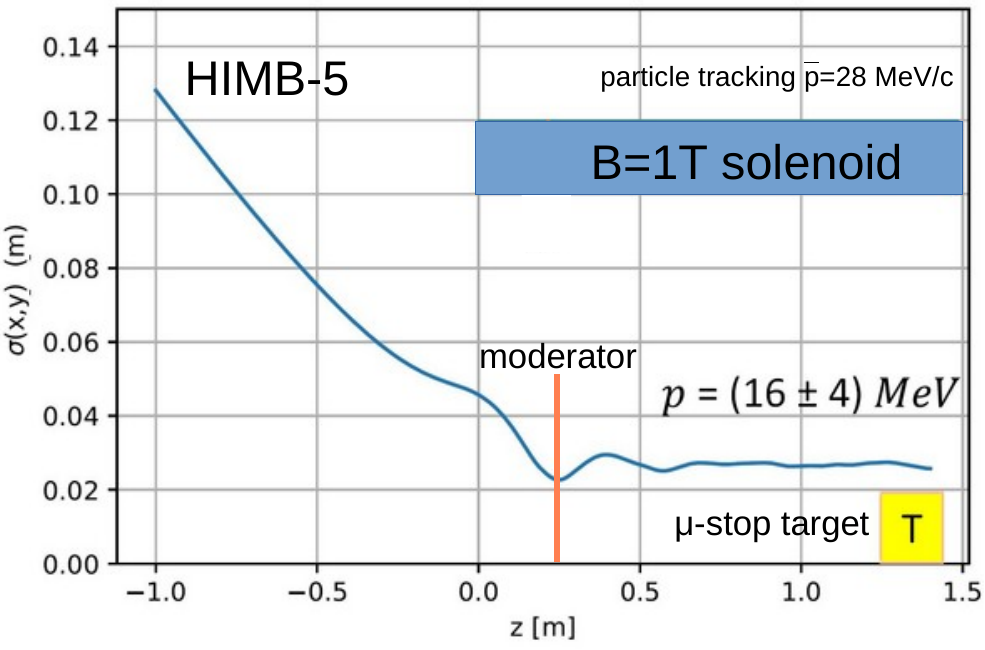}
     \caption{\SI{1}{\sigma} beam envelopes for the \acs{HIMB}-5 scenario (see \autoref{sec:scenarios}) and a solenoid field
       of $B=\SI{1}{\tesla}$ for an unmatched beam optics (left) and a matched beam optics (right) with
       a \SI{600}{\mu m} mylar moderator placed inside the
       solenoid entrance region.
       The \SI{1}{\sigma} beam envelope is determined by transport matrices (left) and using particle tracking
       including energy loss and multiple scattering in the moderator (right). The right plot is
       taken from \cite{Grefkes2021} and has been modified.}
    \label{fig:Mu3e_beam_envelop}
\end{figure}

Consequently, one of the design challenges for Mu3e at \ac{HIMB} is the minimisation
of the matched beam size which can be achieved, according to \eqref{eq:mu3e_matched_beam}, by three
measures:
\begin{itemize}
\item reduction of the muon beam momentum $p$,
\item increase of the magnetic field strength $B$ of the experiment,
\item special optics with oscillating beam envelope.
\end{itemize}
Simulation studies show that the initial muon beam momentum of about
\SI{28}{MeV}/$c$ can be decreased using a moderator to about \SI{16}{}
$\pm$ \SI{4}{MeV}/$c$
without significantly losing muons.
In order to profit from the lower beam momentum, the moderator must be placed
in a region of small beam waist where the beam dispersion is large and 
the additional contribution
from multiple scattering in the moderator is negligible.
Several designs with different placements of the moderator,
different type of moderators (solid and gaseous), and
with and without pre-solenoids have been
simulated by the method of single particle tracking.
The simulation also considers the fringe field of the solenoid which
significantly influences the trajectories of the very low momentum muons.
For the example of a design with a solid moderator in the entrance region of the Mu3e solenoid, the $\SI{1}{\sigma}$ envelope is shown in \autoref{fig:Mu3e_beam_envelop}~(right).
The most critical parameter for the muon transmission efficiency is the
size of the beam pipe inside the experimental magnet.
Simulation studies show that the transmission efficiency scales approximately
linear with the beam-pipe radius in the region $R=30$ - \SI{50}{mm}.
Whether the beam-pipe radius can be increased with respect to the phase~I design
(\SI{30}{mm}) needs to be answered by detailed technical design studies.

The second option is to increase the experimental solenoid field strength.
According to \eqref{eq:mu3e_matched_beam} the beam size  would shrink
$\propto 1/\sqrt{B}$.
At the same time, however, particles from the muon decay would be
bent more strongly, leading to a larger acceptance loss of low-momentum tracks,
if the detector system is not resized accordingly.
Considering the already very tight space constraints of the Mu3e phase I design, a simple
miniaturisation with two inner and two outer pixel layers
would be very challenging.
A better option is to add a third inner pixel layer. With three closely stacked pixel layers
low-momentum tracks can be reconstructed which do not reach
the outer pixel layers due to the strong magnetic field. The higher
redundancy also helps to reduce the hit combinatorics in the track reconstruction.
Ideally, one of the three pixel layers would also provide very precise time
information (timing layer).
Designs with three inner pixel layers are discussed in the context of
a combined Mu3e and MEG search in \autoref{sec:MEGMU3E}.

The third option is to use a special beam optics where the final focus of the muon beam is intentionally not matched
with the experimental solenoid field.
In that case the beam envelope inside the experiment
oscillates with a wave length of $\approx \SI{60}{cm}$ for $B=\SI{1}{\tesla}$ and $p=\SI{28}{MeV}/c$, see \autoref{fig:Mu3e_beam_envelop}~(left).
This beam setup provides periodic minima which are
significantly smaller than the equilibrium
beam size.
Such a design, however, is difficult to realise for several reasons:
A) the beam pipe must be large to include all periodic beam waists,
B) the minima are smeared out due to dispersion effects especially after placing a moderator,
and C) the region around the minimum is relatively narrow, $\approx \SI{10}{cm}$, thus
excluding a long muon stopping target which is favoured to reduce accidental \acs{BG}.

\subsubsection{Muon stopping target}
\begin{figure}[tb!]
  \centering
     \includegraphics[width=0.95\textwidth]{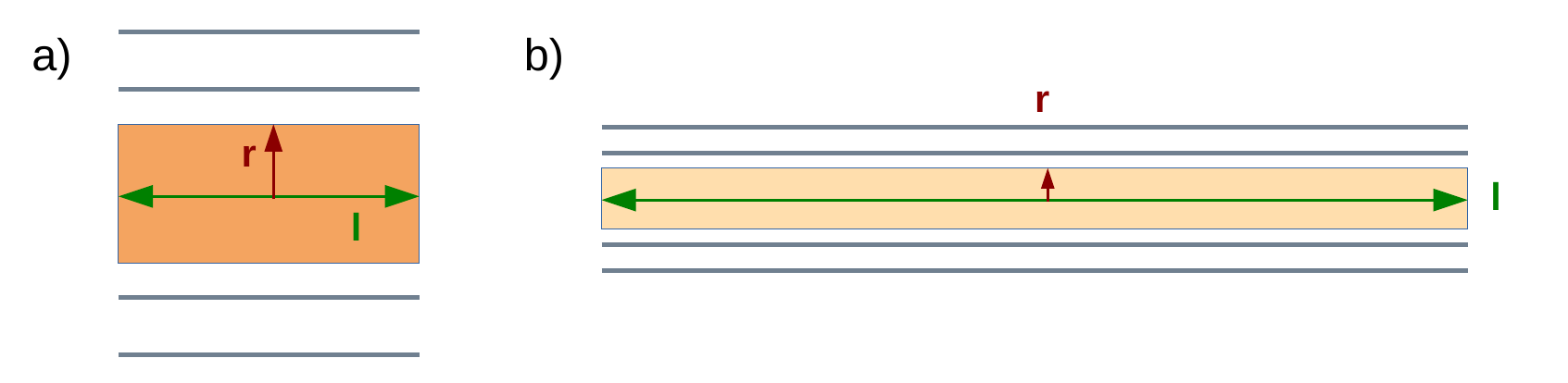}
    \caption{Longitudinal view of the Mu3e muon stopping target region for a)
      a short and wide, and b) a long and narrow geometry.
      The different brightnesses represent different average densities of the
      stopping material. The inner pixel tracking layers are shown in grey.}
    \label{fig:Mu3eTargetRegionLength}
\end{figure}

Accidental \acs{BG} can be best suppressed by using a long muon stopping target and by ensuring excellent vertex resolution,
so that two muons decaying at the same time can be distinguished by tracking (vertexing).
For a given beam momentum, the longitudinal radiation length $X_L$ is fixed by
the required stopping power and given by $X_L \propto l \rho$, with $\rho$ being
the average density of the muon stopping target.
The transverse direction is the preferred detection plane.
The average transverse radiation length is given by the product of target
density and radius which should ideally match the beam size,
$X_T \propto r \rho = X_L \, r/l $.
Since $X_T$ affects the signal particles (e.g. multiple scattering)
and the generation of accidental \acs{BG} (Bhabha), 
the primary design goal\footnote{The relation between longitudinal and
  transverse radiation length holds, irrespective of the detailed stopping target
  geometry and material (gas, solid, etc.).}
is a long muon stopping target and a small beam size,
see \autoref{fig:Mu3eTargetRegionLength}.

\begin{figure}[tb!]
  \centering
     \includegraphics[width=0.9\textwidth]{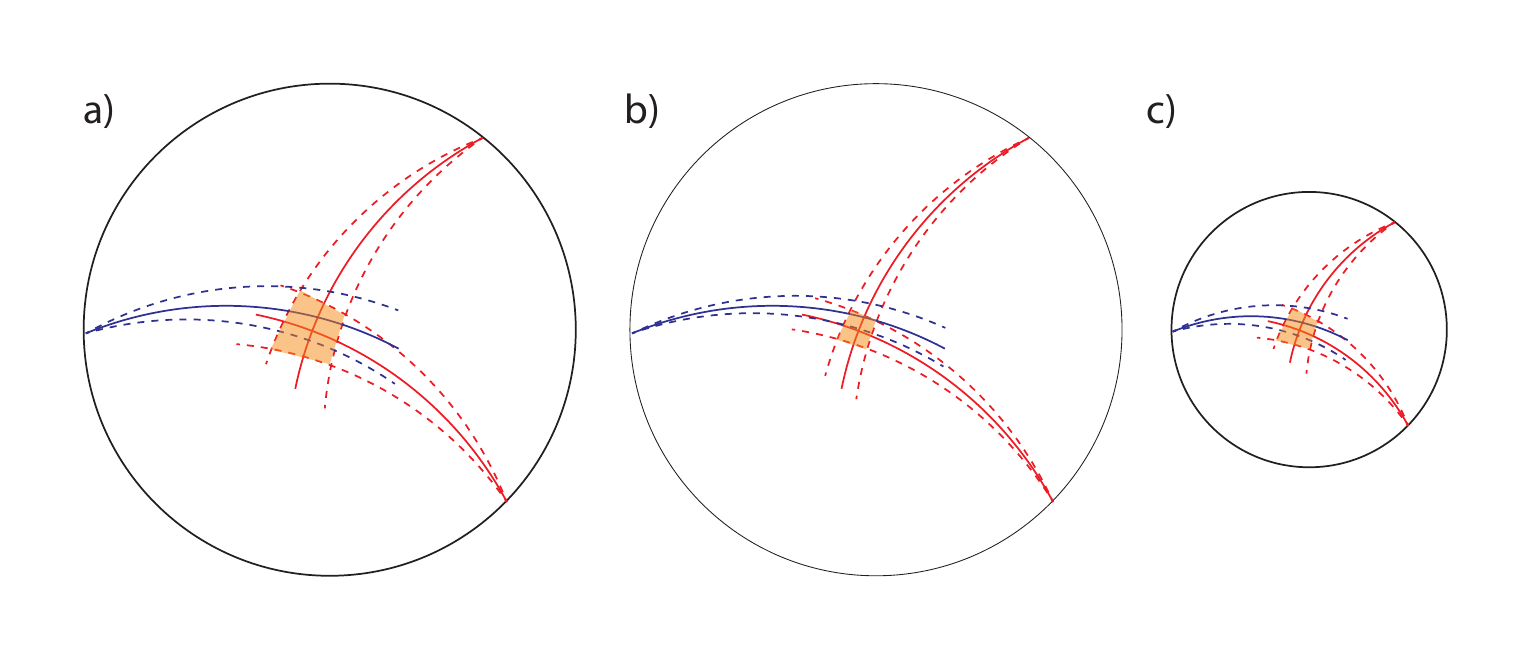}
    \caption{Transverse view of the Mu3e target region. Multiple scattering in the first detector layer leads to extrapolation uncertainties, defining the region susceptible to accidental \acs{BG} (a). This region can be made smaller by reducing material in the first layer (b) or by reducing the transverse extent of the target region (c). }
    \label{fig:Mu3eTargetVertexing}
\end{figure}

The impact of the radial target size on the
directional momentum resolution is illustrated in \autoref{fig:Mu3eTargetVertexing}.
The accidental \acs{BG} rate depends on the vertex resolution which is
mainly determined by multiple scattering in the first tracking layer.
The vertex resolution can be improved by reducing the material in
the first tracking layer or by using a smaller target radius, thus reducing the
extrapolation uncertainties.

For Mu3e phase~II, a hollow double cone structure similar to phase I is considered as well as new ideas like silicon aerogel or gaseous targets. 
Gas is favourable for constructing a long muon stopping target and has also the advantage of a very small transverse radiation length.
Active (instrumented) targets, which provide direct information about the muons stop/decay position, have the potential to improve the tracking, and therefore momentum resolution.
However, the potential improvement is very small and probably not worth the effort\footnote{Active targets could also be used to reject accidental \acs{BG} by
  identifying and rejecting close muon stops in space and time, which however also
  reduces the effective muon stopping rate.
}.

Several target designs and materials have been simulated in a $B=\SI{1}{\tesla}$ field and optimised with the goal to 
equally distribute the muon stops along the beam direction~\cite{Grefkes2021}.
For a muon stopping target radius of \SI{19}{mm}, stopping rates of about
$\SI{2e9}{\mu/s}$ were calculated for solid state targets with different
geometries (double cone, tilted plane, helicoid).
The flattest stopping distribution on the longitudinal axis was achieved with
a \SI{30}{cm} long gaseous ethane target at atmospheric pressure.
The corresponding total stopping rate is about $\SI{2.5e9}{\mu/s}$ and slightly higher as for the solid targets.
These estimates were obtained for a rather large beam-pipe radius of
\SI{50}{mm}.
Which target design eventually is chosen will also depend on the technical
feasibility to build a long ($\gg$\SI{10}{cm}) stopping target with low density.

\subsubsection{Mu3e signal and background}
\label{sec:Mu3e_bg}

Detailed studies of signal efficiencies and \acs{BG} suppression have been performed in the context of Mu3e phase~I
and are documented in the technical design report \cite{Arndt:2020obb}.
Assuming a muon stopping rate of $\SI{1e8}{\mu/s}$ a single event sensitivity of \num{2E-15} is expected.

In order to reach the ultimate sensitivity goal of \num{E-16} at \ac{HIMB}, higher muon stopping rates and an improved detector are required. In the following, some of the challenges for almost completely suppressing \acs{BG} whilst keeping a reasonable signal efficiency in a much harsher environment are discussed.

For \ac{HIMB}, we assume that a surface muon beam with $\SI{2E9}{\mu /s}$ is stopped in the experiment and the muons decay at rest. The signal signature is then two positrons and one electron originating from a common vertex, coincident in time and with a four-momentum sum corresponding to a stopped muon. All \acs{BG} processes that can mimic this signal topology need to be suppressed to an expectation of less than one \acs{BG} event in the signal region over the lifetime of the experiment. The two main \acs{BG} classes, namely the rare muon decay \mtenn and accidental combinations of two positrons and an electron are discussed below.

Besides an excellent \acs{BG} suppression, a decent efficiency for the signal process is required in order to achieve the sensitivity goal in a reasonable amount of time. The efficiency is on the one hand limited by the detector geometry, which does not cover the area around the beam and defines a lower limit on reconstructable transverse momenta (roughly \SI{10}{MeV}/$c$ transverse momentum for the phase~I Mu3e design). On the other hand, any gaps or deficient parts in the detector will affect the efficiency.
As all three tracks have to be fully reconstructed,
detector inefficiencies enter the final efficiency with a high power. The exact signal efficiency depends on the signal kinematics, which in turn depends on the type of new physics.
For the phase~I design it ranges from about 10\% for pure dipole operators to about 19\% for four-fermion operators \cite{Perrevoort:2018okj, Perrevoort:2018ttp, Arndt:2020obb}.
With the planned improvements  for the phase~II detector design, see
\autoref{sec:Mu3e_PhaseII_design}, the signal efficiency is expected to increase by at least 50\% compared to phase~I.

The process \mtenn is allowed in the \ac{SM} and occurs with a \ac{BR} of \num{3.4E-5} \cite{SINDRUM:1985vbg}. It is indistinguishable from the signal process except for the energy and momentum carried away by the neutrinos.
The invariant mass of the $e^+e^-e^+$ system has been calculated to next-to-leading order \cite{Pruna:2016spf}.
Close to the endpoint, the spectrum falls roughly with the sixth power of the \emph{visible mass} and
the branching fraction falls below \num{E-16} about $\SI{1}{MeV}/c^2$ below the endpoint, driving the requirements for the invariant mass resolution.

Combinations of different processes that in total have two positrons and one electron in the final state can mimic the signal signature if their origins in space and time are not resolvable by the detector and their combined kinematics match the signal.
These accidental \acs{BG}s scale with powers of the muon decay rate $R$.
The ordinary muon decay is a very rich source of positrons.
So when studying accidental \acs{BG}s, sources of electrons have to be investigated, where those that appear in coincidence with a positron are particularly dangerous.
Since most electron production processes scale with the amount of material seen by positrons leaving the target region transverse to the beam $X_T$, see \autoref{fig:Mu3eTargetRegionLength}, the best \acs{BG} mitigation is to reduce all material to the minimum.
Furthermore,
accidental \acs{BG}s can be reduced by good timing ($\sigma_t$) and vertex ($\sigma_v$) resolution as well as good resolution for the kinematics, e.g.~the three-particle invariant mass resolution $\sigma_{M_3}$, and the centre-of-mass-system momentum resolution $\sigma_p$.

Detailed simulation studies for the phase~I Mu3e experiment have shown that the most frequent accidental \acs{BG} is from the combination of a Michel positron which undergoes Bhabha scattering in the detector material (leading to an $e^+e^-$ pair) with another Michel positron.
In order to match the signal kinematics, both Michel positrons must have energies close to the maximum allowed energy\footnote{Note that this \acs{BG} has its kinematic endpoint \num{1} electron mass \emph{above} the muon mass, as the electron originates from the detector material and is not created in the muon decay.}.
The scattering of a positron at $\approx \SI{53}{MeV}/c$  with an electron at rest creates an $e^+e^-$ pair with an invariant mass
of $\approx \SI{7}{MeV}/c^2$ which can be used to suppress this \acs{BG} with little signal loss if the two particle invariant mass resolution $\sigma_{M_2}$ is sufficiently small.

A similar \acs{BG} topology is produced by a photon conversion combined with a Michel positron. The rarity of $\approx \SI{53}{MeV}$ photons combined with the small amount of material in the muon stopping target region strongly suppresses this \acs{BG} and obviates the need for a two-particle invariant mass cut (which would remove the expected signal kinematics in case of dominating dipole operators).

The third variety of accidental \acs{BG} with two coincident particles combining with a Michel positron takes the $e^+e^-$ pair from a rare muon decay \mtenn. A detailed study of this \acs{BG} was performed for the phase~I Mu3e experiment \cite{HughesPhD}, a simple rate extrapolation produces 17 events in the signal region for running the phase~I apparatus under \ac{HIMB} conditions, giving a target for the necessary improvements in kinematic, vertex and timing resolution.  

Backgrounds with three unrelated particles can arise from combinations of any of the above processes which include one electron, or from Compton scattering with two Michel positrons.
The \acs{BG} rates scale with the muon rate to the third power, but are also
suppressed by the vertex and timing resolution squared. Extrapolations from
the phase~I simulation show that at muon rates of a few \num{e9} per second
this kind of \acs{BG} is completely negligible if the detector performance is
sufficient to deal with the ``2+1'' accidental \ac{BG}.

\subsubsection{Mu3e phase~II detector design} \label{sec:Mu3e_PhaseII_design}
\begin{figure}[tb!]
  \centering
     \includegraphics[trim=0 0 0 0,clip,width=0.85\textwidth]{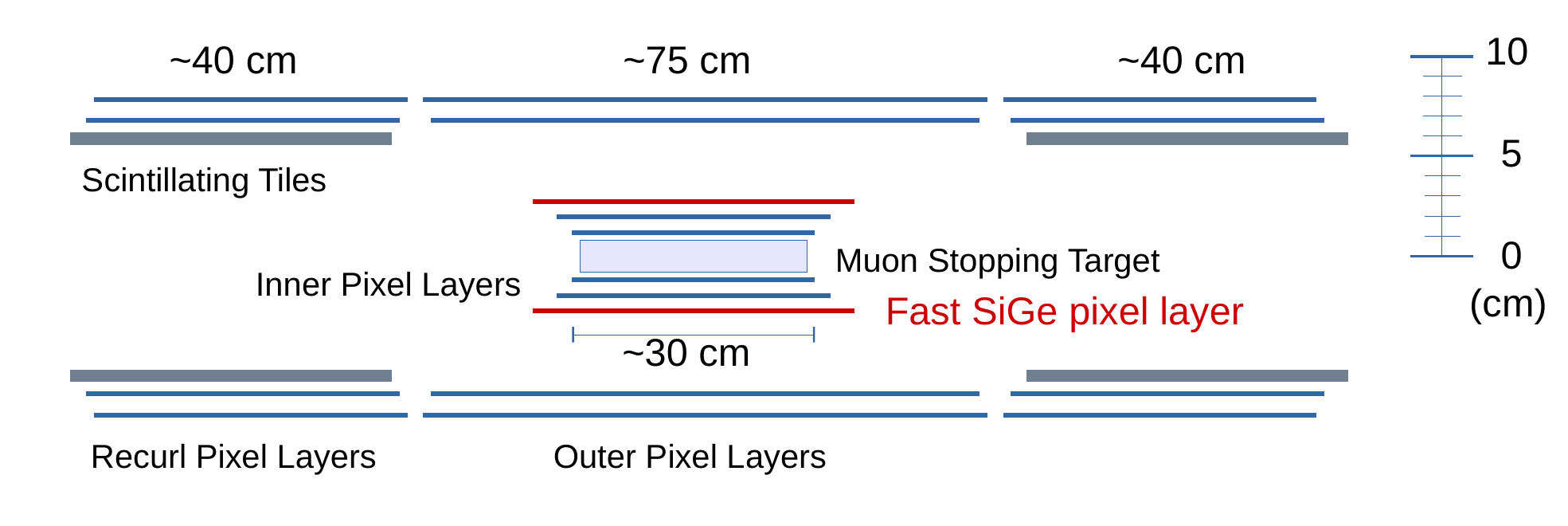}
    \caption{Sketch of an elongated detector design for Mu3e phase~II at \ac{HIMB}.}
    \label{fig:Mu3e_PhaseII_design}
\end{figure}

To further optimise the signal sensitivity it is planned to expand the geometrical
acceptance for tracks with respect to Mu3e phase I design.
This can be achieved by extending the pixel sensor instrumentation in up- and downstream directions.
With the experience from Mu3e phase I, we believe that significantly larger tracking modules can be produced.
For Mu3e phase~II, we consider a vertex detector length of about \SI{30}{cm},
see \autoref{fig:Mu3e_PhaseII_design}, or even longer.
Currently, the longest pixel modules for Mu3e phase~I have a length of \SI{36}{cm}.
Twice as long pixel modules  could be produced with a new pixel sensor generation with improved readout capabilities, like daisy chaining of the data.
With such a design the requirements  will be less stringent for the flexprint, where the individual
pixel sensors are glued onto.

The detector technologies of Mu3e phase~I were specifically chosen to prove the high-rate capability of the experiment for phase~II and to pave the road for extremely high
beam stopping rates in excess of $10^9\,\mu$/s.
Nonetheless, for Mu3e phase~II several improvements with respect to the phase I detector are necessary to further suppress \acs{BG} and to exploit the full \ac{HIMB} potential.
In addition to the extended vertex region and the increase of the overall
acceptance, the scintillating fibre detector, see \autoref{fig:Mu3e_PhaseI_design},  needs to be exchanged due to
occupancy limitations.
The alternative technology has to provide a significantly higher 
granularity and a time resolution of about \SI{100}{ps}, or better.
Therefore, the development of a monolithic silicon pad detector (PicoPix)
based on the \SI{130}{\nano\metre} SiGe BiCMOS process from IHP\footnote{Institute for High Performance Microelectronics in Frankfurt/Oder, Germany}  was started.
This technology was proven in test beams to provide time resolutions as good as $\mathcal{O}(\SI{100}{ps})$ for a design with sub-millimeter pixel size~\cite{Paolozzi_2019}.
Replacing the scintillating fibre detector by a layer of PicoPix sensors would
provide 4-dimensional information (time and spatial coordinates).
The high granularity of these sensors will allow to place them at small radii close to the muon stopping target, see \autoref{fig:Mu3e_PhaseII_design}.
The PicoPix detector
can be regarded as a fifth tracking layer, in addition to the four standard pixel layers, for which 
a new generation of MuPix sensors with improved time resolution and readout capabilities will be used.
Running Mu3e at \ac{HIMB} with about 20 times higher beam rates compared to phase~I will substantially increase the hit combinatorial problem in the online track finding on the filter farm, see \autoref{sec:mu3e_online}.
The generally much improved time resolution of the tracking detectors will help to mitigate the combinatorial problem.

The final layout of the phase~II design will be subject of further simulation studies.
They will address the possibility to increase the magnetic field and
to change the radius of the outer pixel layer to further improve the momentum resolution, see also \autoref{sec:MEGMU3E}.
During the workshop also an idea was discussed to use the most inner pixel layer to stop the muons.
With such a concept the rate of muon stops could be further increased.
The feasibility of this idea needs to be experimentally studied.

\subsubsection{Online data processing at high rates}
\label{sec:mu3e_online}

The data rate produced by the Mu3e detector at \ac{HIMB} is three to four orders of magnitude larger than what can technologically and economically be saved to mass storage. The experiment thus relies on a fast and efficient on-line reconstruction, which identifies interesting events in real time. For the phase~I experiment it was shown \cite{DissVomBruch2017, vomBruch:2015kla, vomBruch:2017fqw} that a compact and affordable farm of a dozen PCs with \acp{GPU} is capable of finding and fitting all relevant particle tracks and identifying three-track vertices, all using algorithms developed specifically for the multiple scattering dominated regime in which Mu3e operates \cite{Kozlinskiy:2017wyl, Berger:2016vak, Kozlinskiy:2014rwa, Schenk2013}.
Assuming a four-layer tracking on the filter farm,
the computing power needed scales roughly with the third power of the rate; even with optimistic assumptions about the evolution of \acp{GPU} until the start of data taking at \ac{HIMB}, this factor 8000 cannot be compensated simply by more and newer hardware. One key to tackling the online reconstruction challenge is improved time resolution of the pixel sensors, in particular in the vertex region. If a resolution of the order of \SI{1}{ns} can be reached, the \ac{GPU} only based approach is likely viable even at \ac{HIMB} rates. Alternatively, large associative memories can be used for matching with pre-computed patterns; this option introduces additional complexity into the system, but we can profit from experience gained in the ATLAS hardware track trigger projects \cite{ATLAS:2021tfo,Dittmeier:2020cct}.

\subsubsection{Mu3e-Gamma}
\label{sec:MEGMU3E}

\begin{figure}[tb!]
  \centering
     \includegraphics[trim=0 0 0 0,clip,width=0.999\textwidth]{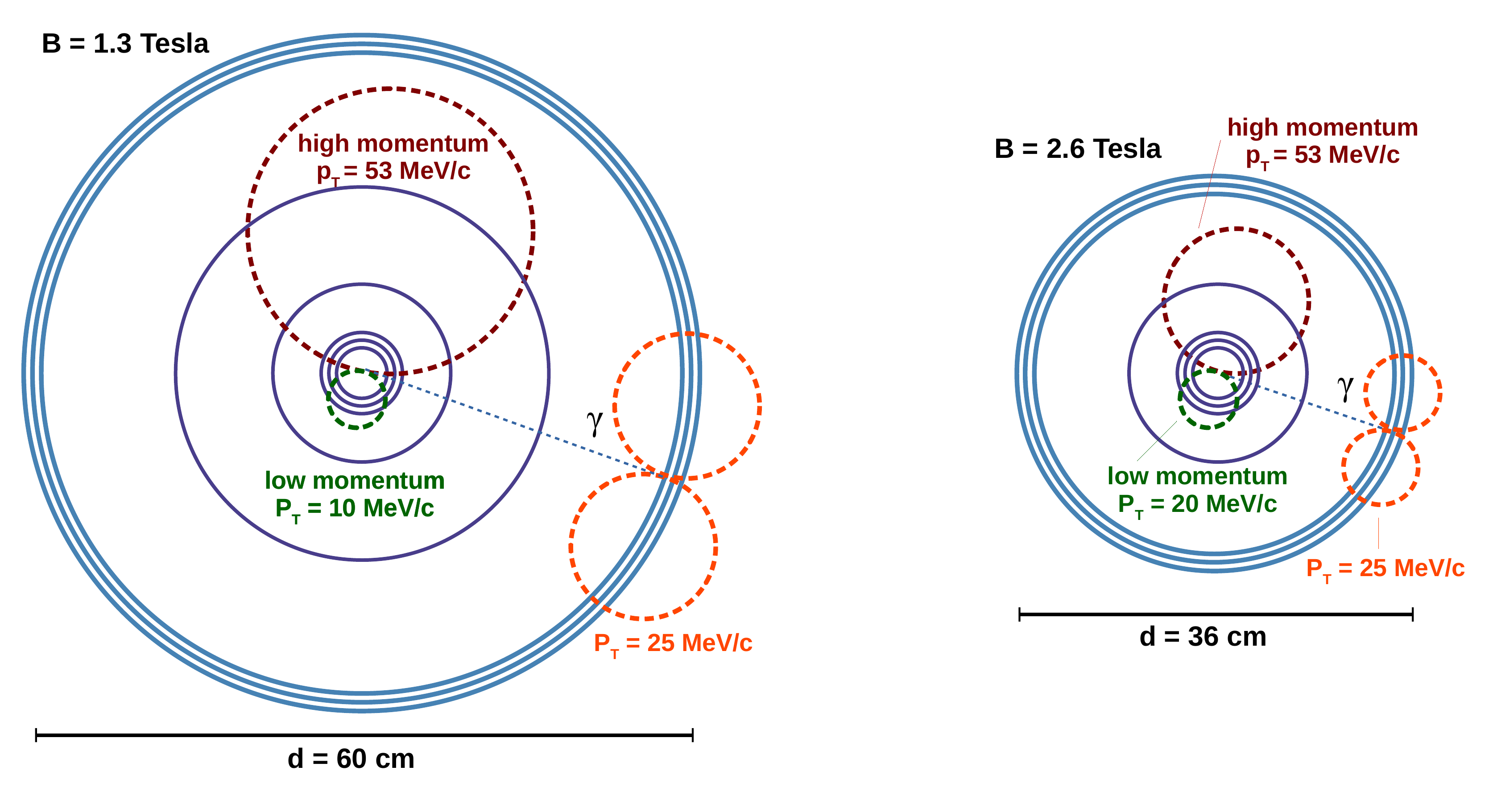}
     \includegraphics[trim=350 0 0 0,clip,width=0.49\textwidth]{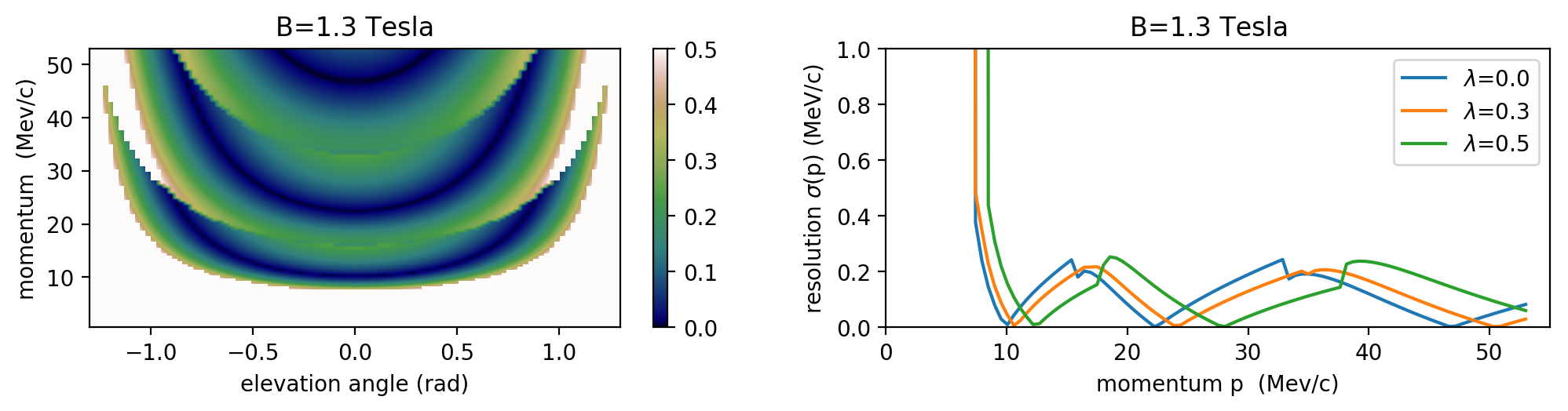}
     \includegraphics[trim=350 0 0 0,clip,width=0.49\textwidth]{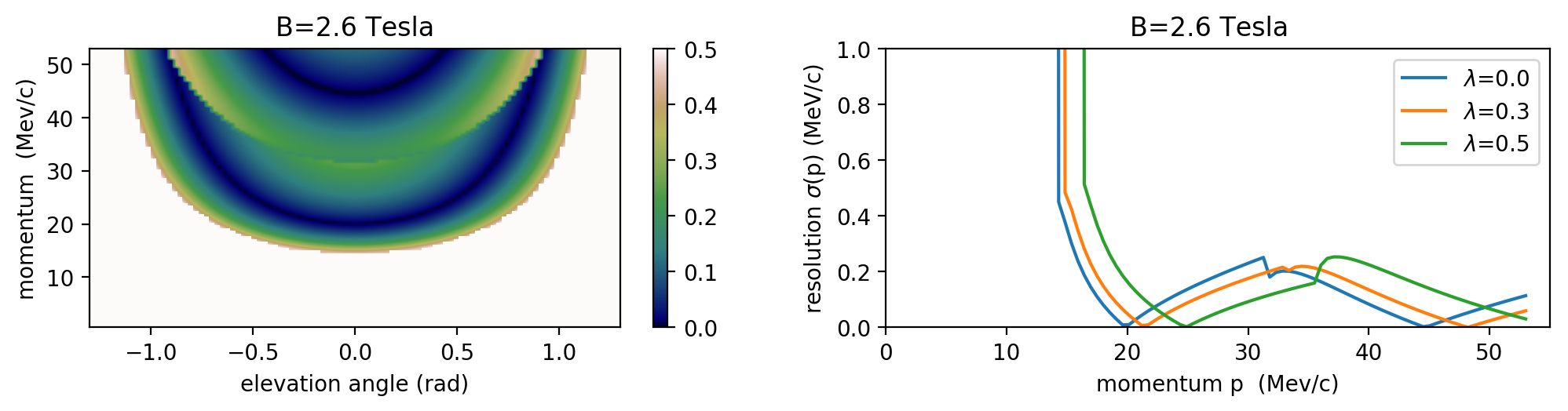}
    \caption{Transverse cross sections of conceptual Mu3e-$\gamma$ designs (top) and calculated momentum
      resolution of charged particles from the muon decays as function of the momentum for three example elevation angles $\lambda$ (bottom).
      The designs are optimised for the detection of $\mathrm{\mu^+ \to e^+ \gamma}$ for a solenoid
      field of $B=\SI{1.3}{T}$ (left) and $B=\SI{2.6}{T}$ (right). 
      Photons are detected in a  dedicated three-layer active converter pixel spectrometer
      which measures the ionisation loss of the $e^+e^-$ pair in silicon. See text for more details.}
    \label{fig:MEGMu3e}
\end{figure}

Considering the very high rates of the HIMB with up to $\SI{1e10}{\mu/s}$, a $\gamma \rightarrow e^+ e^-$ converter can be considered for photon detection, despite the in general low conversion (detection) efficiency.
By tracking the $e^+ e^-$ pair, a photon converter allows  for a precise measurement of the photon position, a good measurement of the photon direction and a very precise measurement of the photon energy.
This makes the converter option very interesting for the search $\mu^+ \to e^+ \gamma$, see also section \autoref{sec:photon conversion spectrometer}.
In particular an active converter is of high interest as it further improves the excellent energy resolution of a photon converter tracker by directly measuring the energy loss of the $e^+e^-$ pair in the converter, which is the main limitation for the energy resolution.

The feasibility of a passive photon converter in the Mu3e phase~I experiment has been studied in \cite{Leuschner2019} using a GEANT4 simulation.
In the studied design the converter consists of \SI{0.1}{mm} gold and the $e^+e^-$ pair is reconstructed in a two-layer \ac{HVMAPS} spectrometer with a pixel size of $\SI{80}{\mu m}$.
For photons with energy $E_\gamma=\SI{53}{MeV}$, a resolution of about \SI{250}{keV} (peak) was found, to be compared with the in average \SI{220}{keV} energy loss of a single MIP in \SI{0.1}{mm} gold.
A significant contribution to the uncertainty comes from the unknown position of the conversion point and the therefore unknown path length of the $e^+e^-$ pair in the converter material.
Also energy straggling significantly contributes to the energy resolution which is more severe for large Z materials due to the larger possible energy transfer in collisions.

Both contributions to the energy resolution can be significantly reduced if thick silicon sensors are used as active converter which measures directly the energy loss in the depleted region.
To make this concept work the sensor should be thick and the depleted region which is relevant for charge collection should be large.
A charge collection region of about $\SI{600}{\mu m}$ can be achieved if a
high ohmic substrate (e.g. $\gtrsim \SI{15}{k \ohm cm}$) is depleted at about \SI{200}{V}.
Using three such active converter layers results in a total radiation length of about 2\%.
The resulting photon conversion (detection) efficiency is on one hand much smaller than for photon calorimeters, on the other hand such a detector provides and excellent photon  energy resolution of about \SI{100}{keV}, more than an order of magnitude better than the upgraded MEG~II xenon calorimeter.
For not too small pixel sizes the active converter serves as spectrometer for the reconstruction of the $e^+e^-$ without the need of any additional tracking detectors, thus largely simplifying the design.

Two possible designs of a Mu3e detector combined with a three-layer active converter pixel spectrometer are shown in \autoref{fig:MEGMu3e} (Mu3e-$\gamma$).
The  designs are optimised for different solenoid fields and provide an excellent momentum resolution of about $\sigma_p \lesssim \SI{250}{keV}/c$ over a large momentum range.
This is achieved by choosing a spacing of the tracking layers that ensures that for all particle momenta at least one section of the track (a section is defined between two detection layers) approximately fulfils the condition of a half turn where the multiple scattering uncertainty on the track momentum vanishes, see discussion in~\cite{Berger:2016vak}.
Because of the excellent momentum resolution of the Mu3e-$\gamma$ designs, this concept is very well suited for a combined search of $\mu^+ \rightarrow e^+ e^+ e^-$ and $\mu^+ \rightarrow e^+ \gamma$.

The $B=\SI{1.3}{T}$ design in \autoref{fig:MEGMu3e} is optimised to provide a large kinematic acceptance\footnote{Assuming standalone tracking in the three innermost pixel layers.} ($p_T \gtrsim \SI{8}{MeV}/c$). With such a design even very exotic processes like $\mu \rightarrow eeeee \nu\nu$ can be detected with good efficiency.
This design requires the instrumentation of large detector areas with pixel sensors which is costly and creates a high power consumption.
The instrumented areas are almost a factor two smaller for the $B=\SI{2.6}{T}$ design, however for the price of a smaller kinematic acceptance ($p_T \gtrsim \SI{15}{MeV}/c$).
The larger solenoid field has also the advantage of better muon beam focusing which leads to an increase of the muon stopping rate of almost a factor $2$ compared to the
$B=\SI{1.3}{T}$ design, see the discussion in \autoref{sec:Mu3eFF}.

In all designs, charged particles from the muon decays remain inside a radius of $r \le \SI{26}{cm}$ and cannot reach the  active converter pixel spectrometer.
The \acs{BG} rate for photon detection is therefore very small, thus making it together with the excellent photon energy resolution a promising instrument to search for
$\mu \rightarrow e^+ \gamma$ or any other decay involving photons in the final state.
Examples are the search for exotic LFV decays such as $\mu^+ \rightarrow e^+ X \gamma$ with $X$ being a pseudo-scalar axion, and
the search for dark photons $\mu^+ \rightarrow e^+ \nu_e \bar{\nu_\mu} A'$ with $A'$ being a long living dark photon which weakly mixes with SM particles and decays in flight 
$A' \rightarrow e^+ e^-$ after a few picoseconds.

For the $\mu \rightarrow e^+ \gamma$ search an excellent timing resolution is crucial.
This could be achieved by using a dedicated timing layer in one of the inner three pixel layers for measuring the $e^+$ timing and by integrating
a $\sigma_t \lesssim \SI{100}{ps}$ \ac{TDC} in the active converter pixel layers.
For monolithic silicon pixel detectors time resolutions of $\mathcal{O}(\SI{100}{ps})$ have only been  achieved with small prototypes, so far~\cite{Paolozzi_2019}.
If such a time resolution can be achieved on a large scale chip in a large scale tracker system will be an important research topic for the coming years.
Alternatively, a dedicated timing detector (e.g. scintillator) could be added in front to the converter station\footnote{The timing layer would measure the time
of the $e^+ e^-$ pair after a half turn.} to measure the photon timing.


\subsection{\texorpdfstring{$\mu \rightarrow e\gamma$}{MEG}} \label{sec:mueg}

\newcommand*{\meg}{\mu \to e \gamma}
\newcommand*{\megsign}{\mu^+ \to e^+ \gamma}
\newcommand*{\mueee}{\mu \to e e e}
\newcommand*{\mueeesign}{\mu^+ \to e^+ e^+ e^-}

\subsubsection{General introduction}
The state-of-the-art of the $\mu^+ \to e^+ \gamma$ search is represented by the MEG~II experiment~\cite{Baldini:2018nnn} at PSI. 

The best current upper limit on the BR of the $\mu^+ \to e^+ \gamma$ decay has been set by the MEG experiment at PSI as $\mathrm{BR}(\mu\to e\gamma) < 4.2\cdot 10^{-13}$~\cite{MEG:2016leq}. An upgrade of the experiment to MEG~II has been carried out. MEG~II preserves the concept of the previous experiment while improving the detector performances roughly by a factor 2 for all the kinematic variables and aiming at running up to $7 \cdot 10^7 \mu^+$/s. The expected final sensitivity is $\mathrm{BR}(\mu\to e\gamma)=
6\cdot10^{-14}$ for a data-taking period of 3 years. The MEG~II experiment has successfully completed its engineering phase and it has just entered into the physics run mode (run2021). For all technical aspects we refer to the most recent document~\cite{Baldini2021symmetry}.

The signature of a $\mu^+ \to e^+ \gamma$ decay at rest is a back-to-back, mono-energetic, time coincident $\gamma$ and $e^{+}$. 
There are two main \ac{BG} sources, the dominant being the accidental coincidences between a high-energy positron from the principal decay $\mu^+ \to e^{+} \nu \overline{\nu}$  (Michel decay) and a high-energy photon from positron annihilation-in-flight or bremsstrahlung or from the \ac{RMD} $\mu^+ \to e^{+} \nu \overline{\nu} \gamma$. The other source comes from the \ac{RMD} itself, when neutrinos take off only a small amount of energy.

\begin{figure}[t!]
  \centering
    \includegraphics[trim=0 0 0 0, clip,width=0.8\textwidth]{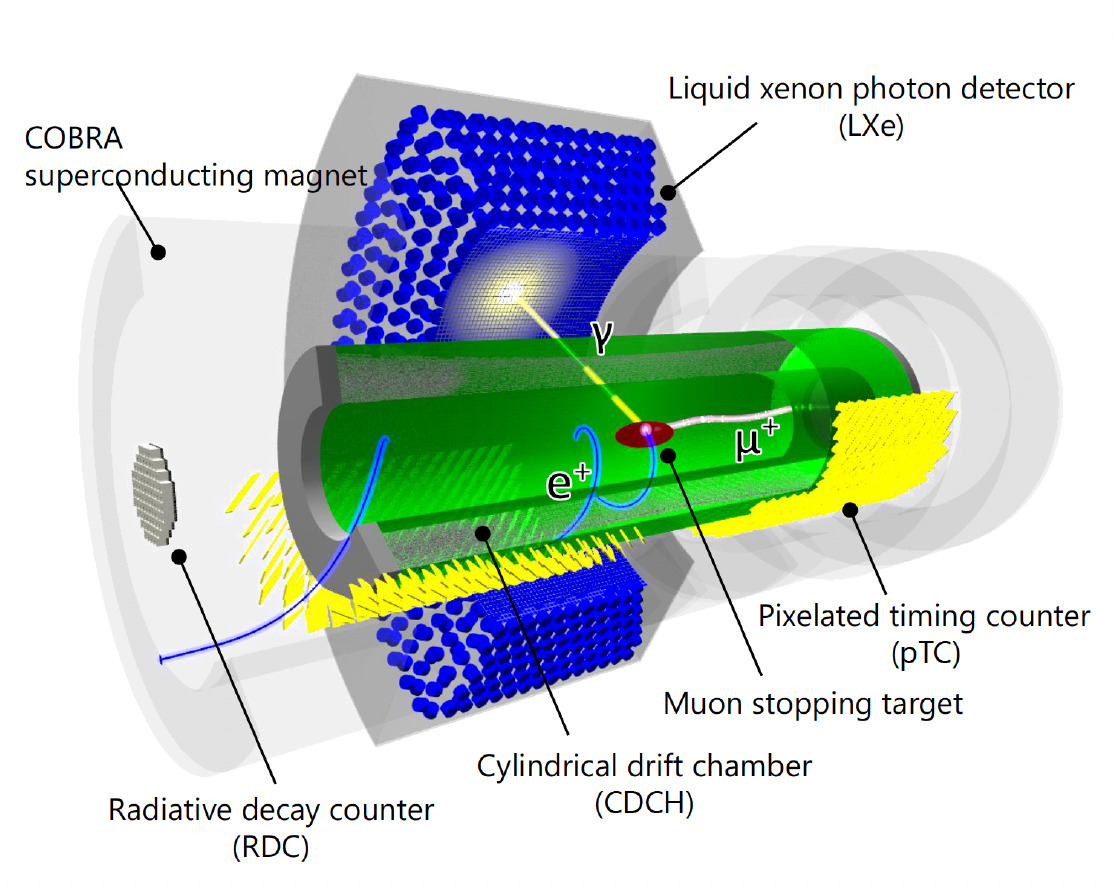}
    \caption{A sketch of the MEG~II experiment with all the key elements.}
    \label{fig:meg_MEGII}
\end{figure} 
\begin{table}[t]
    \caption{Comparison of the $e^{+}$ and $\gamma$ kinematic variable resolutions (in $\sigma$) with the MEG (measured) and MEG~II (expected) apparatus.}
    \label{tab:MEGvsMEGII}       
    \begin{center}
    \begin{tabular}{lll}
        \toprule
         & MEG & MEGII  \\
        \midrule
        $E_{e}$(core) [keV] & 306 & 130  \\
        $\theta_e$(core) [mrad] & 9.4 & 5.3 \\
        $\phi_e$ (core) [mrad] & 8.7 & 3.7 \\
        $t_e$ [ps]& 70 & 35 \\
        $u_{\gamma}$ [mm] & 5 & 2.4 \\
        $v_{\gamma}$[mm] & 5 & 2.2 \\
        $w_{\gamma}$ [mm] & 6 & 3.1 \\
        $E_{\gamma}$ (w $<$ 2 cm) [$\%$]& 2.4 & 1.1 \\
        $E_{\gamma}$ (w $>$ 2 cm) [$\%$]& 1.7 & 1.0 \\
        $t_{\gamma}$ [ps]& 67 & 60 \\

        $t_{e \gamma}$ [ps]& 122 & 84 \\

        Tracking efficiency [$\%$]& 65 & 78 \\
        CDCH-pTC matching efficiency [$\%$] & 45 & 90 \\
        Gamma efficiency  [$\%$]& 63 & 69 \\
        Trigger efficiency [$\%$]& 99 & 99 \\
        \bottomrule
    \end{tabular}
    \end{center}
\end{table}

In MEG~II, positive surface muons with a momentum of 28 $\mathrm{MeV}/c$ are stopped in a thin slanted polyethylene target (thickness 140 $\mathrm{\mu m}$; angle $15$ deg), located at the centre of the apparatus.

All kinematic variables of the $\gamma$ (energy $E_{\gamma}$, time $t_{\gamma}$ and interaction point $X_{\gamma}$) are measured using a \ac{LXe} calorimeter. All kinematic variables of the $e^{+}$ are measured with a spectrometer made of a \ac{CDCH} combined with plastic scintillators coupled to \ac{SiPM} - the so-called \ac{pTC} mounted inside a gradient magnetic field. The performance of the experiment is continuously monitored by a variety of calibration methods. All signals are recorded with custom designed waveform digitisers up to 5 Gsample/s with the DRS4 chip. A flexible trigger system allows to select $\mu^+ \to e^+ \gamma$ candidate events together with pre-scaled calibration data.

A sketch of the MEG~II experiment with all the key elements is shown in \autoref{fig:meg_MEGII}. \autoref{tab:MEGvsMEGII} summarises the performances of
MEG (measured) and MEG~II (expected).

Pushing down the sensitivity of the $\mu^+ \to e^+ \gamma$ search requires both beam rate increase and improved background rejection capability. The \ac{SES} scales as
\begin{equation}
    \mathrm{SES} = \frac{1}{R\cdot T \cdot A \cdot \epsilon}\, ,
\end{equation}
where $R$ is the beam rate, $T$ is the acquisition time, $A$ is the geometrical acceptance and $\epsilon$ the product of all the efficiencies (detection efficiency, selection efficiency, etc.).

The \ac{BR} of the accidental \ac{BG} $B_{acc}$ scales as
\begin{equation}
    B_{acc} = R \cdot \Delta E_e \cdot (\Delta E_{\gamma})^2 \cdot \Delta T_{e\gamma} \cdot (\Delta \Theta_{e \gamma})^2 \, ,
\end{equation}
where $\Delta E_e$, $\Delta E_{\gamma}$, $\Delta T_{e\gamma}$ and $\Delta \Theta_{e \gamma}$ are the positron energy, gamma energy, relative positron and gamma timing and relative positron and gamma angle resolutions.

These formulas suggest that, given a very good detector geometrical acceptance and detector efficiency, the most effective parameter to improve sensitivity is the beam rate. On the other hand the \ac{BR} of the accidental \ac{BG} -- the dominant \ac{BG} -- increases linearly with the beam rate. In order to really benefit from a beam-rate increase therefore relies on detectors able to simultaneously sustain higher beam intensity and provide better kinematic variable resolutions. 

While the measurement of the positron kinematic variables is conceptually delineated and the spectrometer option is the most competitive one, leaving just room for the specific technology to be selected, the measurement of the gamma kinematic variables is in principle open to two different approaches: either detecting the gamma directly via a calorimeter or converting it into an electron-positron pair and then measuring it via a spectrometer.

\begin{figure}[h!]
  \centering
    \includegraphics[trim=0 0 0 0, clip,width=0.8\textwidth]{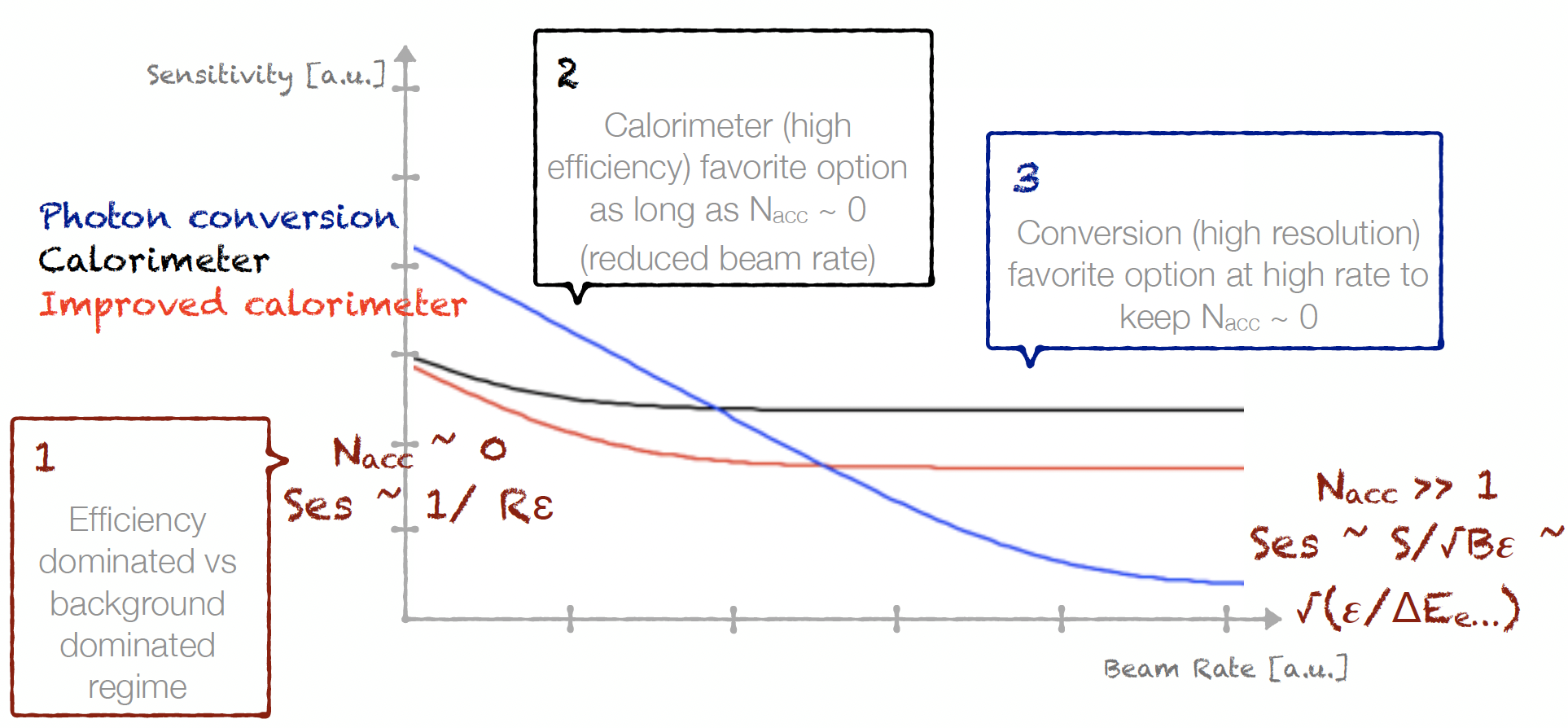}
    \caption{A sketch of the sensitivity as a function of the beam rate for three different scenarios: a) calorimeter (black line) - used as a reference; b) Improved performing calorimeter (red line); c) photon conversion (blue line).}
    \label{fig:meg_SESVSBeam}
 %
\bigskip
\includegraphics[trim=0 0 0 0, clip,width=0.8\textwidth]{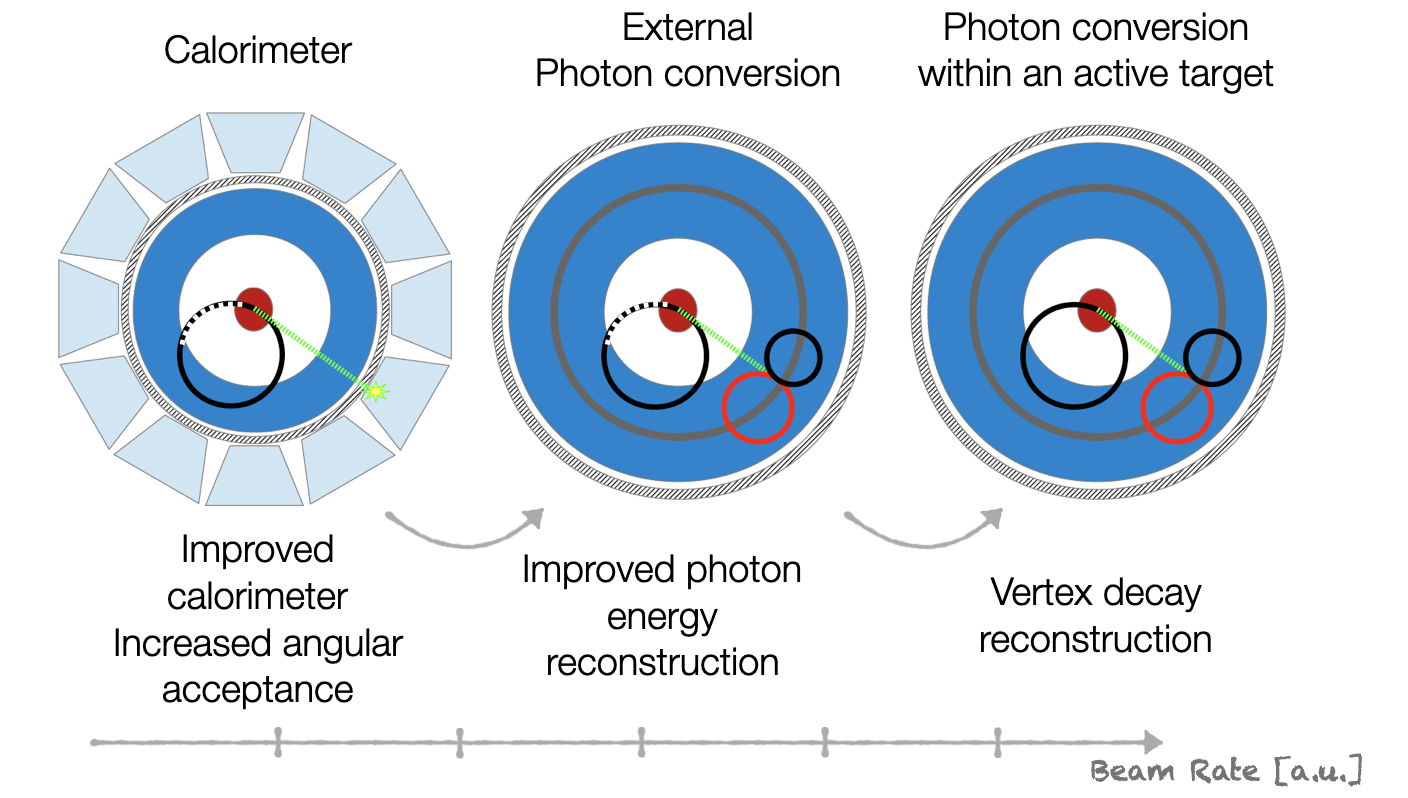}
    \caption{A sketch of the different experimental approaches as a function of the beam rate.}
    \label{fig:meg_ExperimentalApproachesVSBeam}
\end{figure}

The calorimeter option offers a higher detection efficiency compared to the conversion one. Nonetheless, the latter could be the favourite choice at a very high beam rate in order to keep the \ac{BG} under control, due to the better kinematic resolutions. The capability of sustaining a higher beam rate compensates for the drop in terms of efficiency. The best option as a function of the beam rate is defined by the \ac{BG} regime. \autoref{fig:meg_SESVSBeam} shows it in a schematic way. 

In the next sections the different experimental approaches summarised in 
\autoref{fig:meg_ExperimentalApproachesVSBeam} will be presented. While a calorimeter 
with improved performance could still be appealing, based on the current and foreseen
technologies external photon conversion seems to be more promising  at very high beam 
intensities, as expected with \ac{HIMB}. Photon conversion can be further
improved if complemented with an active target.

\subsubsection{Gaseous positron tracker}

A precise reconstruction of the kinematics of 52.8~MeV/$c$ positrons can only be achieved with a very light tracker in a magnetic field. Indeed, at such low momentum, the contribution of the multiple Coulomb scattering to the tracking resolutions tends to become dominant, so that an extremely good position resolution is not necessary if it is compensated by a reduced material budget. Despite the rapid evolution of ultra-thin silicon detectors, this aspect makes gaseous detectors still competitive for this application.

The state of the art is represented by the \ac{CDCH} of the MEG~II experiment~\cite{Tassielli:2020wap}. With a gas mixture of helium and isobutane in 90:10 volume concentrations and cells made of 20~$\mu$m gold-plated tungsten sense wires and 50~$\mu$m silver-plated aluminium field wires, with a minimum cell size of $3.5 \times 3.5$~mm$^2$ and a full-stereo geometry, this detector is designed to provide 100~$\mu$m single-hit resolution in the plane transverse to the beam axis, with an average radiation length of about 350~m. The resulting momentum and angular resolutions are expected to be around 100~keV and 5~mrad, respectively. In consideration of the technical difficulties for building a chamber with thinner wires, and the intrinsic limitations to the achievable single-hit resolution in a drift chamber, we cannot expect any dramatic improvement of these performances, although there are proposals to refine the track reconstruction exploiting the detection of single ionisation clusters~\cite{Chiarello:2017yft} and to use carbon monofilaments to build wires with reduced density~\cite{Tassielli:2021D1}.

For these reasons, in terms of performances, the MEG~II \ac{CDCH} can be considered the benchmark for the next generation of gaseous trackers for $\mu \to e \gamma$. There is anyway a strong limitation to the use of such a kind of detector at beam intensities much higher than $10^8~\mu$/s: the MEG~II \ac{CDCH} is expected to undergo a 25\% gain loss per 
year in the inner and most illuminated region at $7 \times 10^7~\mu$/s, due to wire ageing effects~\cite{Baldini:2018nnn}. While in MEG~II an increase of the sense wire voltage can be applied to compensate for
this loss, with a much higher beam intensity the ageing rate would clearly become unmanageable. 

As a possible solution, a radical change in the geometry of the drift chamber, with wires lying in planes transverse to the beam axis, would reduce the positron rate per unit length in the inner wires, so suppressing the ageing rate by a factor of about 10. This design requires the wire support structures and the readout electronics to be distributed in between the tracker and the outer detectors (positron timing and photon detectors), whose performances could deteriorate by the consequently higher material budget.

As an alternative approach, the replacement of the drift chamber with a \ac{TPC} can be considered, so to strongly reduce the material budget (thanks to the absence of wires in the tracking volume) and exploit the high particle flow capabilities of modern \acp{TPC} readout by \acp{MPGD}. The typical geometry of a $\mu \to e \gamma$ experiment requires anyway a relatively long detector along the beam axis (2~m for the MEG~II \ac{CDCH}). If a conventional \ac{TPC} is used, with the electrons drifting along the beam axis, the deterioration of the longitudinal position resolution due to the electron diffusion would spoil the detector performances. Moreover, simulations show that the current density in the readout detector and the space charge density in the tracking volume would be prohibitive also with the most advanced \acp{MPGD}. We considered, as a possible alternative, a radial \ac{TPC}, i.e. with electrons drifting radially, toward the external surface of the cylindrical detector. This surface could be instrumented with a cylindrical \ac{MPGD} (e.g. a cylindrical GEM~\cite{Bencivenni:2007zz} or $\mu$RWell~\cite{Bencivenni:2014exa}). The shorter drift distance would allow to reduce the readout current and space charge densities, down to values that are comparable with the ones expected in the GEM-\ac{TPC} of the upgraded ALICE experiment~\cite{Ketzer:2013laa}. It supports the feasibility of such a solution, although big challenges need to be faced. Beside the ones already experienced in ALICE, there is the specific need of keeping the material budget on the external surface of the detector very low, which would require a dedicated effort to integrate the front-end electronics with its cooling in the structure of the readout \ac{MPGD}.

\subsubsection{Photon calorimeter}

The photon measurement is the most difficult and crucial part of the $\meg$
experiment. A significant improvement of the photon detector performance would be a key to higher sensitivities of future $\meg$ experiments.   
Photon measurements with a \ac{LXe} scintillation detector 
have been carried out in the MEG/MEG~II experiment.
This calorimetric approach has been nicely working with the muon beam intensity
up to $7\times 10^{7}\,\mu/\mathrm{s}$ in the MEG~II experiment. 
The performance of the calorimeter would, however, be limited at higher beam intensities
above $10^{8}\,\mu/\mathrm{s}$ foreseen at future $\meg$ experiments aiming at $\mathcal{O}(10^{-15})$ sensitivity.

The challenges for new calorimetry for incoming experiments at the intensity frontier is to provide detectors with ultra-precise time resolution and supreme energy resolution. Two very promising materials on the market are BrilLanCe (cerium doped lanthanum bromide, $\mathrm{LaBr_{3}(Ce)}$) and \ac{LYSO} $\mathrm{Lu_{2(1-x)}}$ $\mathrm{Y_{2x}SiO_5(Ce)}$, supported by recent developments aiming at providing relatively large crystals.

Cerium doped lanthanum bromide stands out due to its ultra-high light yield (1.65 $\times$ NaI(Tl)) and by a more than an order of magnitude faster decay time compared to NaI(Tl). With these properties together with its high density,  $\mathrm{LaBr_{3}(Ce)}$ is the ideal medium for calorimetry limited only by the currently available crystal sizes on the market.

Due to recent developments, larger crystals up to a radius $R$ = \SI{4.45}{cm} and a length $L$ = \SI{20.3}{cm} can be produced commercially. A calorimeter built
from such a large crystal is a potential candidate
for the detection of photons at energies from few tens up one hundred MeV. This
corresponds to the interesting energy range of current cLFV experiments. Thus
$\mathrm{LaBr_{3}(Ce)}$ may be a suitable candidate for future experiments in this sector.

\ac{LYSO} on the other hand exhibits a very high density, comparable to BGO and thus features short radiation length ${X_0}$ and Moli\`ere radius ${R_M}$. Despite the fact that the light yield  is only roughly \SI{70}{\percent} of NaI and the decay time roughly three times longer compared to $\mathrm{LaBr_{3}(Ce)}$, its density makes \ac{LYSO} an attractive candidate as well, especially considering that the available crystal size is one of the limiting factors.

\autoref{tab:Brillaproperties} summarises the main scintillation properties compared to the widely used scintillation media. For a quick comparison a figure of merit (F.o.M.) is defined as the square root of the ratio of the scintillation decay time $\mathrm{\tau}$ and the product of the light yield (LY) and the density $\mathrm{\rho}$. 
\begin{table}
\caption{Main scintillation properties for widely used scintillating media. A F.o.M. is given as defined in the text.}
\begin{center}
\scalebox{0.8}
{
\begin{tabular}{ccccc}
\toprule
Scintillator & Density & Light yield & Decay time & F.o.M. \\
                    & $\mathrm{\rho}$ ($\si{g/cm^3}$) & LY (ph/keV) & $\mathrm{\tau}$ (ns) &  $\mathrm{\sqrt{\tau/(\rho \cdot LY)}}$ \\
\midrule
 $\mathrm{LaBr_{3}(Ce)}$ & 5.08 & 63 & 16 & 0.22 \\
 LYSO & 7.1 & 27 & 41 & 0.46 \\
 YAP & 5.35 & 22 & 26 & 0.47 \\
 LXe & 2.89 & 40 & 45 & 0.62 \\
 NaI(Tl) & 3.67 & 38 & 250 & 1.34 \\
 BGO & 7.13 & 9 & 300 & 2.16 \\
 \bottomrule
\end{tabular}
}
\end{center}
\label{tab:Brillaproperties}
\end{table}

\begin{figure}[p]
  \centering
    \includegraphics[trim=0 0 0 0, clip,width=0.65\textwidth]{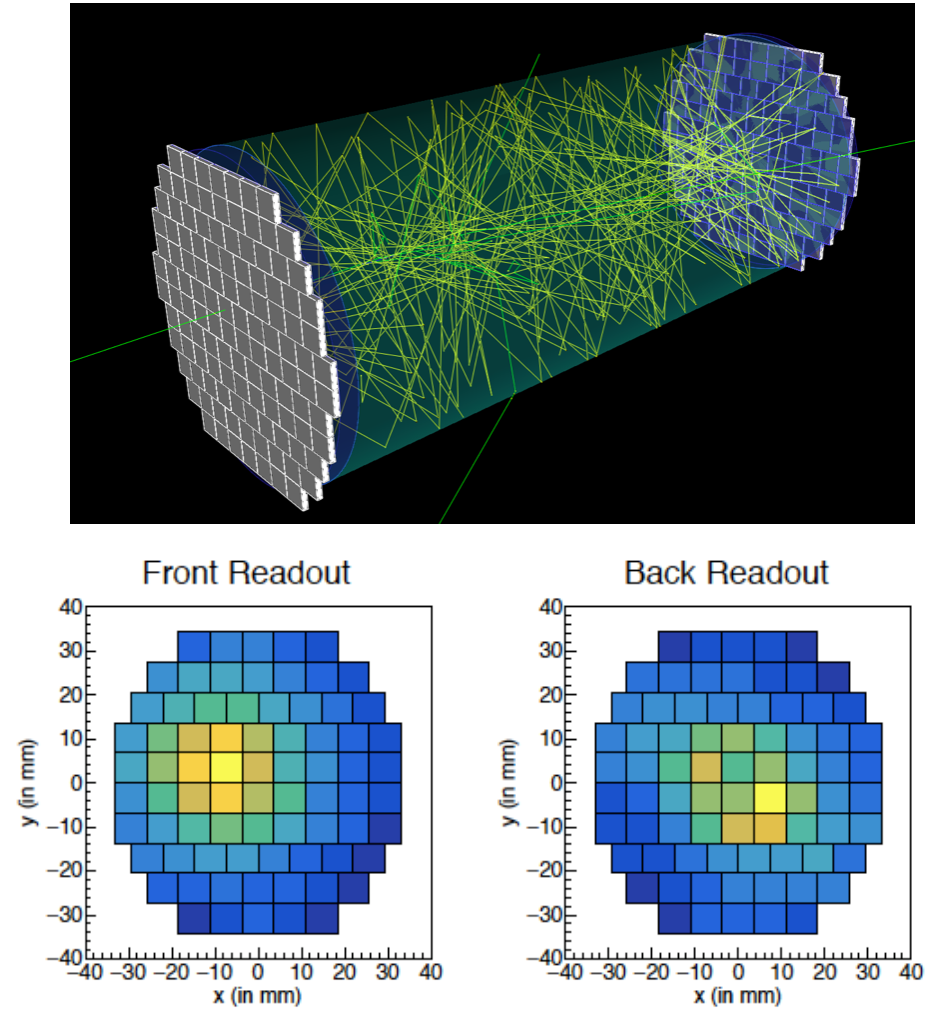}
    \caption{A sketch of the scintillation light distribution associated to a 55 MeV gamma event
impinging on ($R$=3.5\,cm, $L$=16\,cm) \ac{LYSO} crystal coupled to 
Hamamatsu S13360-6025PE \acp{MPPC} (top) and the typical collected charge distribution on the front and back face (bottom).}
    \label{fig:meg_DoubleReadout}
    \includegraphics[trim=0 0 0 0, clip,width=0.9\textwidth]{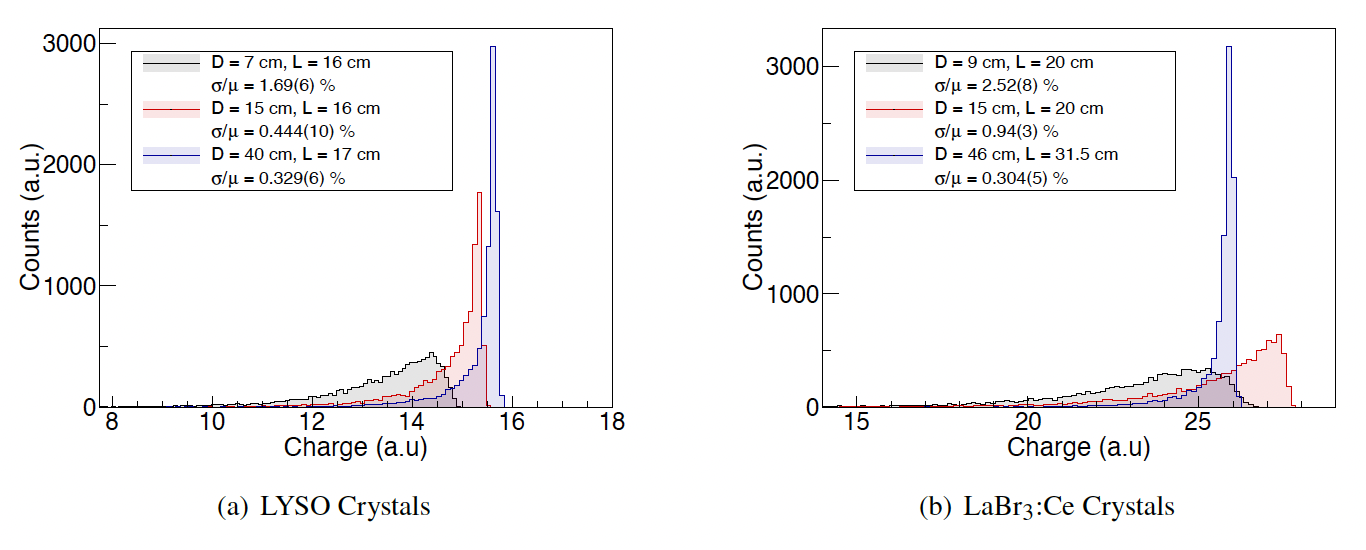}
    \caption{Energy resolution for different crystal sizes, where $\sigma$ and $\mu$ are the standard deviation and the mean . "Available" crystals are shown in black, "Large" sized
crystals in red and the "Ultimate" crystals in blue.}
    \label{fig:meg_calorimeter_LYSOandLaBr3}
\end{figure}

The result is a detector with a high photosensor granularity, high rate sustainability, maximal photosensor coverage area, optimal geometrical acceptance and insensitivity to magnetic fields. Due to the small thickness of modern photosensors of a few mm, even the radiation impinging area can be covered with minimal impact on photons passing through. This feature allows for a double readout scheme, where the \acp{MPPC} are mounted also on the front/entrance face.  The smaller the \ac{MPPC} pixel size the smaller the saturation effect. In addition, the granularity due to the \acp{MPPC} allows some geometrical reconstruction of the event. A sketch of a detector assembly is shown in \autoref{fig:meg_DoubleReadout}.

\autoref{fig:meg_calorimeter_LYSOandLaBr3} shows the results obtained with different  crystal sizes of both $\mathrm{LaBr_{3}(Ce)}$ and \ac{LYSO} up to the so-called ultimate dimension, not yet available on the market. The detectors are exposed to 55\,MeV gammas impinging on the entrance face. Two types of \ac{MPPC} were selected to be used as a model in the simulation: Hamamatsu S13360-6025PE and sensL MicroFj-60035TSV. For crystals
of 15\,cm diameter and 20\,cm and 16\,cm of length for $\mathrm{LaBr_{3}(Ce)}$ and \ac{LYSO} respectively, an energy resolution up to around $1\%$ for
$\mathrm{LaBr_{3}(Ce)}$ and  $0.4\%$ for \ac{LYSO} is obtained. In order for
$\mathrm{LaBr_{3}(Ce)}$ to fully benefit from its higher light yield, one has to
go for even larger crystals of about 40\,cm diameter. In such a configuration, resolutions around $0.3\%$ and better has been quoted also for the  $\mathrm{LaBr_{3}(Ce)}$.

\begin{figure}[t]
  \centering
    \includegraphics[trim=0 0 0 0, clip,width=0.4\textwidth]{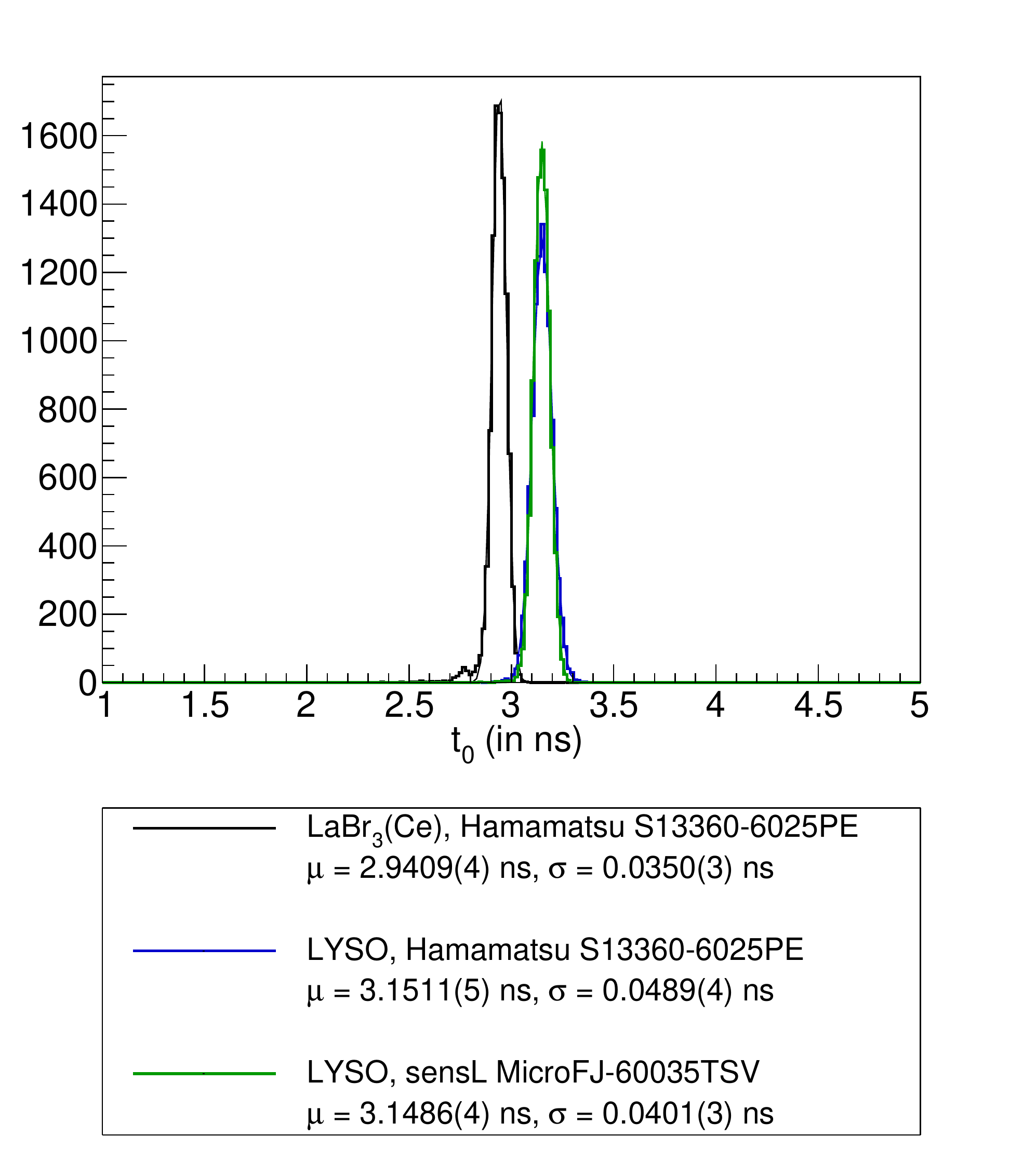}
    \caption{Timing resolutions for the available ($R$=4.45\,cm, $L$=20.32\,cm) $\mathrm{LaBr_{3}(Ce)}$ and
($R$=3.5\,cm, $L$=16\,cm) \ac{LYSO} crystals. Here the $\mathrm{LaBr_{3}(Ce)}$ is
coupled to the Hamamatsu S13360-6025PE. For the \ac{LYSO} both \ac{MPPC} options
(Hamamatsu S13360-6025PE and sensL MicroFJ-60035TSV) are displayed.}
    \label{fig:meg_timing_LYSOandLaBr3}
\end{figure}

Examples of time distributions are shown in \autoref{fig:meg_timing_LYSOandLaBr3}. 
Timing resolutions below 50\,ps are in the reach. The Monte-Carlo simulations are 
based on the Geant4 libraries
with dedicated code that includes the \ac{MPPC} response and the whole
electronic chain up to the DAQ, based on waveform digitisers with a
sampling frequency up to 5 GSample/s. The reconstructed
algorithms are based on waveform analysis. The simulations are supported
by measurements done with available LaBr3(Ce) crystals with
sizes of ($R$=3.81\,cm, $L$=7.62\,cm) and ($R$=1.27\,cm, $L$=10.16\,cm) coupled
to either photomultiplier tubes or \acp{MPPC} and the characterisation
of the \ac{MPPC} response.

In conclusion, new detectors have been considered here based
on either large LaBr3(Ce) or \ac{LYSO} crystals coupled to \acp{MPPC} showing
very promising results for high-energy $\mathcal{O}(50)$\,MeV photon calorimetry.
Independent of the specific detector assembly, simultaneous energy,
timing and position resolutions below 0.5\,MeV, 50\,ps and a few mm appear
to be feasible. Such results put this new calorimetry at the detector
forefront for particle physics research at beam intensity frontiers.

\subsubsection{Photon conversion spectrometer}
\label{sec:photon conversion spectrometer}

Photon measurements based on a pair conversion spectrometer would be an alternative and viable option with several advantages 
over the calorimetric approach with higher beam intensity.
In the pair conversion spectrometer, the incident photon is converted to an
electron-positron pair in a thin converter and 
the conversion pair is then measured with a tracker.
More precise measurements of the energy and position as well as higher rate capabilities are expected.
In addition, the photon direction can be measured with the conversion spectrometer
since the emission of the conversion pair is highly boosted along the incident
photon direction. 
This would provide a possibility to further reduce  the accidental background 
together with a widely spread distribution of the stopping muon.
It must be noted that the precise measurement of the photon direction 
is not possible with the calorimetric approach.
On the other hand, a major challenge for the conversion spectrometer 
is the low detection efficiency due to the low conversion efficiency 
with a thin conversion layer.  
It can be mitigated to some extent with multiple conversion layers.  
The pair spectrometer was employed in the previous $\meg$ experiment, MEGA (1985--1999)\cite{MEGA:2001cwl}.
It consisted of three photon conversion layers, each of which was based on two 250\,$\mu$m thick lead converters, with the overall detection efficiency of 5\%.
The signal statistics can be improved by increasing the beam intensity, limited
by the rate capability of the detectors.  

\begin{figure}[tb!]
\centering
\begin{minipage}[t]{0.4\linewidth}
\includegraphics[width=\linewidth, clip]{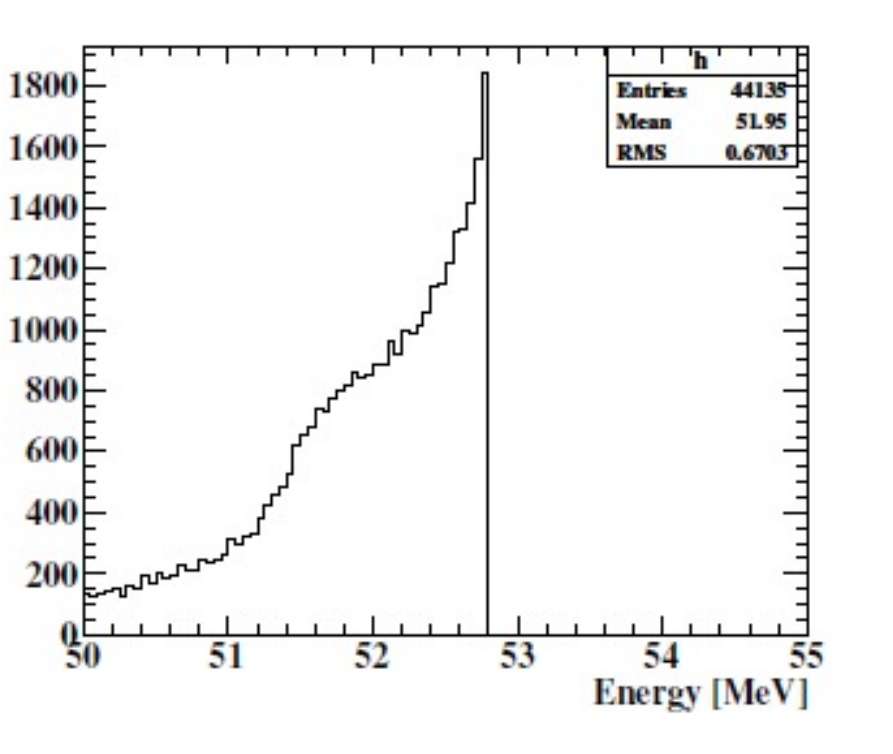}
\caption{Energy sum distribution of the conversion pair emitted from 560\,$\mu$m thick lead converter with signal photon injection.}
\label{fig:meg_energyLossInConverter}
\end{minipage}
\hspace{0.01\linewidth}
\begin{minipage}[t]{0.45\linewidth}
\includegraphics[width=\linewidth, clip]{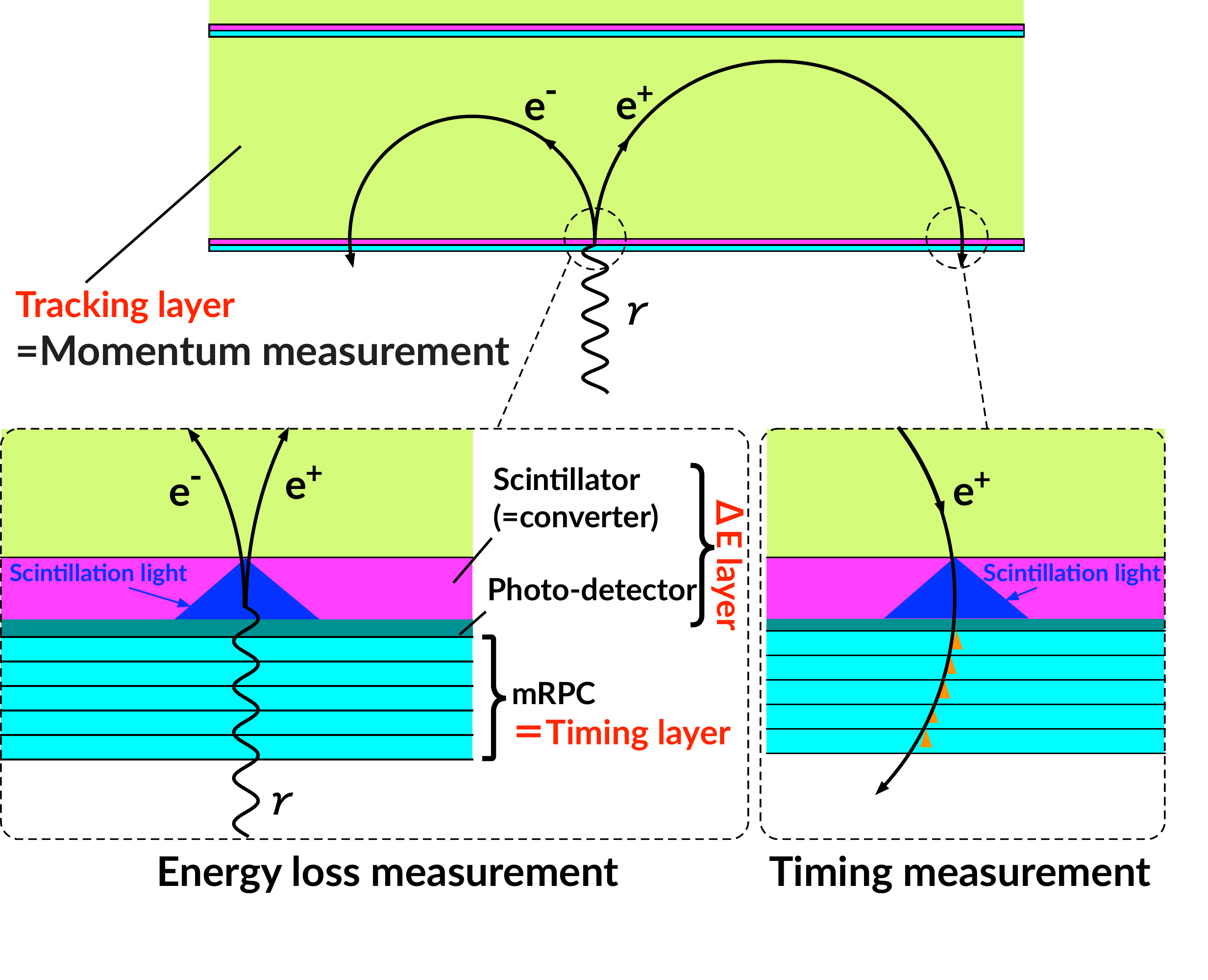}
\caption{Possible structure of the active conversion spectrometer.}
\label{fig:meg_activeConverter}
\end{minipage}
\end{figure}

Another crucial limiting factor of the conversion spectrometer is the energy loss of the conversion pair inside the converter.
\autoref{fig:meg_energyLossInConverter} shows the simulated result for the energy
sum of the conversion pair emitted from a 560\,$\mu$m thick lead converter 
with signal photon injection,
where a significant tail due to the energy loss in the converter can be seen. This leads to an inefficiency for the signal photon and/or 
deterioration of the energy resolution.

It is proposed here to use a so-called ``active converter'' 
where the energy loss of the conversion pair in the converter is measured by the converter itself. 
\autoref{fig:meg_activeConverter} shows a possible structure of 
the conversion spectrometer composed of an active converter, 
a tracking layer, and a timing layer.  
The incident photon is converted in the active converter layer 
which measures the energy loss of the conversion pair in the converter.
The possible technology for the active converter is a
scintillator coupled with a photo-detector or silicon detector.
The tracks of the conversion pair are measured in the tracking layer.
It is not a harsh environment since the Michel positrons do not reach 
the conversion spectrometer placed outside the positron spectrometer 
in the proposed design as shown later. 
A gaseous detector such as a drift chamber or a \ac{TPC} can therefore be used as a tracker, although thin silicon detectors can be an attractive candidate technology as well.
The timing layer placed in front of the active converter measures 
the timing of the returning conversion pair, with which the photon timing can 
be extracted after correcting for the time-of-flight of the conversion pair.   
The timing layer is required to be low-mass to avoid the photon conversion before the active converter, although it can basically be vetoed by placing a thin scintillator layer between the active converter and the timing layer.
The use of \ac{mRPC} detectors is considered as a candidate technology.
The timing can also be measured by the active converter if its timing resolution is good enough.

\begin{figure}[tb]
  \centering
    \includegraphics[width=0.9\textwidth]{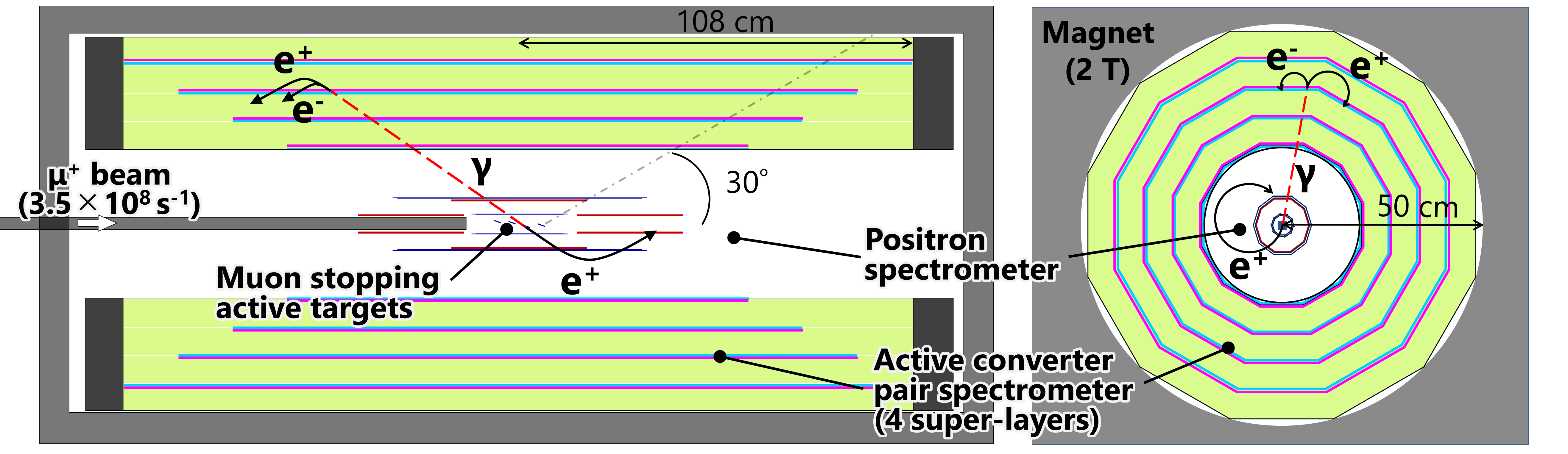}
    \caption{Possible layout of $\meg$ experiment with photon conversion spectrometer.}
    \label{fig:meg_designWithPairSpectrometer}
\end{figure}

\autoref{fig:meg_designWithPairSpectrometer} shows a possible layout of the $\meg$ experiment
with the photon conversion spectrometer. The photon spectrometer with four layers of the active converters is placed 
outside the positron spectrometer.
Silicon detectors are envisaged as a technology option 
for the positron spectrometer 
because of their high rate capability.
The design of the positron spectrometer is inspired by the Mu3e detector, which 
would provide a possibility of a concurrent search for $\mueee$ in this setup. One of the most important improvements in the experimental design compared to the MEG~II experiment is 
the enhanced signal acceptance especially for the zenith angle, 
which is $\pm 60^\circ$, while it is $\pm 20^\circ$ for the MEG~II experiment.
The enhanced zenith-angle acceptance would enable us to measure the angular distribution 
of $\meg$ with polarised muon beam after a possible discovery of $\meg$ decay
in the MEG~II experiment.
Together with the measurement of the \ac{BR}, we could pin-down 
the underlying new physics behind $\meg$.

The materials for the active converter are being investigated from the viewpoint 
of the expected performance of the active converter such as conversion efficiency and 
the detector resolutions.
One of the crucial parameters is the critical energy, above 
which the energy loss by bremsstrahlung dominates over ionisation energy loss.
Since the active converter can basically measure the ionisation energy loss only, 
the critical energy should ideally be higher than the energies of the conversion pair, 
which range from 0 to 52.8\,MeV. 
The critical energy is known to have an empirical dependence on the atomic number as 
$610\,\mathrm{MeV}/(Z+1.24)$~\cite{ParticleDataGroup:2020ssz}.
The critical energies of \ac{LYSO} and YAP as good candidates for scintillator materials 
from the viewpoint of the detection efficiency are 12\,MeV and 23\,MeV, respectively, 
which are not sufficiently high compared to the energies of the conversion pair.
On the other hand, the critical energy for  plastic scintillator, 
which is not ideal from the viewpoint of the detection efficiency, is 93\,MeV and thus
sufficiently high.
The comparison among the candidate materials is not straightforward since some part of the energy loss by  bremsstrahlung can be absorbed and measured by the converter.
Simulation studies are in progress to compare the performance of different materials as the active 
converter. The energy loss measurement can be ruined when the returning conversion pair hits the converter.
A segmentation of the scintillator into cells is therefore required 
for the active converter.

The \ac{GasPM} and \ac{SiPM} are under consideration 
as the photodetector for the scintillator converter.
The \ac{GasPM} consists of a photocathode and \ac{RPC} as electron multiplier.
The technology of the ultra-low mass \ac{RPC} with diamond-like carbon electrodes developed 
for the MEG~II radiative decay counter~\cite{Oya:2021iqx}
can be used for the \ac{RPC} of the \ac{GasPM}.
The development of the large-area photocathode sensitive 
to visible scintillation light is, however, extremely challenging.
The combination of the \ac{VUV} scintillator and the CsI photocathode sensitive to \ac{VUV} light is 
also under investigation because CsI photocathodes are much easier to handle.
The scintillation readout by \ac{SiPM} is also viable option.
A good light collection efficiency and its small position dependence have 
been already demonstrated by the CALICE scintillator calorimeters\,\cite{Tsuji:2019zuj}. 

Further studies on the design and the performance of the photon conversion spectrometer are in progress, including detailed simulation studies and prototype tests for the candidate technologies.

\subsubsection{\texorpdfstring{Silicon pixel sensor for $\mu\rightarrow e\gamma$}{Silicon pixel sensor for MEG}}\label{sec:pixelMEG}

With the advent of monolithic silicon pixel sensors, which can be thinned to a
few tens of micrometers, silicon detectors became an attractive alternative
for tracking low-energy particles, especially for high-rate
applications, where gaseous detectors suffer from sparks and ageing effects.
Ultra-light mechanical designs combined with gaseous helium cooling of the sensors
allow for the construction of silicon tracking layers with a thickness of about
1 per mill of radiation length, as demonstrated by the Mu3e Phase~I experiment.
A similar design could also be used for precisely measuring the trajectory of
the positron in the $\mu\rightarrow e\gamma$ search.
Since positrons from this decay are mono-energetic a further optimisation of
the Mu3e spectrometer can be done and applied to MEG to achieve a momentum resolution of about 100\,keV/$c$.
This optimisation aims at detecting the decay positrons after curling half a
turn (\SI{180}{\degree} in the solenoid field).
This can be achieved by operating the Mu3e pixel detector in a magnetic field
of about \SI{2.5}{\tesla} (instead of  \SI{1.0}{\tesla}) or by increasing the
radius of the outer pixel layer with respect to the Mu3e phase I design.
In order to fully profit from the high rate capability of the silicon detector
technology (in terms of radiation hardness) the sensor is required to provide
a very good time resolution as well.
The \ac{HVMAPS} sensor used for the Mu3e phase~I experiment has a
time resolution of about \SI{5}{ns} and would perfectly fulfil the requirements of a  $\mu\rightarrow e\gamma$ search at \ac{HIMB}.
For the $\mu\rightarrow e\gamma$ search the first tracking layer must be as thin as possible in terms of radiation length.
The thinner the sensor the less multiple scattering affects the measurement 
of the initial positron direction, which is derived from the hit positions in
the first two tracking layers and is crucial for the topological \ac{BG}
rejection (back-to-back signature of signal).

\begin{figure}[tb!]
  \centering
     \includegraphics[width=0.69\textwidth]{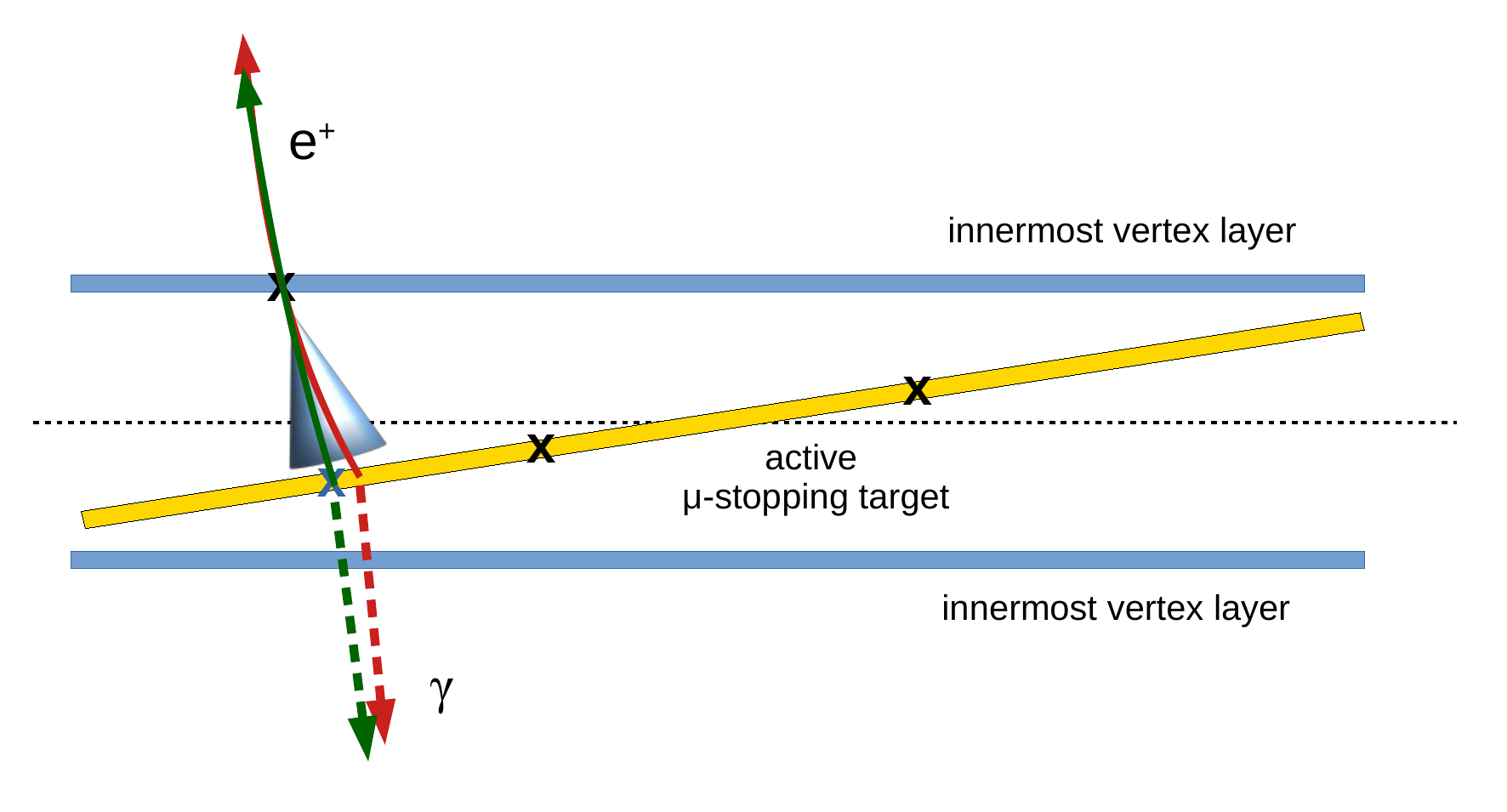}
     \caption{Sketch of the active muon stopping target concept for the reconstruction of $\meg$ events.
       Extrapolated positron tracks are assigned to hits (muon decay) in the active target to improve the vertex information 
       (direction of the positron and photon).}
    \label{fig:active_target}
\end{figure}

The experimentally proven radiation hardness of the \ac{HVCMOS} process \cite{Schoning:2020zed}
also allows to use \ac{HVMAPS} for beam monitoring applications and the usage as
active muon stopping target such that 
the muon decay position is directly measured. The achievable resolution is a
few tens of microns and much smaller than the uncertainty of a few hundreds of
micrometer by extrapolating the reconstructed positron trajectory back to
the muon stopping target, see \autoref{fig:active_target}.
By directly measuring the muon decay vertex, the direction of  the
signal positrons and photons can be more precisely determined and
\ac{BG} not exhibiting the back-to-back topology can be more efficiently
suppressed.
This concept, however, only works if the active muon stopping target is not too
thick since multiple scattering inside the material would otherwise compromise
the measurement of the positron direction.
Studies are ongoing where the signal response of thinned \ac{HVMAPS} is measured
and a small reduction of the hit detection efficiency has been observed with
the MuPix10 sensor for substrate thickness below 50\,$\mu$m.
This efficiency loss, however, might be recovered with an improved amplifier
design in future sensor versions.
Furthermore, in order to make this concept viable, the density of muon stops
must not be too high as otherwise ambiguities in the matching of positron
tracks with the muon decay positrons will reduce the track linking efficiency. 
This requirement defines a limit on the maximum muon stop rate, depending on the maximum
distance (radius) of the first tracking layer and the thickness of
the first tracking layer in case of a silicon tracker.

\begin{figure}[tb!]
  \centering
    \includegraphics[width=0.49\textwidth]{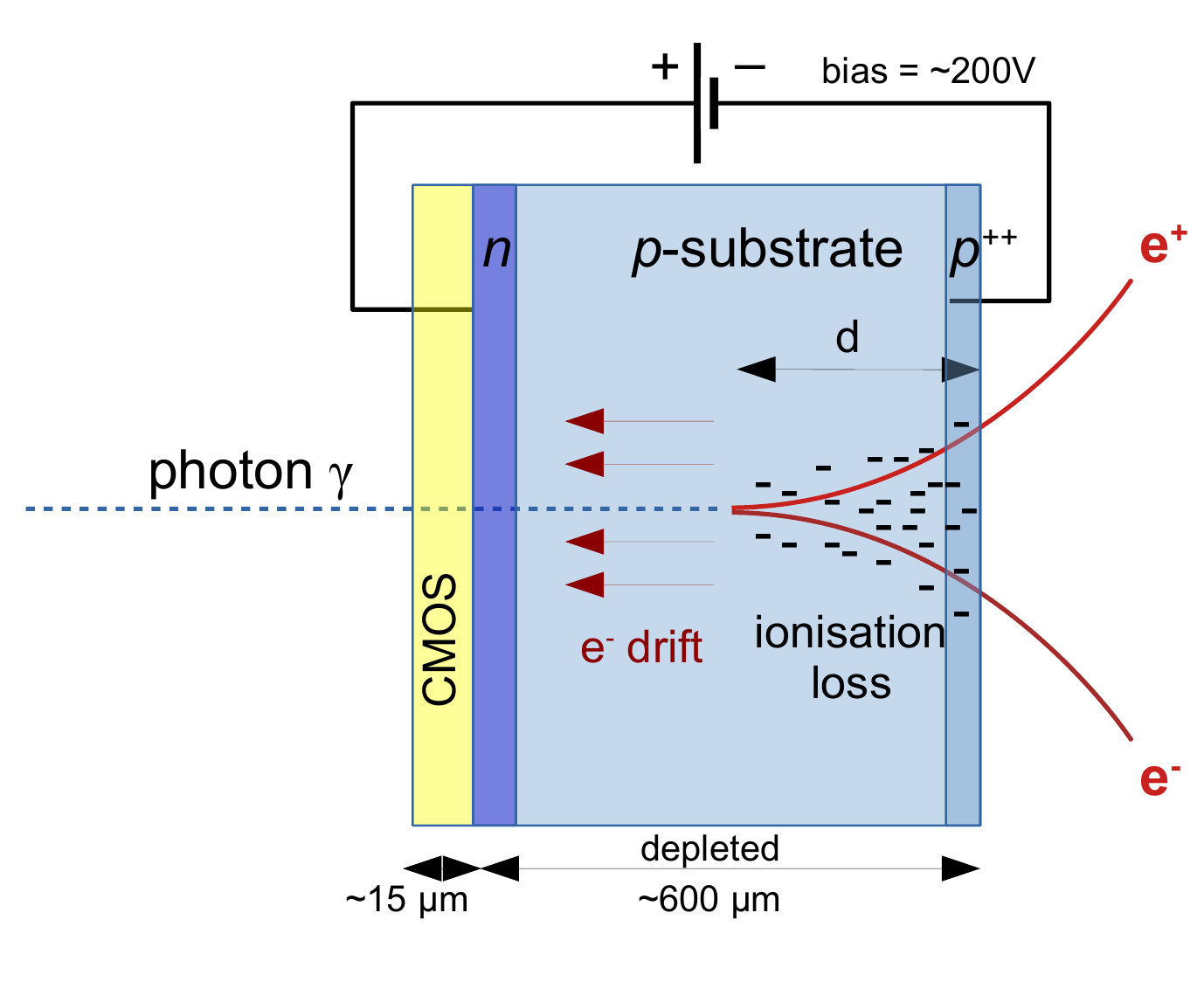}
     \caption{Sketch of an active converter silicon sensor implemented as \acs{HVMAPS}. The ionisation loss is proportional to the measured signal amplitude.
     The active zone where the charge (electrons) is collected is about 40 times larger than the inactive zone where the CMOS circuits are placed for readout.
}
    \label{fig:active_converter_silicon}
\end{figure}

The possibility to implement an active conversion target (see \autoref{sec:photon conversion spectrometer})
in a depleted silicon detector as photon detector
was already discussed in the context of a combined Mu3e-MEG experiment in \autoref{sec:MEGMU3E}.
A sketch of such a detector is shown in \autoref{fig:active_converter_silicon}.
Photons with an energy $\gtrsim \SI{10}{MeV}$ convert in the sensor material and create an $e^+e^-$ pair which undergoes ionisation loss.
Electron-hole pairs are produced in the silicon. The electrons drift to the anode (n-well) and are registered by the readout circuitry which is implemented in CMOS. 
An interesting feature of this concept is that the active converter can also be used as high precision $e^+e^-$ spectrometer if the pixel size is not too large.
The goal is to reach an energy resolution of about $\sigma_E \approx \SI{100}{keV}$ for \SI{53}{MeV} photons.

For the development of the active converter pixel sensor the following challenges need to be tackled.
Assuming a substrate resistivity of about $\gtrsim \SI{15}{k \ohm cm}$ a depletion voltage of about \SI{200}{V} is required to fully deplete $\SI{600}{\mu m}$ of silicon.
The pixel and guard-ring design must therefore allow for a breakdown voltage $\ge \SI{200}{V}$.
The pixel size should not be larger than $\SI{150}{\mu m}$ for the $B=\SI{2}{\tesla}$ setup so that spatial hit uncertainties do not deteriorate the momentum resolution.
The pixel sensor circuitry has to provide a time resolution of $\sigma_t  \lesssim \SI{100}{ps}$.
The \ac{TDC} and the readout electronics has to be capable to provide a time resolution of a $\sigma_t \lesssim \SI{100}{ps}$.
The amplitude of the signal should be measured with a relative accuracy of 10\% or better.
Furthermore all active converter layers must not have any significant amount of support material, making the detector design very challenging.
Significant R\&D effort will be required to realise such an active converter pixel sensor.

In summary, monolithic silicon detectors are a very promising detector technology for search for $\meg$.
They provide excellent tracking in a high-rate experiment at HIMB and can also be used as active muon stopping target and active photon converter.


\subsection{Muon moments}\label{sec:moments}

Measurements of magnetic and electric dipole moments have historically provided unique insights into new physics as well as stringent tests of the \ac{SM}\@. There is huge potential for a future muon ($g-2$) experiment at PSI, using re-accelerated muons from a new low-energy, high-brightness source (muCool) coupled to the future \ac{HIMB}. This complements ongoing efforts at Fermilab and J-PARC, potentially further increasing the precision.  
Similarly, there is potential to improve the sensitivity of a search for a muon \ac{EDM} to better than \SI{1e-23}{\ecm}. 
A null result sets a stringent limit on a currently poorly constrained Wilson coefficient and a discovery of a muon \ac{EDM} would prove the existence of physics beyond the \ac{SM}.  

\subsubsection{\texorpdfstring{${g-2}$}{g-2} and muon electric dipole moment}
\label{sec:Motivation}

The \ac{AMM} of the electron and muon have a long tradition as precision tests of the \ac{SM}, going back to Schwinger's famous prediction $a_\ell=(g-2)_\ell/2=\alpha/(2\pi)$~\cite{Schwinger:1948iu,Kusch:1948mvb}. 
For the electron, the precision at which \ac{SM} prediction and direct measurement can be confronted is approaching $10^{-13}$, with~\cite{Hanneke:2008tm}
\begin{equation}
\label{aedirect}
 a_e^\text{exp}=1\,159\,652\,180.73(28)\times 10^{-12}.
\end{equation}
Regarding the \ac{SM} prediction, 
the mass-independent $4$-loop \ac{QED} coefficient is known semi-analytically~\cite{Laporta:2017okg}, so that the dominant uncertainties, both at the level of $10^{-14}$, arise from the numerical evaluation of the $5$-loop coefficient~\cite{Aoyama:2019ryr} and hadronic contributions~\cite{keshavarzi:2019abf}.\footnote{  
For the $5$-loop \ac{QED} coefficient there is a $4.8\sigma$ tension between \cite{Aoyama:2019ryr} and \cite{Volkov:2019phy} in the evaluation of diagrams without closed lepton loops.} 
The present limiting factor thus concerns independent input for the fine-structure constant $\alpha$, with the 
current most precise measurements in atom interferometry~\cite{Parker:2018vye,Morel:2020dww} differing by $5.4\sigma$,
\begin{align}
a_e^\text{SM}[\text{Cs}]&=1\,159\,652\,181.61(23)\times 10^{-12},\notag\\
 a_e^\text{SM}[\text{Rb}]&=1\,159\,652\,180.25(10)\times 10^{-12},
\end{align}
i.e., $2.5\sigma$ above and $1.6\sigma$ below \eqref{aedirect}, respectively. 

The value for the muon provides an intriguing hint of new physics given the $4.2\sigma$ discrepancy between the theoretical predication and current world’s best measurement. The experimental value is
\begin{equation}
\label{amuexp}
 a_\mu^\text{exp}=116\,592\,061(41)\times 10^{-11},
\end{equation}
with a combined precision of \SI{0.35}{ppm} after averaging the Run-1 result from the Fermilab experiment~\cite{Abi:2021gix,Albahri:2021ixb,Albahri:2021kmg,Albahri:2021mtf} and the Brookhaven measurement~\cite{Bennett:2006fi}. The resulting $4.2\sigma$ tension 
with the \ac{SM} prediction~\cite{Aoyama:2020ynm}
\begin{align}
\label{amuSM}
a_\mu^\text{SM}=116\,591\,810(43)\times 10^{-11}
\end{align}
will be scrutinised at subsequent runs at Fermilab~\cite{Grange:2015fou} and at J-PARC~\cite{Abe:2019thb}, with a precision goal of \SI{0.14}{ppm} and \SI{0.45}{ppm}, respectively. In particular, the J-PARC experiment will pioneer a new experimental technique that does not rely on the so called ``magic momentum'' in a storage ring, see also~\cite{Gorringe:2015cma}, with a set-up similar to the experiment proposed here, see \autoref{sec:gm2prospectsPSI}. 

The uncertainty in the \ac{SM} prediction~\eqref{amuSM}
is completely dominated by hadronic contributions, with \ac{HVP} and \ac{HLbL} at \SI{0.34}{ppm} and \SI{0.15}{ppm}
producing a total precision of \SI{0.37}{ppm}. This value reflects the current recommendation from the muon $g-2$ Theory Initiative~\cite{Aoyama:2020ynm}, which is based on major input from~\cite{aoyama:2012wk,Aoyama:2019ryr, czarnecki:2002nt,gnendiger:2013pva,davier:2017zfy,keshavarzi:2018mgv,colangelo:2018mtw,hoferichter:2019gzf,davier:2019can,keshavarzi:2019abf,kurz:2014wya,melnikov:2003xd,masjuan:2017tvw,Colangelo:2017fiz,hoferichter:2018kwz,gerardin:2019vio,bijnens:2019ghy,colangelo:2019uex,Blum:2019ugy,colangelo:2014qya}. 
While the current precision suffices for the comparison to the precision of the current measured value, further improvements both on \ac{HVP} and \ac{HLbL} are required to match the experimental projections,  
including new $e^+e^-\to \text{hadrons}$ data, consolidated lattice-QCD calculations of \ac{HVP}, and direct input on space-like \ac{HVP} from the proposed MUonE experiment~\cite{MUonE:LoI,Banerjee:2020tdt}. A first 
lattice calculation of \ac{HVP} at sub-percent precision~\cite{Borsanyi:2020mff} has been reported recently, indicating a tension with the data-driven determination~\cite{Lehner:2020crt,Crivellin:2020zul,Keshavarzi:2020bfy,Malaescu:2020zuc,Colangelo:2020lcg}. Other developments include new data 
for $e^+e^-\to \pi^+\pi^-$ from SND~\cite{Achasov:2020iys} and improved radiative corrections~\cite{Campanario:2019mjh}. For \ac{HLbL}, a recent evaluation in lattice 
QCD agrees with the phenomenological result at a similar level of precision~\cite{Chao:2021tvp}, but the sub-leading contributions to \ac{HLbL} ~\cite{Hoferichter:2020lap,Bijnens:2020xnl,Bijnens:2021jqo,Zanke:2021wiq,Danilkin:2021icn,Colangelo:2021nkr} need to be better understood to meet the target precision set by experiment.  

The current discrepancy between the measured and theoretical value is larger than the electroweak contribution and so requires some form of enhancement mechanism, but there are several well-motivated examples that provide this.
One example concerns new light, weakly coupled models that provide an explanation via the small mass of the new particle~\cite{Holdom:1985ag,Pospelov:2008zw,Chen:2015vqy,Davoudiasl:2018fbb,Darme:2020sjf,Cadeddu:2021dqx,Buen-Abad:2021fwq,Darme:2021qzw}. 
Alternatively, solutions with new heavy particles above the electroweak scale are possible if the chirality flip originates from a large coupling to the \ac{SM} Higgs, instead of the muon Yukawa coupling in the \ac{SM}, leading to a chiral enhancement that allows for viable solutions for particle masses up to tens of TeV~\cite{Czarnecki:2001pv,Stockinger:1900zz,Crivellin:2018qmi,Capdevilla:2021rwo}. 

The most-studied theoretical framework of \ac{BSM} physics above the electroweak scale that can explain $g-2$ of the muon is the \ac{MSSM}. Here, the necessary chiral enhancement is provided by $\tan\beta \equiv v_u/v_d$, where
$v_u$ and $v_d$ are the vacuum expectation values of the two Higgs doublets $H_u$ and $H_d$, which give mass to up-type and down-type fermions, respectively. A large value of $\tan \beta\approx 50$ is suggested by top--bottom Yukawa coupling unification~\cite{Ananthanarayan:1991xp,Carena:1994bv}, which can thus provide a natural explanation~\cite{Lopez:1993vi,Chattopadhyay:1995ae,Dedes:2001fv}.

Another possible explanation, also motivated by the hints for lepton flavour universality violation in semi-leptonic $B$ decays, are leptoquarks. In fact, two scalar leptoquark representations can provide a chiral enhancement factor of {$m_t/m_\mu \approx 1600$}~\cite{Djouadi:1989md,
Chakraverty:2001yg,Cheung:2001ip,Bauer:2015knc,ColuccioLeskow:2016dox,Crivellin:2020tsz}. This allows for a \ac{BSM} explanation with perturbative couplings that is not in conflict with direct \ac{LHC} searches. It is furthermore very predictive as it involves only two free couplings whose product is determined by $g-2$. Thus, correlated effects in $h\to\mu^+\mu^-$~\cite{Crivellin:2020tsz}, and, to a lesser extent, in $Z\to\mu^+\mu^-$~\cite{ColuccioLeskow:2016dox,Crivellin:2020mjs} arise.

Alternatively, there exist many heavy BSM explanations such as composite or extra-dimensional
models~\cite{Das:2001it,Xiong:2001rt,Park:2001uc} or models with vector-like leptons~\cite{Stockinger:1900zz,Giudice:2012ms,Dermisek:2013gta,Falkowski:2013jya,Altmannshofer:2016oaq,Kowalska:2017iqv,Crivellin:2018qmi,Arnan:2019uhr,Crivellin:2021rbq}, including in addition a second Higgs doublet~\cite{Ferreira:2021gke,Chun:2020uzw,Frank:2020smf}(2HDM). Note that also a pure 2HDM can provide a solution, either via Barr--Zee diagrams in the 2HDM-X~\cite{Cao:2009as,Broggio:2014mna,Wang:2014sda,Ilisie:2015tra,Abe:2015oca,Crivellin:2015hha} or through a lepton-flavour-violating $\tau\mu$ coupling~\cite{Crivellin:2019dun,Iguro:2020rby,Wang:2021fkn,Hou:2021qmf}, which is, however, strongly constrained from $h\to\tau\mu$ searches.


Unlike the magnetic dipole moment of the muon, the \ac{EDM} is predicted to be well below the reach of current experiments and so any measurement would be an indication of new physics.
In comparison to other \ac{EDM} bounds~\cite{Chupp:2017rkp},  the limit on the muon \ac{EDM} is particularly weak: the current direct limit
$d_\mu <\SI{1.5e-19}{\ecm}$~\cite{Bennett:2008dy} is ten orders of magnitude higher than the one for $d_e<\SI{1.1e-29}{\ecm}$~\cite{Andreev:2018ayy}. 
One reason for this is that in the past, the muon \ac{EDM} was often discarded as a valuable BSM probe by rescaling the limits on $d_e$, assuming \ac{MFV}~\cite{Chivukula:1987fw,Hall:1990ac,Buras:2000dm,DAmbrosio:2002vsn} (by the ratio $m_\mu/m_e$), with a resulting limit of $d_\mu <\SI{2.3e-27}{\ecm}$, which is orders of magnitude below the direct limit. However, \ac{MFV} is, at least to some extent, an ad-hoc symmetry imposed to allow for light particle spectra, in particular within the \ac{MSSM}, where this reduces the degree of fine-tuning in the Higgs sector while respecting at the same time flavour constraints. 
However, since the \ac{LHC} did not discover any new particles directly~\cite{Butler:2017afk,Masetti:2018btj}, the concept of naturalness is challenged. Furthermore, LHCb, Belle, and BaBar discovered significant tensions in semi-leptonic $B$ decays~\cite{Aaij:2014pli,Aaij:2014ora,Aaij:2015esa,Aaij:2015oid,Khachatryan:2015isa,ATLAS:2017dlm,Sirunyan:2017dhj,Aaij:2017vbb,LHCb:2020zud,LHCb:2021zwz} implying a discrepancy significantly above the $5\sigma$-level within a global analysis~\cite{Alguero:2021anc,Altmannshofer:2021qrr,Alok:2020bia,Hurth:2020ehu,
Ciuchini:2020gvn}. These hints for \ac{BSM} physics point towards a significant violation of lepton flavour universality and are therefore not compatible with \ac{MFV} in the lepton sector~\cite{Cirigliano:2005ck}. 

Further, the $4.2\sigma$ tension in the muon $g-2$ sets the expected size for the muon \ac{EDM} if the $CP$-violating phase in the respective Wilson coefficient is sizeable. That is, even though the value of $g-2$ is not directly related to the \ac{EDM}, any \ac{BSM} contribution would result from the real part of the Wilson coefficient whose imaginary part determines the \ac{EDM}, with an $\mathcal{O}(1)$-phase (in case of a chirally enhanced explanation) leading to a muon \ac{EDM} of the order of $\SI{e-22}{\ecm}$. Moreover, while $g-2$ by itself does not conflict the \ac{MFV} paradigm, solutions with chiral enhancement mentioned above can violate the \ac{MFV} scaling~\cite{Giudice:2012ms,Crivellin:2018qmi}, and particularly in the case of leptoquarks they even must do so, in order to respect the bounds from $\mu\to e\gamma$~\cite{ColuccioLeskow:2016dox}. At the same time, such scenarios automatically provide an a priori free phase, leading to a large \ac{EDM} unless the phase happens to be small.   

Therefore, it is well-motivated that the BSM flavour structure goes beyond MFV, a notion sometimes contested on grounds of naturalness arguments. However, this does not mean that a conflict with $\mu\to e\gamma$ arises, as in the limit of vanishing neutrino masses lepton flavour is conserved, and thus it is possible to completely disentangle the muon from the electron \ac{EDM}. This can even be achieved via a symmetry, such as $L_\mu-L_\tau$ symmetry~\cite{He:1990pn,Foot:1990mn,He:1991qd}, which, even after its breaking, protects the electron \ac{EDM} and $g-2$ from BSM contributions~\cite{Altmannshofer:2016oaq}.
Also from an \ac{EFT} point of view~\cite{Pruna:2017tif,Crivellin:2018qmi,Crivellin:2019mvj}, it is clear that the muon \ac{EDM} can be large and that its measurement is the only way of determining the imaginary part of the associated Wilson coefficient. 

For these reasons, both a high-precision measurement of the muon $g-2$ and a dedicated muon \ac{EDM} experiment are highly motivated, and would be valuable contributions in the search for \ac{BSM} physics in low-energy precision observables.


\subsubsection{A \texorpdfstring{${g-2}$}{g-2} / EDM measurement at PSI}  \label{sec:moments_measurement}

Currently the most precise measurement of the muon $g-2$ is being performed at Fermilab\@, with an expected relative precision of \SI{0.14}{ppm}. The second experiment proposed at J-PARC~\cite{Abe:2019thb} will have a  precision of about \SI{0.45}{ppm}. Both experiments also intend to use the same data to search for the muon \ac{EDM} with a precision of about \SI{1e-21}{\ecm}~\cite{Chislett:2016jau,Abe:2019thb}.

The possible experiment at PSI would provide an independent measurement of the muon $g-2$ with a different experimental technique and systematic uncertainties, as well as providing insight into uncharted terrain by improving on the sensitivity to a muon \ac{EDM}\@.

Both measurements observe the spin precession $\vec{\omega}$ of a muon in a storage ring with an electric field $\vec{E}$ and magnetic field $\vec{B}$ given by
\begin{equation}
	\vec{\omega}=\frac{q}{m}\left[a\vec{B}-\left(a+\frac{1}{1-\gamma^2}\right)
	\frac{\vec{\beta}\times\vec{E}}{c}\right] +
	\frac{q}{m}\frac{\eta}{2}\left(\vec{\beta}\times\vec{B}+\frac{\vec{E}}{c}\right).
\label{eq:omegaMu1}
\end{equation}
Certain choices of the muon momentum and the combination of electric and magnetic fields permit measurements with strongly reduced systematic effects.

\begin{figure}[t]
	\centering
			\includegraphics[width=0.75\columnwidth]{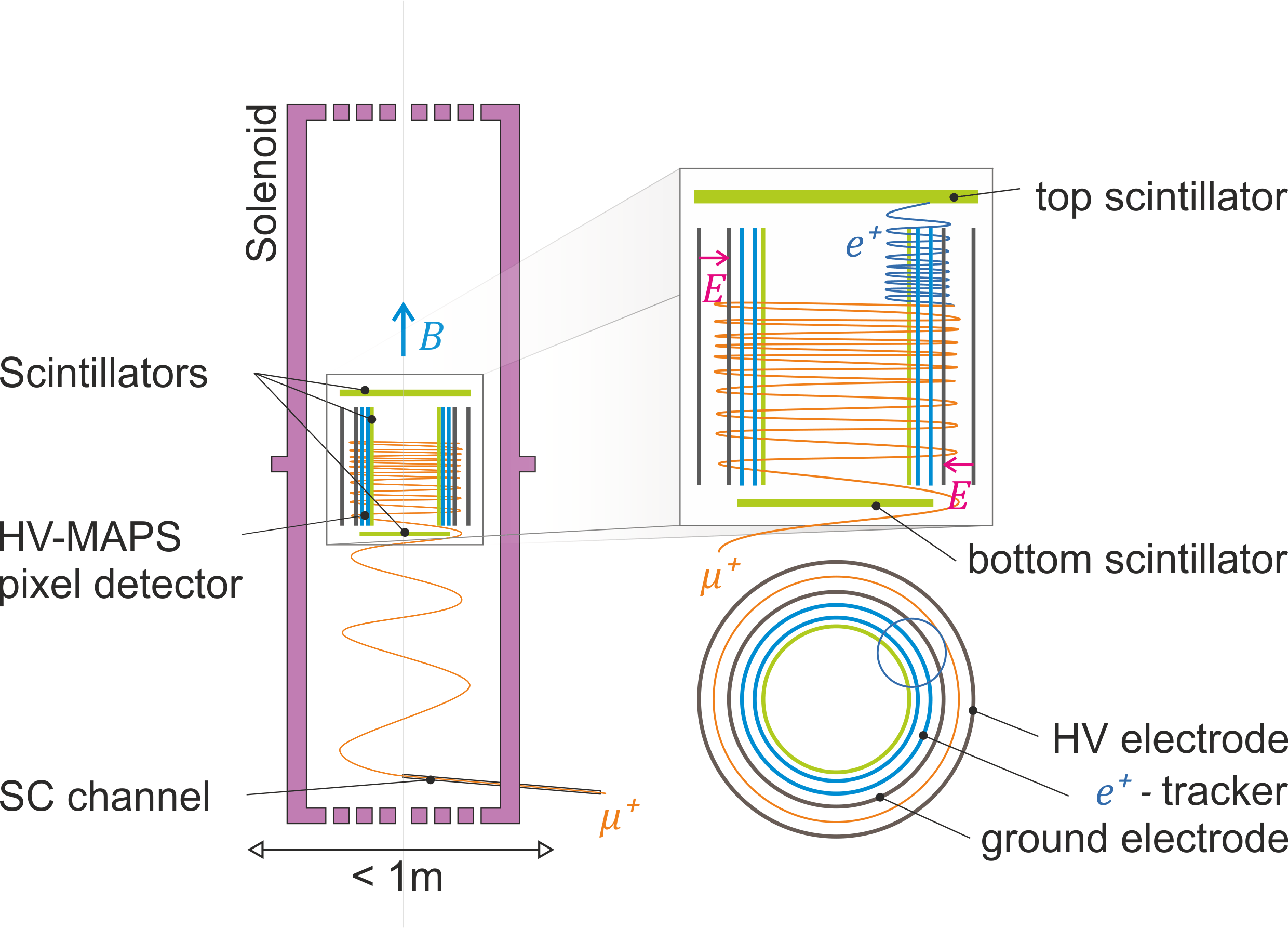}
		\caption{Sketch of the helix muEDM instrument, not to scale. }%
		\label{fig:ComparisonStorageVsHelix}%
\end{figure}

The muEDM collaboration at PSI proposes a search for the muon \ac{EDM} with a sensitivity of about \SI{6e-23}{\ecm} for a year of data taking using muons with a momentum of $|\vec{p}|=\SI{125}{MeV/}c$, $|\vec{\beta}|=|\vec{v}|/c = 0.77$, and an average polarisation of better than $P=93\%$ from the $\mu$E1 beam line at PSI with a particle flux of up to $2 \times 10^8$~$\mu^+$/s.
The concept is based on the frozen-spin technique~\cite{Farley:2003wt,Adelmann:2010zz} combined with a spiral injection into a magnetic field of $B=\SI{3}{T}$, similar as in the J-PARC~$(g-2)$/muEDM experiment~\cite{Iinuma:2016zfu,Abe:2019thb}.
A sketch of the experiment is shown in \autoref{fig:ComparisonStorageVsHelix}.

The search profits from the large electric field in the rest frame of the muon $|E^{\ast}|=|\gamma c \vec{\beta}\times\vec{B}|=\SI{1.1}{GV/m}$. 
A radial electrode system provides an electric field $\vec{E}_{\mathrm{f}}$ perpendicular to the motion of the muon and the magnetic field $\vec{B}$, hence $\vec{\beta}\cdot\vec{B}=\vec{\beta}\cdot\vec{E}_{\mathrm{f}}=0$, and $\vec{B}\cdot\vec{E}_{\mathrm{f}}=0$.
By adjusting the strength of the electric field $\vec{E}_{\mathrm{f}}$ such that
\begin{equation}
		a\vec{B} = \left(a-\frac{1}{\gamma^2-1}\right)\frac{\vec{\beta}\times\vec{E_{\mathrm{f}}}}{c},
\label{eq:FrozenSpinCondition}
\end{equation}
it is possible to cancel the anomalous precession term in \eqref{eq:omegaMu1}, which simplifies the spin precession to
\begin{equation}
	\vec{\omega}=	\frac{q}{m}\frac{\eta}{2}\left(\vec{\beta}\times\vec{B}+\frac{\vec{E_{\mathrm{f}}}}{c}\right),
\label{eq:omegaMu2}
\end{equation}
known as the frozen-spin technique.

The statistical sensitivity of this proposal is limited by the maximum muon momentum and the large lateral phase at $\mu$E1. As we outline in the following, by using a dedicated beam line with sub millirad divergence and a momentum of $P=\SI{210}{MeV}/c$, this concept has the potential to improve the sensitivity of the muEDM search to better than $\SI{1e-23}{\ecm}$. Without the electric field, i.e., with both electrodes grounded, this same setup measures the anomalous magnetic moment and could reach a relative statistical precision of the muon $g-2$ of about \SI{0.1}{ppm}.

\subsubsection{Prospects for a search of the muon electric dipole moment}

The sensitivity for a muon \ac{EDM} experiment deploying the frozen-spin technique\,\cite{Farley:2003wt} with $E\approx a B c\beta \gamma^2$ is given by
\begin{equation}
		\sigma(d_\mu) = \frac{\hbar}{2 \beta c P \gamma B \sqrt{N}\alpha \tau }.
\label{eq:EDMSensitivity}
\end{equation}

\begin{figure}%
\centering
\subfloat[]{
\includegraphics[width=0.475\columnwidth]{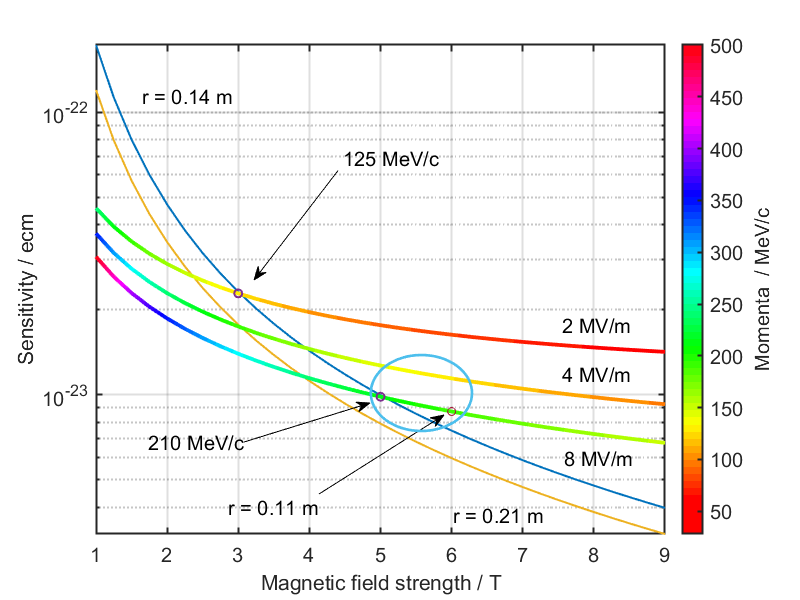}
\label{subfig:muEDMSensSingle}}%
\hfill
\subfloat[]{
\includegraphics[width=0.475\columnwidth]{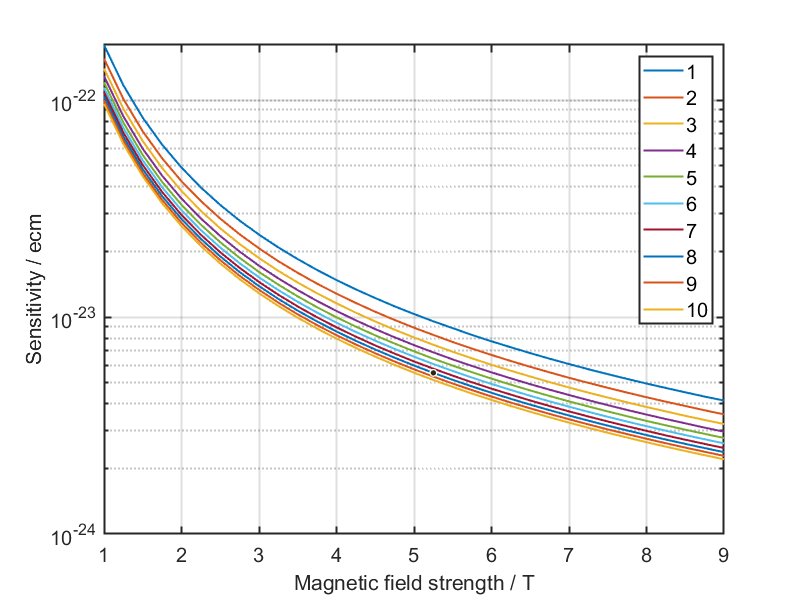}
\label{subfig:muEDMSensMultiple}}
\caption{Sensitivity landscape for muon \ac{EDM} searches. (a)~Single muon-on-demand: Sensitivity increases with magnetic field strength keeping the radius constant. Electric fields above \SI{8}{MV/m} are very difficult to obtain for large  electrodes with extremely thin material thickness (positron transmission). The region within the circle indicates the possible parameter space for a future muEDM search using re-accelerated muons from muCool. (b)~Increase in sensitivity with multiplicity of muons per injection for the case $r=\SI{0.14}{m}$, as function of the magnetic field.} %
\label{fig:muEDMsens}%
\end{figure}

\autoref{fig:muEDMsens} illustrates expected sensitivities to a muon \ac{EDM} in the case that a new dedicated beam could deliver muons with a phase space optimally adapted for injection. We assume an overall positron detection efficiency of 70\%. 
As a benchmark case we consider the solenoid based layout described in the letter of intent~\cite{Adelmann:2021udj} with a muon orbit of $r=\SI{0.14}{m}$. 
Note that the sensitivity at $B=\SI{3}{T}$, the nominal field value for the experiment proposed for the $\mu$E1 beamline, will be improved by a factor two if the next muon can be injected on request, whenever the previous positron decay was confirmed or a time-out of five times the laboratory lifetime is reached.
In this single-muon-at-a-time scenario it is possible to improve the sensitivity to about \SI{1e-23}{\ecm}.
\autoref{fig:muEDMsens}b shows that with a mean muon multiplicity above three, i.e. in average three muons are injected at a time, and a magnetic field above $B=\SI{5}{T}$ a sensitivity of better than \SI{8e-24}{\ecm}, c.f.\ current limit $d_\mu <\SI{1.5e-19}{\ecm}$~\cite{Bennett:2008dy}, in a year of measurement can be reached. 
This could be accomplished with a muon rate from muCool of \SI{5e5}{\per\second} and a repetition rate of \SI{100}{kHz} (see \autoref{tab:muEDMComparison} for a comparison of potential sensitivities). 

\begin{table}%
\caption{Comparison of two future scenarios on a re-accelerated high-brightness muon beam  with the concept proposed for $\mu$E1. A measurement with a sensitivity better than \SI{1e-23}{\ecm} can only be realised with a larger magnetic field and higher momentum, which in turn also requires a higher electric field. Note that injection of more than one muon per measurement also requires a longer time-out, which partially compensates for the increase in injected muons.}
    \begin{center}
        \begin{tabular}{lccc}
        \toprule
         & $\mu$E1 & \multicolumn{2}{c}{ \ac{HIMB} muCool} \\
         & \SI{125}{MeV}/$c$ &  \SI{125}{MeV}/$c$ & \SI{210}{MeV}/$c$ \\
        \midrule
        $E$-Field (MV/m) & 2 & 2 & 8 \\
        $B$-Field (T) & 3 & 3 & 5 \\
        radius  (m) & 0.14 & 0.14 & 0.14 \\
        ${\mathrm{e}^+}$/year (1 muon) & \num{7.3e11}& \num{4e12} & \num{3e12}   \\
        \midrule
        Sensitivity/ year (1 muon)& 6 & 2.3 & 1 \\
        (\SI{1E-23}{\ecm}) & & &\\
        Sensitivity/ year  (3 muons) &- & 1.9 & 0.8 \\ 
        (\SI{1E-23}{\ecm}) & & & \\
        \bottomrule
        \end{tabular}
    \end{center}
\label{tab:muEDMComparison}
\end{table}

\subsubsection{Prospects for a high-precision measurement of the muon \texorpdfstring{${g-2}$}{(g-2)}}
\label{sec:gm2prospectsPSI}
The \ac{HIMB} in combination with a muon cooling stage and a re-acceleration beamline means that the same experimental setup, without the electric field applied, can be used to make a dedicated measurement of the magnetic moment of the muon. In this \ac{AMM} measurement in the muEDM setup (AMMiED) no electric field will be applied and \eqref{eq:omegaMu1} will reduce to
\begin{equation}
	\vec{\omega}=\frac{q}{m}\left(a\vec{B}\right)
\label{eq:omegaMu3}
\end{equation}
in the absence of an \ac{EDM}. In the case of using the same apparatus as for the search for an \ac{EDM} the central electrode should be removed, to reduce multiple scattering, while the outer electrode is simply grounded.

The relative sensitivity of the anomalous frequency measurement is given by
\begin{equation}
		\sigma(a_\mu)_{\mathrm{rel}} = \frac{\sqrt{2}\quad m_\mu}{P\sqrt{N}\gamma\tau\alpha~ e a_\mu B}.
\label{eq:SensAMM}
\end{equation}

\subsubsection*{Measurement of \texorpdfstring{$\boldsymbol{g-2}$}{(g-2)} in muEDM setup}
\begin{figure}%
\centering
\includegraphics[width=0.75\columnwidth]{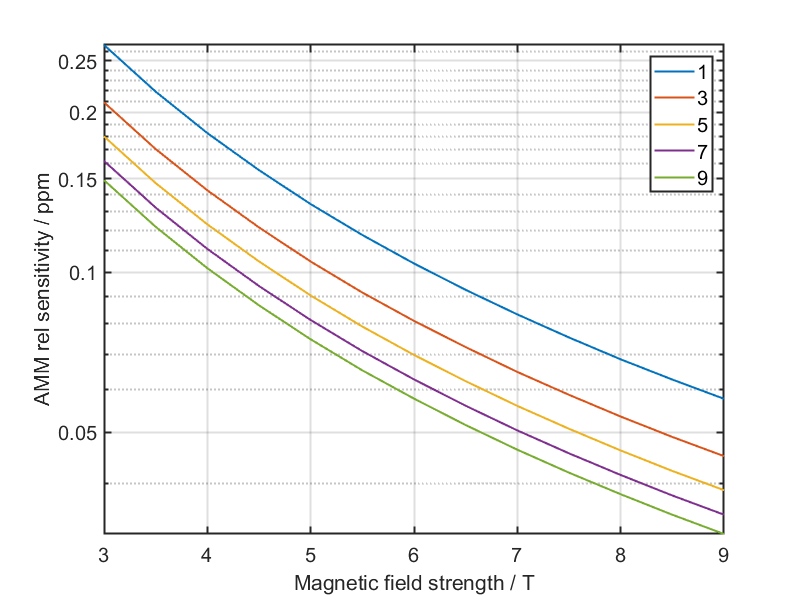}%
\caption{Statistical sensitivity in ppb for a measurement of the muon $g-2$. For this calculation the radius was fixed to $r=\SI{0.14}{m}$ and $P=qBr$.}%
\label{fig:AMMSens}%
\end{figure}

For the sensitivity calculation, shown in \autoref{fig:AMMSens}, we assume the same values for $P$ and $\alpha$ as in the muEDM scenario and a muon on request operation scheme with an integral positron detection efficiency of 70\%. It can be seen that for a field strength of $B=\SI{6}{T}$ a statistical sensitivity of \SI{0.1}{ppm} can be reached, matching the expected final precision of the experiment at Fermilab. An improvement can be seen when the muon multiplicity is increased with the potential to achieve about \SI{0.06}{ppm}, or to match the Fermilab sensitivity with a lower field strength of $B=\SI{4}{T}$. 

In terms of the systematic uncertainties, the experimental design benefits from the more compact magnet and hence a high field homogeneity is expected, as demonstrated in \ac{MRI} technologies. Further investigations need to be undertaken in order to fully understand the exact levels of these uncertainties. 

Both previously discussed scenarios for a muon \ac{EDM} and \ac{AMM} measurement are based on the assumption of a highly efficient re-acceleration of muons to at least \SI{125}{MeV}/$c$ from the future low-energy, high-brightness source muCool. The caveat of such a scenario is the difficulty of a low loss extraction of muons from the high magnetic field region in muCool to the first acceleration stage and subsequent further acceleration. Although the initial phase space density and lateral extension within the muCool source are excellent, significant losses will appear due to muon decay and a mismatch of final beam and the experimental acceptance phase space, resulting in the likely situation of one single-muon-on-request-scenario.

\subsubsection*{Tiny all magnetic \texorpdfstring{$\boldsymbol{g-2}$}{(g-2)} storage ring}
\begin{figure}%
\centering
\includegraphics[width=0.75\columnwidth]{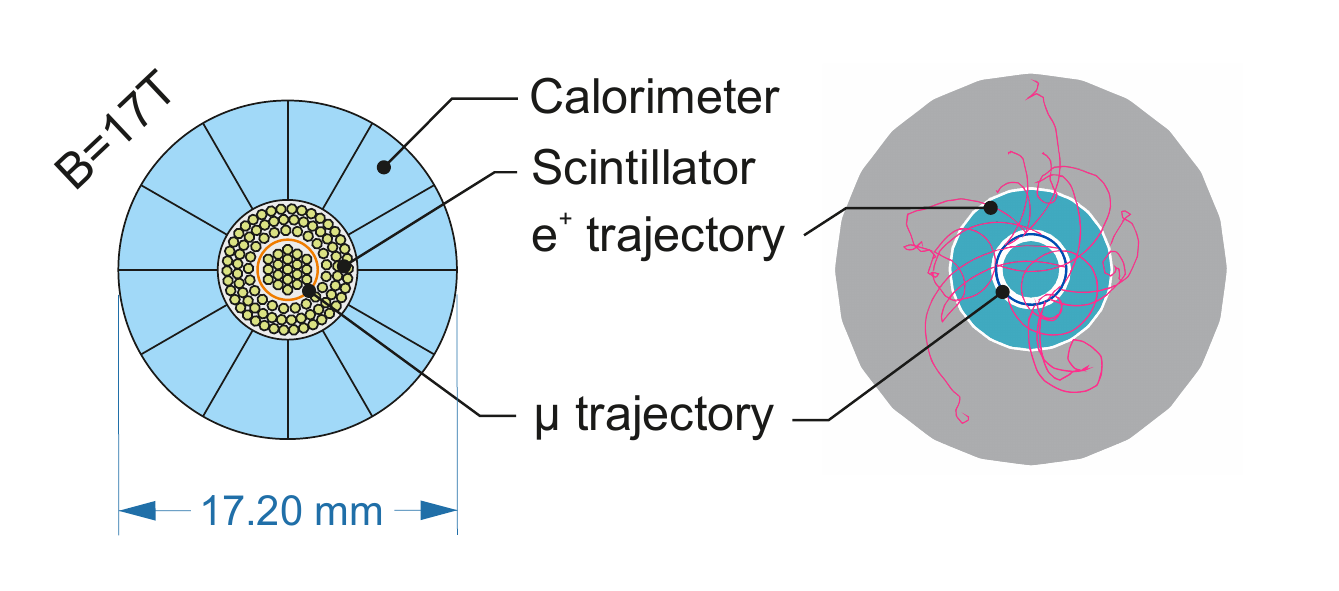}%
\caption{Sketch of a hyper compact $g-2$ storage ring experiment, TiAMMo, using muons with $E=\SI{1}{MeV}$ from a moderately re-accelerated low-energy, high-brightness muon source in a \SI{17}{T} magnetic field.}%
\label{fig:TinyAMM}%
\end{figure}

An alternative concept, similar to first ideas presented during the PSI workshop 2010 by D.~Taqqu, is to use non-relativistic muons, with an energy of $1\,\mathrm{MeV}/c$, in a tiny storage ring as depicted in \autoref{fig:TinyAMM}.
This scenario circumvents the challenges of a long re-acceleration beam line by proposing to use only one \ac{RFQ} acceleration stage within the same high magnetic field of muCool and transporting the muons into a high magnetic solenoid field of \SI{17}{T}.  
The muons longitudinal motion along the field lines is stopped by a short quadrupolar pulse similar as in the PSI muEDM concept.
The muon is then confined to an orbit of about \SI{6}{mm} diameter by using a weakly focusing magnetic field configuration. 
A combination of scintillating fibres and a segmented calorimeter detects the decay positron. 
Although, sub nanosecond timing resolution for scintillating fibres has been reported, a direct measurement of the muon spin precession frequency to extract $g$ and not $g-2$ seems still too ambitious as it would require a resolution of  better than \SI{40}{ps}. Instead we propose to use a central scintillating fibre bundle within the tiny storage orbit to distinguish between in and outwards going positrons for a measurement of the anomalous precession frequency.

\begin{table}%
\caption{Comparison of principle parameters of Fermilab E989~\cite{Grange:2015fou}, J-PARC~\cite{Abe:2019thb} and our proposals (AMMiED and TiAMMo).}
    \begin{center}
    {\footnotesize
        \begin{tabular}{lcccc}
        \toprule
        & Fermilab E989 & J-PARC & AMMiED & TiAMMo \\
        \midrule
        Muon momentum & $\SI{30.9}{GeV}/c$ & $\SI{300}{MeV}/c$ & $\SI{125}{MeV}/c$ & $\SI{14.6}{MeV}/c$ \\
        Lorentz $\gamma$ & 29.3& 3 & 1.56 & 1 \\
        Polarisation & 100\% & 50\%& 95\% & 95\% \\
        Magnetic field $B$ &  \SI{1.45}{T} &\SI{3}{T} & \SI{6}{T} & \SI{17}{T} \\
        Focusing field & E-quadrupoles & weakly focusing (wf) & wf & wf \\
        Cyclotron period & \SI{149}{ns} & \SI{7.4}{ns}  & \SI{3.8}{ns} & \SI{0.4}{ns} \\
        AMM precession period & \SI{4.37}{\micro\second} &\SI{2.11}{\micro\second} &\SI{1.05}{\micro\second} & \SI{0.37}{\micro\second} \\
        Number of detected $e^+$ & \num{1.6e11} & \num{5.7e11} & \num{7e12}& \num{1.7e12} \\
        \midrule
        AMM precision (stat.) & \SI{0.1}{ppm} & \SI{0.45}{ppm} & \SI{0.1}{ppm} &  \SI{0.08}{ppm} \\ 
        AMM precision (sys.) & \SI{0.1}{ppm} & \SI{0.07}{ppm} &
        --   & -- \\
        \bottomrule

        \end{tabular}
    }
    \end{center}
\label{tab:CompTable}
\end{table}

\subsubsection{Conclusion}
There is potential to investigate the tantalising hints of new physics by measuring the magnetic dipole moment of the muon using a different method and to reach four orders of magnitude below the current muon \ac{EDM} limit using a novel low-energy, high-brightness muon source coupled to the proposed \ac{HIMB}\@. In this section we discussed possible prospects for a muon \ac{EDM} measurement, which will ideally require a highly efficient re-acceleration to muon momenta of about \SI{210}{MeV}/$c$. For a measurement scheme based on a single-muon-on-request we show that a sensitivity of $\sigma(d_{\mu})=\SI{1e-23}{\ecm}$ may be expected within one year of data taking, with a potential improvement if multiple muons are stored at the same time.
Similarly, we presented two possible scenarios for a high-precision measurement of the muon $g-2$ with a statistical sensitivity better than 0.1\,ppm in one year of data taking, matching or even surpassing the prospected sensitivity of the currently ongoing effort at Fermilab (see \autoref{tab:CompTable}). Again higher sensitivity could be achieved if the source provided sufficient muons to permit a storage of more than one muon at a time within the storage ring. The tiny non-relativistic storage ring proposal has the advantage to avoid a long re-acceleration beam line. The price for this is a re-acceleration scheme within a solenoid guiding field and a high magnetic field for storage of $B=\SI{17}{T}$ resulting in an orbit of less than \SI{6}{mm} diameter.


\subsection{Muonium} \label{sec:muonium}

\subsubsection{Development of novel muonium sources at PSI} \label{sec:Msources}

A high-intensity, low-emittance atomic muonium $M=(\mu^+ e^-)$ beam is being developed at PSI, which would enable improving the precision of $M$ spectroscopy measurements, and may allow a direct observation of the $M$ gravitational interaction. 

Improvements in measuring the 1S-2S transition frequency using state-of-the art laser spectroscopy techniques is strongly motivated by recent experiments measuring the muon \ac{AMM}~\cite{Abi:2021gix}, and other searches for \ac{BSM} physics. Measurement of the free fall of $M$ atoms would be the first test of the weak equivalence principle using elementary antimatter ($\mu^+$) and a purely leptonic system. 

Both experiments rely on the high-intensity, continuous muon beams at PSI with significant benefits expected from the proposed \ac{HIMB} upgrade. The especially challenging gravity experiment additionally relies on a proposed novel atomic $M$ source, with cold atoms converted in the surface of \ac{SFHe}. 

Present state-of-the-art vacuum muonium sources are room temperature, porous materials that allow combination of the muon with an electron from the bulk, and a following quick diffusion inside the nanoscopic pores, with laser ablated silica aerogel being one of the best room temperature converters. Such sources provide $\sim 3\%$ muon-to-vacuum $M$ conversion using surface $\mu^+$ beams of 28~MeV/$c$ momentum \cite{beer2014enhancement}.

Mesoporous materials have been shown to convert $\mu^+$ to vacuum $M$ with efficiencies of 40\% (20\%) at room temperature (down to 100 K) when using highly moderated, keV energy muons from the \ac{LEM} beamline at PSI~\cite{Antognini:2011ei,Khaw:2016ofi}; this has an intensity four orders-of-magnitude lower than a surface muon beam. Future developments of moderated low-energy muons (the muCool experiment described in \autoref{sec:facilities:mu_cool}) may reach one to two orders of magnitude higher rates.  However, such converters produce a $M$ beam with broad (thermal) energy and angular ($\sim \cos\theta$) distributions. Improving the source quality by cooling these samples results in lower emission rates, with no observable emission below $\sim 50$~K due to the decreased diffusion constant, and to the sticking of $M$ to the pore walls that occurs unavoidably with any conventional $M$ converter \cite{Antognini:2011ei, kim2015}.

A newly proposed experiment at PSI plans to tackle the above difficulties by using \ac{SFHe} as a vacuum $M$ converter, due to its inert nature that rejects impurities from its bulk even at the lowest temperatures \cite{taqqu2011ultraslow, kirch2014testing, soter2021development}. Based on the calculated and measured chemical potentials of hydrogen isotopes in \ac{SFHe} ~\cite{saarela1993hydrogen, marin1998atomic}, $M$ atoms are expected to have a high chemical potential in the liquid compared to the thermal energies, implying that they would be ejected from the \ac{SFHe} surface with a well defined longitudinal velocity of $v_{M}\sim 6300$\,m/s only broadened by the thermal spread which is expected to fall below the Landau velocity ($v_{t} \approx 50$\,m/s).

Assuming high muon-to-vacuum-muonium conversion rates on the order of 10-20\% in \ac{SFHe}, the present $\pi$E5 beamline at PSI could provide a flux of $1-5\times10^5$~$M$/s from the cold atomic beam, while the \ac{HIMB} beamline could offer up to an order of magnitude higher rates.

\subsubsection{Spectroscopy}

Pure leptonic systems, such as positronium and muonium are free of finite-size effects and therefore are ideal systems to test bound state \ac{QED} \cite{Savely}. They are essential to extract fundamental constants such as the muon mass and magnetic moment. Those systems can also be used to search for new physics, including dark-sector particles and new muonic forces \cite{Frugiuele:2019drl}, as well as testing Lorentz/CPT symmetries \cite{2014-Vargas} or measure the effect of gravity on antimatter via the gravitational redshift  \cite{Karshenboim:2008zj}.

Past $M$ spectroscopy experiments were conducted between $1980-2000$ at TRIUMF, RAL and LAMPF (see \cite{2016-Jung} for a recent review). As a result of the
difficulty in obtaining a high flux of $\mu^+$ and the necessity to
slow down the muons so that $M$ into vacuum can be formed efficiently, all those  experiments were essentially limited by statistics, or statistics-related systematic effects \cite{2016-Jung}. With its
intense $\mu^+$ beam, PSI harbours tremendous opportunities for
improving $M$ spectroscopy experiments. Higher statistics makes it
possible to implement experimental techniques that are systematically
more robust. In this respect the \ac{LEM}
beamline at PSI \cite{Prokscha:2008zz} plays a crucial role as recently demonstrated by the Mu-MASS collaboration  \cite{Crivelli:2018vfe} which improved the determination of the $M$ Lamb shift (LS) by one order of magnitude \cite{Ohayon:2021qof}.

 A recent review of the ongoing measurements of the 1S-2S transition and the LS of muonium in the context of the Mu-MASS  experiment at PSI is given in \cite{Ohayon:2021dec}. A description of the progress of the Muonium Spectroscopy Experiment Using Microwave (MuSEUM) that is ongoing at J-PARC aiming to improve the muonium \ac{HFS} can be found in \cite{2020-MUSEUM}. Here we point out the impact that \ac{HIMB} would have on muonium spectroscopy. \ac{HIMB} would increase by orders of magnitude the available statistics of these experiments. This will also allow for accurate studies of the systematic effects and for the implementation of new measurement schemes, such as the employment of an enhancement cavity with a larger laser beam to reduce AC-stark shift in the 1S-2S measurement or the use of separated oscillatory fields spectroscopy to measure Lamb shift, fine structure and \ac{HFS}.

\ac{HIMB} would allow to push the experimental projected accuracy of the ongoing MuMASS and MUSEUM experiment by an order of magnitude, below 1 kHz for the 1S-2S transition and to the few Hz level for the \ac{HFS}.
Assuming the theoretical accuracy could also be improved to match the experimental one,  combining those measurement will result in a very stringent test of bound state \ac{QED} \cite{Eides:2018rph}. It will also allow to improve the determination of fundamental constants such as the muon mass, the muon magnetic dipole moment, and  a determination of the Rydberg constant independent of finite size effects as well as the determination of the fine structure constant at a level comparable with the current best determinations \cite{Morel:2020dww,Parker:2018vye} .

\begin{figure}[t]
    \centering
    \includegraphics[width=0.7\columnwidth]{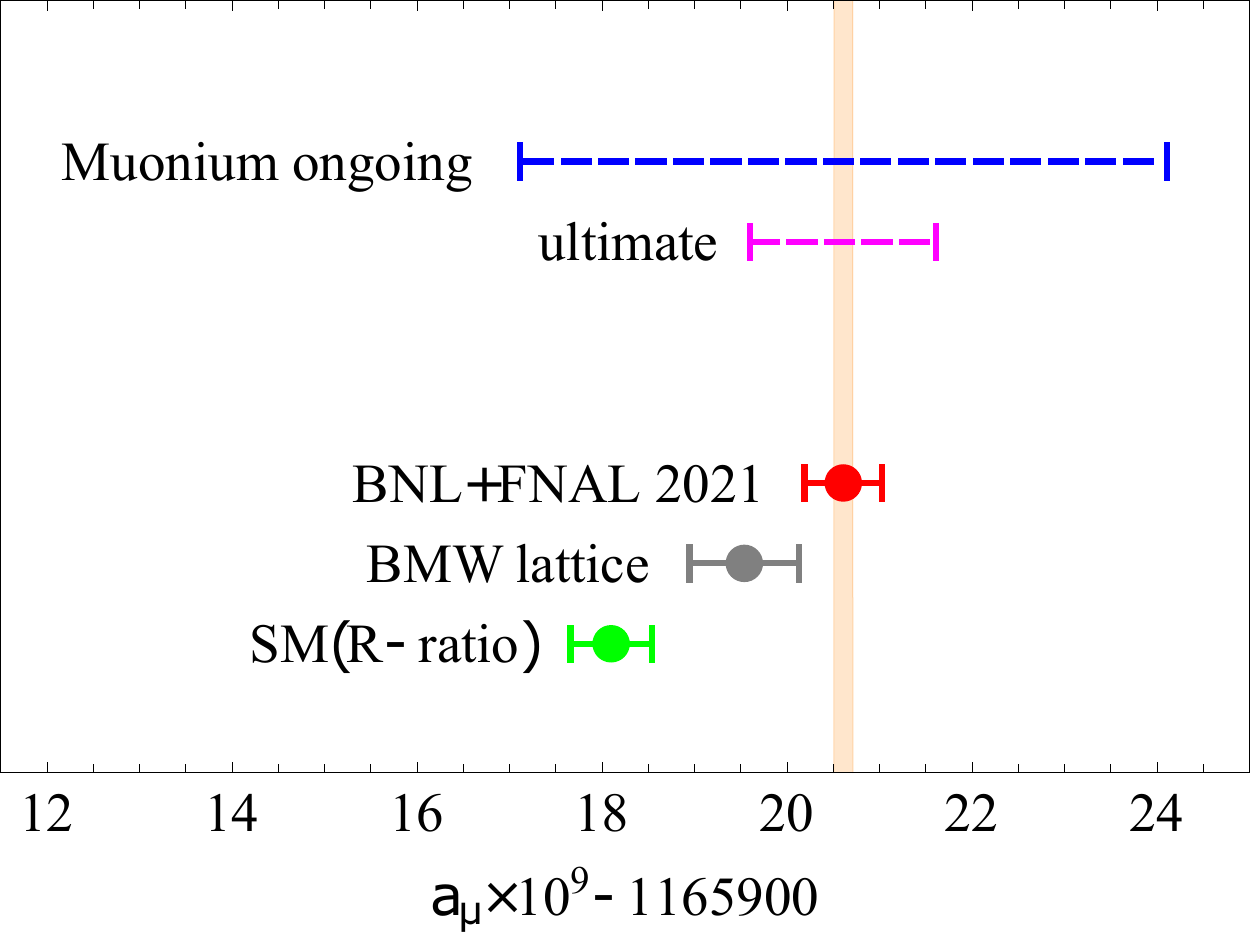}  
    \caption{Muon \acs{AMM} determined from the combined measurements at BNL and  \acs{FNAL} (red)~\cite{Abi:2021gix}, \ac{SM} calculations with leading-order \acs{HVP} evaluated from $e^+e^-\to$ hadrons data (green)~\cite{Aoyama:2020ynm} or using a recent lattice QCD result (gray)~\cite{Borsanyi:2020mff}. Also shown are the projected sensitivities in muonium (dashed) with the currently planned spectroscopy improvements (blue) and its ultimate improvement which could be reached at \acs{HIMB} (magenta), centred around the current experimental average. The orange band shows the four-fold improved \acs{FNAL} standard deviation expected in the near future. Adapted from \cite{Delaunay:2021uph}.}
    \label{fig:amu}
\end{figure}

Moreover, as pointed out recently \cite{Delaunay:2021uph}, 
$M$ spectroscopy at this level of precision would provide an independent determination of the muon \ac{AMM} with an uncertainty at the level of the current tension. This is illustrated in \autoref{fig:amu}.

$M$ spectroscopy offers also the possibility to search for new light bosons coupled to electrons and muons \cite{Frugiuele:2019drl}.   This is very interesting since those could provide an explanation in terms of new physics of the muon \ac{AMM}. 
In this case, a  dark force between the electron and the antimuon could be mediated for example by a new scalar or a new vector gauge boson giving rise to Yukawa-like attractive potentials~\cite{Fadeev:2018rfl}
\begin{align}
V_{ss}(\vec{r}) = -&g_e^s g^s_\mu\, \frac{e ^{-m_s r}}{4 \pi r} \\
V_{vv}(\vec{r}) =-&g^v_e g^v_\mu \frac{e ^{m_v r}}{4 \pi r} \nonumber \\
+&\frac{g^v_e g^v_\mu}{4}\bigg(\vec{\sigma}_e\cdot\vec{\sigma}_\mu
\Big(\frac{1}{r^3}+\frac{m_v}{r^2}+\frac{m_v^2}{r}- \frac{8\pi}{3}\delta(\vec{r})\Big)\nonumber  \\
& \qquad 
-(\vec{\sigma}_e\cdot\hat{r})(\vec{\sigma}_\mu\cdot\hat{r})
\Big(\frac{3}{r^3}+\frac{3m_v}{r^2}+\frac{m_v^2}{r}\Big)\bigg)
    \, \frac{e ^{-m_v r}}{4 \pi m_e m_\mu}
\end{align}
where $V_{ss}$ and $V_{vv}$ are for scalar and vector potentials, respectively, $m_s$ is the scalar boson mass, $m_v$ is the vector boson mass, and $g^{s,v}_e, g^{s,v}_\mu$ are the coupling strengths to electrons and antimuons. 
The effect of such new forces on the measured transitions can be estimated by applying first order perturbation theory.

\begin{figure}[tb]
\centering
\includegraphics[width=0.9\textwidth]{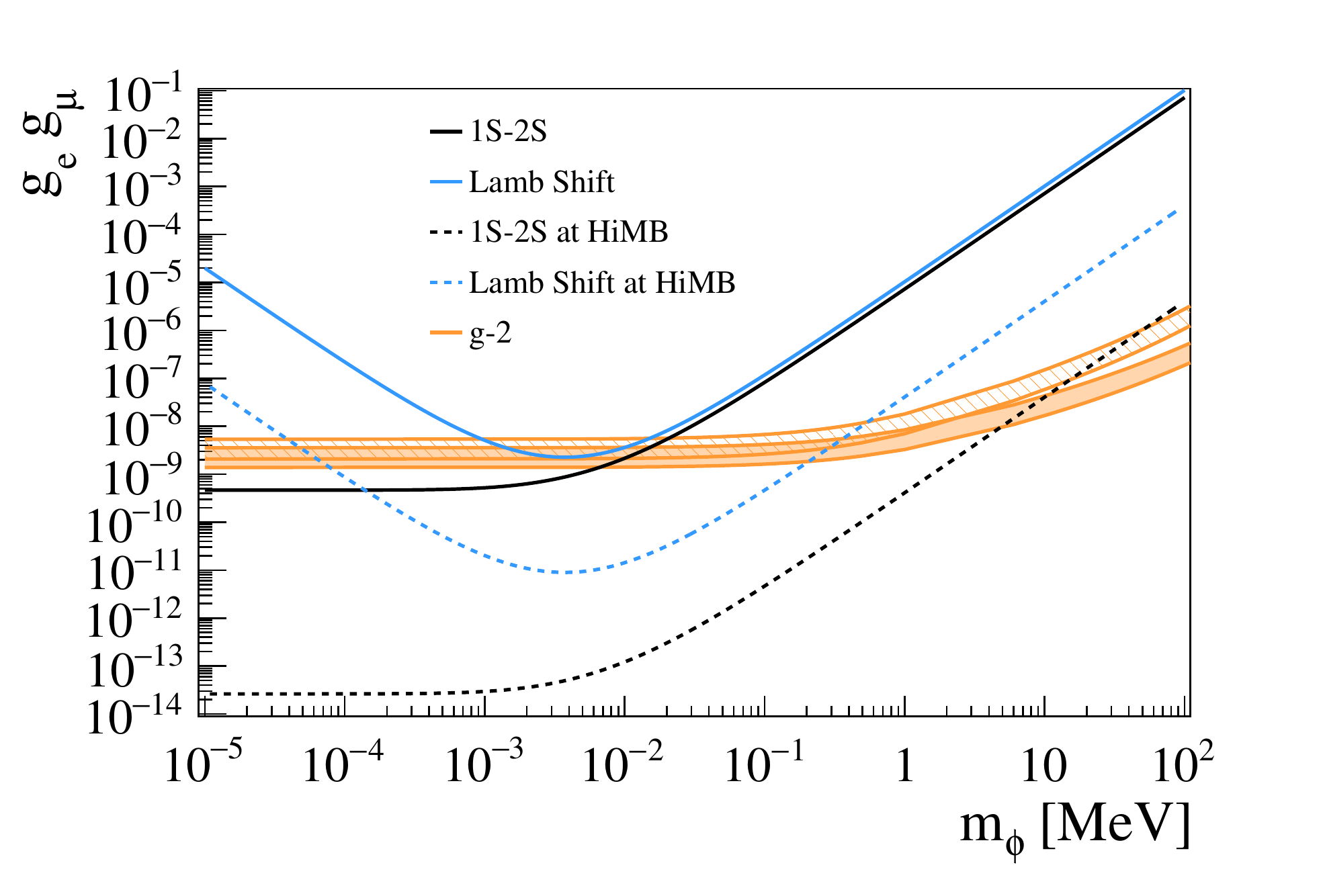}
    \caption{Constraints from $M$ spectroscopy on $g^{s}_e, g^{s}_\mu$ as a function of the scalar/vector mass. The solid black line is the constraint from the $M$ 1S$-$2S measurement \cite{2000-1S2S}  while the blue line is from the recent Lamb shift measurement at PSI \cite{Ohayon:2021qof}. The dashed lines are the projections of the uncertainty that would be enabled by \ac{HIMB}. The orange band represents the region suggested by the $g-2$ muon anomaly considering the lower bound from the measurement of the electron gyromagnetic factor for the scalar case, while the hatched region is for the vector one.}
\label{fig:scalar}
\end{figure}

In \autoref{fig:scalar}, we present the current and projected sensitivity with \ac{HIMB} of muonium spectroscopy to new physics. The constraints on $g^{s}_e,\,g^{s}_\mu$ as a function of the scalar/vector mass, which are nearly
identical in the mass range considered here, are compared to the region favoured
by the $g-2$ muon anomaly, considering the bounds from the electron gyromagnetic
factor \cite{Hanneke:2008tm}. However, as pointed out in
\autoref{sec:Motivation}, the situation for the electron \ac{AMM} should be
clarified since the two latest determinations of the fine structure constant
using atom interferometry differ from each other by 5\,$\sigma$
\cite{Parker:2018vye,Morel:2020dww}.  
We do not present results from experiments at the intensity frontier since those
typically depend on assumptions on the decay
channels. An example is the recent results of the NA64 experiment placing
stringent bounds on new bosons with the assumption that those would decay invisibly \cite{NA64:2021xzo}.
The combination of the Lamb shift and 1S$-$2S $M$ measurements provides the most stringent laboratory constraint excluding that a new scalar/vector boson with a mass $<10\,$keV could contribute to the muon $g-2$ anomaly. 

Another interesting application is that $M$ spectroscopy is sensitive to Lorentz and CPT violating effects. In the context of the Standard Model Extension (SME) this would improve the current limits  \cite{2014-Vargas}. 
Moreover, a test of the gravitational behaviour of second generation leptons via the gravitational redshift \cite{Karshenboim:2008zj} will become possible. 

The realisation of \ac{HIMB} would thus greatly expand the physics reach of $M$ spectroscopy.

\subsubsection{Gravitational free fall experiment of muonium}

\begin{figure}[t]
\centering
  \includegraphics[width = 0.9\textwidth]{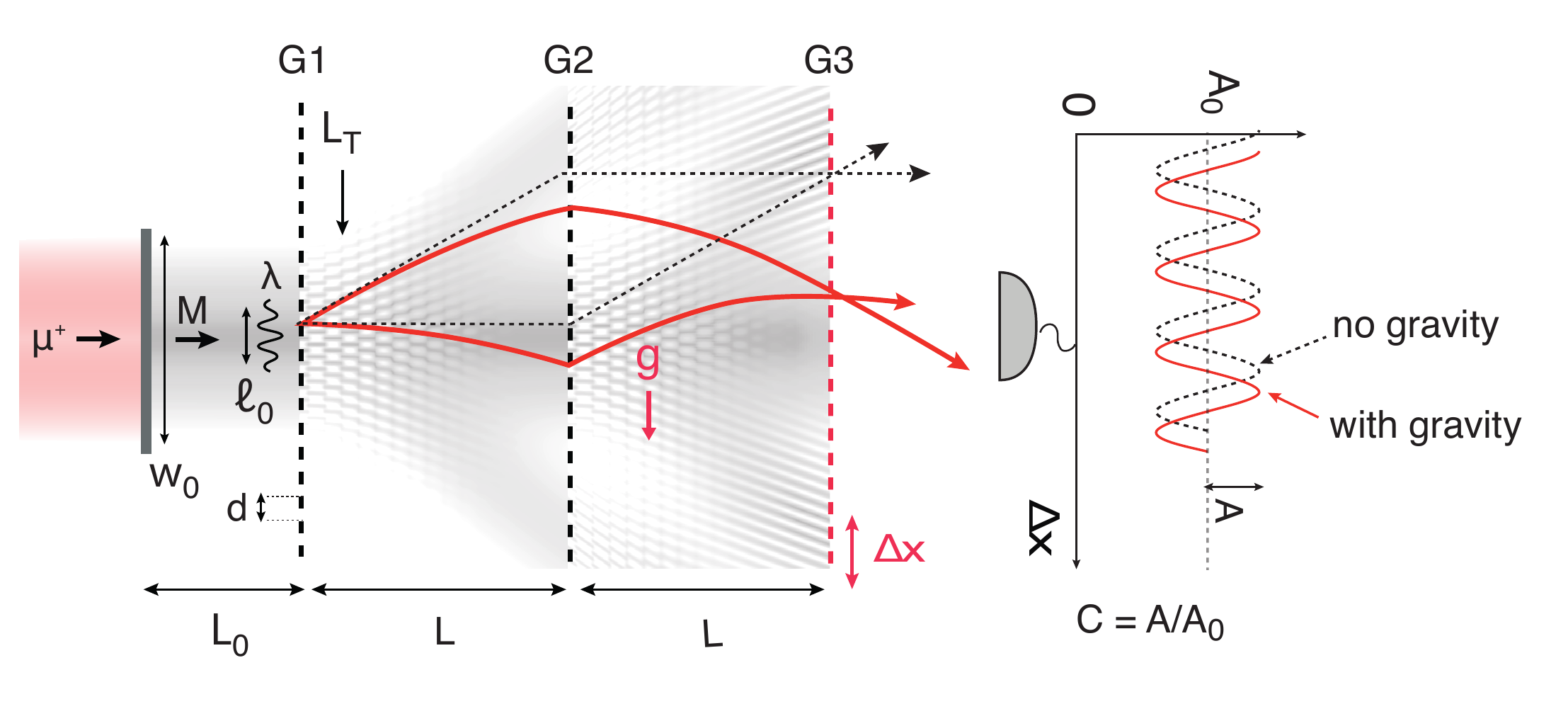}
  \caption{A three-grating interferometer used to measure the gravitational
    interaction of $M$ atoms.  The quantum diffraction pattern caused by the
    gratings G1 and G2 with a fully coherent beam is given in grey.
    Classical trajectories (red and dashed lines) are shown to
    illustrate the effect of gravity on the measured interference
    pattern appearing at G3. The vertical shift of the interference
    pattern caused by the gravitational acceleration $g$ is detected by
    measuring the transmitted $M$ rate while scanning G3 in vertical
    direction. See details in text.}
  \label{principle}
\end{figure}

The novel cold atomic beam under development would enable the measurement of the gravitational fall of $M$ atoms by means of atom interferometry. However, such an experiment is inherently challenging due to the short lifetime or the muon ($\tau \approx 2.2 ~\mu$s) which allows a mere $\Delta x=\frac{1}{2}gt^2=600$~pm gravitational fall in a time of $t=5\tau$. Measuring small deflections like this needs a precise knowledge of the initial momentum of the atoms, and requires strict momentum selection. Two periodic gratings (G1 and G2) with horizontal slits of pitch $d$ and spaced by a distance $L$ could be used to achieve this momentum selection as shown in \autoref{principle}.

The classical and quantum regime of this device is characterised by the de Broglie wavelength of the atoms, $\lambda = h/p $, and grating pitch $d$ in terms of the Talbot length, $L_T=d^2/ \lambda$, which is approximately 18 microns for thermal $M$ atoms with $\lambda_M \approx 0.56$~nm. Trajectory selection for both classical and quantum regimes of this device by G1 and G2 will result in an intensity pattern with the same periodicity $d$ at a distance $L$ after G2. Gravitational acceleration and deflection of the atoms causes a phase shift $\delta \phi$ of this pattern in the vertical direction as $\delta \phi =2\pi gT^2/d$, where $T=L/v_{M}$ is the $M$ time of flight between each pair of gratings. 

Direct observation of this sub-micron pattern and sub-nanometer shifts needed for measuring $M$ gravity would be extremely hard. It is possible however to carry out an indirect measurement using a third grating (G3) of the same pitch $d$, placed at distance $L$ from G2. By counting the total rate of $M$ atoms transmitted through G3 as a function of the G3 vertical position $\Delta x$ the phase shift can be measured. 

The contrast of the intensity pattern $C$ is defined by the ratio of the amplitude and the average yield $C =A/A_0$ as shown in \autoref{principle}. This contrast strongly depends on the transverse coherence length of the beam, $\ell_{0}$, that determines how many slits of G1 are illuminated with a coherent wavefront. This coherence length in relation to the beam width $w_{0}$ and the interferometer parameters (the grating periodicity $d$ and distances $L$) together with the de Broglie wavelength ($\lambda$) of the atoms is sufficient to estimate the interferometer performance in the first approximation. In analogy to statistical optics (Van Cittert-Zernike theorem \cite{goodman2015statistical}), we can relate the transverse coherence length of the $M$ beam to the transverse momentum distribution of the atoms as $\ell_{0} = \frac{1}{2}\frac{\lambda}{\alpha}\approx 16$~nm, where $\alpha$ is the angular spread of the cold $M$ source. 

The sensitivity in measuring the gravitational acceleration $g$ is given by \cite{oberthaler1996inertial}
\begin{equation}
\Delta g = \frac{1}{2\pi T^2}\frac{d}{C \sqrt{N}} \; ,	
\label{eq:sensitivity}
\end{equation}
where $N$ is the number of $M$ atoms transmitted through G3 and measured by the detector given by
\begin{equation}
N = N_0 \; \varepsilon_0 \, e^{-(t_0+2T)/\tau} \; (T_G)^3 \;   \varepsilon_\mathrm{det} \,  ,
\label{eq:number}
\end{equation}
with $N_0$ being the number of $M$ atoms produced at the $M$ source, and $\varepsilon_0$ the $M$ transport efficiency from the source to G1. The $M$ decay is accounted for by the third term $e^{-(t_0+2T)/\tau}$, where $t_0$ is the time of flight from the source to G1. The number of detected $M$ atoms is further reduced by the $M$ detection efficiency $\varepsilon_\mathrm{det}$, and by the limited transmission $T_G$ of a single grating.
The short lifetime of the muon necessitates a gain in sensitivity by using a small grating pitch $d$. Maximal sensitivity, as a tradeoff between phase shift $\delta \phi$ and statistics $N$, is obtained for $T\approx 6-8~\mu$s corresponding to an interferometer length of 40-50~mm. 

A calculation of the interferometer parameters to extract the contrast $C$, uses an approximation of the $M$ source with a Gaussian Schell-model beam \cite{mcmorran2008model}, and adapted mutual intensity functions that are widely used to describe the propagation of partially coherent light \cite{goodman2015statistical}. Using realistic parameters on the initial beam size and quality expected from the superfluid source above, the fringe contrast of $C\approx 0.3$ at the exact position of G3 can be achieved, with a few $\mu$m high-contrast region along the beam axis. Such a measurement thus requires precise G3 positioning with $\mu$m-accuracy in the beam axis, and below-nm-accuracy in the vertical direction.

From \eqref{eq:sensitivity} we see that determining the sign of $g$ (more precisely to reach $\Delta g/g=1$) in about one day, requires the detection of 3.2~$M$/s, assuming a contrast $C=0.3$. Following \eqref{eq:number}, and taking pessimistic estimates from Monte-Carlo simulations and initial detector and grating studies studies by using $T_G = 0.3$, $\varepsilon_0=0.75$ and $\varepsilon_\mathrm{det}=0.3$, at the source we need $N_0 \approx 1.4\times 10^4$~$M$/s. As a comparison the $\pi$E5 beam line at PSI can presently deliver $3.6 \times 10^{6}$~$\mu^+$/s at a momentum of 10~MeV/$c$ within a transverse area of about 400~mm$^2$. At this muon momentum we may expect a muon-to-vacuum-M conversion efficiency of about 0.1-0.3 in the best case, when fast diffusion of the atoms can be assumed. This will result in $M$ rates of up to $\sim 1.1 \times 10^{6}$~$M$/s. These high rates may allow a further collimation of the $M$ beam to a $5\times 1$~mm area, which would put less strain on grating production and alignment and would cut the number of useful $M$ atoms conservatively by a factor 5 mm$^2$/400~mm$^2 =0.013$. Using these parameters where there is room for contingency, we expect to produce the necessary rate of $\sim 5 \times 10^4 $~$M$/s in an small area of $\sim 5\times 1$~mm$^2$, and reach the goal sensitivity of $\Delta g =\frac{9.8 \mathrm{~m/s^2} }{\sqrt{\mathrm{\#\; days}}}$ with present $\mu^+$ sources. An increase by two orders of magnitude in $\mu^+$ rates expected by the proposed \ac{HIMB} project at PSI will allow shorter measurement times and further improvements on the sensitivity to $g$.


\subsection{Further particle-physics applications} \label{sec:pool}

\subsubsection{Muon decays} \label{sec:mudec}

LFV muon decays $\mu\to e e e$ and $\mu\to e \gamma$ are flagship
cases for HIMB. In this subsection we consider other SM and BSM muon
decay processes that allow to further scrutinise the operator causing
the decay of the muon, but also allows for a more general approach to
look for low-mass BSM particles coupling e.g. to electrons or
neutrinos. As we will describe there, has been a lot of activity in the
past decades to use muons as a more general laboratory for BSM
physics. Many of the results described here can in principle be
improved upon with an increased rate provided by HIMB.

The Michel decay of muons $\mu^+\to e^+ (\nu_e\bar\nu_\mu)$ played an
important role in many aspects of particle physics. Its description
through the Fermi theory with a $V\!-\!A$ four-fermion interaction was
an early application of effective theories and the determination of the
Fermi constant via the muon lifetime~\cite{Webber:2010zf} is a crucial
ingredient for electroweak precision tests. 

A more general analysis of muon decays is typically done allowing for
right-handed neutrinos, but neglecting their
mass~\cite{Fetscher:1986uj, Danneberg:2005xv}. If no information on
the positron polarisation is available, the energy spectrum of the
decay positron can be expressed in terms of four decay
parameters~\cite{Fetscher:2021ldh}, usually denoted by $\rho$,
$\delta$, $\eta$ and $P_\mu\,\xi$, with $P_\mu$ the polarisation of
the muon. The presence of BSM physics results in deviations of these
parameters from their SM value. Earlier searches for right-handed
currents~\cite{Jodidio:1986mz} have been
generalised~\cite{TWIST:2011aa} and lead to stringent constraints on
$\rho$, $\delta$, $\eta$ and $\xi$.

The most recent and most precise measurements of the muon decay parameters $\rho$, $\delta$, and $\mathcal{P}_\mu \xi$ were obtained by the TWIST experiments stopping a highly polarised beam in a silver target located inside a 2 T solenoid and surrounded by tracking and timing detectors \cite{TWIST:2011jfx, TWIST:2011egd}. These measurements were not statistically limited, but relied on a precise control of all relevant systematic effects. In that sense, such measurements will not profit from the increase in muon rates offered by \ac{HIMB}. This is especially true when measuring $\mathcal{P}_\mu \xi$ where a very precise knowledge of the muon beam polarisation is paramount and which is almost impossible to achieve at a high-acceptance beamline such as \ac{HIMB}. However, the measurements of $\rho$ and $\delta$ could profit from a beam such as \ac{HIMB}-muCool (see \autoref{sec:facilities:mu_cool}), which would allow to stop the beam in an extremely thin target thus greatly reducing systematic uncertainties related to the passage of the Michel positron through the target and would allow a very compact detector arrangement potentially reducing the systematic uncertainties related to detector response and calibrations.

Interpreting the muon decay parameters in terms of Wilson coefficients
of a general EFT containing SM particles and some number of sterile
neutrinos provides a framework to combine muon decay constraints with more general BSM explorations. Contrary to \cite{Fetscher:1986uj, Danneberg:2005xv} this
requires to allow for non-negligible sterile neutrino masses, as e.g. in
\cite{Forero:2011pc,Alonso:2012ji,deGouvea:2015euy}. In addition to
the Michel decay, also the \ac{RMD} $\mu\to e (\nu\bar\nu) \gamma$
and the rare muon decay $\mu\to e (\nu\bar\nu) (e e)$ can be used. The
additional visible final-state particles can help to analyse potential
effects in more detail.  On the other hand it has to be kept in mind
that the low energy scale of the process severely limits the scale of
\ac{BSM} physics that can be tested.

While the rate of the \ac{RMD} and rare decay with HIMB is no issue
at all, analysing these processes would require to collect a certain
amount of data disregarding the kinematic cuts used in \ac{cLFV}
searches. For the rare decay in connection with Mu3e this is probably
more realistic than for the \ac{RMD} in connection with a
MEG-like experiment.

Regarding more general searches for low-mass BSM, in recent years
theorists have actively explored new physics prospects below $\sim$
GeV~\cite{Beacham:2019nyx}, and various experiments have found hints
for new particles below the muon mass.  For instance, there are neutrino
experiments which individually find evidence for an additional massive
neutrino,
and as a possible explanation for the $^8$Be anomaly in nuclear decays~\cite{Krasznahorkay:2015iga}
a 17\,MeV boson has been suggested.  The fact that a 17\,MeV boson coupled
to electrons is difficult to confirm or rule out motivates the search
for deviations in SM muon decays.

There are also numerous theoretically well motivated scenarios with
nearly massless \acp{ALP}. They appear as Goldstone
bosons and possibly have even LFV couplings. This can lead to muon
decays where there are BSM particles among the
invisibles~\cite{Agrawal:2021dbo, Beacham:2019nyx,Bauer:2021mvw}.  Searches for the
two-body decay $\mu\to e X$ with $X$ an \ac{ALP} were carried out at
TRIUMF~\cite{Jodidio:1986mz, Bayes:2014lxz}. For sufficiently large
$m_X$ such a process would lead to a bump in the positron energy
spectrum. For nearly massless $X$ the endpoint of the spectrum is
modified and a careful comparison to the SM prediction is required. In
this context there is also the possibility to use a forward detector
to increase the sensitivity, since the SM background is strongly
suppressed in the forward
direction~\cite{Pilaftsis:1993af,Hirsch:2009ee,Calibbi:2020jvd}. 
The sensitivity of $\mu A \to e A$ experiments to emission of an \ac{ALP}
is explored in \cite{GarciaTormo:2011jit}. Several decades ago Cristal
Box looked for $\mu\to e X \gamma$~\cite{Goldman:1987hy} and $\mu\to e
\gamma\gamma$~\cite{Bolton:1988af}. Regarding the latter, the two photons could arise from an \ac{ALP} decay $X\to\gamma\gamma$ or a contact interaction (of
dimension 8). More recently, MEG
has looked for $\mu\to e X\to e
\gamma\gamma$~\cite{Baldini:2020okg} and improved the limits for the branching ratio  $X\to\gamma\gamma$ in the range $m_X=20-45$\,MeV.

Another prominent example of a BSM particle that can be looked for in
muon decays is a dark photon.  For Mu3e there are first
studies~\cite{Perrevoort:2018ttp} for dark photon and LFV two-body
decays. While they can be carried out in phase I of Mu3e, a dedicated
experimental search with HIMB would improve the sensitivity.

Astrophysics can constrain particles with masses
$\stackrel{<}{{}_\sim} $ few MeV that are feebly coupled to the first
generation, and energy frontier experiments exclude new particles with
large couplings --- but for new particles coupled to the first and/or
second generation, there remains an allowed
triangle~\cite{Beacham:2019nyx}, beyond astrophysics, the energy
frontier and current intensity frontier bounds, which precision $\mu$
decays could explore.  

An $X$ boson coupled to neutrinos could alleviate
\cite{Kreisch:2019yzn} the current $H_0$ tension \cite{Bernal:2016gxb,
  Planck:2018vyg} in cosmology, and be produced via $\mu\to e \nu
\bar{\nu} X$.  Some current experimental constraints on this scenario
have been discussed~\cite{Lyu:2020lps}, but the theoretical study
\cite{Lessa:2007up} suggests that meson decays could be more sensitive
to such an $X$ than $\mu\to e \nu \bar{\nu} X$, because $X$ emission
removes the chiral suppression of the meson decays.

To summarise, there are many low-mass BSM physics scenarios that can
affect muon decay processes.  A general \ac{EFT} extension as well as a
collection of simplified low-mass BSM models are being implemented for
muon decay processes in the framework
McMule~\cite{Banerjee:2020rww}. Together with the precise prediction
of SM muon decays this gives a theoretical basis for a detailed
investigation of the potential impact of future measurements with
\ac{HIMB}.

\def\mup    {\ensuremath{\mu^+}}
\def\mun    {\ensuremath{\mu^-}}
\def\Mu     {\ensuremath{M}}
\def\epmu   {\ensuremath{e^+_\mu}}
\def\epA    {\ensuremath{e^+_{\text{A}}}}
\def\enmu   {\ensuremath{e^-_\mu}}
\def\enA    {\ensuremath{e^-_{\text{A}}}}
\def\aMu   {\ensuremath{\overline{\Mu}}}
\def\MuaMu  {\ensuremath{\Mu\overline{\Mu}}}
\def\pmuamu {\ensuremath{P_{\MuaMu}}}
\def\gmuamu {\ensuremath{G_{\MuaMu}}}
\def\gf     {\ensuremath{G_{\text{F}}}}
\def\sioo   {\ensuremath{\text{SiO}_2}}
\def\eatom  {\ensuremath{e_\text{atom}}}
\def\emichel{\ensuremath{e_\text{Michel}}}
\def\ekinmu {\ensuremath{E^{\text{kin}}_{\Mu}}}
\def\nmudec {\ensuremath{N_{\Mu}^{\text{decay}}}}

\def\eV   {\ensuremath{\,\text{eV}}}
\def\usec   {\ensuremath{\,\mu\text{s}}}
\def\mm   {\ensuremath{\,\text{mm}}}

\def\ie   {\textit{i.e.\/}}

\def\Kz    {\ensuremath{K^0}}
\def\Kbar  {\kern 0.2em\bar{\kern -0.2em K}{}}
\def\Kzb   {\ensuremath{\Kbar^0}}
\def\KzKzb {\ensuremath{\Kz \kern -0.16em \Kzb}}

\subsubsection{Muonium - antimuonium oscillations}
Bruno Pontocorvo predicted there might be
\MuaMu\ oscillations~\cite{Pontecorvo:1957cp}, in analogy to
\KzKzb\ mixing, even before \Mu\ was
discovered~\cite{Hughes:1960zz}. In the past, \MuaMu\ oscillations
were searched for in Ar gas~\cite{Amato:1969cp}, at
TRIUMF~\cite{Marshall:1981mk,Huber:1989by}, with foils at
LAMPF~\cite{Bolton:1981ug} and also with \sioo\ powder targets at
LAMPF~\cite{Ni:1987iza,Ni:1989zb,Matthias:1991fw}, and most recently
with the same approach at PSI~\cite{Abela:1996dm,Willmann:1998gd}. The current best upper limit
for the probability of spontaneous muonium to antimuonium oscillation
is $P_{\MuaMu} < 8.3\times 10^{-11}$ (90\%\,CL) and was obtained with
the MACS apparatus at PSI.

Muonium has a lifetime of $2.2\usec$ and decays into a fast positron
($\epmu$) from the muon decay and a slow electron ($\enA$) from the
bound electron. The decay of an antimuonium is signalled by a final
state with the opposite electrical charges, \ie, a fast electron
($\enmu$) and a slow positron ($\epA$).

\begin{figure}[tb]
  \begin{centering}
   \unitlength1.0cm 
   \begin{picture}(15.0,5.5)
     \put( 0.2, 0.6){\includegraphics[height=4.0cm]{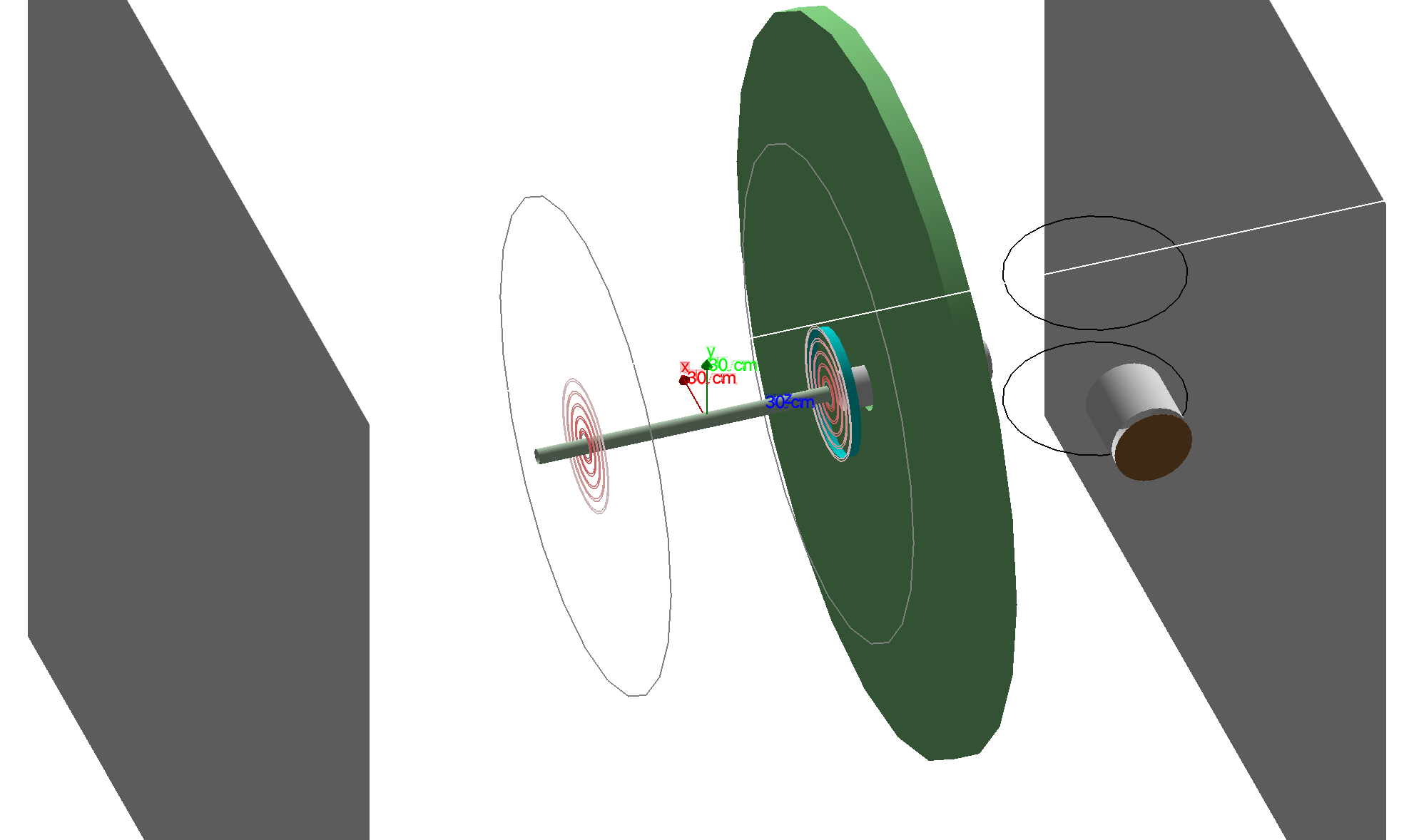}}
     \put( 7.0, 0.0){\includegraphics[height=5.0cm]{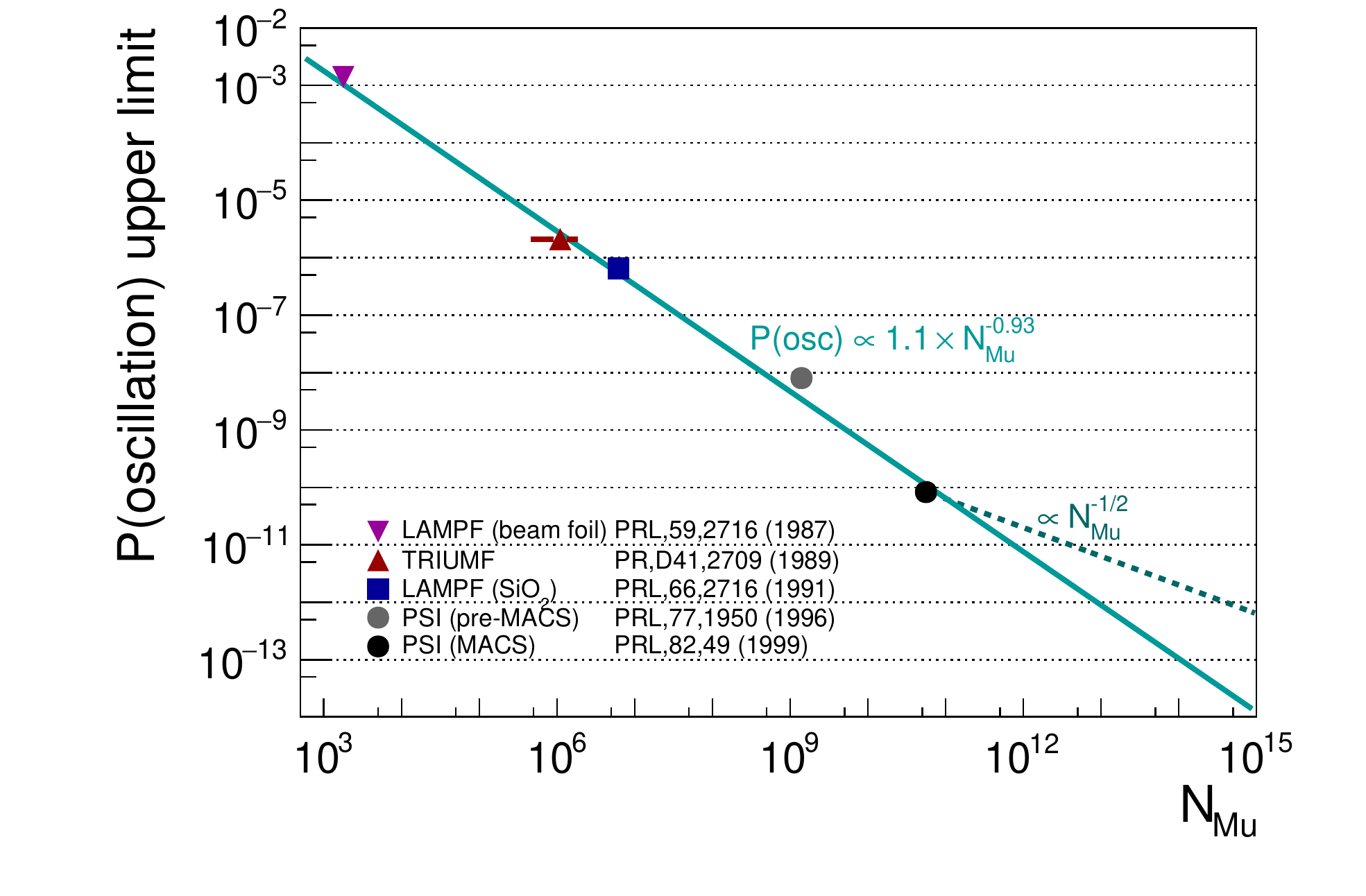}}
   \end{picture}
 \caption{(left) Straw-man GEANT4 setup emulating the MACS
   apparatus. The \mup\ beam enters from the left and hits a target
   (hidden in the beam pipe) inside the 5-layer tracker. (right)
   Development of the upper limit of the \MuaMu\ oscillation
   probability vs.~the number of produced $N_\Mu (=N_{\text{Mu}})$ in the acceptance of the experiments. The results are taken verbatim from the
   publications and consist of upper limits at 90\%CL and 95\%CL (no
   re-scaling has been performed as the difference is not visible on
   this plot). The lines are to guide the eye. }
   \label{f:muamuhistory}
  \end{centering}
\end{figure}

To study the different beam setups in more detail, a straw-man
GEANT4~\cite{GEANT4:2002zbu} simulation setup emulating the MACS
apparatus was developed as shown in \autoref{f:muamuhistory} (left).
A target is embedded inside a five-layer tracker with total length of
1m. In a region extending over 2\,cm along the \mup\ beam direction,
an electric field of 1.5\,kV/cm  provides acceleration for $\epA$,
and a solenoidal magnetic volume bends the accelerated $\epA$ onto a
sensitive volume.  This simulation allows the modelling and estimation
of (1) the decay characteristics of the \Mu, (2) the acceptance
(defined as the fraction of \Mu\ decays inside the active detector
volume), and (3) the detection efficiency as a function of the
\Mu\ kinetic energy (\ekinmu) and production divergence. For a real
detector, a configuration allowing the simultaneous measurement of
\Mu\ and \aMu\ (e.g. with the accelerating electric field
perpendicular to the \mup\ direction) would be chosen.

In all setups, the simulated detector setup was kept
constant. Therefore, the fraction of \Mu\ decays inside the active
volume ($f_D$) decreases strongly with increasing
\ekinmu. \autoref{t:muamubeams} illustrates this behaviour.

\begin{table}[tb]
 \begin{center}
  \caption{Expected \Mu\ decays per second (\nmudec) in vacuo for different beam
    intensities $N_{\mup}$ and $\mup\to\Mu$ (in vacuo) conversion
    efficiencies $\eta_{\Mu}$. The factor $f_D$ specifies the fraction
    of \Mu\ decaying inside the detector volume corresponding to the
    \Mu\ kinetic energy \ekinmu.  }
  \medskip
  \begin{tabular}{llllll}
   \toprule
   Target &$N_{\mup}\, [\text{s}^{-1}]$  &$\eta_{\Mu}$ &$\ekinmu [\eV] $ &$f_D$ &\nmudec$\,[\text{s}^{-1}]$ \\
   \midrule
   \acs{HIMB}-3/Al$^{(a)}$       &$1.0\times10^{10}$ &0.001 &20000  &0.07  &$7.0\times10^5$  \\
   \acs{HIMB}-3/\sioo{}$^{(b)}$ &$1.0\times10^{10}$ &0.028 &0.026 &1.00  &$2.8\times10^8$ \\
   \acs{HIMB}-3/aerogel &$1.0\times10^{10}$   &0.015 &0.026  &1.00  &$1.5\times10^8$  \\
   \bottomrule
  \end{tabular}\\
  \smallskip
  \hspace*{-2cm}${}^{(a)}$Assuming $\eta_{\Mu}$ as in~\cite{Bolton:1981ug} \quad${}^{(b)}$Target not stable over time.\\
  \label{t:muamubeams}
 \end{center}
\end{table}

To significantly improve on the current best upper limit $P_{\MuaMu} <
8.3\times 10^{-11}$ (90\%\,CL) of~\cite{Willmann:1998gd}, a minimum of
$\nmudec = 10^{12}$ is required, cf.~\autoref{f:muamuhistory}
(right). This is reachable with aerogel targets with relatively short
measurements, and may also be achieved with foils.

\subsubsection{Atomic parity violation in muonic atoms} \label{sc:apv}

A muonic atom is formed when a negative muon comes to rest in a material and subsequently gets captured by a nearby atom. After this atomic capture, the muon quickly cascades down to the $1s$ atomic orbital, emitting Auger electrons and X-rays. Due to its relatively large mass, there is significant overlap between the muon wave function and the atomic nucleus. This makes this system particularly attractive to study short range interactions between the muon and a nucleus/nucleon.

Muonic atoms are being used to determine absolute nuclear charge radii, most recently by the CREMA collaboration for low Z nuclei~\cite{Antognini:1900ns,CREMA:2016idx,Krauth:2021foz} and by the muX collaboration for high-Z radioactive elements~\cite{Wauters:2021cze}.

Another compelling physics case for muonic X-rays is \ac{APV} which arises from the mixing of the opposite parity $2s_{1/2}$ and $2p_{1/2}$ atomic levels, resulting in an E1 admixture in the otherwise pure M1 $2s_{1/2}-1s_{1/2}$ transition~\cite{Missimer:1984hx}. The most straightforward \ac{PV} observable is the angular correlation between the $2s-1s$ X-ray and the electron from the $\mu^-$ decay in orbit.

Observing the single photon $2s-1s$ transition is a challenge however. For Z$\backsimeq$30 nuclei, this transition has a \ac{BR} of $\mathcal{O}(10^{-4})$. At lower Z two-photon and Auger transitions completely dominate the depopulation of the 2s level. A clean detection of this X-ray is hampered by a dominant background from Compton scattered $(n>2)p-1s$ X-rays. In addition, the \ac{SM} amplitude of a \ac{PV} observable is expected to be smaller than $10^{-3}$. For boron (Z=5) or neon (Z=10) the \ac{PV} $e^- \gamma$ correlation is expected to be at the 10$^{-2}$ or 10$^{-3}$ level, respectively. Additional challenges arise however due to the Auger depopulation of the $2s$ level, necessitating a low pressure gas target to avoid repopulating the electric orbitals~\cite{PhysRevLett.78.4363}. Furthermore, the small energy difference between the $2s$ and $2p$ levels, responsible for the large \ac{APV}, makes it hard to resolve the $2p-1s$ and $2s-1s$ transitions with traditional X-ray detectors.

\begin{figure}[tb]
\center
\includegraphics[width=\textwidth]{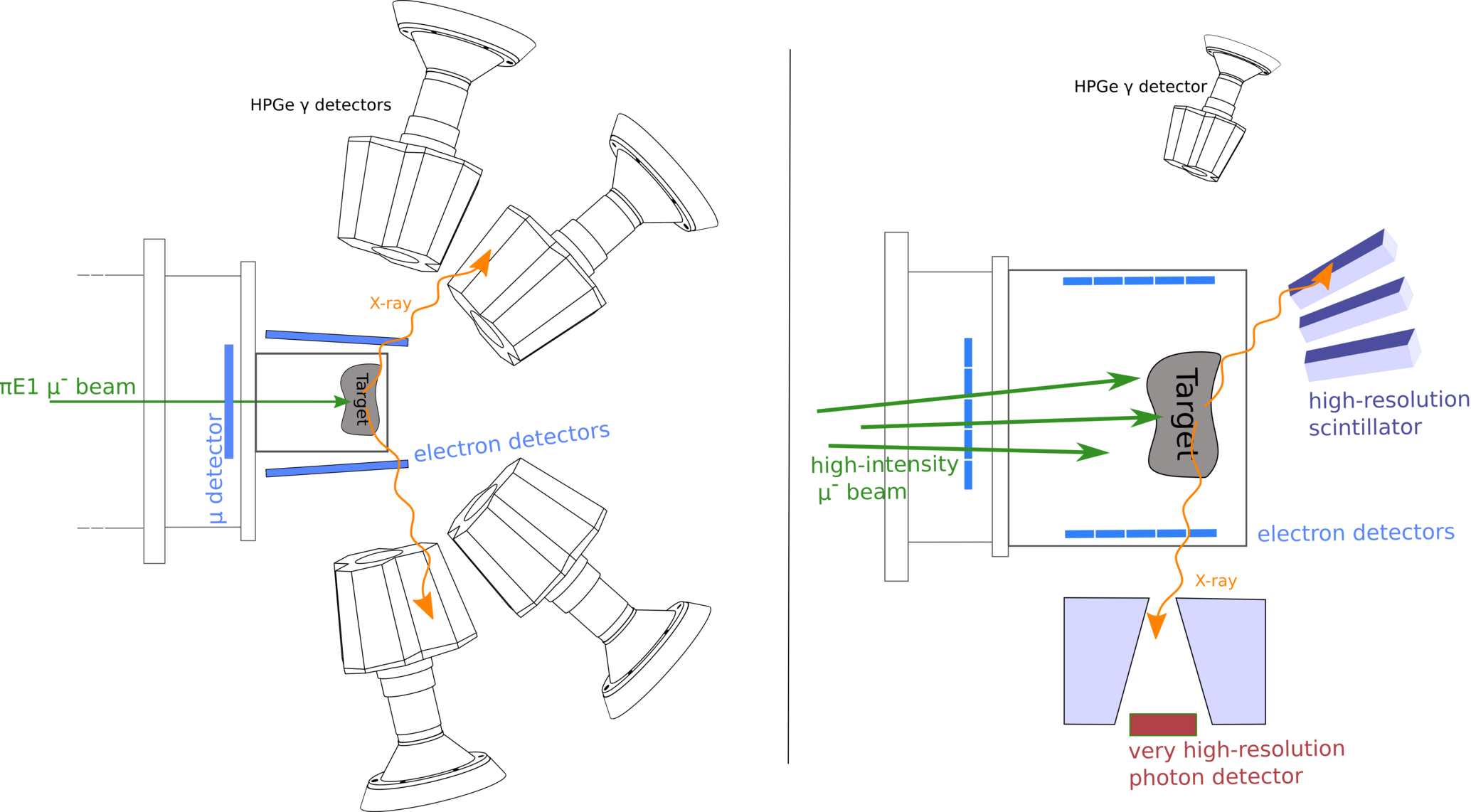}
\caption{Left: Sketch of the muonic X-ray setup currently used at the $\pi$E1 beam area, with a thin muon counter, a small target surrounded by Michel electron detectors, and a \acs{HPGe} detector array. Right: Accepting higher incoming muon rate requires the development of larger targets and/or a confining magnetic field, and the use of novel high-resolution photon detectors such as LaBr$_3$ detectors and microcalorimeters.}
\label{fig:apv:setup}
\end{figure}

The muX collaboration has been investigating the reach of a $2s-1s$ \ac{APV} experiment for Z$\backsimeq$30 nuclei. Deploying a \ac{HPGe} detector array at the $\pi$E1 beamline, the transition was observed in Kr and Zn for the first time~\cite{Wauters:2021cze}. Despite being the photon detector of choice for many applications in the \numrange[]{0.1}{10}~\SI{}{\mega \electronvolt} energy range, \ac{HPGe} detectors have long charge collection times and are susceptible to neutron damage, limiting the instantaneous and integrated rate, which in turns limits a muX-style \ac{APV} experiment~(\autoref{fig:apv:setup}) to a sensitivity of $\mathcal{O}(10^{-3})$ on a \ac{PV} observable such as the $e^- \gamma$ correlation.

To fully profit from the high $\mu^-$ rates at \ac{HIMB}, or even from the current intensities available at the $\pi$E5 beam area, alternative high-resolution photon detector technologies have to be considered. One promising technology are high-resolution inorganic scintillators such as doped LaBr$_3$ detectors, where resolutions of better than 2~\% at \SI{1}{\mega \electronvolt} are on the horizon. A rough estimate with 50 $1\times 1$~inch detectors placed \SI{25}{\centi\metre} from a muon target shows that a statistical precision on the $e^- \gamma$ correlation of better than $10^{-3}$ is possible.

For Z$\backsimeq$30 nuclei this would mean an $\mathcal{O}(1)$ \ac{SM} test. Searching for muon-specific dark forces, as suggested in~\cite{Karshenboim:2014tka}, this level 
of sensitivity can be competitive when only taking the low-energy $\mu$ two-loop 
constraints from normal \ac{APV}. 
Taking into account \ac{LHC} bounds from non-resonant di-lepton 
data~\cite{Crivellin:2021bkd}, the sensitivity of a muon \ac{APV} experiment to $Q_W$ 
needs to be similar as the $^{133}$Cs \ac{APV} measurement, which is at the 1\% level. 
As a consequence, from this perspective only a measurement with low-Z nuclei is worth 
pursuing as the \ac{APV} amplitude is experimentally accessible due to
the near degeneracy of the $2s_{1/2}$ and $2p_{1/2}$ states. This energy split quickly 
increases with  a Z$^4$ dependence, yielding \ac{APV} amplitudes $< 10^{-3}$ for 
Z$\geq$30 nuclei. To realize such \ac{APV} measurement with sufficient sensitivity, 
R\&D efforts need to be put into novel detector technologies, as well as in efficient 
low-pressure gas-targets.

The \ac{HIMB} $\mu^-$ intensities also open up the opportunity to apply ultra-high resolution $\gamma$ detectors such as crystal spectrometers~(see e.g. \cite{KESSLER2001187}) and state of the art cryogenic microcalorimeters (see e.g.~\cite{Kraft-Bermuth:2018mgz, WINKLER2015203}) to muonic X-rays spectroscopy. A first possible physics case are bound-state \ac{QED} tests using exotic atoms as suggested in~\cite{Paul:2020cnx}. Such a measurement can probably be done with a fairly simple target station.

\subsubsection{Muon conversion} \label{sec:muconv}

Searches for the conversion of the muon into an electron in the field of the nucleus have been performed at PSI in the past and indeed the current best limit on this so far unobserved process is given by the SINDRUM II collaboration in~\cite{SINDRUMII:2006dvw}. Large efforts are currently being undertaken by the Mu2e and COMET collaborations \cite{Mu2e:2014fns,COMET:2018auw} to perform these measurements at Fermilab and J-PARC, respectively, aiming at improving the sensitivity of the SINDRUM II experiment by a factor $10^4$. The search for muon conversion profits from the pulsed structure of the Fermilab and J-PARC proton accelerators as after the formation of the muonic atom a certain amount of time can be waited until the background events stemming from the proton pulse on the muon production target has significantly dropped.

\begin{figure}
\center
\includegraphics[width=\textwidth]{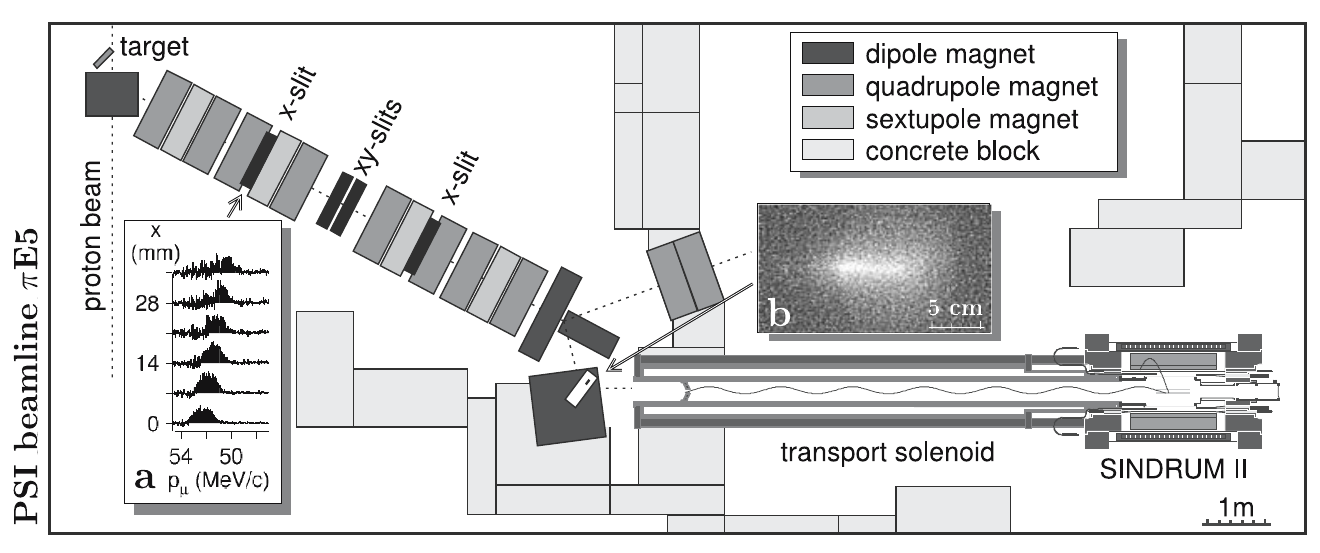}
\caption{Experimental setup of the SINDRUM II experiment. Muons and pions are produced by the proton beam at Target E and are transported by the $\pi$E5 beamline. A degrader located at the beginning of the transport solenoid separates the pions and muons and guides the muons to the SINDRUM II spectrometer located at its end. Picture reproduced from \cite{SINDRUMII:2006dvw}, where also more details can be found.}
\label{fig:setup_sindrum}
\end{figure}

The experiments at PSI needed to adopt a different method to reduce such backgrounds, as the quasi-continuous \SI{50}{\MHz} structure of the proton accelerator at PSI does not allow to implement this strategy. Instead the experiment relied on making the muon beam as pure as possible by strongly reducing the amount of pions transported by the beamline responsible for the largest backgrounds. \autoref{fig:setup_sindrum} shows the setup employed by the SINDRUM II experiment. It consisted of the actual spectrometer shown at the end of the beamline used to accurately track and measure the particles emitted from the muon conversion gold target placed at its centre. A transport solenoid connected the spectrometer to the $\pi$E5 beamline. At the beginning of the transport solenoid  an \SI{8}{\mm} thick CH$_2$ degrader was placed. The degrader was used to separate the surviving pions in the beam at the momentum of \SI{52}{\MeV}/$c$ through their much shorter range in material compared to muons. The thickness was chosen to stop all pions while most of the muons continued further downstream -- guided by the transport solenoid -- to the spectrometer. Figure~1 in \cite{SINDRUMII:2006dvw} shows the difference in range for muons and pions in the degrader (however keep in mind that this simulation was performed for monoenergetic beams thus exaggerating the separation power).

\begin{figure}
\center
\includegraphics[width=0.7\textwidth]{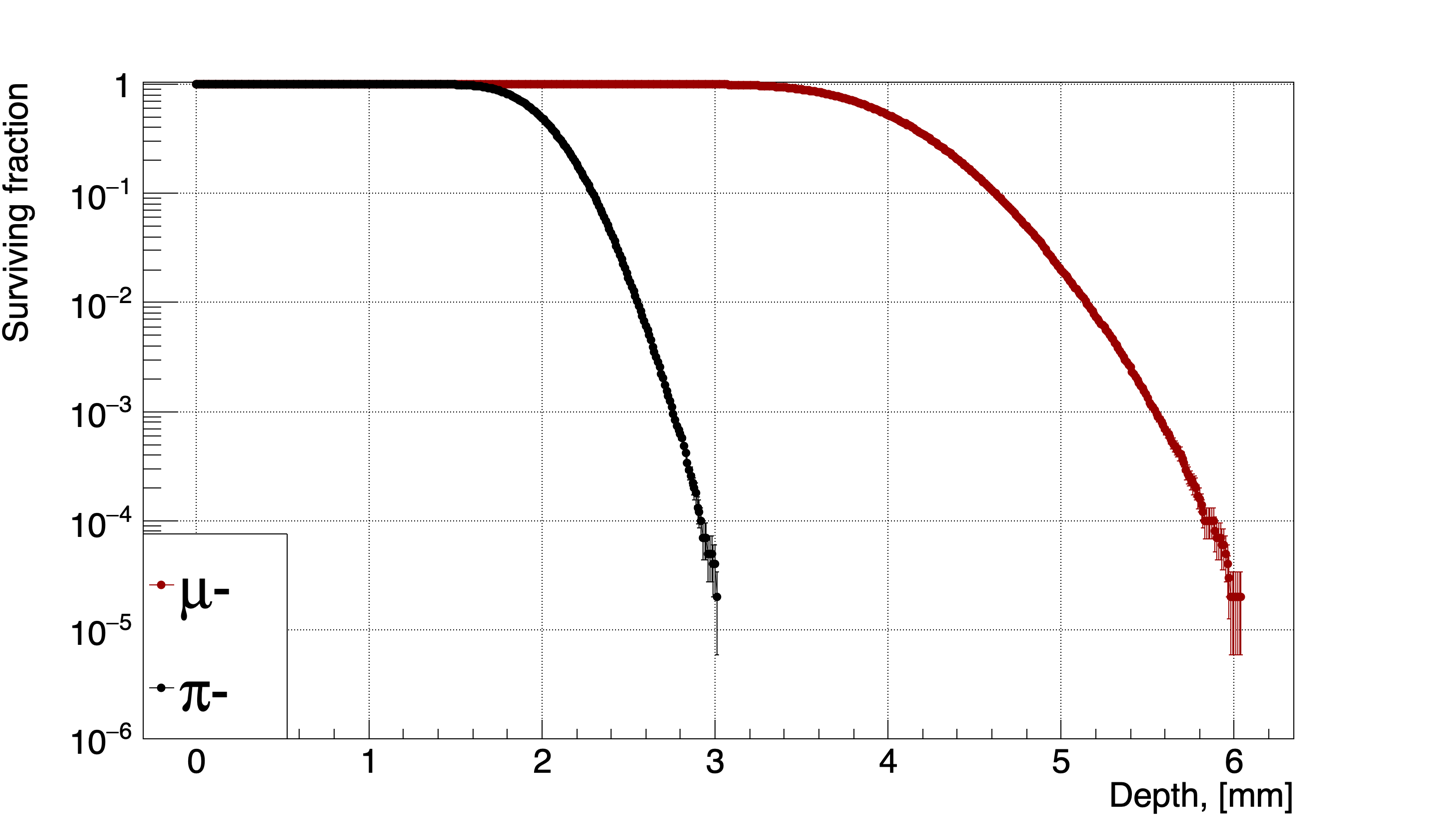}
\caption{Range of negative pions and muons in a CH$_2$ degrader at \SI{40}{\MeV}/$c$ and with a momentum resolution of 3\%. A degrader with a thickness a bit above \SI{3}{\mm} would be appropriate to nicely separate pions and muons.}
\label{fig:muon_pion_range}
\end{figure}

In order to see how such an experiment could be improved by \ac{HIMB} at PSI, let us examine what the improvement in muon rate could be and what strategies could be employed to combat backgrounds at the same time:
\begin{itemize}
\item \textbf{Muon rate:} \\
SINDRUM II was running at a beam momentum of \SI{52}{\MeV}/$c$ in the $\pi$E5 area of PSI with an estimated rate of around $2\times 10^6$~$\mu^-$/s available at that time. Due to the limited focusing power of the solenoids used in the \ac{HIMB} beamline compared to quadrupoles, the momentum reach of the \ac{HIMB} muon beamlines are limited. Good transmission should be achieved up to a momentum of about \SI{40}{\MeV}/$c$, at which point it is expected to reach about $10^8$~$\mu^-$/s. At this momentum the muon/pion separation separation through a degrader is obviously not as good as in the case of SINDRUM II -- see \autoref{fig:muon_pion_range}. However, due to the lower momentum the pion content of the beamline is approximately an order of magnitude lower to start with. 

\item \textbf{Backgrounds:}\\
\acp{BG} in the SINDRUM II experiment originated mostly from pions decaying in front of or stopping in the degrader. High-energy electrons generated in these decays could reach the spectrometer and mimic the electron produced in the muon conversion process. Different strategies can be employed to reduce the \ac{BG} to levels much below the ones achieved by SINDRUM II to fully exploit the increased muon rate. i) A spectrometer with modern detector technology as, e.g., employed by the Mu3e experiment (see \autoref{sec:mu3e}) will much improve the identification of the electron emitted in the muon conversion process and thus allow for tighter cuts and correspondingly reduced backgrounds. ii) Developments on active targets for such kind of experiments will also provide an additional handle to reduce \acp{BG} by uniquely identifying events originating from the conversion target. iii) Using a curved transport solenoid with the degrader at its entrance would eliminate a direct line-of-sight between the degrader and the spectrometer thus stopping any high-energy electrons originating at the degrader from reaching the spectrometer. Similar concepts are also employed by the Mu2e and COMET experiments  \cite{Mu2e:2014fns,COMET:2018auw}.
\end{itemize} 

Based on the above considerations an improvement of around two orders of magnitude over the sensitivity of the SINDRUM II experiment seems feasible by performing a dedicated and new experiment at \ac{HIMB}. Given the fact that both Mu2e and COMET aim at reaching four orders of magnitude improved sensitivity, the reach of such an experiment at PSI is certainly very limited. The situation could of course change if either these experiments are not able to reach their ultimate sensitivity or detect the muon conversion process with a relatively high branching ratio. At this point, an experiment at \ac{HIMB} could become interesting -- especially as it is hard for the experiments at pulsed beams to work with high-Z conversion targets, which would reveal some insights into the underlying physics of the conversion process \cite{Kitano:2002mt}.



\section{Muon spin spectroscopy and material science with HIMB}\label{sec:musr}
%

Muon spin spectroscopy typically uses positive muons as highly sensitive local magnetic probes to study a broad range of research topics in solid-state physics, chemistry and materials science \cite{Yaouanc:2011Book}. The experimental techniques referred to as \ac{muSR} (for Muon Spin Rotation, Relaxation, Resonance or Research) are universally applicable since muons can be implanted in any material. The spin-$\frac{1}{2}$ muons are produced with 100\% spin polarisation, providing \ac{muSR} with a great advantage compared to other local probe methods such as \ac{NMR} or \ac{ESR} 
which typically rely upon a tiny thermal equilibrium spin polarisation. The muon is a very sensitive probe of local static and dynamic magnetic properties of materials. Its short mean lifetime of 2.2~$\mu$s and  relatively large gyromagnetic ratio of $2\pi \times 135.5$~MHz/T makes it suitable for measurements of magnetic fields ranging from $\sim \mu$T to several Tesla, and fluctuating on a time scales from pico- to microseconds. The muon decays into two neutrinos and an energetic positron which is emitted preferentially along the direction of the muon spin, thus providing information on the time evolution of the muon spin polarisation in the investigated material. 

Applications of \ac{muSR} include a large variety of topics in condensed matter research with a focus on the investigation of novel and unconventional magnetic and superconducting materials. As a local probe, \ac{muSR} is a powerful tool in this context since it allows studying the detailed nature of magnetic and superconducting phases on a microscopic scale and as a function of control parameters such as doping, temperature, applied magnetic field or pressure. One of the particular advantages of \ac{muSR} in such measurements is its sensitivity to both the superconducting and magnetic volume fractions, and to their respective order parameters. In the case of superconductors, fundamental microscopic parameters such as the magnetic penetration depth can be determined absolutely. In magnetic materials, \ac{muSR} is extremely sensitive to small magnetic moments and short range magnetic order. In addition, it is possible to discriminate between static and dynamic magnetism and to characterise magnetic fluctuations within a unique time window, which falls between that of neutron scattering and \ac{NMR}. Thanks to the volume fraction sensitivity of \ac{muSR}, the technique is often used to complement scattering probes like X-ray diffraction and neutron scattering where phase volume information is difficult to obtain. 

In addition to the common applications of \ac{muSR} in magnetism and related subjects, it is used to investigate molecular dynamics, charge transport phenomena such as polaron motion in conducting polymers, electron and spin transport as well as ion conduction in technologically relevant battery materials. As a "light isotope" of the proton the $\mu^+$ can form the hydrogen-like exotic atom muonium [$M = (\mu^+\,e^-)$] which may substitute for hydrogen in insulators, semiconductors and organic materials and provide accurate spectroscopic information on hydrogen levels in these materials and their potential for practical applications. As a "light proton isotope", the muon can also serve as a test model for light particle diffusion, radical formation, and chemical reactions with the largest known kinetic isotope effect \cite{Fleming:2011Science}.

The majority of the \ac{muSR} experiments use one of the following experimental setups, which vary in the orientation of the initial muon spin direction with respect to the external field and the physical information that can be obtained. Very roughly speaking, \ac{ZF} \ac{muSR} allows studying the static magnetic order parameter, the volume fraction and magnetic correlation times. Longitudinal-field (LF, a magnetic field is applied along the initial muon spin direction) \ac{muSR} is used to discriminate between static and dynamic ground states and to investigate magnetic fluctuations, as well as spin or ion diffusion rates. Transverse-field (TF, a magnetic field is applied perpendicular to the initial muon spin direction) \ac{muSR} is often used to determine fundamental length scales in superconductors and to measure the q-integrated local susceptibility (Knight shift).

In time-integrated measurements the time evolution of the polarisation is not determined, and the measurements can be carried out at very high muon rates (also at continuous mean beams, because one does not need to have only one muon at a given time in the sample as in time-differential measurements, see \autoref{musr:different_muon_source}). Time-integrated experiments provide information about the energy-level scheme of the muon spin coupled to other spins species (electrons and/or nuclei). Examples are avoided-level-crossing spectroscopy (ALC) or radio-frequency $\mu$SR (RF-$\mu$SR).

\subsection{Different muon sources around the world}\label{musr:different_muon_source}
Condensed matter research using \ac{muSR} requires intense muon beams which at present are available at four large scale facilities: the ISIS facility at the Rutherford Appleton Laboratory (UK), the MUSE facility at J-PARC (Japan), the CMMS facility at TRIUMF (Canada) and the \ac{SmuS} facility at PSI. The first two are pulsed muon sources with 50 Hz and 25 Hz repetition rates, respectively, while those at TRIUMF and at PSI deliver continuous muon beams. Pulsed and continuous muon sources complement each other allowing for different kinds of experiments. At pulsed sources it is possible to determine very small muon spin relaxation rates due to the virtually zero background between the pulses, but these sources are strongly limited in time resolution due to the 50 -- 80~ns width of the muon pulses. In contrast, at continuous sources the time resolution can be up to a factor of thousand better since it is solely determined by the detector construction and the read-out electronics. However, the incoming muon rate has to be reduced compared to a pulsed beam facility to minimise pile-up events. In a traditional \ac{muSR} setup at a continuous beam, an incoming muon is registered in a muon counter, which opens a data gate of typically 10~$\mu$s length. A valid \ac{muSR} event is given after the detection of the decay positron in one of the scintillators of the positron spectrometer. In case of a pileup event, where a second muon or a second positron is observed within the data gate, it is not possible to determine which muon belongs to which positron, and the event is discarded. At a beam rate of $4\times 10^4$/s, the accepted rate of $1.8\times 10^4$/s is at maximum for a data gate of 10~$\mu$s. At beam rates $>4\times 10^4$/s, the increasing pile-up probability causes a reduction of the accepted rate to $<1.8\times 10^4$/s. In contrast, at pulsed beams, the accepted rate of incoming muons can easily exceed $10^5$/s. 

The rate limitation at a continuous muon beam could be overcome, if the stopping position of the incoming muon in the studied sample could be detected and the corresponding emitted positron tracked. In this case, each decay positron could be assigned unambiguously to its parent muon by vertex reconstruction, thus allowing for multiple muons to stop in the sample at any given moment. This makes \ac{muSR} measurements with at least ten times higher incoming muon rates feasible. This paradigm shift for \ac{muSR} measurements at a continuous source will be discussed in detail in \autoref{musr:new_opportunities}.

\subsection{\texorpdfstring{Current status of \acs{muSR} at PSI}{Current status of muSR at PSI}}\label{musr:current_status}
The Laboratory for Muon Spin Spectroscopy (LMU) develops and operates six state-of-the-art instruments of the \ac{SmuS} facility, the worldwide most powerful muon source for condensed-matter research. 
%
%
The six instruments are permanently installed at five separate secondary muon beamlines of the \ac{HIPA} cyclotron complex. The different instruments provide the users with a variety of experimental capabilities with respect to temperature, magnetic field, pressure, time resolution, measurement geometry, probing depth and minimal sample size to fulfil the various requirements for the broad scientific spectrum addressed by the Swiss and international user community (see \autoref{musr:table1}). To maintain the leading international research position, the \ac{muSR} instruments need to be kept at the state-of-the-art. This requires a continuous development of new experimental capabilities to be able to address topical and novel research questions.
%
%
\begin{table}[tb!]
    \centering
    \begin{tabular}{lrrrp{4.5cm}}
    \toprule
    Instrument  &  $\mu^+$ energy & T-range & B-range & Comments \\
    \midrule
     DOLLY    & 4 MeV       & 0.25 -- 300 K & 0 -- 0.6 T &  Unique device for uniaxial strain up to 1~GPa. \\
     FLAME    & 4 MeV       & 0.02 -- 300 K & 0 -- 3.5 T & Flexible and Advanced \ac{muSR} Environment.  Begin operation in 2022.\\
     GPD      & 15 -- 60 MeV & 0.30 -- 500 K & 0 -- 0.6 T & Unique hydrostatic pressure up to 2.8 GPa.\\
     GPS      & 4 MeV       & 1.60 -- 1000 K & 0 -- 0.78 T& General Purpose Instr.  Unique MORE capability (Muons On REquest).\\
     HAL-9500 & 4 MeV       & 0.01 -- 300 K & 0 -- 9.5 T & Unique combination of low T and very high B.  Record time resolution in \ac{muSR} of 60 ps.\\
     \ac{LEM}      & 1 -- 30 keV & 2.30 -- 600 K & 0 -- 0.3 T & Unique \ac{LEmu} beam and \ac{LEmuSR} spectrometer. Thin films \& heterostructures.  $\mu^+$ range up to 200 nm.\\
    s-Ne      & 1 -- 20 keV &               &            & 23 kHz moderated $\mu^+$.\\
    s-Ar      &             &               &            & 14 kHz moderated $\mu^+$.\\
    \bottomrule
    \end{tabular}
    \caption{The instruments of \acs{SmuS} with several unique features. 
    T-range and B-range denote the available range of temperatures and magnetic fields, respectively. 
    At all instruments, the muon spin can be rotated between 10$^\circ$ and 90$^\circ$ by so-called spin-rotators ($\vec{E}\times\vec{B}$ fields), or special asymmetric quadrupole settings of 
    the muon beam line (GPD instrument). This allows zero-field-, LF-, and TF-\ac{muSR} measurements. 
    \ac{LEM}, s-Ne: solid neon moderator; s-Ar: solid argon moderator. The \ac{LEmu} rates are for a
    proton current of 2~mA, "slanted" target E \cite{Berg:2015wna}.}
    \label{musr:table1}
\end{table}

The research programme at \ac{LMU}/\ac{SmuS} includes the ``classical'' applications of \ac{muSR} in magnetic materials, superconductors, and semiconductors, as described at the beginning of
\autoref{sec:musr} but also goes well beyond this scope. Frontier research topics of recent years include high-spin molecules \cite{Shafir:2005prb,Tesi:2018ChemComm,Kiefl:2016acsnano}, low-dimensional magnets
\cite{Forslund:2019SciRep,Lancaster:2013PhysScrip}, quantum spin liquids \cite{Tustain:2020NPJQM,Gao:2019NatPhys}, organic superconductors \cite{Pratt:2016JSPS}, conducting polymers \cite{Wang:2016JPSJ,Pratt:JPhysCondMatt2004}, liquid crystals \cite{McKenzie:2013PRE}, topological materials \cite{Guguchia:2020NatComm,Guguchia:2019NPJQM,Krieger:2020PRL,Krieger:2019PRB}, defect regions in novel solar-cell materials \cite{Alberto:2018PhysRevMat,Curado:2020ApplMatTod}, and defects in semiconductor device heterostructures and near-surface regions \cite{Woerle:2019PRB,Woerle:2020PRAppl,Prokscha:2020PRAppl}. 

\begin{figure}[tb!]
  \centering
    \includegraphics[trim=0 -10mm 0 0,clip,width=0.49\textwidth]{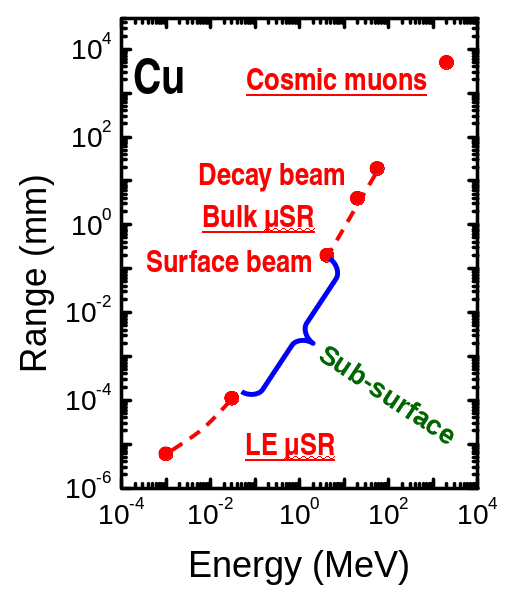}
    \includegraphics[trim=0 0 0 0,clip,width=0.46\textwidth]{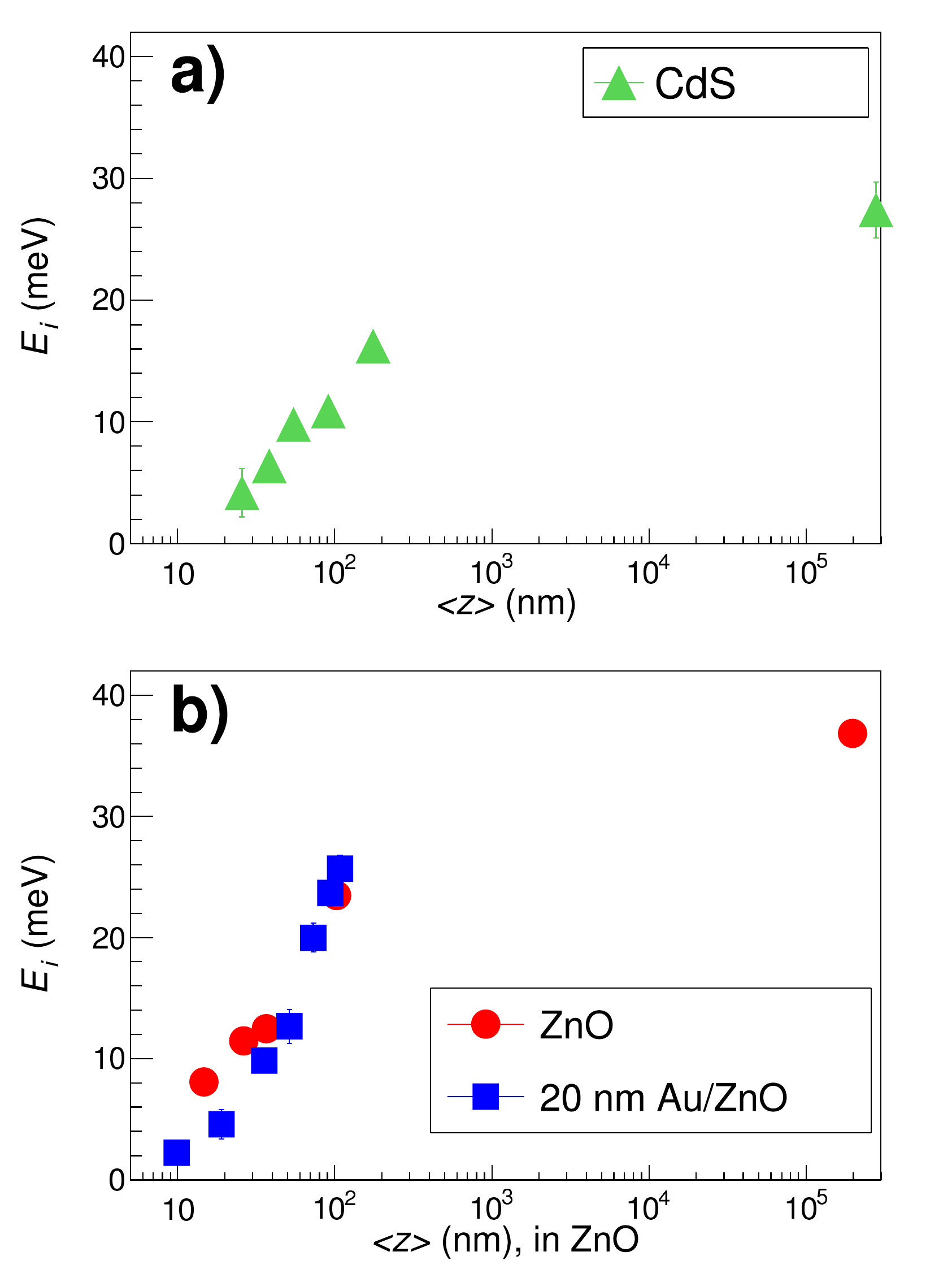}
    \caption{Left: calculated muon stopping ranges in Cu as a function of muon energy. Right: example for
    the range gap in the determination of the activation energy $E_i$ of shallow muonium as a function of
    mean depth $\langle z \rangle$ in CdS and ZnO (Reprinted figure with permission from~\cite{Prokscha:2014DepthDependence}. Copyright 2014 by the American Physical Society.).}
    \label{musr:figure_muonstopping}
\end{figure}

Despite the very successful research program at \ac{SmuS}, there are limitations of \ac{muSR} for the further evolution of the technique towards novel applications and research directions. These limitations, summarised below, are determined by the muon beam and detection characteristics of the \ac{muSR} spectrometers.
\begin{itemize}
\item \textbf{Large sample area}: The sample area  must be $> 4\times 4$~mm$^2$ due to the large phase space of typical muon beams, the limited incoming muon rate and the difficulty in determining whether the muon stopped in the sample or not. For example, this makes it difficult to contribute to the field of some novel quantum materials, which often can be grown with high quality or single crystalline form only in small amounts and small cross sections ($\leq 1\times 1$~mm$^2$).
\item \textbf{Limited muon rate:} As discussed in \autoref{musr:different_muon_source}, the maximum incoming muon rate is limited to about $4\times 10^4$/s due to the pileup problem at continuous muon beams, and the inability of detecting the muon/positron vertex. The accepted muon rate, i.e. those events without pileup, is then limited to only $1.8\times 10^4$/s for a 10-$\mu$s data gate of a typical \ac{muSR} experiment at a continuous beam. This limits the statistical precision of an experiment, compared to pulsed muon sources with $> 10^5$/s accepted muon rate, and it makes it often impossible to investigate very subtle effects like very slow magnetic dynamics or time reversal symmetry breaking in unconventional superconductors.
\item \textbf{Limited data gate:} Due to the continuous beam, uncorrelated muon-positron events produce a small time-independent background in the \ac{muSR} spectra. Therefore, the signal to noise ratio gets progressively worse for longer measurement times limiting the available data gate length. This obstructs the measurement of long relaxation times and reduces the frequency resolution for ultra precise Knight shift measurements. These problems can partially be avoided by the use of the so-called Muons On REquest system (MORE) \cite{Abela1999} which however reduces the count rate.
\item \textbf{Limited pressure:} S$\mu$S has the world leading \ac{muSR} high-pressure facility. Due to the limitations in beam spot size and sample size -- causing problems with background of muons stopping in the pressure cells, or problems with the mechanical stability of large samples in the device for applying uniaxial pressure -- the maximum pressures are 1~GPa and 2.8~GPa for uniaxial and hydrostatic pressures, respectively.
\item \textbf{Inaccessible depth range:} Due to available beam energies at reasonable rates, the depth range of muons is limited to either $< 200$~nm [\ac{LEmu}], or $\geq 100$~$\mu$m by $\mu^+$ with energies $> 4$~MeV, provided by the surface- or decay-muon beams. The currently accessible depth ranges are shown in \autoref{musr:figure_muonstopping}. There is a big inaccessible gap in the so-called sub-surface muon beam region, which is at energies $E_\mu$ between $30$~keV and $\sim 3.0$~MeV, corresponding to muon momenta $p_\mu$ of $2.5$~MeV/$c$ and  $25$~MeV/$c$, respectively.
\end{itemize}

The reason for the range gap is the strong dependence on momentum of sub-surface muon beam intensity $I(p)$ \cite{Pifer:1976ia}
\begin{equation}\label{musr:equation1}
    I(p) \propto p^{3.5}.
\end{equation}
Already at 25~MeV/$c$, the beam intensity drops by a factor of two, making it unfeasible to go to even lower beam momenta/energies for \ac{muSR} applications. Additional complications arise due to the fact that in a standard \ac{muSR} spectrometer, the muons have to cross a 200-$\mu$m-thin scintillator to detect the incoming muon, plus additional thin metal windows in the radiation shield of the cryostat where the sample is located. In these materials, the muons lose tens to hundreds of keV, which requires an incoming beam energy in the MeV range.  
Note, that this momentum dependence of beam intensity does not apply to the low-energy muon beam at $E_\mu < 30$~keV: these muons are generated by the moderation of a highest-intensity surface muon beam with a $I > 2\times 10^8$/s in a cryogenic solid moderator (solid argon or solid neon) to generate \ac{LEmu} with a rate up to $2.3\times 10^4$/s at the moderator target, see Table~\ref{musr:table1}, and \cite{Prokscha:2008zz, Bakule:2004ContPhys,Morenzoni:2004JPCondMatt}.
%
\subsection{New opportunities for \texorpdfstring{\acs{muSR}}{muSR} at \acs{HIMB}: key science drivers}\label{musr:new_opportunities}

\begin{figure}[tb!]
  \centering
     \includegraphics[trim=0 0 0 0,clip,width=0.55\textwidth]{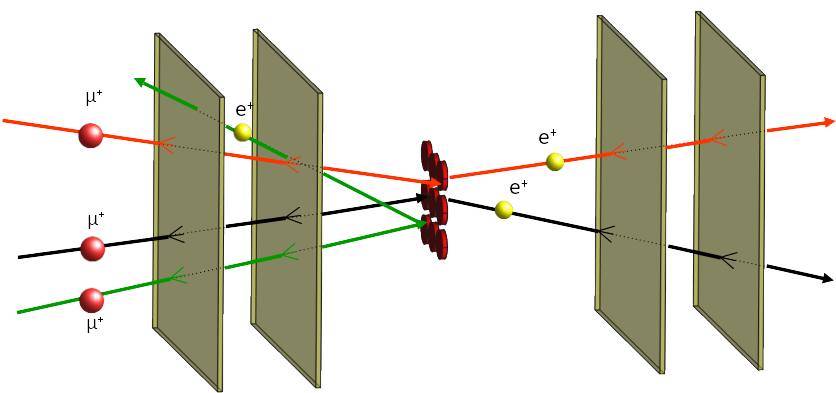}
     \includegraphics[trim=0 0 0 0,clip,width=0.4\textwidth]{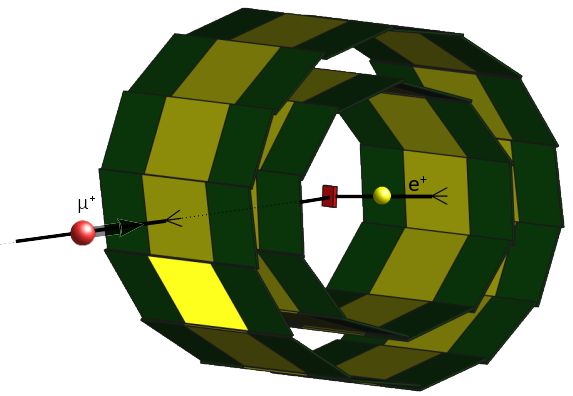}
    \caption{Sketch of the vertex detection scheme for future \ac{muSR} instruments.}
    \label{musr:figure_vertex_detection}
\end{figure}

\ac{HIMB} in combination with the novel vertex reconstruction (see \autoref{musr:figure_vertex_detection}) -- based on the development of a new generation of Si-Pixel detectors (see \autoref{sec:facilities:detector_developments}) -- 
will lead to a "quantum leap" of the \ac{muSR} technique, enabling new research directions/science for \ac{muSR} by much faster, higher statistics and more efficient measurements, extending the depth range of \ac{muSR}, introducing lateral spatial resolution, and by being able to apply external stimuli with unprecedented levels. The envisaged determination of the muon decay vertex with a precision of $\lesssim 1$~mm will
\begin{itemize}
    \item enable to measure small samples of cross section $\leq 1$~mm$^2$, at least twenty times smaller than currently possible. It will also allow measuring multiple samples in parallel, thus, for example, determining the phase diagram as a function of doping in one measurement. This is extremely important for the efficient and fast characterisation of novel quantum materials, which can be often grown in small amounts only. 
    \item enable to avoid the pile-up problem by tracking each muon and its corresponding decay positron. The incoming muon beam rate can be significantly increased by at least one order of magnitude compared to the current limit of $\sim 4\times 10^4$/s. Experiments can be carried out much faster, therefore more efficiently using precious beam time at the two to three times oversubscribed instruments. Due to the significantly enhanced statistics, subtle effects can be detected, for example to measure very slow magnetic fluctuations in quantum materials or detect ultra weak magnetic fields in time reversal symmetry breaking superconductors.
    \item enable to avoid uncorrelated background in the \ac{muSR} spectra which allows extending the accessible data gate by at least a factor two. This opens the possibility to measure significantly longer relaxation times and to increase the frequency resolution for high field measurements (Knight shift). 
    \item enable the mapping of imposed lateral variation with $\sim$~mm resolution of sample parameters under chemical, thermal or pressure gradients. This allows measuring extended portions of phase diagrams in one sample or stimulating gradient induced properties like time reversal symmetry breaking in superconductors by thermal gradients.
    \item enable new opportunities for \ac{muSR} under extreme conditions. With smaller samples of $\sim$~mm size, the application of about ten times larger uniaxial strain (10~GPa) and hydrostatic pressures (30~GPa) will become possible, allowing the exploration of so far inaccessible regions in the phase diagrams of novel materials.
    
    Other external stimuli such as illumination or the study under \ac{DC} or \ac{RF} electromagnetic fields will become possible with at least one order of magnitude higher intensities/fields. The use of small samples will ease the mitigation of sample heating by more efficient cooling, and it will allow building very compact and mechanically stable and reliable setups. Possible applications are, e.g., novel investigations of charge carrier lifetimes and dynamics in semiconductors by photo-generated carriers, vortex dynamics in superconductors under \ac{DC} currents, skyrmion dynamics under \ac{DC} or \ac{RF} fields, or quantum information processing using molecular spins or skyrmions in pump-probe experiments (see \autoref{musr:new_opportunities:pump-probe}). 
\end{itemize}
Besides the exciting new opportunities offered by the combination of \ac{HIMB} and vertex reconstruction, \ac{HIMB} will enable i) novel applications in energy materials by closing the range gap between 200~nm and 200~$\mu$m (\autoref{musr:new_opportunities:devices} and \autoref{musr:table2}), ii) new pump-probe experiments at continuous muon beams (\autoref{musr:new_opportunities:pump-probe}), iii) low-energy muon applications with an order of magnitude intensity gain (\autoref{musr:new_opportunities:low-energy-muons}), and iv) significantly improved applications of negative muons in non-destructive, depth-selective elemental analysis studies (\autoref{musr:new_opportunities:elemental-analysis}).

\subsubsection{Novel quantum materials}\label{musr:novel_quantum_materials}

Core activities in condensed matter physics and materials science are the search for, understanding and tuning of phases of quantum materials. \ac{muSR} as a powerful local probe technique can provide unique insights in the properties of unconventional magnetic and superconducting materials. A large variety of new materials, including those with topologically non-trivial band structures, offer the possibility of realising interesting phenomena, where \ac{muSR} is a very sensitive experimental technique for their characterisation.

The possibility of measuring ten to hundred times smaller samples, at ten to hundred times higher rates, will have a huge impact on the use of \ac{muSR} for the investigation of novel quantum materials. In the following, we give a
 short overview of possible \ac{muSR} applications with \ac{HIMB}, which currently can only be carried out to a limited extent or are even unfeasible.
\begin{itemize}
    \item Higher uniaxial pressure up to 10~GPa due to smaller sample size, which means higher mechanical stability of sample and setup and better homogeneity of samples as well as higher hydrostatic pressure of up to 30 GPa by using high-pressure anvil cells with small sample volumes.
    \begin{itemize}
        \item Realisation and detection of quantum spin liquids by systematic and symmetry-lowering control of the magnetic interactions by uniaxial strain in kagome, triangular and Kitaev systems.
        \item Tuning of flat-band superconductivity in kagome-lattice systems, for example in LaRu$_3$Si$_2$, XV$_3$Sb$_5$ (X=K, Rb, Cs) and investigation of the interplay of \ac{TRSB}, \ac{CDW} order and superconductivity in these systems.
        \item Investigation of 3D \ac{CDW} order beyond 1~GPa uniaxial strain in YBa$_2$Cu$_3$O$_7$, where only small high quality crystals are available.
        \item Hydrostatic and uniaxial pressure tuning of the competition of \ac{CDW} and superconductivity, e.g. in NbSe$_2$.
        \item Uniaxial pressure tuning of stripe order and pair density wave superconductivity in La$_{2-x}$Ba$_x$CuO$_4$.
        \item Tuning of semiconductor to metal transitions in magnetic layered transition metal dichalgoneides (TMC), e.g. 2H-MoTe$_2$.
        \item Pressure tuning of superconducting Weyl semi-metal 1T-MoTe$_2$.
        \item Stabilisation and control of topological fermions by tuning of magnetic competition, e.g. in the ferromagnetic Weyl semi-metal Co$_3$Sn$_2$S$_2$.
        \item Novel low-temperature topological magnetic states in TbMn$_6$Sn$_6$: pushing topological properties towards room temperature.
        \item Strain-induced topological magnon transition in ferromagnetic kagome systems.
        \item Tuning of mechanically soft quantum magnets, e.g. CsFeCl$_3$ and TlCuCl$_3$ which is only possible with small specimens.
        \item Lifting degeneracy of multi-component unconventional superconductivity, e.g. study of \ac{TRSB} in Sr$_2$RuO$_4$ under strain.
        \item Elucidating the source of unconventional high temperature magnetism in Sr$_2$RuO$_4$ using high pressure.
        \item Local probe investigation of the magnetic and superconducting phase diagram of FeSe under hydrostatic pressures larger than 2 GPa.
        \item Strain-tuning of nematicity in Fe-based superconductors.
        \item Investigation of two-dome superconductivity in the temperature-pressure phase diagram of K$_{0.8}$Fe$_{1.7}$Se$_2$.
        \item Black phosphorus as an emerging 2D material: Superconducting phases at pressures up to 20 GPa.
        \item Pressure induced superconductivity in elemental Fe.
    \end{itemize}
    \item Small samples allow application of stronger external stimuli (higher current densities, electric fields or luminosity).
    \begin{itemize}
        \item Current induced tuning of magnetic properties of iridates.
        \item Current induced modification of charge  order in the quantum chain systems, e.g. in (La,Sr,Ca)$_{14}$Cu$_{24}$O$_{41}$.
        \item Microwave / RF induced Skyrmion motion.
        \item Current induced vortex lattice dynamics.
        \item Manipulation of spin states in molecular magnets using light and microwave irradiation.
        \item Spectroscopic investigation of muonium states in semiconductors and insulators.
    \end{itemize}
    \item Simultaneous measurements of multiple and high quality samples.
    \begin{itemize}
        \item Much more efficient exploration of phase diagrams (transition temperatures and order parameters) of unconventional superconductors and novel magnets, while thermal and magnetic history are identical for all samples.
        \item Precise measurements of Knight shifts and line shapes in superconductors, spin liquids and topological materials, using a reference sample (e.g. Ag) for {\em in-situ} calibration of magnetic field and detection of field drifts.
    \end{itemize}
    \item High statistics measurements and low positron background (allows long data gates of $20$~$\mu$s).
    \begin{itemize}
        \item Small changes in relaxation rates and internal magnetic fields to investigate \ac{TRSB} in superconductors.
        \item Subtle features in \ac{muSR} depolarisation function at long times: quantum entanglement and quantum decoherence in F-Mu-F systems.
        \item Measurement of decoherence times of spin states in single molecule/ion magnets.
        \item Volume fraction of spontaneous magnetic fields in \ac{TRSB} in unconventional superconductors.
        \item Thermal gradients as a source of \ac{TRSB} fields in unconventional superconductors.
    \end{itemize}
     \item Laterally resolved \ac{muSR} measurements.
    \begin{itemize}
        \item Thermal gradient effects on materials, e.g. spin Seeback effect.
        \item Homogeneity of phases in magnetic and superconducting sample.
        \item Measurement of the effect of light irradiation intensity on properties of materials in a single shot, e.g. by irradiating with a gradient of light intensity.
    \end{itemize}
\end{itemize}

This list -- although by far not complete -- demonstrates the versatile future opportunities for \ac{muSR} with \ac{HIMB} combined with vertex reconstruction, allowing a significantly increased access to sample phase space of quantum materials by \ac{muSR} and thus enabling the exploration of new research avenues by \ac{muSR}.

\subsubsection{Energy materials}\label{musr:new_opportunities:devices}

In recent years, \ac{LEmuSR} has been successfully extended to the study of semiconductor interfaces to provide valuable information -- with unprecedented depth resolution of a few nanometres -- about charge carrier profiles/gradients and their manipulation by illumination, as well as defect regions in technologically relevant Si and SiC and its oxide interfaces, and defect regions in \acf{CIGS} based solar cell heterostructures \cite{Alberto:2018PhysRevMat,Curado:2020ApplMatTod,Woerle:2019PRB,Woerle:2020PRAppl,Prokscha:2020PRAppl,woerle:low-energy2020}. The goal of these studies is to obtain a better microscopic characterisation of device interfaces with a depth resolution superior to standard characterisation techniques, in order to be able to adapt the growth conditions and achieve optimum device efficiency.

Another important extension of \ac{muSR} demonstrated its capability to measure the Li$^+$ ion diffusion coefficient, $D$, in battery materials \cite{sugiyama:li-ion2015,sugiyama:spin2019}. In contrast, conventional \ac{NMR} is unable to provide useful information due to the dominant effect of localised magnetic moments on the \ac{NMR} spin-lattice relaxation. With the knowledge of $D$, other crucial parameters relevant for the performance of battery materials can be derived, such as the reactive surface area, diffusion pathway, and density of mobile ions.

So far, these \ac{muSR} studies are either limited to thin layers/heterostructures with a thickness $< 200$~nm, or to bulk material. In many cases, semiconductor and solar cell devices as well as real life battery devices have thicknesses in the range of hundreds of nanometres to tens of $\mu$m. Therefore, it would be highly appealing to use such a sensitive local probe technique as \ac{muSR} to study interface and bulk regions in real devices in-operando, to gain unparalleled insights into their function and to enable the development of better devices in future. With the expected intensities of low momentum sub-surface muons at \ac{HIMB}, as shown in \autoref{musr:table2}, the study of buried layers in so far inaccessible depth ranges will become possible, allowing the extension of \ac{muSR} to become a new in-operando technique for contributing to the solution of important societal problems in energy research. 
\begin{table}[ht]
    \caption{Expected sub-surface muon rates at \ac{HIMB} for \ac{muSR} applications, where we assume
    a surface muon rate of $10^8$/s being available in the new $\mu$H3 beamline, where
    the GPS and FLAME instruments are located. This rate is expected downstream of the $\mu$H3 spin-rotator. Beam rates calculated using \eqref{musr:equation1}. Such low beam energies will require
    the development of new muon entrance detectors.}
    \label{musr:table2}
    \begin{center}
    \begin{tabular}{lrrr}
    \toprule
    $\mu^+$ momentum  &  $\mu^+$ energy & beam rate & mean implantation depth \\
    \midrule
    2.5 MeV/$c$ & 30 keV      & $1.7\times 10^4$/s & $\sim 180$ nm\\
    3.2 MeV/$c$ & 50 keV      & $4.0\times 10^4$/s & $\sim 350$ nm\\  
    3.8 MeV/$c$ & 70 keV      & $7.4\times 10^4$/s & $\sim 500$ nm\\  
    4.6 MeV/$c$ & 100 keV     & $1.5\times 10^5$/s & $\sim 700$ nm\\  
    \bottomrule
    \end{tabular}
    \end{center}
\end{table}

\subsubsection{Pulsed beam and pump-probe}\label{musr:new_opportunities:pump-probe}
\begin{figure}[tb!]
  \centering
    \includegraphics[trim=0 -10mm 0 0,clip,width=0.49\textwidth]{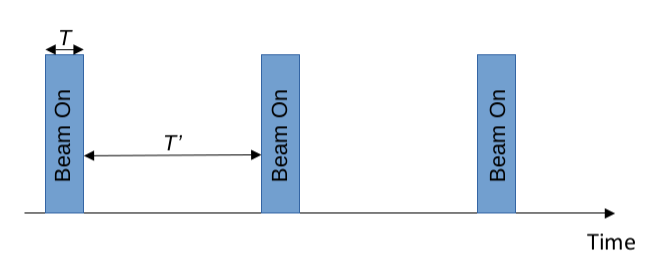}
    \includegraphics[trim=0 0 0 0,clip,width=0.49\textwidth]{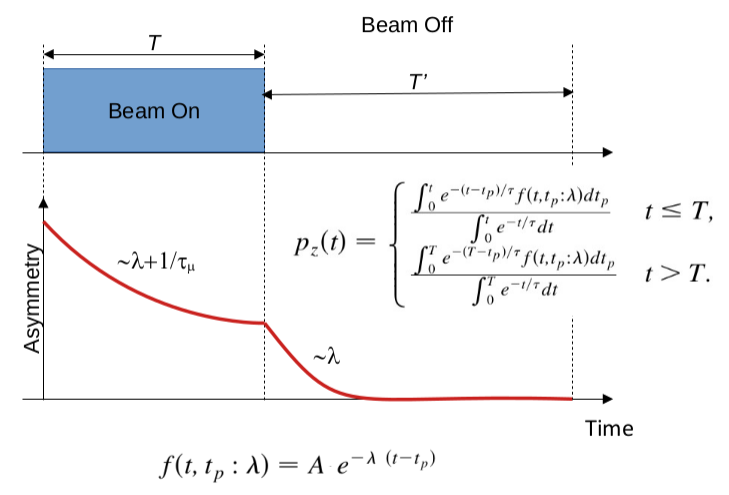}
    \caption{Left: pulsing scheme for chopping the surface muon beam. A possible timing scheme, adjusted to the muon life time, is T = 1~$\mu$s, and T$^\prime$ = 19~$\mu$s, resulting in a 50~kHz repetition rate. During the pulse of length T, on average about 100 $\mu^+$ will stop in the \ac{muSR} spectrometer. Right: expected polarisation functions $p_{z}(t)$~\cite{Salman:2006PRL}.}
    \label{musr:figure_pulsing}
\end{figure}

%
%
The new $\mu$H3 beamline at \ac{HIMB} is expected to deliver a surface muon rate of $>10^8$/s downstream of the $\mu$H3 spin-rotator (see also \autoref{musr:table2}). The spin-rotator will significantly cut the initially large phase space, allowing for pulsing/chopping of the continuous surface muon beam. A possible pulse scheme is sketched in \autoref{musr:figure_pulsing}.  
\begin{figure}[tb!]
  \centering
    \includegraphics[trim=0 0 0 0,clip,width=0.49\textwidth]{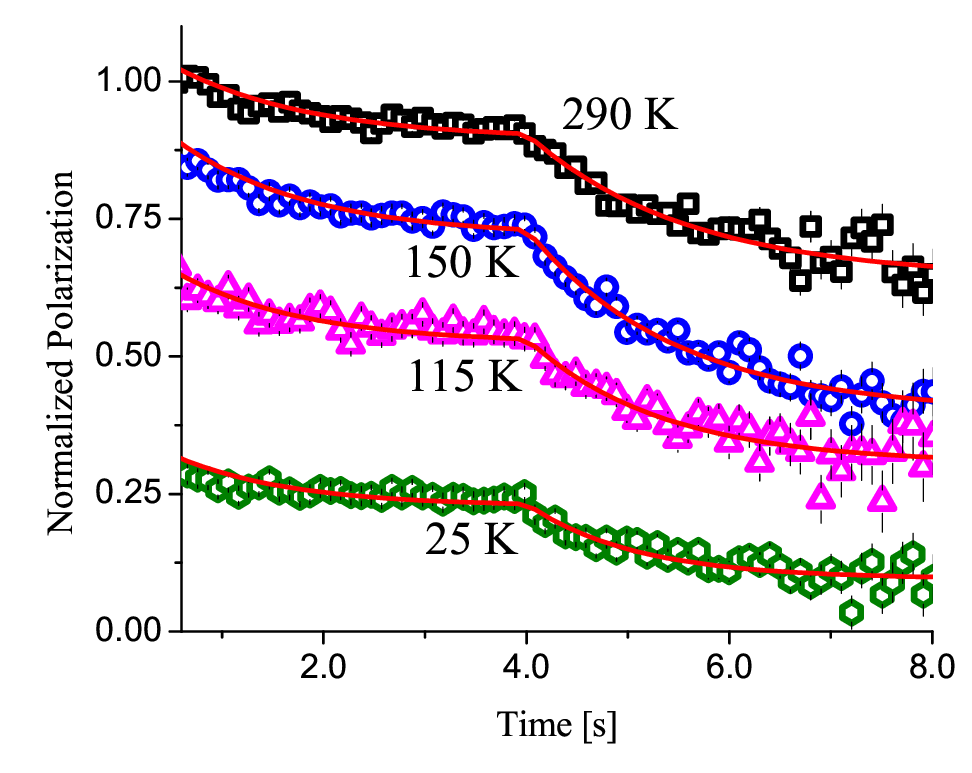}
    \includegraphics[trim=0 -40mm 0 0,clip,width=0.49\textwidth]{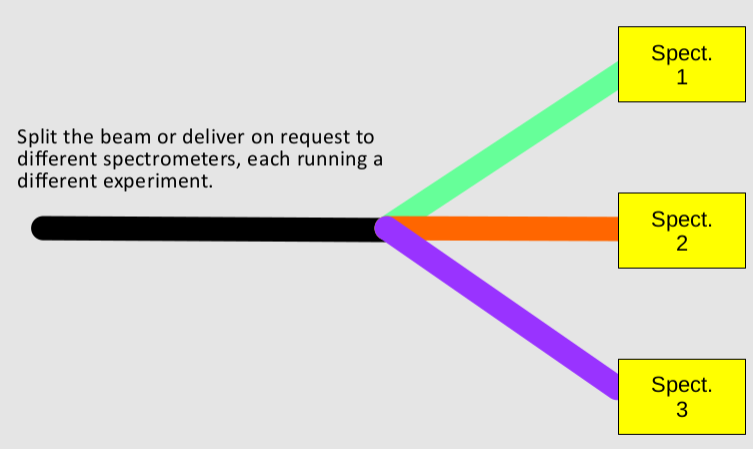}
    \caption{Left: Example for the measurement of spin-lattice relaxation in pulsed $\beta$-\acs{NMR} using radioactive $^8$Li$^+$ implanted in SrTiO$_3$ (Reprinted figure with permission from~\cite{Salman:2006PRL}. Copyright 2006 by the American Physical Society.).
    Right: Scheme of multiple instruments on the same muon beamline.}
    \label{musr:figure_pulsing_1}
\end{figure}
During the pulse of a length of T = 1~$\mu$s, about 100 $\mu^+$ will enter the \ac{muSR} spectrometer on average at an incoming rate of $10^8$/s. Between the pulses, the measurement is free of accidental background, similar to a pulsed muon beam facility. This technique is already in use at the $\beta$-\acs{NMR} facility at TRIUMF, where a beam of radioactive $^8$Li$^+$ ions with tunable energies between 1 and 30~keV is used for depth-resolved condensed matter physics applications in thin films or near-surface regions, see \autoref{musr:figure_pulsing_1}, similar to \ac{LEmuSR}. Compared to \ac{muSR}, $^8$Li$^+$ $\beta$-\acs{NMR} is sensitive to much longer times scales due to its orders of magnitude longer lifetime of 1.2~s. For a much more efficient use of beamtime in this scheme, the beam could be split to serve different spectrometers at the same time, as shown on the right side of \autoref{musr:figure_pulsing_1}.

Such a scheme of pulsed muon beams, with an average muon rate of $5\times 10^6$/s, will enable new research directions at a continuous muon beam facility:
\begin{itemize}
    \item The possibility of measuring the effect of external stimuli synchronised with the muon pulse will allow novel \ac{muSR} applications in quantum information processing with molecular spin systems, magnetic skyrmions etc. For example, quantum states can be prepared in molecular magnet systems and the decoherence time of these states can be measured by \ac{muSR}. Magnetic skyrmions have the potential to be used as information carriers in future spintronics devices, susceptible to manipulation by very small electric currents or magnetic/electric field pulses. In a magnetic skyrmion system, collective magnon modes can be resonantly stimulated by GHz microwave excitation -- \ac{FMR} -- and probed by its effect on the \ac{muSR} polarisation and the local internal field at the muon site. This combination of \ac{FMR}-\ac{muSR} has the potential to examine magnon modes while accessing the internal fields and fluctuations by \ac{muSR}.
    \item In general, such a pump-probe scheme can be applied to many systems to gain information about the dynamics in the MHz to GHz range by using light or microwave stimuli, and probe with the muon as a very sensitive local magnetic field sensor. For example, excess carrier lifetimes in semiconductors can be measured by photoexcited \ac{muSR} \cite{yokoyama:photoexcited2017}.
    \item With this kind of a pulsed beam -- similar to a pulsed muon beam facility but with much better time resolution -- a long data gate of 20~$\mu$s width can be employed due to the very low positron background to measure very small relaxation rates, low muon spin precession frequencies and better resolved split frequencies (e.g. in semiconductors with shallow muonium states). High-statistics measurements would become possible -- without vertex reconstruction -- to measure very small changes of \ac{muSR} parameters (see also \autoref{musr:novel_quantum_materials}).
    \item In this pulsing scheme with an effective muon rate in the MHz range, a \ac{muSR} spectrum can be recorded with sufficient statistics within seconds. This will allow to extend \ac{muSR} to a new field of applications: the study of transient states with seconds lifetimes. Examples are slow magnetic dynamics, slow chemical reactions and diffusion processes, and vortex creep in superconductors. 
\end{itemize}

\subsubsection{Low-energy muons}\label{musr:new_opportunities:low-energy-muons}

The current techniques for generating low-energy $\mu^+$ with energies in the eV- and keV- range at continuous muon beam facilities are based on the moderation of a 4-MeV surface muon beam in either a thin film of a solid rare gas (argon or neon) \cite{Prokscha:2008zz,Bakule:2004ContPhys} or in low-pressure He gas placed in superimposed electric and magnetic fields (muCool, see \autoref{sec:facilities:mu_cool} and~\cite{Belosevic:2018fnj}). The conversion efficiencies to the eV range are typically small, between $10^{-5}$ and $10^{-4}$ per incoming $\mu^+$. This requires very high surface muon beam intensities of $> 10^8$/s to be able to generate $10^3$/s -- $10^4$/s low-energy $\mu^+$. This order of magnitude of \ac{LEmu} rates is at the lower limit for practical \ac{LEmuSR} applications. Obviously, an increase of the 4-MeV surface muon beam rate on the moderator target by one or even two orders of magnitude would have an enormous impact on the entire research program with low-energy muons. 

The PSI \ac{LEM} facility is the leading facility for low-energy $\mu^+$ applications, delivering up to 23~kHz moderated muons. However, taking into account beam transport and detection efficiencies of the \ac{LEM} facility, the maximum recorded \ac{LEmuSR} event rate is less than 2.7~kHz at the moment, far below the pile-up limit at a continuous muon beam.  This reduces the available statistics of a typical \ac{LEmuSR} measurement to $10^6$ to $10^7$ recorded muon decays in the \ac{muSR} histograms, and thus the investigation of small \ac{muSR} signals or subtle changes in depolarisation rates and local magnetic fields is not feasible. An additional limitation is the large \ac{LEmu} beam spot with a \ac{RMS} width of 6 -- 7~mm. This requires a minimum sample cross section of $10\times 10$~mm$^2$. For optimum usage of the beam, samples with sizes of $25\times 25$~mm$^2$ are required. Using a collimator in front of the \ac{LEmuSR} start detector with 10~mm or 15~mm diameter reduces the beam spot size to \ac{RMS} values of 3 -- 4~mm, while losing 45 -- 60\% of beam rate. 
\begin{itemize}
    \item An increase of \ac{LEmu} rate by one order of magnitude obviously leads to significantly improved \ac{LEmuSR} experiments with much better statistics and faster measurements. Small samples with $5\times 5$~mm$^2$ cross section or even smaller will become possible using beam collimation, while compensating the beam loss in the collimator by the increased initial rate after moderation. Such small samples will enable new \ac{LEmuSR} applications on novel, small size quantum materials and device structures. In addition, experiments under external stimuli such as illumination, electric fields or electric currents can be carried out with much higher light intensities and current densities without excessive heating, therefore expanding the parameter range of \ac{LEmuSR} to new regions.
    Current experiments to directly measure the local magnetic field generated by current-induced spin accumulation at the surface/interface (spin Hall effect) are limited by sample heating in the large cross section samples \cite{aqeel:2017probing}. In samples with 5 times smaller area, the current densities can be increased by a factor 5 (in-plane current) or even a factor 25 (out-of-plane current), therefore bringing the local magnetic field generated by spin accumulation closer to the detection limit of \ac{LEmuSR}. 
    %
    Furthermore, the quality and homogeneity of electrical contacts can be much better controlled with small sample areas $\ll 25\times 25$~mm$^2$, which is the ideal sample size at \ac{LEM} at the moment. This is important for experiments on new device structures, such as reversible spin storage in metal/metal-oxide/fullerene heterojunctions \cite{moorsom:reversible2020}, but also for future applications to study charge carrier dynamics and accumulation/depletion in the presence of defects at technologically important semiconductor interfaces \cite{woerle:low-energy2020} and solar cell devices \cite{Alberto:2018PhysRevMat,Curado:2020ApplMatTod}. For the latter, \ac{LEmuSR} is used as a local probe technique to characterise these interfaces with a resolution of a few nanometre with the overall goal to obtain important insights in optimising the growth conditions of these devices.
    \item
    With a significant increase of \ac{LEmu} rate, about twenty times lower temperatures of $\sim 100$~mK will become feasible at the \ac{LEM} facility. At the moment, the lowest temperature of $\sim 2.3$~K is achieved with a home-made $^4$He cryostat, where the 6-cm-diameter opening in the radiation shield of the cryostat\footnote{\ac{LEmu} would stop in any of the typically used $\mu$m-thick radiation shield windows.} leads to a thermal load of $\sim 200$~mW on the sample. A \ac{DR}, needed 
    to reach $\leq 100$~mK, has a cooling power of only $\sim \mu$W. This requires a cooling of all surfaces ``seen" by the sample to 4~K. Since one needs an opening of several cm diameter in the radiation shield, 
    all the beam pipes with direct view on the sample have to be cooled with liquid $^4$He. This means, that the
    existing \ac{LEmu} beam at \ac{LEM} needs to be extended by a section with another bend, where from the bend towards the \ac{DR} the entire beam section will be cooled. This extension of the beamline will cause a significant loss of at least 50\% in transmission, which can be compensated by an increase of the initial \ac{LEmu} rate. The possibility of achieving twenty times lower temperatures down to 100~mK or less  will open new fascinating applications of \ac{LEmuSR} in various fields of condensed matter physics, for example:
    \begin{itemize}
        \item Direct determination of the magnetic penetration depth and coherence length in low temperature superconductors, e.g.~the type-I superconductor Al with large $\kappa$ and a T$_{c}$ of 1.6 K.
        \item Magnetic and superconducting properties of 2D electron gases at interfaces, where coexistence of magnetic and superconducting order at LaAlO$_3$/SrTiO$_3$ interfaces (T$_{c} \sim 200$~mK) was observed \cite{li:coexistence2011,bert:direct2011}. \ac{LEmuSR} can measure the vortex broadening of the magnetic field distribution in the superconducting state to determine the magnetic penetration depth as a function of temperature. This allows determining the symmetry of the superconducting order parameter.
        \item Direct probing of \ac{TRSB} magnetic fields at the surface of unconventional superconductors with low T$_{c}$, e.g.~SrRuO$_4$ with T$_{c} \sim 1.5$~K \cite{luke:time-reversal1998}.
        \item Extension of \ac{LEmuSR} to determine the magnetic penetration depth in unconventional heavy-fermion superconductors. For example, UTe$_2$ with T$_{c} \sim 1.6$~K is showing signs of spin-triplet pairing, \ac{TRSB}, multiple order parameters and chiral surface states \cite{jiao:chiral2020}.
        \item Proximity effects with low-temperature superconductors.
        \item Low-temperature magnetism: frustrated magnetic systems, spin-ice and quantum spin liquid systems, and its properties near surfaces or interfaces.
    \end{itemize}
    \item A beam of \ac{LEmu} can be used to efficiently generate muonium in vacuum, which is essential for performing high-precision muonium spectroscopy to provide stringent tests of bound state \ac{QED} and to determine fundamental parameters of the Standard Model (such as muon mass, muon magnetic moment, etc.), as described in \autoref{sec:muonium}. At the \ac{LEM} facility, the efficient generation of thermal muonium in vacuum has been demonstrated \cite{Antognini:2011ei, Khaw:2016ofi}, which will enable a new muonium 1s-2s experiment with up to hundred times improved precision \cite{Crivelli:2018vfe}.
    Recently, the efficient generation of a muonium beam in the metastable 2s state \cite{janka:intense2020} has been achieved in the \ac{LEM} facility, which enabled a ten times more precise measurement of the 2s-Lamb shift in muonium \cite{Ohayon:2021qof}. It is obvious, that the precision of muonium spectroscopy would tremendously benefit from an order of magnitude increase in \ac{LEmu} rate.
    \item A very exciting future opportunity -- also pursued at the MUSE facility at J-PARC \cite{miyake:ultra2013} -- is the generation of muon microbeams by re-acceleration of the small phase space of the moderated muon beams to hundreds of keV or even MeV. The moderated muons from muCool might be better suited for this purpose compared to the moderation in a solid argon or neon layer due to the smaller phase space of muCool. However, the extraction of muons from the muCool apparatus and routine operation at a large phase space surface muon beam still need to be demonstrated. 
    A re-accelerated beam of moderated muons would have a much smaller phase space compared to the existing sub-surface and surface muon beams. This would allow focusing the beam on spot sizes of $< 1$~mm$^2$, tunable implantation depths from hundreds of nanometre to the $\mu$m range, and the realisation of a transmission muon microscope \cite{Yamazaki:2020muontransmission} -- a novel extension of muon applications to material science. Obviously, this is a further route to completely new muon applications offered by \ac{HIMB}.
\end{itemize}
An upgrade of the $\mu$E4 beamline is currently being studied, where the use of solenoids instead of quadrupole magnets can yield an improved transmission and focusing on the moderator target. With this upgrade, an increase of about a factor of two of muons on moderator \cite{Zhou:2021simulation} can be expected. However, for full exploitation of the capabilities of \ac{LEmu} experiments, \ac{LEmu} rates of $> 10^5$/s are highly desirable.

\subsubsection{Elemental analysis and \texorpdfstring{\acs{muSR}}{muSR}}\label{musr:new_opportunities:elemental-analysis}

The use of negative muons for non-destructive, elemental analysis by the detection of element-specific muonic X-rays has been tested some decades ago \cite{kohler:application1981,daniel:application1984}, while its routine use has been initiated since 2010 at the pulsed muon sources at J-PARC \cite{ninomiya:development2010} and ISIS \cite{hillier:probing2016}. A main drawback of these studies at pulsed muon sources to date is the use of \ac{HPGe} detectors, which have long dead times of 5 - 10~$\mu$s. This means, that per muon pulse of width 50 - 100~ns, on average only one muonic X-ray can be detected, resulting in an event rate determined by the repetition rate of the muon facility (50~Hz at ISIS, and 25~Hz at J-PARC). This pile-up problem does not exist at a continuous muon beam with a rate of even 100~kHz, where the average distance in time between two muons is 10~$\mu$s. A feasibility study at PSI in 2018 has demonstrated (for the setup, see \autoref{fig:apv:setup}, that at the existing $\pi$E1 beamline with $\mu^-$ rates of about 30~kHz and beam momentum of $\sim 30$~MeV/$c$, the statistics can be increased by more than three orders of magnitude compared to the pulsed muon beam facilities. In addition, active collimation is possible at a continuous muon beam leading to a significant reduction of background. This enables faster measurements with much better signal-to-background ratio, higher sensitivity to low-concentration elements, and the resolution of fine structure splittings and a much better separation of adjacent muonic X-ray lines from different elements.

\ac{HIMB} has the potential to lead to a significant boost of $\mu^-$ applications in elemental analysis:
\begin{itemize}
    \item With the higher rates, even faster measurements will be possible, and in-operando measurements of devices such as Li-batteries will become feasible, where the diffusion of Li ions through charging/discharging cycles can be monitored as a function of depth.
    \item Lower beam momenta $< 10$~MeV/$c$ will become feasible, extending the range of muon stopping depths from $\sim 1~\mu$m to $> 1$~mm.
    \item Due to the high rates, it will become possible to perform tomography, a very exciting novel extension to image the distribution of elements inside bulky samples and artefacts.
    \item With \ac{HIMB}, $\mu^-$ could potentially be used in ``parasitic mode": the extraction of muons from the new production target H is made by solenoids, transporting $\mu^+$ and $\mu^-$ in the same way. At the first bending magnet, a "parasitic beamline" can be installed to transport $\mu^-$ -- deflected by the opposite angle with respect to the $\mu^+$ -- into a separate beamline, where the momentum can be adjusted by a second bending magnet. In this way, a dedicated experimental area for $\mu^-$ research would become available without the need to share the beamtime with one of the overbooked $\mu^+$SR instruments. 
\end{itemize}

With higher intensity $\mu^-$ beams, a new kind of $\mu^-$SR experiments will become possible: "X-ray triggered $\mu^-$SR" with routing of histograms for individual elements, i.e. "element-specific" $\mu^-$SR. This will open new possibilities to measure element specific diffusion rates and local magnetic fields at element specific lattice sites. 

%
%
%
%

%
%


\section{The \acs{HIMB} project \& related beam and detector developments}\label{sec:facilities}

\subsection{The \acs{HIMB} project}\label{sec:scenarios}
\begin{figure}
\center
\includegraphics[width=0.9\textwidth]{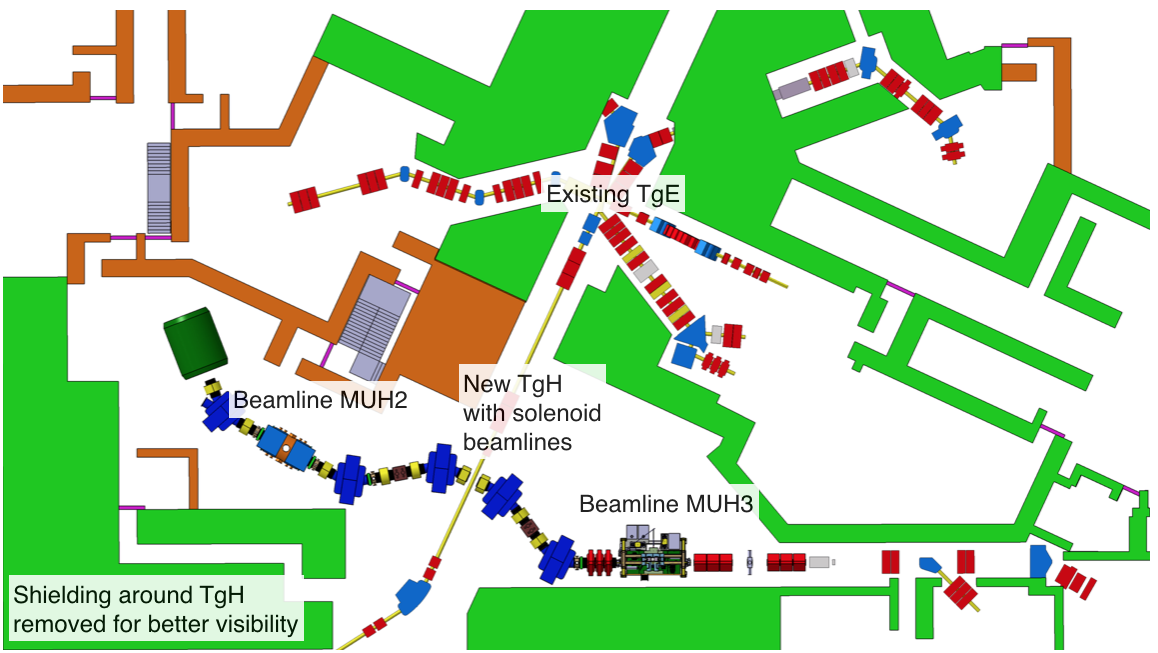}
\caption{Layout of the new solenoid beamlines in the experimental hall of PSI. The new 20-mm long target TgH is located at the same place as the previous 5-mm long target TgM. The two new beamlines MUH2 and MUH3 replace the two existing beamlines $\pi$M1 and $\pi$M3. }
\label{fig:himb_layout}
\end{figure}


The \ac{HIMB} project aims at delivering $\mathcal{O}(10^{10})$ surface muons per second (positive muons stemming from pion decays at rest close to the surface of the production target \cite{Pifer:1976ia}) to experiments. To do so, the existing TgM station located in the experimental hall at PSI on the high-energy proton beamline of the HIPA accelerator and its connected beamlines $\pi$M1 and $\pi$M3 will be completely rebuilt. \autoref{fig:himb_layout} shows how the layout of the new target and beamlines will look like. While the existing graphite target features a length of \SI{5}{mm}, this will be increased to \SI{20}{mm} in order to achieve the required rate. Additionally, the target is slanted at 10~degrees with respect to the proton beam such that the overall length of the graphite slab can be increased to \SI{100}{mm}, while keeping the effective length as seen by the proton beam at \SI{20}{mm}. This optimisation of the target geometry increases the rate of emitted surface muons by about a factor of two.

\begin{figure}
\center
\includegraphics[width=0.7\textwidth]{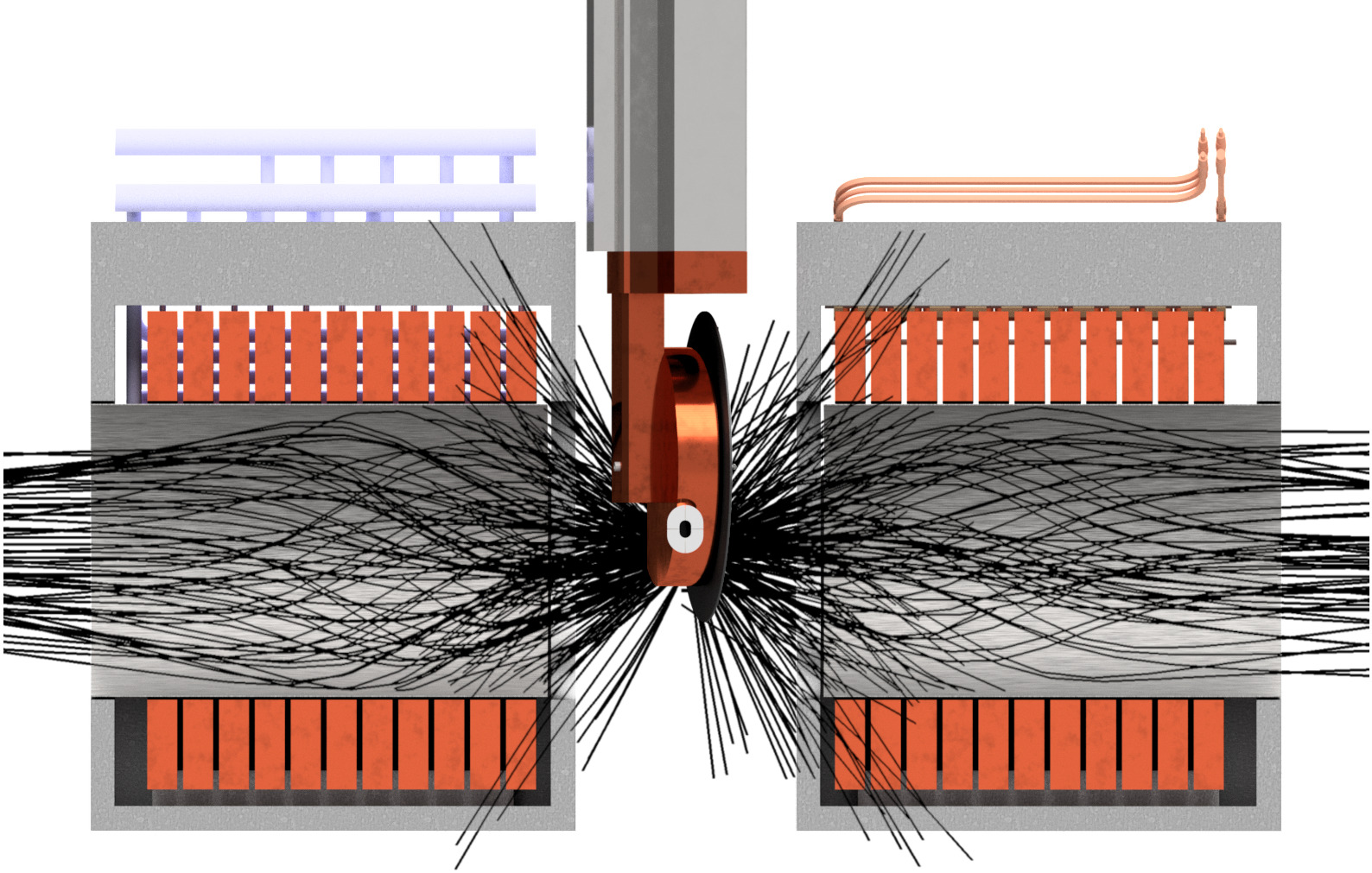}
\caption{Cross-sectional view of the two capture solenoids together with the new target TgH in the foreground. The black lines show trajectories of surface muons emitted from the target.}
\label{fig:muon_capture}
\end{figure}

The key to reach the intensity goal is to use capture solenoids with a 500-mm aperture located only \SI{250}{mm} away from the target. \autoref{fig:muon_capture} shows a cross-section through the two capture solenoids overlaid by the trajectories of surface muons emitted from the target. Already from this picture alone it is clear that a large fraction of the emitted surface muons can be captured by this arrangement and transported further. The capture solenoids will be radiation-hard and are based on the design of two existing solenoids built for the $\mu$E4 beamline at PSI \cite{Prokscha:2008zz}. In order to transport the large phase space accepted by the capture solenoids the rest of the beamline also relies on solenoids for focusing combined with large-aperture dipoles. Overall a combined capture and transport efficiency of around 10\% is achieved compared to the typical efficiencies of much less than 1\% found in the other secondary beamlines of PSI.

\ac{HIMB} is embedded in the larger IMPACT project that combines two major upgrades of the PSI facilities: \ac{HIMB} and TATTOOS \cite{TATTOOS}. The IMPACT project team consisting of people from the PSI divisions NUM (Research with Neutrons and Muons), NES (Nuclear Energy and Safety), BIO (Biology and Chemistry), GFA (Large Research Facilities), and LOG (Logistics) is currently preparing its Conceptual Design Report to be completed by the end of 2021 with the aim of having the project accepted to the 2023 Swiss Roadmap for Research Infrastructures. If accepted, its implementation would occur during the period 2025-2028 with regular user operations starting in 2029.

Based on the transported phase space, two scenarios for the beam spot at the final focus were formulated for the \ac{HIMB} science case workshop depending on the final divergence achievable at the end of the beamline.

\ac{HIMB}-5:
\begin{itemize}
\item $10^{10}$~$\mu^+$/s at 28~MeV/$c$ and  $\sim$10\% momentum-bite (FWHM)
\item beam spot $\sigma_{x,y} \sim 50$~mm
\end{itemize}

\ac{HIMB}-3:
\begin{itemize}
\item $10^{10}$~$\mu^+$/s at 28~MeV/$c$ and  $\sim$10\% momentum-bite (FWHM)
\item beam spot $\sigma_{x,y} \sim 30$~mm
\end{itemize}

These scenarios and especially the achievable rate of $10^{10}$~$\mu^+$/s formed the basis for the different science cases developed for the workshop and described in this document.

\begin{figure}
\center
\includegraphics[width=0.8\textwidth]{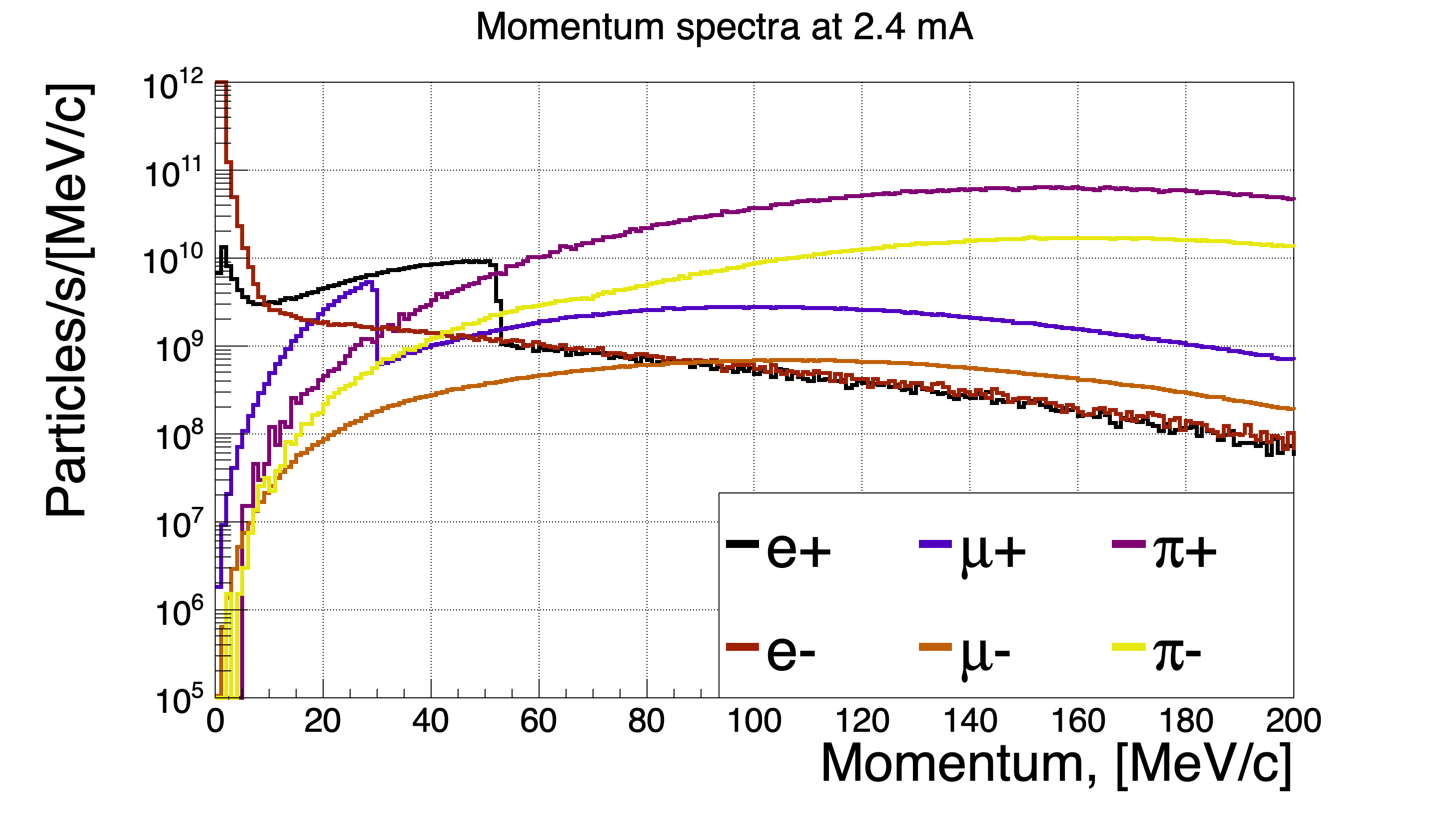}
\caption{Momentum spectra of positive and negative particles registered at the entrance of the capture solenoid.}
\label{fig:momentum_spectrum}
\end{figure}

Of course the \ac{HIMB} beamlines do not only transport surface muons, but also other particles such as negative muons, electrons, positrons, and pions. Figure~\ref{fig:momentum_spectrum} shows the momentum spectrum of all these particles at the entrance to the capture solenoid. The beamlines are designed for a good transport efficiency up to about 40~MeV/$c$ with the dipoles capable of reaching 80~MeV/$c$. This allows, e.g., the use of pion beams for calibration purposes albeit not at the highest intensities.

At 28~MeV/$c$ the rates at the end of the beamline are expected to be $2\times 10^8$/s for negative muons, $2\times 10^{10}$/s for positrons, and $7\times 10^{9}$/s for electrons.

One interesting suggestion raised during the workshop was the notion to study the option of creating a parasitic $\mu^-$ beam by splitting at the first dipole of the channel the positive and negative muons and thus creating the possibility to simultaneously run experiments with positive and negative muons in two legs of the beamline (see also \autoref{musr:new_opportunities:elemental-analysis}). The feasibility of such an approach is currently under investigation.


\subsection{muCool beam development}\label{sec:facilities:mu_cool}

Experiments with muons at the high-intensity frontier, as experiments
searching for lepton flavour violation (see \autoref{sec:CLFV}), typically make use of secondary
beam lines with large acceptance tuned to transport surface muons with
momentum $p=28$~MeV/$c$ (equivalent to 4~MeV kinetic energy).
These muons are copiously created by $\pi^+$-decay from pions stopping
close to the surface of the pion production target (proton target).
Muons of lower momenta, from $\pi^+$ decaying below the target surface, can also be extracted  and
transported by tuning the secondary beam lines for
the corresponding momentum.
However, because of the momentum straggling in the target, the
intensity of these sub-surface muon beams decreases rapidly with
momentum ($p^{3.5}$-dependence~\cite{Pifer:1976ia}).
The muon scattering in the production target, combined with the large area
of the production target and the large acceptance of the secondary
beamline, results in muon beams with poor phase space quality.
For example the $\mu$E4 beam line which has presently the largest muon
flux of about $5\times 10^8 \; \mu^+$/s at the moderator is having
$\sigma_{x,y}\approx 20$~mm, $\theta_{x,y}\approx 150$~mrad at a
momentum of 28 MeV/$c$  and a momentum spread $\Delta p/p\approx
7$\% (at FWHM)~\cite{Prokscha:2008zz}. Other beamlines have slightly smaller phase
space at the cost of muon flux~\cite{Adam:2013vqa, Berg:2015wna}.

The muCool project aims to transform these beams of 28~MeV/$c$ momentum
to a beam with momentum of about 1.5~MeV/$c$ (10~keV kinetic energy)
with 0.1~keV energy spread and a phase space of
$\sigma_x\sigma_{\theta_x}=20$~mm\,mrad with an efficiency ranging
from $2\cdot10^{-5}$ to $2\cdot10^{-4}$.
Because such a beam can be easily focused to a sub-mm size
it is well suited for precision experiments requiring a
target with small transverse area and small stopping power or
requiring muons to be stopped close to the surface.
Moreover, such a beam is well suited for re-acceleration: beams from
few keV to 100 MeV energy can be thus obtained by combining the muCool device
with a suited acceleration stage opening various possibilities
including storage experiments such as the muon \ac{EDM} and the muon
$g-2$ where a small phase space is required.

\subsubsection*{The proposed muCool compression scheme}
\label{muC:scheme}

\begin{figure}[tb]
  \begin{center}
    \includegraphics[width=0.60\textwidth]{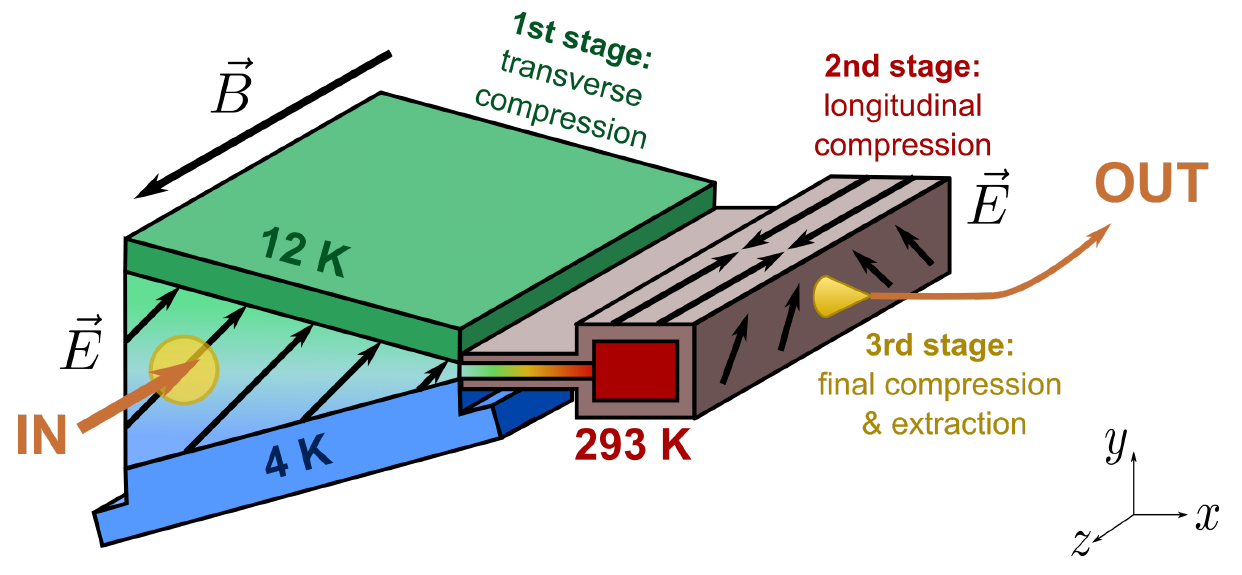}
	\caption{\label{muC:fig1} Schematic diagram of the muCool device. A
          surface muon beam is stopped in a cryogenic He gas target
          with a vertical temperature gradient inside a 5-T field. The
          extent of the stopped muons is reduced first in the
          transverse ($y$), then in the longitudinal ($z$) direction
          using a complex arrangement of E-field and gas density
          gradient. The compressed muon beam is then extracted
          through an orifice into vacuum and re-accelerated along the
          $ -z $-direction.  }
        \end{center}
\end{figure}

In the proposed muCool scheme~\cite{Taqqu:2006mv}, a surface muon beam propagating in the
$-z $-direction is slowed down in a He gas target featuring a strong electric
($E$) field inside a strong magnetic ($B$) field as shown in \autoref{muC:fig1}.
In the slowing-down process, the muon energy is rapidly reduced to the eV range where the E-field becomes important.
The E-field, in conjunction with the B-field and gas density
gradients, leads to drifting of the slowed-down muons drastically
reducing their initially large spatial extent.
In this drift process in the gas, the muons are guided 
to a mm-sized spot.
%
The drift velocity of the $\mu^+$ in a gas with  E- and B-fields is given by~\cite{Blum2008}
\begin{align}
  \vec{v}_D = \frac{\mu |\vec{E}|}{1 + \omega^2 / \nu^2} \left[ \hat{E} + \frac{\omega}{\nu} \hat{E} \times \hat{B}+ \frac{\omega^2}{\nu^2} \left(\hat{E} \cdot \hat{B} \right) \hat{B} \right]. 
\label{muC-Eq1}
\end{align}
In \eqref{muC-Eq1} $\mu$ is the muon mobility, $\omega =
eB/m$ the cyclotron frequency of the muon, $\nu$ the average
$\mu^+-$He collision rate, and $\hat{E}$ and
$\hat{B}$ the unit vectors of the electric and magnetic
fields, respectively.

The spatial extent of the muon stop distribution decreases by making
$\vec{v}_D$ position-dependent, so that $\mu^+$ stopped at different
locations in the target drift in different directions, and converge to
a small spot.
This can be achieved by applying a complex E-field pointing in
different directions at different positions, and by making the collision
frequency $\nu$ position-dependent through  gas 
density gradients.

The muCool setup has been conceived as a sequence of stages having various
density and electric field conditions.
In the first stage, which is at cryogenic temperatures, the muon beam
is stopped and compressed in $y$-direction (transverse compression).
In the second stage, which is at room temperature, the muon beam is
compressed in $z$-direction (longitudinal compression).
In the third stage, the muons are extracted from the gas target into vacuum.
This extraction is followed by re-acceleration in $-z$-direction to
keV energy, and extraction from the B-field.

\subsubsection*{Demonstration of transverse, longitudinal and mixed compressions}
\label{muC:transverse}
The technology and principle at the core of the muCool cooling scheme,
i.e., the ability to move the muon in the He gas with B-field and complex
arrangements of E-field and gas density, has been validated in various
experiments~\cite{Bao:2014xxa, Belosevic:2018fnj, Antognini:2020uyp}.
The time evolution of the number of decay-positrons hitting the plastic
scintillators placed around the targets were used in
these experiments to reveal the average motion of the muon ensemble with 
increasing (decreasing) number of counts  indicating muons
approaching (leaving) the acceptance region of the
considered scintillator.
\begin{figure}[tb]
\begin{center}
  \includegraphics[width=0.30\textwidth]{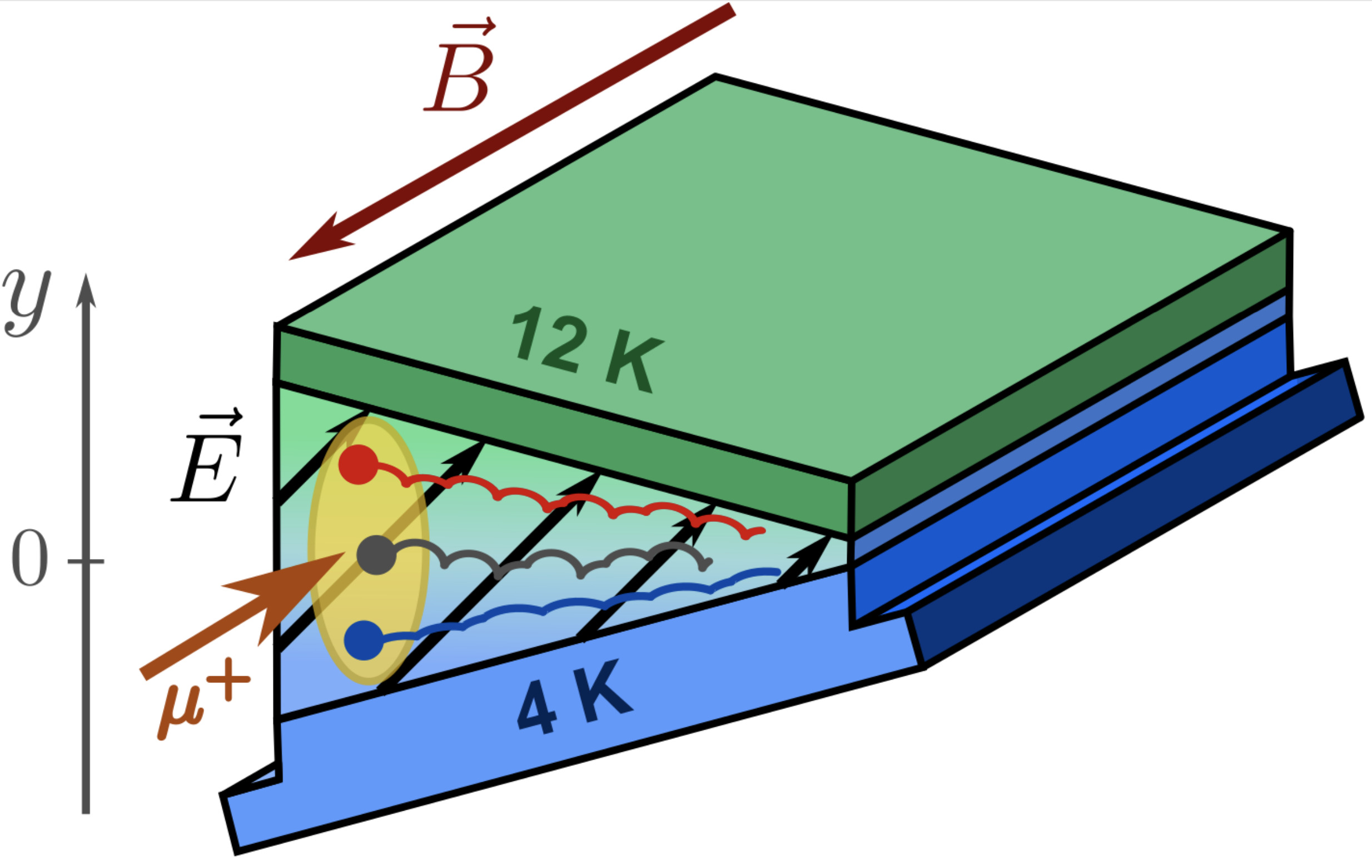}
    \hfill
  \includegraphics[width=0.35\textwidth]{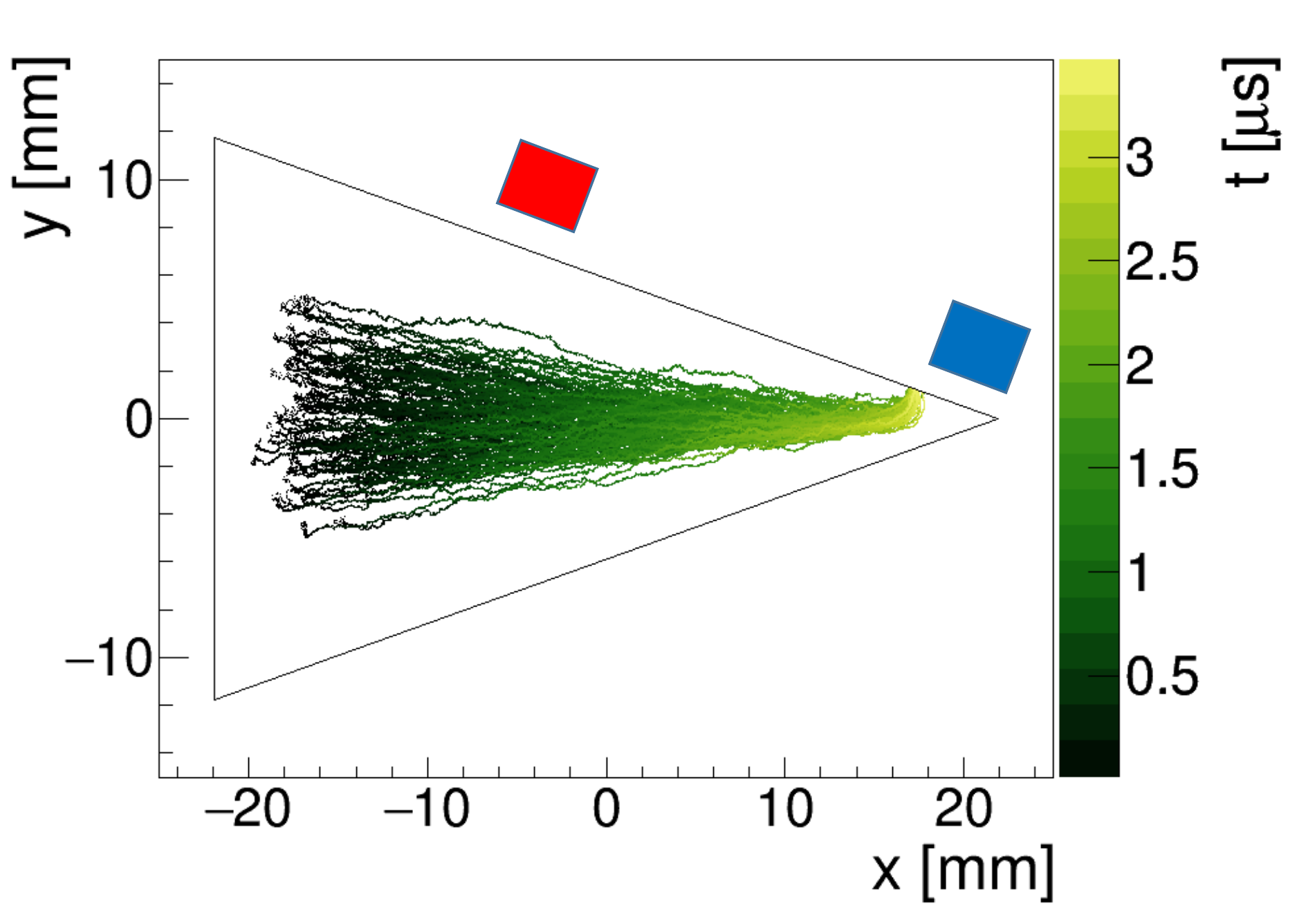}
  \hfill
  \includegraphics[width=0.28\textwidth]{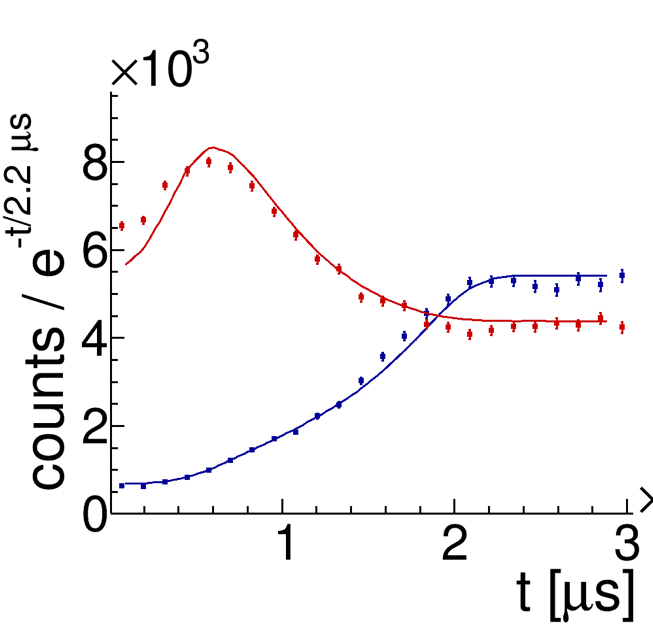}
\caption{\label{muC:fig2} (Left) Sketch of the target used to test the
  transverse compression at cryogenic temperature with temperature 
  gradient. (Middle) Geant4 simulation of muon trajectories starting
  at $x \approx -15$~mm and drifting with time in $+x$-direction while
  compressing in the $y$-direction. The approximate positions of two
  plastic scintillators (red, blue) used to measure decay positrons
  are indicated. (Right) Measured and simulated time spectra for the
  two plastic scintillators of middle panel. The time
  zero is given by a counter detecting the muon entering the
  target. The counts are lifetime compensated.
}
\end{center}
\end{figure}
\begin{figure}[tb]
\begin{center}
  \includegraphics[width=0.28\textwidth]{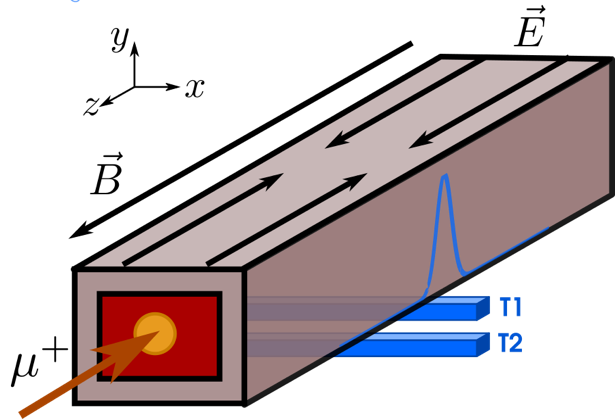}
    \hfill
  \includegraphics[width=0.30\textwidth]{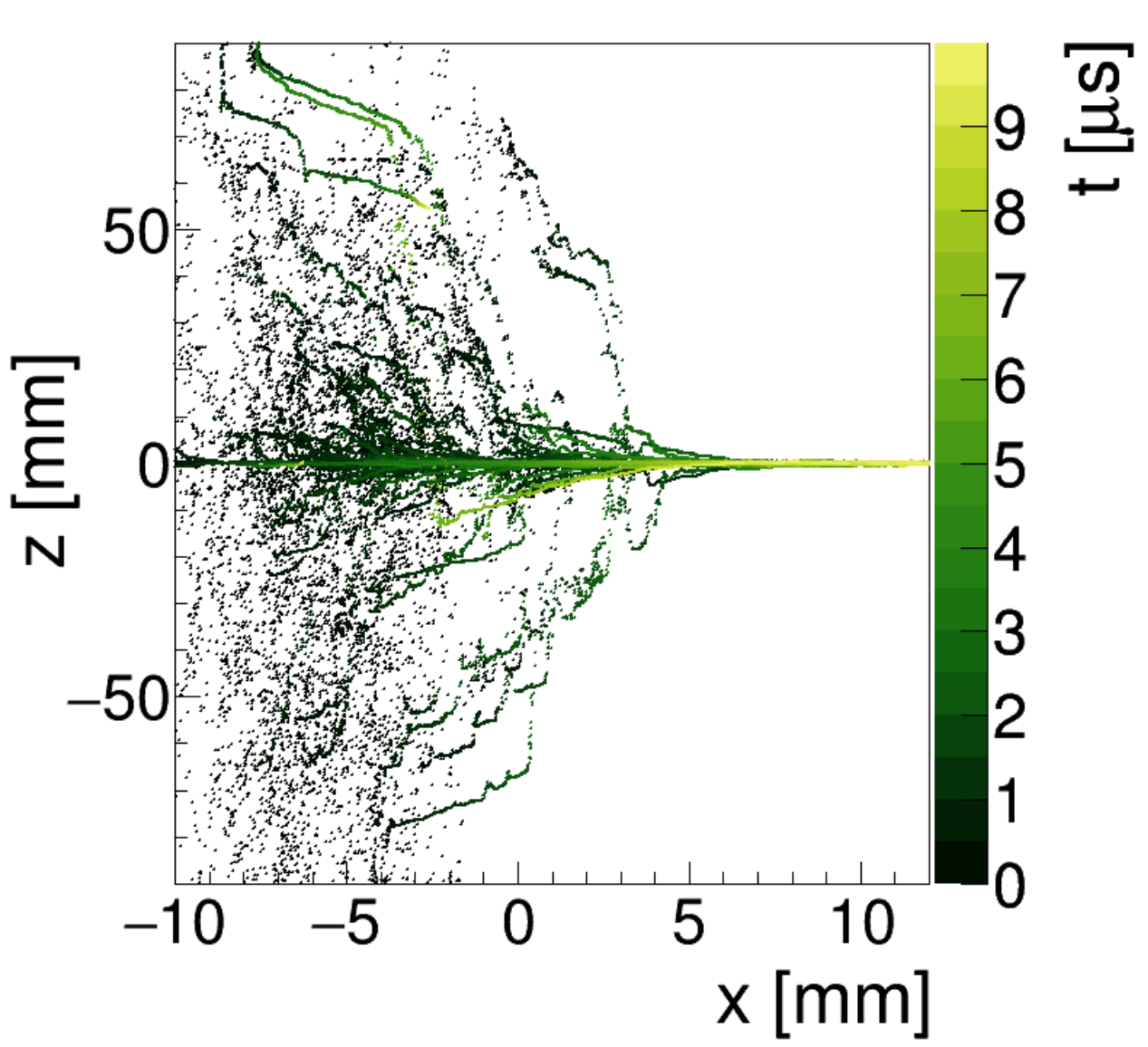}
  \hfill
  \includegraphics[width=0.28\textwidth]{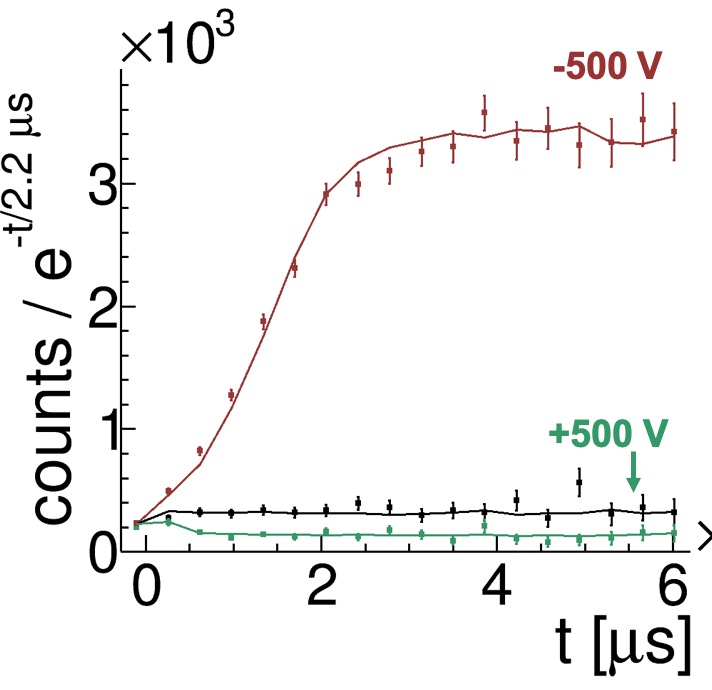}
\caption{\label{muC:fig3} (Left) Sketch of the setup used to test the
  longitudinal compression at room temperature. The scintillators T1 and T2 in coincidence
  detect the $\mu^+$ accumulating around $z\approx 0$. The blue curve indicates the region of acceptance for coincident events. (Middle) 
  Simulated $\mu^+$ trajectories. (Right) Measured and simulated
  time spectra for various HV at the target mid plane: negative HV (red) attracting the muons to the central plane, positive HV (green) repelling them and no HV
  (black).} 
\end{center}
%
%
\begin{center}
  \includegraphics[width=0.60\textwidth]{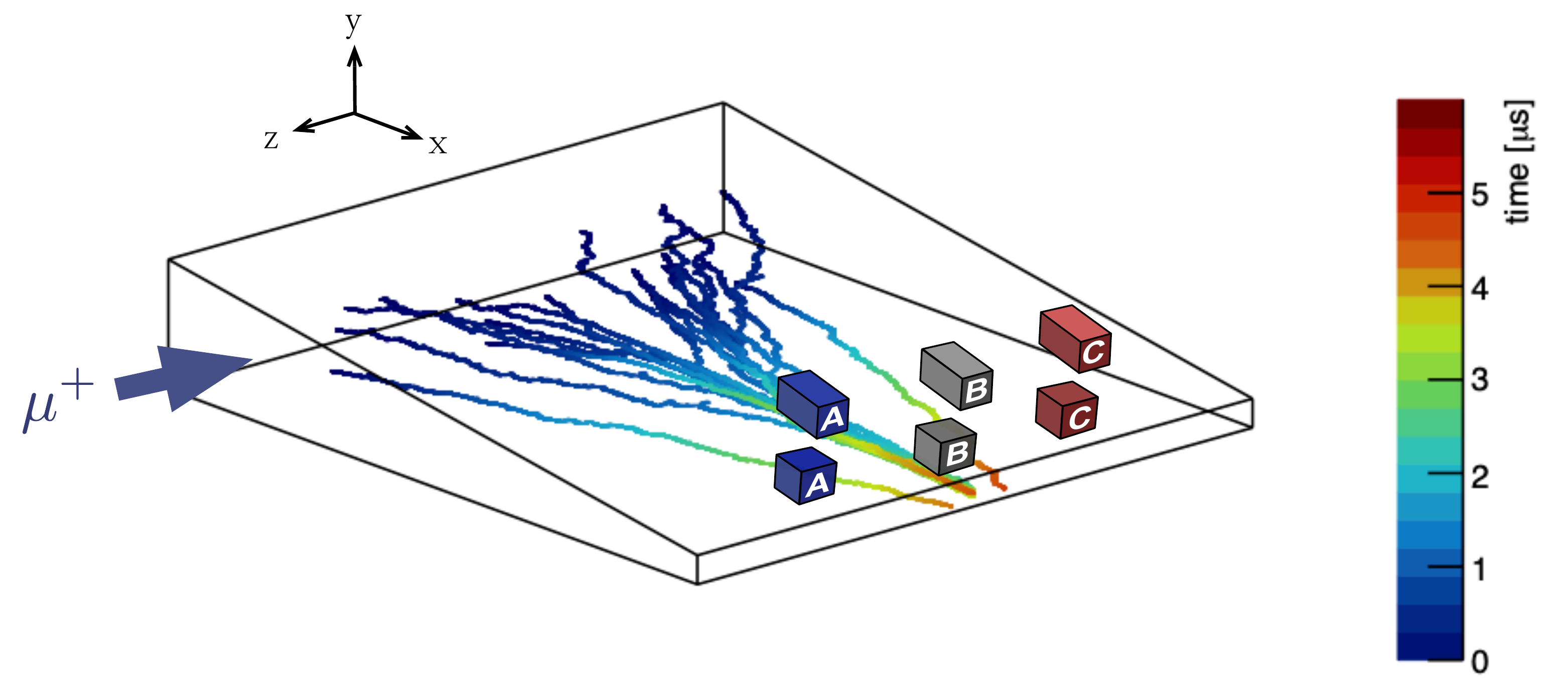}
  \hfill
  \includegraphics[width=0.35\textwidth]{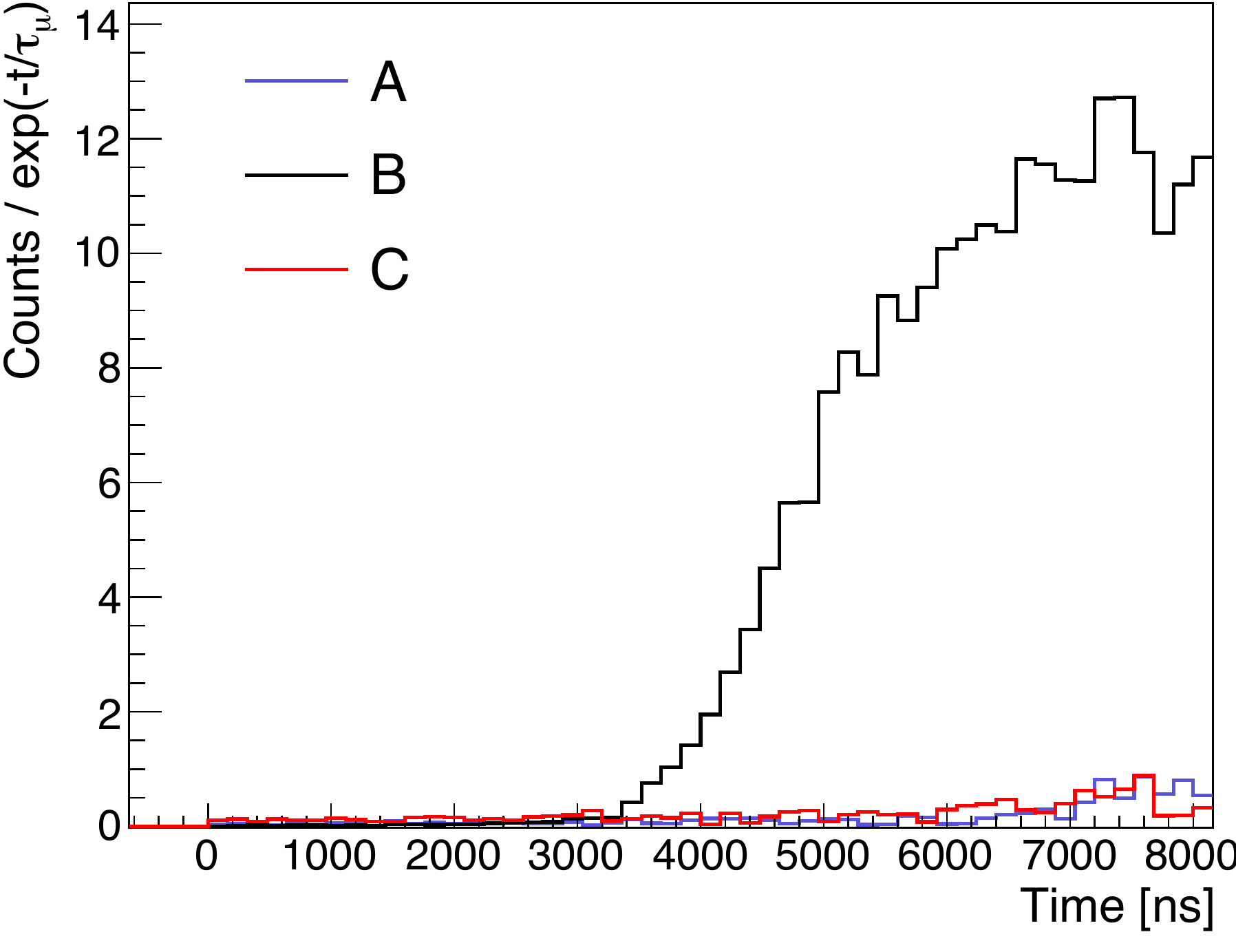}
  \caption{\label{muC:fig4} (Left) Simulated muon trajectories in the
    mixed transverse-longitudinal compression target with a vertical
    density gradient at cryogenic temperatures, $ E_x$ and $ E_y$ components as in the transverse compression target, and an $ E_z $
    component pointing to the target mid-plane at $ z=0 $.  The positions of the  plastic scintillators pairs (A,B,C)
    used to expose the compression are also shown. (Right) Measured time
    spectra in the  scintillator pairs in coincidence. }
\end{center}
\end{figure}
\autoref{muC:fig2} shows a schematic of the setup, a simulation of muon trajectories in the target, and two measurements of the transverse
compression~\cite{Antognini:2020uyp}.
Good agreement between simulated and measured time spectra was
observed confirming the validity of our simulation tools and target
technology.
Similarly \autoref{muC:fig3} demonstrates longitudinal compression.

To avoid the challenging connection between transverse (at cryogenic
T) and longitudinal (at room T) stages, a target has been developed and
commissioned in which both transverse and longitudinal compressions occur
simultaneously (see \autoref{muC:fig4})~\cite{Belosevic:2019bnw}.
Data analysis is still ongoing, but the measured time spectra indicate
that a muon stop distribution with volume $\Delta x \times \Delta
y\times \Delta z=10\times 10\times 60$~mm$^3$ can be transformed in
5\,$\mu$s into a beam drifting in $x$-direction with 10~eV kinetic
energy capable of passing an aperture of $\Delta y\times\Delta
z=1\times 1$\,mm$^2$ size with an efficiency of about 50\% (excluding
muon decay losses).

\subsubsection*{Vacuum extraction and re-acceleration}
\label{muC:next}

\begin{figure}[t]
\begin{center}
  \includegraphics[width=0.7\textwidth]{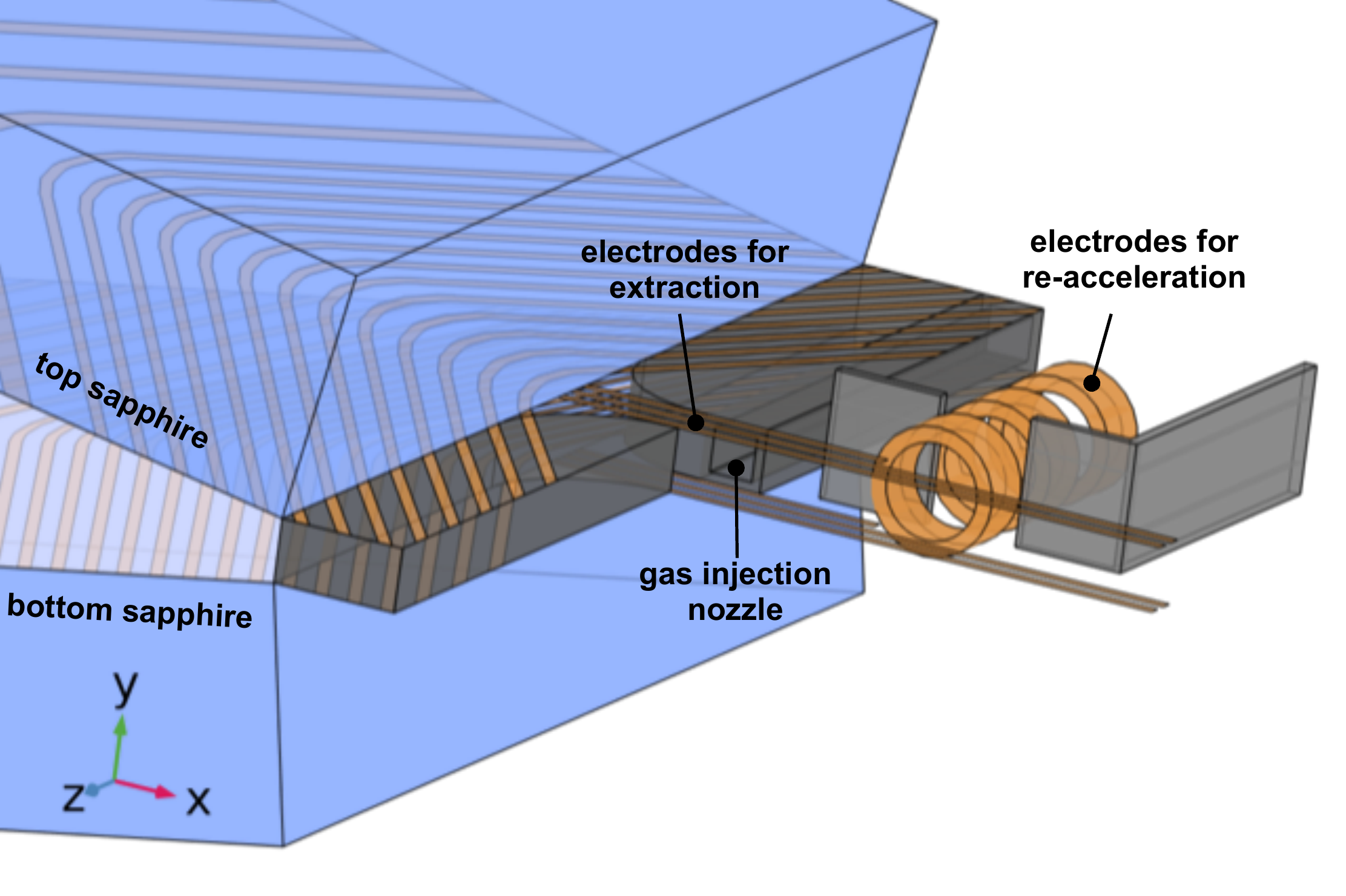}
\caption{\label{muC:fig5} 3D-rendering of the mixed-compression target
  modified to include an orifice, a He gas inlet,
  strips of electrodes for muon extraction and drift in $+x$-direction
  and ring-electrodes for re-acceleration in $-z$-direction. In blue
  the sapphire plates are shown defining the temperature of the top
  and bottom target walls, in orange the electrodes defining the
  electric fields, and in grey the plastic frame of the target. }
\end{center}
\end{figure}

A possible extraction scheme that we are presently simulating is shown
in \autoref{muC:fig5}.
It is based on a mixed-compression target modified to allow $\mu^+$
extraction from the gas target into ``vacuum'' (low density gas
region) through an orifice of $1\times 1$\,mm$^2$ aperture.
To compensate for the He atoms leaving the target through the same
orifice, new He gas is continuously injected at the orifice
perpendicular to the $\mu^+$ motion which acts as a gas barrier for the
target gas.
The injected gas needs to be efficiently evacuated through a system of
differentially pumped regions to maintain a good beam quality while
the muons are drifted in $+x$-direction using suited electrodes.
These muons are eventually coupled  into the re-acceleration
stage where acceleration in $-z$-direction using
ring-shaped electrodes biased at decreasing potentials is occurring.
%
%

\subsubsection*{Projected performance of the complete muCool beam}  
\label{muC:new-beam}

This section briefly discusses the estimated performance of the
\ac{HIMB}-muCool scheme assuming a muCool target with a single active
region for mixed-compression with a performance based on the
commissioned mixed-compression target. The
efficiencies of the various stages are also summarised in
\autoref{muC-table1}.

\begin{itemize}
\item{\textbf{Coupling the beam into the solenoid}} \\ The \ac{HIMB} beam has an
  average momentum of 27.7~MeV/$c$, a large transverse phase space
  $\sigma_x\sigma_{\theta_x}=1910$~mm\,mrad and a large momentum
  spread of 3.5~MeV/$c$ (FWHM) corresponding to 12.6\% momentum
  bite. Geant4 simulations show that 36\% of the muons
  are back-reflected when injecting on-axis the \ac{HIMB} beam into a 5~T
  solenoid of 60~mm inner coil diameter. This fraction becomes 44\%
  for $14^\circ$ tilt between solenoid and beamline axis to
  inject the muons 50~mm off the solenoid axis.

\item{\textbf{Coupling the beam into the target}} \\ Geant4
  simulations show that the fringe field of a 5~T solenoid with 60~cm
  inner-coil diameter focuses the \ac{HIMB} beam to
  $\sigma_x\approx\sigma_y\approx 11$~mm.  Hence, only a fraction of
  48\% of the muons entering the solenoid would impinge on the
  entrance face (end-cap) of the already commissioned mixed-compression
  target.  For a mixed target with a three times larger side wall (but
  same target extent in $x$-direction) this fraction increases to
  83\%. The feasibility to realise this larger target with the
  adequate electric field has still to be demonstrated by R\&D.
  
\item{\textbf{Stopping in the active region of the gas target}} \\ The
  large momentum spread and transverse momentum distribution of the
  \ac{HIMB} beam, in conjunction with the short length of the active region
  and low density of the target leads to an overall small stopping
  probability in the region of interest.
  Geant4 simulations predict that with an optimised moderation at the
  target entrance, 0.41\% of the muons impinging on the entrance face
  (end-cap) of the target eventually stop in the target's active
  region. Here we assumed a 50~mm long active region as realised in
  the commissioned mixed-compression target with a pressure of 10~mbar
  and a temperature gradient from 6~K to 22~K. We should be able to
  improve this stopping probability by at least a factor of 1.7 by
  operating the target at colder temperatures (4 to 16 K gradient) and
  by elongating the extent of the active region  from 50~mm to 60~mm while
  keeping the same compression efficiency. Moreover, we are presently evaluating a possible reduction of the momentum spread by introducing some dispersion in the \ac{HIMB} beam in conjunction with a position-dependent moderation.

\item{\textbf{Compression and drift to the orifice region}} \\ Preliminary
  analysis of data taken with the commissioned mixed-compression
  target demonstrates that muons drift and compress with an average
  kinetic energy smaller than 10~eV from the stopping region of 50~mm
  length and 10~mm height to the orifice region with an efficiency of
  80\% in about 5~$\mu$s.  When including lifetime losses  this
  efficiency is reduced to 8\%.

\item{\textbf{Extraction from the gas target through the orifice}}\\ From
  preliminary electric fields layouts, Geant4 simulations predict that
  the compressed muon beam can be extracted from the target with an
  efficiency of about 50\% (neglecting lifetime losses) in about
  0.7~$\mu$s though an orifice with $1\time 1$ mm$^2$ aperture.
   Including lifetime-losses, this
  probability becomes thus 40\%. 

\item{\textbf{Coupling to the re-acceleration region}} \\ Preliminary
  simulations of the muon motion from the orifice to the
  first accelerating ring  (where the electrostatic re-acceleration is starting)
  show that a transport with  90\% efficiency (without decay-losses)
  in 0.5~$\mu$s can be obtained. This  efficiency is thus about 70\%,
  when accounting for the muon lifetime. The performance of this transport
  strongly depends on the electric field that can be applied, and the
  distance muons have to travel before a gas density suited for
  re-acceleration is reached.  Note that muons are injected in the first ring
  with transverse energies of $\mathcal{O}$(20~eV) and sub-mm size.
  
\item{\textbf{Re-acceleration to 10 keV energy}} \\ In the static
  re-acceleration process guided by a series of ring electrodes at
  decreasing (positive) HV, the only losses are the ones related to
  the decay during the time of flight.  They have been estimated to be
  about 20\%. Some degradation of the beam quality might occur in
  this stage due to collisions with the rest gas (to be investigated).

\item{\textbf{Extraction from the B-field}} \\The extraction of the 10~keV
  muons from the solenoid is obtained by transporting the muons from
  the central region of the solenoid at 5~T to a region of 0.01~T
  field where a grid with 70\% transmission terminates abruptly the
  magnetic field lines. In this transport from high-field to low-field,
  the transverse beam energy is decreased by a factor of 500 while the
  beam radius increases by $\sqrt{500}$. At the grid the abrupt
  termination of the B-field produces a  radial field component 
  that impart an azimuthal momentum to the muons of about 3~keV/c
  (assuming a grid with 2~mm wide apertures) resulting effectively in
  a beam of 2~mrad divergence. Hence the beam is extracted into a
  field-free region with a phase space of 
  $\sigma_x\sigma_{\theta_x}\approx 20$~mm\,mrad (unnormalised) at 10~keV.
\end{itemize}
\begin{table}[t]
    \caption{The first column shows estimated baseline efficiencies of
      the various muCool stages using the commissioned
      mixed-compression target as a reference point. We assume here a
      target with only a single active region (10 mbar, 6-22 K, 50 mm long). The second column shows
      the room for possible improvements after dedicated R\&D, still
      assuming a single active compression region.   All
      numbers include muon-decay losses.  \label{muC-table1}}
  \begin{center}
    {\footnotesize
    \begin{tabular}{lll}
      \toprule
      Baseline   & Possible         & Description\\
      Efficiency & Improvements      & \\
      \midrule
      $5.6\cdot 10^{-1}$ &           & Coupling to the 5~T solenoid with 60 mm coil diameter  \\
      $4.8\cdot 10^{-1}$ & $\times 2$   & Impinging on the target entrance-face   \\
      $4.1\cdot 10^{-3}$ & $\times 1.6$   & Stopping probability in active region of the target   \\
      $8\cdot 10^{-2}$ & $\times 1.5$ & Compression towards the orifice  (within 5 $\mu$s)\\
      $4\cdot 10^{-1}$ & $\times 1.3$ & Extraction from the orifice\\
      $7\cdot 10^{-1}$ & & Drift from orifice to re-acceleration region (in  $\sim 0.5$ $\mu$s) \\
      $8\cdot 10^{-1}$ & & Re-acceleration and transport to the iron grid \\
      $7\cdot 10^{-1}$ & & Transmission through the iron grid terminating the B-field\\
      \midrule
      $1.4\cdot 10^{-5}$&$\times 6$& Total baseline compression efficiency (and possible improvement)\\
      \bottomrule
    \end{tabular}
    }
  \end{center}
\end{table}
As can be seen from \autoref{muC-table1}, which summarises the baseline
efficiencies of the various muCool stages, a total baseline
compression efficiency of $1.4\cdot 10^{-5}$ is expected from the
muCool setup applied to the \ac{HIMB}.
Still assuming a target with a single active region, we expect some
room for improving this efficiency with dedicated R\&D up to a value
of $1\cdot 10^{-4}$.
Moreover, because the stopping probability in the active region of the
commissioned mixed-compression target is smaller than 1\%, a long muCool
target could be realised having multiple active regions in series
along the $z$-direction, each of them with its own extraction orifice.
With some degradation of the 6D phase space, the various
outputs can be combined to produce a single more intense muon
beam. Alternatively, each of the outputs can be extracted individually
from the magnet and sent to various experiments operating
simultaneously.
The number of active regions in a muCool target in principle can be as
high as 100, but practically the required vacuum quality in the
acceleration stage will probably limit their number to a few.

In summary, the muCool target transforms the \ac{HIMB} input beam of 4~MeV
energy, with 1 MeV  energy spread (at FWHM), and 
$\sigma_x\sigma_{\theta_x}=1910$~mm\,mrad (unnormalised)  into a beam of
10~keV with $<0.1$~keV energy spread (at FWHM) and
$\sigma_x\sigma_{\theta_x}\approx 20$~mm\,mrad (unnormalised) with an efficiency ranging from
$2\cdot 10^{-5}$ to $1\cdot 10^{-4}$ assuming a target with a single
active region.  Such a beam can be focused to sub-mm sized targets and
after re-acceleration is well suited for storage-ring experiments.
Its efficiency can be further improved by implementing multiple active
regions in the same elongated target.

\subsubsection*{Post-acceleration from 1 to 60 MeV energy}

The  muon beam delivered by the muCool setup can be
further accelerated to the energy appropriate for 
experiments such as the muon EDM and the muon $g-2$ measurements,
where the required muon beam energy  ranges from around 1 MeV to
60 MeV or higher. 
One of the most important parameters of the muon accelerator is the
transmission rate: the path length should be minimised because muons
are quickly decaying.
Circular accelerators therefore may not be optimal, at least in our
energy range, due to the fact that the length allocated for
acceleration is quite short in comparison to the total path length.
Conversely, the linear accelerator (linac) is efficient in this aspect
and thus is our baseline choice.
As described above, the initial acceleration up to several tens of keV
energy will be accomplished simply by setting the ground of the
muCool target at positive HV so that the muons leaving the muCool setup 
are at a kinetic energy basically given by this HV.
This electrostatic pre-acceleration provides muons to the linac at a
proper energy in a continuous way.
For some experiments it may be more convenient to have an extraction in
bunches containing several muons.
Accumulating the muons for
several microseconds before letting them enter the re-acceleration
stage is feasible, but 
at the cost of a largely reduced muon flux due to the
2~$\mu$s muon lifetime.

In any case, we design the downstream linacs to be compatible with CW
operation as well as a pulsed operation.
Among various types of
linacs, an \ac{RFQ} would be well suited to take
over the muon beam after the initial acceleration because an \ac{RFQ} acts like an
“all-in-one” cavity, which focuses, bunches and accelerates the beam.
An acceleration of negative muonium atoms using an \ac{RFQ} has been
demonstrated in~\cite{Bae:2018atj}.
Our preliminary investigation arrived at a parameter set shown in
\autoref{muC-table2}.
It is proposed to divide the \ac{RFQ} into two
parts, one for capturing and bunching, and the other for the
acceleration.
This increases flexibility in parameter optimisation
drastically.

Two scenarios with a 352-MHz RF system and a 750-MHz RF system are
currently considered.
The higher frequency system offers a shorter accelerator at the
expense of a smaller acceptance and a slightly higher RF power
consumption.
The transmission efficiency can be sufficiently high for the estimated
emittance of 0.55 mm mrad rms (normalised).
The muon flux degradation is about 10\% for an
accelerator length of 5~m.
\begin{table}[t]
    \caption{RFQ main parameters for two RF frequencies.
      The normalised emittance corresponds to a geometrical
      emittance of 40~mm\,mrad at 10~keV.
      The intermediate energy is the kinetic energy of the muon beam at the end of the capture section.
      \label{muC-table2} }
  \begin{center}
    \begin{tabular}{lll}
      \toprule
      Parameters                                        &  Values for        & Values for \\
                                                        &  RF=352\,MHz        & RF=750\,MHz\\
      \midrule
      Input emittance (rms, normalised)                 & 0.55\,mm mrad             & 0.55\,mm mrad\\
      Input energy, kinetic                             &  20\,keV                  & 20\,keV  \\
      Intermediate energy, kinetic                      & 170\,keV                  & 190\,keV \\
      Output energy, kinetic                            & 2\,MeV                    & 2\,MeV  \\
      Output energy spread                              & 0.75\%                   & 0.55\% \\
      Capture section voltage, vane-to-vane             & 21\,kV                    & 28\,kV    \\  
      Acceleration section voltage, vane-to-vane        & 65\,kV                    & 50\,kV    \\
      Total length (Capturing/bunching + Accel.)          & 2.4 + 3.0\,m              & 2.0 + 2.4\,m \\
      Beam power                                        & Negligible               & Negligible \\
      Power consumption                                 & 300\,kW (CW)              & 340\,kW (CW) \\
      Transmission efficiency $\mu$-decay not included & 96\%                      & 94\% \\
      \bottomrule
  \end{tabular}
\end{center}
\end{table}
A rough estimation of the costs to build
such an accelerator is 250k~CHF/m for the accelerating
structure and 2.3 million CHF for the 352-MHz RF system.
The cost of the RF system includes estimates for a 300-kW solid state power amplifier, waveguides, cooling infrastructure, a local signal source with LLRF-and interlock-system, 
support structures and cabling. 
Not included are shielding, manpower, maintenance and running cost.
The 750-MHz RF system would be as expensive as the 352-MHz RF system.

In case where the above estimated cost is prohibitive,
we may consider to reuse the 500-MHz system of the \ac{SLS} 
which is currently in operation but will be available after the ongoing \ac{SLS} upgrade
provided that the performance of the \ac{RFQ} is similar at 500~MHz.
We need to clarify if this option is cost-effective: how long it can keep up and how much it costs to be refurbished if necessary.
Another option is to reduce the duty factor from CW (100\%) to a lower value
that can be accepted by the experiments.
Obviously, this will save running costs, and we also expect a lower cost for the accelerating structure
since the cooling part can be simplified.

The power dissipation is estimated for CW operation and fully
dominated by the RF wall loss since the muon beam power is
negligible. A consumption of several hundred kW is feasible as already
demonstrated in high power proton linacs~\cite{osti_335199}. It is also possible to
utilise a superconducting cavity, which offers a highly
energy-efficient operation at the expense of higher initial
construction cost and increased engineering complexity.

The obtained parameters are for the final energy of 2~MeV.
Although it is hard to change the output energy once we built the RFQ,
it is straightforward at the design stage.
We simply decrease(increase) the length of the second part of the RFQ cavity
if the desired beam energy is lower(higher) than 2~MeV. The final energy is a key parameter driving the overall costs.

The beam is further accelerated for the higher energy applications. The
Lorentz beta is still $\ll 1$ at the exit of the \ac{RFQ}, hence a drift tube
linac may be employed. The accelerator chain proposed here is indeed
similar to an established layout of proton linac facilities, see
e.g.~\cite{ESS_TechnicalDesignReport}. Other options are, however, to be considered before
finalising the accelerator design.

\subsubsection*{Selected applications for muCool beams (with or without re-acceleration):}
\label{muC:applications}

This new beam opens the way for next generation experiments with muons
and muonium where the reduced phase-space is of great advantage
spanning applications in fundamental particle physics, atomic physics
and materials science through the $\mu$SR technique.\\

\paragraph{Materials science.}
The muCool beam can greatly benefit $\mu$SR investigations by
delivering muons simultaneously to several $\mu$SR instruments at the
maximum rate allowed by pile-up effects with energy tunability and
sub-mm size.
Because the pile-up effects in the typically 10~$\mu$s-long
observation time window become increasingly unsustainable for rates
exceeding $4\cdot 10^{4}$~$\mu/$s (without vertexing, see \autoref{musr:new_opportunities}), the full \acs{HIMB}-muCool potential
could be exploited by distributing the keV-energy sub-mm beam between
several $\mu$SR instruments operating simultaneously.

Even assuming the baseline efficiency of $1.4\cdot10^{-5}$, the muCool
setup injected with $10^{10} \mu$/s from \ac{HIMB} and provided with an
appropriate muon distribution system would be able to deliver muons
simultaneously to several $\mu$SR instruments with keV energies
similar to the present \ac{LEM} beamline~\cite{Prokscha:2008zz} (best performing
low-energy beam-line world-wide) but with a sub-mm transverse size
instead of the presently cm-size.
The number of beams that could be extracted from the muCool setup can
be further increased by improving the overall efficiency of the muCool setup and by realising a long muCool target having multiple
active regions in series along the $z$-direction, each of them with its
own extraction orifice.
Owing to the limited rate acceptance of the $\mu$SR instruments (for
cw beams), there is no need to merge the beams exiting the various
orifices. Each of these beams could be extracted from the solenoid at
slightly different off-axis positions with negligible distortion of
their beam quality (relative to on-axis extraction).

Hence, the \acs{HIMB}-muCool scheme would be able to deliver numerous beams,
with their number limited only by the technical capability of
realising the needed vacuum requirements at the re-acceleration
stages.
Each of these beam could be easily transported to a compact $\mu$SR
instrument in a background-free environment capable of analysing sub-mm
samples at 50~kHz rate and keV energies.
Pile-up effects could be further reduced because the sub-mm low-energy beams can be easily ``kicked away" to prevent ``second muons''
entering the setup in the measurement time window of the ``first muon".
Combined with high-rates capability, this feature allows the
implementation of a muon-on-demand scheme that minimises the dead time
of the $\mu$SR spectrometer: as soon as a decay-positron is detected
in the spectrometer the next muon is allowed to enter.

The small phase space of the muCool beam is also well suited for
efficient re-acceleration to higher energies as discussed previously. Among others, the energy
range between 30~keV to a few hundred keV presently not accessible at
PSI could be covered.

To reduce the number of $\mu$SR instruments, it would also be possible 
to focus several of the extracted beams from the same muCool target
into various samples (or at various locations of a larger sample)
mounted in the same $\mu$SR spectrometer. Using a positron tracking
system (as described in \autoref{sec:musr}) it would then be possible to
distinguish uniquely the muon decays originating from the various
samples in order to disentangle  the corresponding
time spectra.

Summarising, the \acs{HIMB}-muCool scheme would improve by an order of
magnitude the sample throughput of the PSI $\mu$SR facility
especially in the keV-energy regime because the muCool device could
deliver multiple low-energy beams with 50~kHz rates clean of ``second
muons''.
At the same time the transverse beam size would be reduced by more than two orders of
magnitude, tremendously benefiting investigations of small samples
(see \autoref{sec:musr}).
Moreover, it also allows muon re-acceleration to energies that are presently not
accessible.\\ 

\paragraph{High quality muonium sources. }
Muon to vacuum-muonium conversion is very efficient for keV-energy
muons~\cite{Antognini:2011ei}.
Hence, the sub-mm muCool beam at keV-energy could be converted into a
high-brightness muonium source.
This novel muonium source could be exploited to vastly improve
on the precision of muonium spectroscopy~\cite{Crivelli:2018vfe}, and for the investigation of the gravitational interaction of muonium~\cite{MAGE:2018wxk}.
%

As detailed in \autoref{sec:muonium}, the \acs{HIMB}-muCool beam opens the way for improving by an order of
magnitude the ongoing spectroscopy measurements in muonium: below
1~kHz for the 1S-2S transition and down to few Hz for the \ac{HFS}.
The high muonium rates from the \acs{HIMB}-muCool
will allow accurate studies of systematic uncertainties, the
implementation of novel experimental schemes less prone to
it, and  laser spectroscopy of other transitions.

The combination of the various  transition measurements in muonium will lead to:
stringent tests of bound-state \ac{QED} (purely leptonic
systems), the best determination of fundamental constants such as the
muon mass and the muon magnetic moment, an independent determination
of the muon $g-2$ with uncertainty comparable with the present
discrepancy, and searches for possible new (beyond the standard model) muon-electron couplings.

As detailed in \autoref{sec:muonium}, the \acs{HIMB}-muCool beam will also greatly benefit the measurement of the muonium acceleration in the Earth's gravitational field, as a test of the equivalence principle for anti-matter and second-generation particles. 
%

Note in conclusion that several output beams delivered by the muCool setup could be merged
on the same spot of a muon-to-muonium converter yielding a muonium source
with a flux of several $10^{5}\,$muonium/s emitted from a 1 mm spot size.\\

%

\paragraph{Storage ring experiments.}
As detailed in \autoref{sec:moments}, the search for a muon \ac{EDM} and the precise
measurement of the muon $g-2$ represent well motivated channels for
physics beyond the Standard Model.
The currently ongoing experiment at Fermilab aims at a measurement of
 $g-2$ with statistical uncertainty of 0.1 ppm and a similar
uncertainty from systematics~\cite{Chislett:2016jau}.
With the same apparatus, searches of the muon EDM with a final sensitivity of
$1\cdot 10^{-21}$ e\,cm will be accomplished.
A second collaboration at J-PARC also aims at the same quantities with a
combined (statistics + systematic) precision of 0.45~ppm for the $g-2$,
and a sensitivity of $1\cdot 10^{-21}$ e\,cm for the muon EDM.

The muonEDM collaboration at PSI proposes (see \autoref{sec:moments}) a search
for the muon EDM based on the frozen-spin technique applied to a
compact muon storage ring~\cite{Adelmann:2021udj}.
Preliminary studies show that a sensitivity of $6\cdot 10^{-23}$ e\,cm
could be reached in the PSI experiment using the $\mu$E1 beam at
125~MeV/$c$ delivering $2\cdot 10^8$~$\mu^+$/s with an average
polarisation of better than 93\% and assuming a B-field of 3 T.
Because of the small phase space acceptance of the storage ring, the
coupling efficiency for the $\mu$E1 beam is only $2.5\cdot 10^{-4}$ so
that only $5\cdot 10^{4}$ $\mu^+$/s are eventually stored in orbit.

The sensitivity to the muon EDM is given by (see \autoref{sec:moments})
\begin{equation}
  \sigma(d_\mu)=\frac{\hbar}{2\beta\gamma c B P \sqrt{N} \alpha \tau}
\end{equation}
where $B$ is the field strength, $P$ the muon polarisation, $N$ the
number of stored muons, $\alpha$ the average decay-asymmetry, $\tau$ the muon
lifetime, and can be improved with the muCool beam for two reasons:
Firstly, owing to its small phase-space the muCool beam can be
efficiently coupled into the storage ring resulting in a larger rate of
stored muons $N$.
Secondly, owing again to its small phase-space the muCool beam can be
efficiently accelerated to larger momenta to increase $\gamma$ and
$\beta$. With larger $\gamma$ and $\beta$ also the B-field can be
increased to keep the optimal radius of curvature.
For a storage ring with 5~T field the optimal momentum is
around 190~MeV/$c$. 

A muCool target performing approximately at the baseline efficiency followed by a
post-acceleration stage would deliver muons with a rate of about $1.4\cdot
10^{5}$~s$^{-1}$  so that the rate of muons injected in orbit is increased by a
factor of 3 compared to the  experiment at the $\mu$E1 beamline.
R\&D on the muCool target has the potential to significantly improve
the muon rate delivered to the storage ring.
Yet, using a cw beam, i.e., in the scenario where a single muon at a
time is analysed in the storage ring, it is
essentially impossible to improve the sensitivity beyond $1\cdot
10^{-23}$ e\,cm per year of data taking (here we assumed 5 T field and
190~MeV/$c$ momentum).
This limitation is originating from  pile-up effects.

A small improvement can be obtained by bunching the muCool beam: 5
muons per bunch with 100~kHz repetition rate would result for example in a sensitivity of
$8\cdot 10^{-24}$ e\,cm per year of measurement.
While formation of these bunches is technically possible, the
muon-decay losses over the needed accumulation period are severe due to the 2~$\mu$s lifetime of the muon.
These losses can only be compensated by realising a muCool target
with several active regions and by recombining the various outputs at
the cost of some energy spread of $\mathcal{O}$(1 keV).

The same setup used to search for the muon EDM can be used, with minor
modifications, to measure the muon $g-2$.
As detailed in \autoref{sec:moments} with a field strength of 6~T and muon on
request at 125~MeV/$c$ (that requires in concrete terms a cw beam
delivering about $5\cdot10^{5}\mu$/s) a statistical sensitivity of
0.1~ppm can be reached similar to the ongoing Fermilab experiment.
Also in this case, bunching of the muCool beam could be used to
improve the measurement down to 0.06~ppm.

An alternative layout is being investigated in which the $g-2$
measurement is performed by injecting muons of only 1~MeV energy into
a 17 T solenoid that is acting as mini storage ring.
One of the advantages of this scheme is that it requires only the
RFQ-acceleration stage but not the linac, reducing size and cost.

Hence, \acs{HIMB}-muCool has the capability to contribute to two 
flagship quantities of the present particle physics landscape with high potential for new physics. 

In conclusion, the \acs{HIMB}-muCool complex provides a beam (or multiple
beams) that push forward our capabilities where we are already
world-leading such as $\mu$SR, muonium spectroscopy and fundamental particle physics with
muons, while expanding our activities to include storage-ring and gravity experiments.


\subsection{Detector developments}\label{sec:facilities:detector_developments}

To take full advantage of the increased rate of muons provided by \ac{HIMB}, a new generation of silicon pixel detectors is required. Silicon pixel detectors are used in particle physics for the reconstruction of trajectories of charged particles (tracking) and the identification of vertices from the decay of non-stable particles. Tracking is based on pattern recognition and track fitting and thus requires detectors with high granularity and excellent spatial resolution. 

PSI has a long-standing tradition in the design, development and construction of silicon pixel detectors. These detectors have proven their functionality for example in the CMS experiment at the CERN LHC~\cite{CMS:2008xjf, Kastli:2007pe,CMSTrackerGroup:2020edz} and in applications in x-ray photon science at \ac{SLS} and SwissFEL~\cite{Jungmann-Smith:pp5080,doi:10.1080/08940886.2018.1528428,Allahgholi:gb5080}. The pixel detectors used in this context are built using the so-called hybrid technology, in which the particle sensing element and the necessary readout electronics are separate entities, connected using the bump-bonding technique~\cite{Broennimann:2005qv}. 

An alternative approach for pixel detectors is the monolithic technology, where the sensor and readout electronics are parts of the same entity. While CMOS based \ac{MAPS}~\cite{Kenney:1994jjz} are well known in particle physics, only recently commercially available technologies enable depleted substrates and thus \ac{DMAPS}~\cite{Peric:2007zz, Turchetta:2001dy}. The depleted substrate is mandatory for a radiation
resistant and fast signal generation and thanks to the full amplification stage and readout architecture being implemented in the silicon sensor chip, much smaller active thickness can be used as the smaller charge signal can be processed without performance degradation. The material budget per layer is significantly reduced as only a single silicon layer is needed and the reduced leakage current due to the
lower silicon volume results in relaxed cooling requirements. This makes \ac{DMAPS} the most attractive pixel detector concept for future experiments. 

In fact, the pixel detector for the Mu3e experiment is being constructed in an \ac{HVMAPS} technology~\cite{Peric:2013cka}, which allows a high voltage being applied to the silicon substrate to achieve the depletion zone. The MuPix family of chips that has been developed over the past few years~\cite{Schoning:2020zed} allows to build a tracking detector with an unprecedented low-material design~\cite{Arndt:2020obb}. Low-material detectors are key for the reconstruction of the trajectory of the decay positrons to minimise the effect of multiple scattering on the spatial resolution.  

The MuPix chips have an active area of $20\times20$\,mm${^2}$ and are thinned down to a thickness of 50$\,\mu$m. Using thin high-density interconnect circuits, made of two layers of a substrate of polyimide (10$\,\mu$m) and aluminium (12$\,\mu$m), a detector layer ends up with a radiation length of $\approx 1.15\times 10^{-3}X_0$. The pixel readout is triggerless and always-on, i.e., every hit in a chip is read out with a time-stamp having a resolution of 15 ns or better. With a pixel pitch of 80$\,\mu$m, a position resolution of 23$\,\mu$m is achieved.

The next generation of pixel detectors will provide increased granularity in space and time to distinguish particle locations and their interaction times, which is necessary to cope with the high muon flux delivered by \ac{HIMB}. 
The evolution of these detector concepts in view of particle physics experiments at \ac{HIMB} are discussed in \autoref{sec:Mu3e_PhaseII_design}, \autoref{sec:pixelMEG}, and \autoref{sec:moments_measurement}. 

Pixel detectors also enable new scientific approaches in muon experiments in solid-state physics and materials science, in particular \ac{muSR}. The spatial and temporal resolution provided by the pixel detector allows for example the measurement of multiple samples in parallel in the same exposure, the separation of domains with different magnetic properties in custom samples as well as to overcome the current limitation in muon rate (as discussed in \autoref{sec:musr}).

\begin{figure}[tb!]
  \centering
     \includegraphics[trim=0 0 0 0,clip,width=0.5\textwidth]{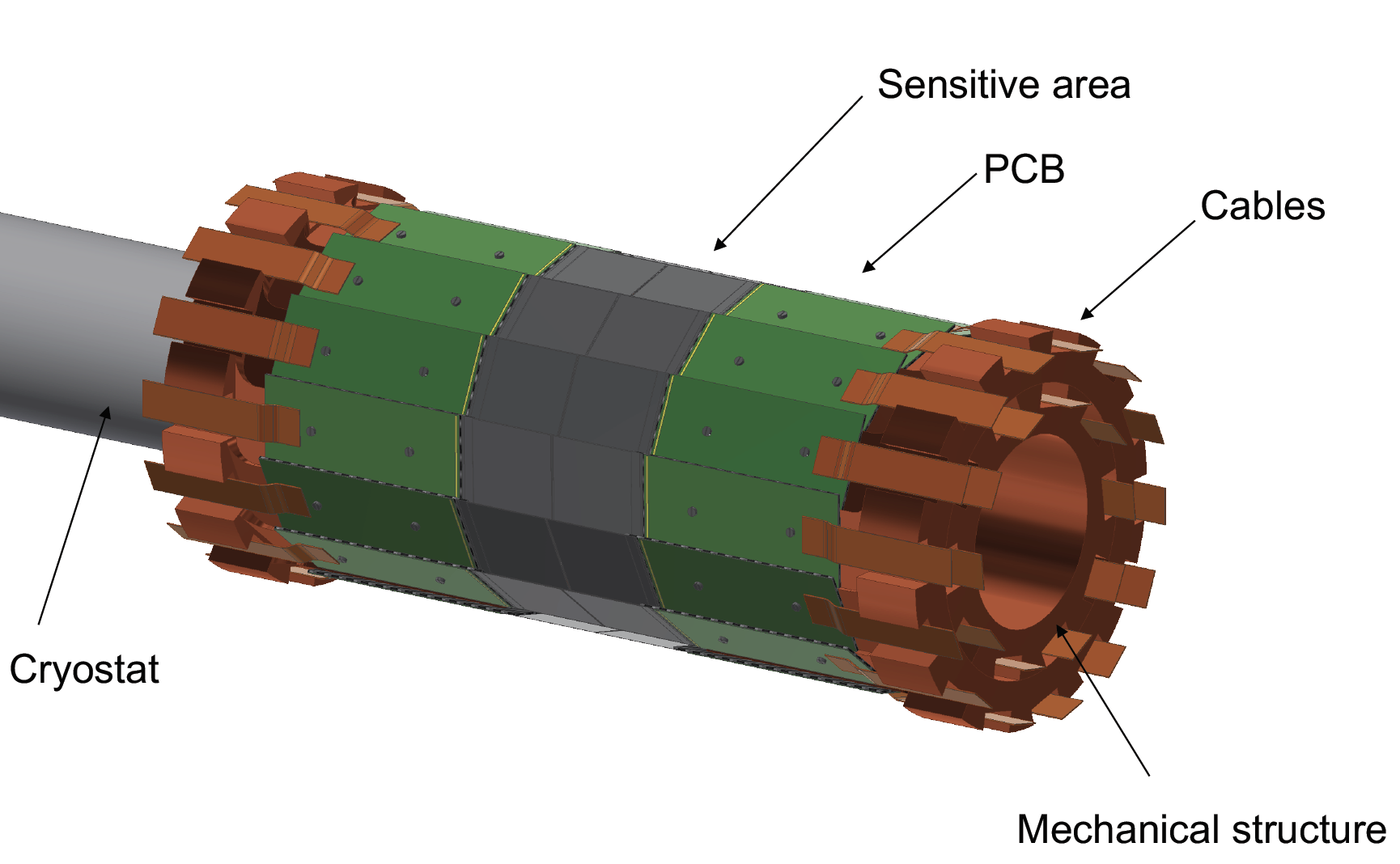}
      \includegraphics[trim=0 0 0 0,clip,width=0.42\textwidth]{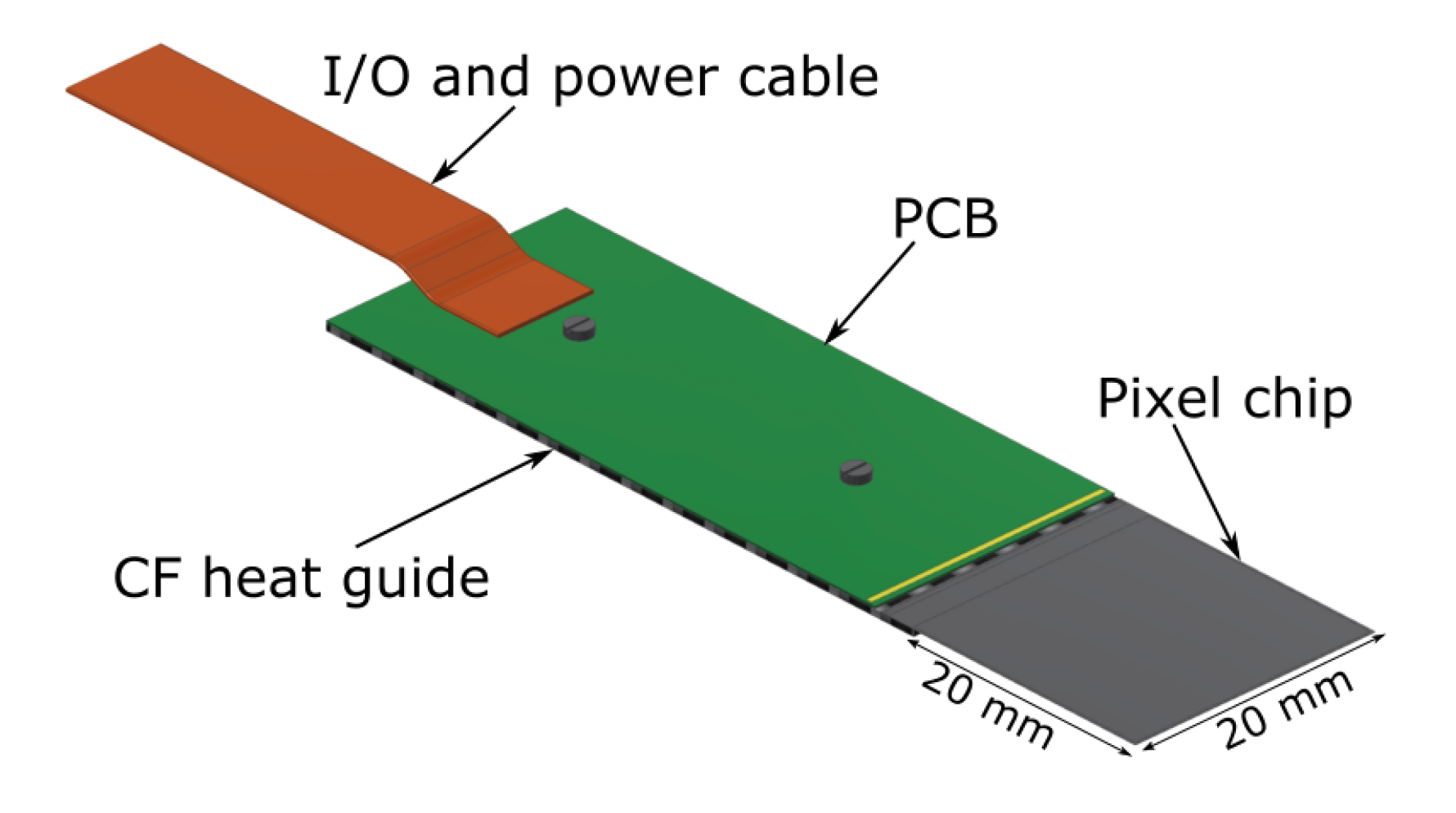}
    \caption{(Left) Drawing of a prototype pixel detector for the general purpose surface muon instrument. The pixel detector is built from two cylindrical layers surrounding the cryostat in which the samples is placed. (Right) Drawing of a pixel detector module. The sensitive element is a \ac{DMAPS} chip of the MuPix family developed in the context of the Mu3e experiment.}
    \label{fig:pixel}
\end{figure}

A concept for a prototype pixel detector for the general purpose surface muon instrument GPS is shown in \autoref{fig:pixel}. The pixel detector is built from two concentric cylinders surrounding the cryostat in which the sample is placed. The cylinder axis is placed perpendicular to the muon beam. The pixel detector is used to measure the trajectory of the incoming muon and the outgoing positron, and to reconstruct the muon decay vertex in the sample. 

Due to the low momentum of the muons ($<30$\,MeV), the spatial resolution of the detector will be dominated by multiple scattering effects and building the detector from minimum material will be crucial. The prototype detector uses a \ac{DMAPS} chip of the MuPix family as a sensitive element and all components needed for readout, power and cooling of the detector are placed outside the active area. The prototype detector is built from 34 modules with a total active area of 136\,cm${^2}$. The radius of the inner and outer pixel layers are 21\,mm and 31\,mm, respectively. The detector acceptance determined for a sample of 100'000 simulated muons with an average momentum of 27\,MeV is 56\%.

The effect of multiple scattering on the resolution has been studied assuming a sensor thickness of 50$\,\mu$m and was found to be about 0.65\,mm at the muon decay point. The possibility of thinning the sensors to thicknesses of less than 50$\,\mu$m is currently being studied. With thinner sensors with a thickness of 30$\,\mu$m the resolution can be improved to less than 0.50\,mm.


\section{Conclusions} \label{sec:conclusions}
Two new high-intensity muon beams HIMB at PSI with up to two orders of magnitude higher and completely unprecedented intensities of surface muons will boost muon science in many ways. Flavour physics is a hot topic in particle physics, with charged lepton flavour searches and muon electromagnetic moment measurements in particular. Such experiments at HIMB will remain at the forefront of research for the next two decades or longer. Other fundamental atomic, nuclear and particle physics experiments will complement these flagship experiments, testing essentially all known interactions with considerable discovery potential for physics beyond the current Standard Model of particle physics. Likewise, in condensed matter physics HIMB will open up completely new regions of experimental phase space, for smaller sample sizes, for faster spatially and time-resolved measurements, and for depth profiling in so far inaccessible regions. All experiments at HIMB will considerably benefit from technological advances in detector technology, target technology and sample environment. Application of novel technologies, in particular of thin and fast tracking detectors and of new muon beam cooling schemes will further push the boundaries of what is presently possible in muon physics. The physics case for HIMB will continue to develop over time and lead to a definition of an initial physics program at this eagerly awaited new facility. The cases presented here already motivate a very strong and unique program for the next two decades.

\section*{Acknowledgements}
\addcontentsline{toc}{section}{Acknowledgements}

The science case presented here is built on the work of numerous
people over many years and is a true community effort. It is a
pleasure to thank all our colleagues at PSI and the external
users for their contributions and support. A special thank you
goes to all speakers and other
\href{https://indico.psi.ch/event/10547/registrations/participants}{participants}
of the \href{https://indico.psi.ch/event/10547}{HIMB Science Case
workshop} of April 2021.

We gratefully acknowledge support by the Swiss National Science Foundation (SNF)
through the grants PCEFP2\_181198, PCEFP2\_181117, PP00P2\_176884, 200441, 197052,
197346; by the European Research Council (ERC) through CoG. 725039, 818053;
by the European Union's Horizon 2020 research and innovation programme through grant 
858199; by the German Research Foundation (DFG) under Project WA 4157/1; 
and by the JSPS Core-to-Core Program, A. Advanced Research Networks
JPJSCCA20180004 and JSPS KAKENHI Grant Number JP21H04991 (Japan).
 

\newpage 
\printbibliography[heading=bibintoc]

@article{Grillenberger:2021kyv,
    author = "Grillenberger, Joachim and Baumgarten, Christian and Seidel, Mike",
    title = "{The High Intensity Proton Accelerator Facility}",
    doi = "10.21468/SciPostPhysProc.5.002",
    journal = "SciPost Phys. Proc.",
    volume = "5",
    pages = "002",
    year = "2021"
}

@article{SciPost:ppPSI,
    editor = "Signer, Adrian and Kirch, Klaus and Hoffman, Cyrus",
    title = "{Particle Physics at PSI}",
    doi = "10.21468/SciPostPhysProc.5",
    journal = "SciPost Phys. Proc.",
    volume = "5",
    year = "2021"
}

@article{RevModPhys.69.1119,
  title = {Heavy-fermion systems studied by \ensuremath{\mu}SR technique},
  author = {Amato, A.},
  journal = {Rev. Mod. Phys.},
  volume = {69},
  issue = {4},
  pages = {1119--1180},
  numpages = {0},
  year = {1997},
  month = {10},
  publisher = {American Physical Society},
  doi = {10.1103/RevModPhys.69.1119},
}

@inbook{doi:10.7566/JPSCP.2.010201,
author = { Elvezio Morenzoni and  Thomas Prokscha and  Hassan Saadaoui
and  Zaher Salman and  Andreas Suter and  Bastian M. Wojek and  Jordan
Baglo and  Ivan Bozovic and  Masrur Hossain and  Robert F. Kiefl 
and  Gennady Logvenov and  Oren Ofer},  
title = {Low-Energy Muons at PSI: Examples of Investigations of
Superconducting Properties in Near-Surface Regions and
Heterostuctures}, 
booktitle = {Proceedings of the International Symposium on Science
Explored by Ultra Slow Muon (USM2013)}, 
chapter = {},
pages = {},
doi = {10.7566/JPSCP.2.010201},
}

@article{AMATO2009606,
title = {Probing the ground state properties of iron-based superconducting pnictides and related systems by muon-spin spectroscopy},
journal = {Physica C: Superconductivity},
volume = {469},
number = {9},
pages = {606-613},
year = {2009},
note = {Superconductivity in Iron-Pnictides},
issn = {0921-4534},
doi = {10.1016/j.physc.2009.03.017},
author = {A. Amato and R. Khasanov and H. Luetkens and H.-H. Klauss},
}

@article{doi:10.1080/08957959.2016.1173690,
author = {R. Khasanov and Z. Guguchia and A. Maisuradze and
D. Andreica and M. Elender and A. Raselli and Z. Shermadini and
T. Goko and F. Knecht and E. Morenzoni and A. Amato}, 
title = {High pressure research using muons at the Paul Scherrer Institute},
journal = {High Pressure Research},
volume = {36},
number = {2},
pages = {140-166},
year  = {2016},
publisher = {Taylor & Francis},
doi = {10.1080/08957959.2016.1173690},
}

@Article{condmat5020042,
AUTHOR = {Z. Guguchia},
TITLE = {Unconventional Magnetism in Layered Transition Metal Dichalcogenides},
JOURNAL = {Condensed Matter},
VOLUME = {5},
YEAR = {2020},
NUMBER = {2},
ARTICLE-NUMBER = {42},
ISSN = {2410-3896},
DOI = {10.3390/condmat5020042}
}

@article{10.3389/fphy.2021.651163,
author={Shang, Tian and Shiroka, Toni},
title={Time-Reversal Symmetry Breaking in Re-Based Superconductors: Recent Developments},
JOURNAL={Frontiers in Physics},
VOLUME={9},
PAGES={270},
YEAR={2021},
DOI={10.3389/fphy.2021.651163},
ISSN={2296-424X},
}

@article{Pifer:1976ia,
    author = "Pifer, A. E. and Bowen, T. and Kendall, K. R.",
    title = "{A High Stopping Density $\mu^+$ Beam}",
    doi = "10.1016/0029-554X(76)90823-5",
    journal = "Nucl. Instrum. Meth.",
    volume = "135",
    pages = "39--46",
    year = "1976"
}

@article{Prokscha:2008zz,
    author = "Prokscha, T. and Morenzoni, E. and Deiters, K. and Foroughi, F. and George, D. and Kobler, R. and Suter, A. and Vrankovic, V.",
    title = "{The new muE4 beam at PSI: A hybrid-type large acceptance channel for the generation of a high intensity surface-muon beam}",
    doi = "10.1016/j.nima.2008.07.081",
    journal = "Nucl. Instrum. Meth. A",
    volume = "595",
    pages = "317--331",
    year = "2008"
}

@article{Adam:2013vqa,
    author = "Adam, J. and others",
    title = "{The MEG detector for ${\mu}^+\to e^+{\gamma}$ decay search}",
    eprint = "1303.2348",
    archivePrefix = "arXiv",
    primaryClass = "physics.ins-det",
    doi = "10.1140/epjc/s10052-013-2365-2",
    journal = "Eur. Phys. J. C",
    volume = "73",
    number = "4",
    pages = "2365",
    year = "2013"
}

@article{Berg:2015wna,
    author = "Berg, F. and others",
    title = "{Target Studies for Surface Muon Production}",
    eprint = "1511.01288",
    archivePrefix = "arXiv",
    primaryClass = "physics.ins-det",
    doi = "10.1103/PhysRevAccelBeams.19.024701",
    journal = "Phys. Rev. Accel. Beams",
    volume = "19",
    number = "2",
    pages = "024701",
    year = "2016"
}

@article{Taqqu:2006mv,
    author = "Taqqu, D.",
    title = "{Compression and extraction of stopped muons}",
    doi = "10.1103/PhysRevLett.97.194801",
    journal = "Phys. Rev. Lett.",
    volume = "97",
    pages = "194801",
    year = "2006"
}

@incollection{Blum2008,
author = {Blum, W. and Riegler, W. and Rolandi, L.},
doi = {10.1007/978-3-540-76684-1_2},
pages = {1--48},
title = {{The Drift of Electrons and Ions in Gases}},
year = {2008}
}

@article{Bao:2014xxa,
    author = "Bao, Yu and others",
    title = "{Muon cooling: longitudinal compression}",
    eprint = "1402.2418",
    archivePrefix = "arXiv",
    primaryClass = "physics.acc-ph",
    doi = "10.1103/PhysRevLett.112.224801",
    journal = "Phys. Rev. Lett.",
    volume = "112",
    number = "22",
    pages = "224801",
    year = "2014"
}

@article{Belosevic:2018fnj,
    author = "Belosevic, I. and others",
    title = "{muCool: A next step towards efficient muon beam compression}",
    eprint = "1811.08332",
    archivePrefix = "arXiv",
    primaryClass = "physics.acc-ph",
    doi = "10.1140/epjc/s10052-019-6932-z",
    journal = "Eur. Phys. J. C",
    volume = "79",
    number = "5",
    pages = "430",
    year = "2019"
}

@article{Antognini:2020uyp,
    author = "Antognini, A. and others",
    title = "{Demonstration of Muon-Beam Transverse Phase-Space Compression}",
    eprint = "2003.11986",
    archivePrefix = "arXiv",
    primaryClass = "physics.acc-ph",
    doi = "10.1103/PhysRevLett.125.164802",
    journal = "Phys. Rev. Lett.",
    volume = "125",
    number = "16",
    pages = "164802",
    year = "2020"
}

@phdthesis{Belosevic:2019bnw,
    author = "Belosevic, Ivana",
    title = "{Simulation and experimental verification of transverse and longitudinal compression of positive muon beams: Towards a novel high-brightness low-energy muon beam-line}",
    doi = "10.3929/ethz-b-000402802",
    school = "Zurich, ETH",
    year = "2019"
}

@article{Bae:2018atj,
    author = "Bae, S. and others",
    title = "{First muon acceleration using a radio frequency accelerator}",
    eprint = "1803.07891",
    archivePrefix = "arXiv",
    primaryClass = "physics.acc-ph",
    doi = "10.1103/PhysRevAccelBeams.21.050101",
    journal = "Phys. Rev. Accel. Beams",
    volume = "21",
    number = "5",
    pages = "050101",
    year = "2018"
}

@article{osti_335199,
title = {CW RFQ fabrication and engineering},
author = {Schrage, D. and Young, L. and Roybal, P.},
doi = {},
url = {https://www.osti.gov/biblio/335199}, 
place = {United States},
year = {1998},
month = {12}
}

@book{ESS_TechnicalDesignReport,
author = {Carlile, Colin and Miyamoto, Ryoichi and Pahlsson, A. and Trojer, M. and Weisend II, J.G. and Ainalem, Marie-Louise and Andersen, Ken and Batkov, K. and Carlsson, Patrik and Ene, D. and Heden, B. and Hedin, K. and Jackson, Andrew and Jacobsson, P. and Kirstein, Oliver and Lanfranco, Giobatta and Lee, Yongjoong and Lindroos, Mats and Rebec, J. and Zugazaga, A.},
year = {2013},
month = {04},
pages = {},
title = {European Spallation Source Technical Design Report},
isbn = {978-91-980173-2-8},
doi = {10.13140/RG.2.1.2040.6483/1}
}

@article{Antognini:2011ei,
    author = "Antognini, A. and others",
    title = "{Muonium emission into vacuum from mesoporous thin films at cryogenic temperatures}",
    eprint = "1112.4887",
    archivePrefix = "arXiv",
    primaryClass = "physics.atom-ph",
    doi = "10.1103/PhysRevLett.108.143401",
    journal = "Phys. Rev. Lett.",
    volume = "108",
    pages = "143401",
    year = "2012"
}

@article{Crivelli:2018vfe,
    author = "Crivelli, P.",
    title = "{The Mu-MASS (MuoniuM lAser SpectroScopy) experiment}",
    eprint = "1811.00310",
    archivePrefix = "arXiv",
    primaryClass = "physics.atom-ph",
    doi = "10.1007/s10751-018-1525-z",
    journal = "Hyperfine Interact.",
    volume = "239",
    number = "1",
    pages = "49",
    year = "2018"
}

@article{MAGE:2018wxk,
    author = "Antognini, A. and Kaplan, Daniel M. and Kirch, Klaus and Knecht, Andreas and Mancini, Derrick C. and Phillips, James D. and Phillips, Thomas J. and Reasenberg, Robert D. and Roberts, Thomas J. and Soter, Anna",
    collaboration = "MAGE",
    title = "{Studying Antimatter Gravity with Muonium}",
    eprint = "1802.01438",
    archivePrefix = "arXiv",
    primaryClass = "physics.ins-det",
    reportNumber = "IIT-CAPP-17-5",
    doi = "10.3390/atoms6020017",
    journal = "Atoms",
    volume = "6",
    number = "2",
    pages = "17",
    year = "2018"
}

@article{Chislett:2016jau,
%    author = "Chislett, Rebecca",
%    editor = "D'Ambrosio, G. and Iacovacci, M. and Passera, M. and Venanzoni, G. and Massarotti, P. and Mastroianni, S.",
%    collaboration = "Muon g-2",
%    title = "{The muon EDM in the g-2 experiment at Fermilab}",
%    doi = "10.1051/epjconf/201611801005",
%    journal = "EPJ Web Conf.",
%    volume = "118",
%    pages = "01005",
%    year = "2016"
%}

@article{Adelmann:2010zz,
%    author = "Adelmann, A. and Kirch, K. and Onderwater, C. J. G. and Schietinger, T.",
%    title = "{Compact storage ring to search for the muon electric dipole moment}",
%    doi = "10.1088/0954-3899/37/8/085001",
%    journal = "J. Phys. G",
%    volume = "37",
%    pages = "085001",
%    year = "2010"
%}

@article{Adelmann:2021udj,
%    author = "Adelmann, A. and others",
%    title = "{Search for a muon EDM using the frozen-spin technique}",
%    eprint = "2102.08838",
%    archivePrefix = "arXiv",
%    primaryClass = "hep-ex",
%    month = "2",
%    year = "2021"
%}

@article{Baldini:2018nnn,
    author = "Baldini, A. M. and others",
    collaboration = "MEG II",
    title = "{The design of the MEG II experiment}",
    eprint = "1801.04688",
    archivePrefix = "arXiv",
    primaryClass = "physics.ins-det",
    doi = "10.1140/epjc/s10052-018-5845-6",
    journal = "Eur. Phys. J. C",
    volume = "78",
    number = "5",
    pages = "380",
    year = "2018"
}

@article{Pontecorvo:1957cp,
    author = "Pontecorvo, B.",
    title = "{Mesonium and anti-mesonium}",
    journal = "Sov. Phys. JETP",
    volume = "6",
    pages = "429",
    year = "1957"
}

@article{Hughes:1960zz,
    author = "Hughes, V.W. and McColm, D.W. and Ziock, K. and Prepost, R.",
    title = "{Formation of Muonium and Observation of its Larmor Precession}",
    journal = "Phys. Rev. Lett.",
    volume = "5",
    pages = "63--65",
    year = "1960"
}

@article{Amato:1969cp,
    author = "Amato, J.J. and Crane, P. and Hughes, V.W. and Rothberg, J.E. and Thompson, P.A.",
    title = "{Search for muonium-antimuonium conversion}",
    journal = "Phys. Rev. Lett.",
    volume = "21",
    pages = "1709--1712",
    year = "1968"
}

@article{Bolton:1981ug,
    author = "Bolton, P.R. and others",
    title = "{Observation of Muonium in Vacuum}",
    doi = "10.1103/PhysRevLett.47.1441",
    journal = "Phys. Rev. Lett.",
    volume = "47",
    pages = "1441--1444",
    year = "1981"
}

@article{Marshall:1981mk,
    author = "Marshall, Glen M. and Warren, J.B. and Oram, C.J. and Kiefl, R.F.",
    title = "{A Search for Muonium to Anti-muonium Conversion}",
    reportNumber = "TRI-PP-81-23",
    journal = "Phys. Rev. D",
    volume = "25",
    pages = "1174",
    year = "1982"
}

@article{Ni:1987iza,
    author = "Ni, B. and others",
    title = "{Search for Spontaneous Conversion of Muonium to Antimuonium}",
    journal = "Phys. Rev. Lett.",
    volume = "59",
    number = "24",
    pages = "2716--2719",
    year = "1987"
}

@article{Ni:1989zb,
    author = "Ni, B. and others",
    title = "{Search for spontaneous conversion of muonium to anti-muonium}",
    journal = "Phys. Rev. D",
    volume = "48",
    pages = "1976--1989",
    year = "1993"
}

@article{Matthias:1991fw,
    author = "Matthias, B.E. and others",
    title = "{New search for the spontaneous conversion of muonium to anti-muonium}",
    journal = "Phys. Rev. Lett.",
    volume = "66",
    pages = "2716--2719",
    year = "1991"
}

@article{Huber:1989by,
    author = "Huber, T.M. and others",
    title = "{Search for Mixing of Muonium ($\mu^+ e^-$) and Anti-muonium ($\mu^- e^+$)}",
    reportNumber = "TRI-PP-89-64",
    journal = "Phys. Rev. D",
    volume = "41",
    pages = "2709--2725",
    year = "1990"
}

@article{Abela:1996dm,
    author = "Abela, R. and others",
    title = "{Improved upper limit on muonium to anti-muonium conversion}",
    eprint = "nucl-ex/9805005",
    archivePrefix = "arXiv",
    reportNumber = "HD-PI--96-04",
    journal = "Phys. Rev. Lett.",
    volume = "77",
    pages = "1950--1953",
    year = "1996"
}

@article{Willmann:1998gd,
    author = "Willmann, L. and others",
    title = "{New bounds from searching for muonium to anti-muonium conversion}",
    eprint = "hep-ex/9807011",
    archivePrefix = "arXiv",
    reportNumber = "UHD-PI-MY-9812",
    journal = "Phys. Rev. Lett.",
    volume = "82",
    pages = "49--52",
    year = "1999"
}

@article{GEANT4:2002zbu,
    author = "Agostinelli, S. and others",
    collaboration = "GEANT4",
    title = "{GEANT4--a simulation toolkit}",
    reportNumber = "SLAC-PUB-9350, FERMILAB-PUB-03-339, CERN-IT-2002-003",
    doi = "10.1016/S0168-9002(03)01368-8",
    journal = "Nucl. Instrum. Meth. A",
    volume = "506",
    pages = "250--303",
    year = "2003"
}

@article{Kastli:2007pe,
    author = "Kastli, H. Chr. and others",
    editor = "Ambrosi, Giovanni and Bilei, Gian M. and Fano, Livio and Passeri, Daniele and Santocchia, Attilio and Zuccon, Paolo",
    title = "{CMS barrel pixel detector overview}",
    eprint = "physics/0702182",
    archivePrefix = "arXiv",
    doi = "10.1016/j.nima.2007.07.058",
    journal = "Nucl. Instrum. Meth. A",
    volume = "582",
    pages = "724--727",
    year = "2007"
}

@article{CMS:2008xjf,
    author = "Chatrchyan, S. and others",
    collaboration = "CMS",
    title = "{The CMS Experiment at the CERN LHC}",
    doi = "10.1088/1748-0221/3/08/S08004",
    journal = "JINST",
    volume = "3",
    pages = "S08004",
    year = "2008"
}

@article{CMSTrackerGroup:2020edz,
    author = "Adam, W. and others",
    collaboration = "CMS Tracker Group",
    title = "{The CMS Phase-1 Pixel Detector Upgrade}",
    eprint = "2012.14304",
    archivePrefix = "arXiv",
    primaryClass = "physics.ins-det",
    doi = "10.1088/1748-0221/16/02/P02027",
    journal = "JINST",
    volume = "16",
    number = "02",
    pages = "P02027",
    year = "2021"
}

@article{Jungmann-Smith:pp5080,
author = "Jungmann-Smith, J. H. and Bergamaschi, A. and Br{\"{u}}ckner, M. and Cartier, S. and Dinapoli, R. and Greiffenberg, D. and Huthwelker, T. and Maliakal, D. and Mayilyan, D. and Medjoubi, K. and Mezza, D. and Mozzanica, A. and Ramilli, M. and Ruder, Ch. and Sch{\"{a}}dler, L. and Schmitt, B. and Shi, X. and Tinti, G.",
title = "{Towards hybrid pixel detectors for energy-dispersive or soft X-ray photon science}",
journal = "Journal of Synchrotron Radiation",
year = "2016",
volume = "23",
number = "2",
pages = "385--394",
doi = {10.1107/S1600577515023541},
}

@article{doi:10.1080/08940886.2018.1528428,
author = {A. Bergamaschi and M. Andrä and R. Barten and C. Borca and M. Brückner and S. Chiriotti and R. Dinapoli and E. Fröjdh and D. Greiffenberg and T. Huthwelker and A. Kleibert and M. Langer and M. Lebugle and C. Lopez-Cuenca and D. Mezza and A. Mozzanica and J. Raabe and S. Redford and C. Ruder and V. Scagnoli and B. Schmitt and X. Shi and U. Staub and D. Thattil and G. Tinti and C. F. Vaz and S. Vetter and J. Vila-Comamala and J. Zhang},
title = {The MÖNCH Detector for Soft X-ray, High-Resolution, and Energy Resolved Applications},
journal = {Synchrotron Radiation News},
volume = {31},
number = {6},
pages = {11-15},
year  = {2018},
publisher = {Taylor & Francis},
doi = {10.1080/08940886.2018.1528428},
}

@article{Allahgholi:gb5080,
author = "Allahgholi, Aschkan and Becker, Julian and Delfs, Annette and Dinapoli, Roberto and Goettlicher, Peter and Greiffenberg, Dominic and Henrich, Beat and Hirsemann, Helmut and Kuhn, Manuela and Klanner, Robert and Klyuev, Alexander and Krueger, Hans and Lange, Sabine and Laurus, Torsten and Marras, Alessandro and Mezza, Davide and Mozzanica, Aldo and Niemann, Magdalena and Poehlsen, Jennifer and Schwandt, Joern and Sheviakov, Igor and Shi, Xintian and Smoljanin, Sergej and Steffen, Lothar and Sztuk-Dambietz, Jolanta and Trunk, Ulrich and Xia, Qingqing and Zeribi, Mourad and Zhang, Jiaguo and Zimmer, Manfred and Schmitt, Bernd and Graafsma, Heinz",
title = "{The Adaptive Gain Integrating Pixel Detector at the~European XFEL}",
journal = "Journal of Synchrotron Radiation",
year = "2019",
volume = "26",
number = "1",
pages = "74--82",
doi = {10.1107/S1600577518016077},
}

@article{Broennimann:2005qv,
    author = "Broennimann, Ch. and Glaus, F. and Gobrecht, J. and Heising, S. and Horisberger, M. and Horisberger, R. and Kastli, H. C. and Lehmann, J. and Rohe, T. and Streuli, S.",
    editor = "Grosse-Knetter, J. and Krueger, H. and Wermes, N.",
    title = "{Development of an indium bump bond process for silicon pixel detectors at PSI}",
    eprint = "physics/0510021",
    archivePrefix = "arXiv",
    doi = "10.1016/j.nima.2006.05.011",
    journal = "Nucl. Instrum. Meth. A",
    volume = "565",
    pages = "303--308",
    year = "2006"
}

@article{Kenney:1994jjz,
    author = "Kenney, Christopher J. and Parker, Sherwood I. and Peterson, Vincent Z. and Snoeys, Walter J. and Plummer, James D. and Aw, Chye Huat",
    title = "{A Prototype monolithic pixel detector}",
    reportNumber = "UH-511-766-93",
    doi = "10.1016/0168-9002(94)91411-7",
    journal = "Nucl. Instrum. Meth. A",
    volume = "342",
    pages = "59--77",
    year = "1994"
}

@article{Peric:2007zz,
    author = "Peric, Ivan",
    editor = "Ambrosi, Giovanni and Bilei, Gian M. and Fano, Livio and Passeri, Daniele and Santocchia, Attilio and Zuccon, Paolo",
    title = "{A novel monolithic pixelated particle detector implemented in high-voltage CMOS technology}",
    doi = "10.1016/j.nima.2007.07.115",
    journal = "Nucl. Instrum. Meth. A",
    volume = "582",
    pages = "876--885",
    year = "2007"
}

@article{Turchetta:2001dy,
    author = "Turchetta, R. and others",
    title = "{A monolithic active pixel sensor for charged particle tracking and imaging using standard VLSI CMOS technology}",
    doi = "10.1016/S0168-9002(00)00893-7",
    journal = "Nucl. Instrum. Meth. A",
    volume = "458",
    pages = "677--689",
    year = "2001"
}

@article{Peric:2013cka,
    author = "Peri\'c, Ivan and others",
    editor = "Unno, Yoshinobu and Toyokawa, Hidenori and Arai, Yasuro and Hatsui, Takaki",
    title = "{High-voltage pixel detectors in commercial CMOS technologies for ATLAS, CLIC and Mu3e experiments}",
    doi = "10.1016/j.nima.2013.05.006",
    journal = "Nucl. Instrum. Meth. A",
    volume = "731",
    pages = "131--136",
    year = "2013"
}

@article{Schoning:2020zed,
    author = {Sch\"oning, A. and others},
    title = "{MuPix and ATLASPix -- Architectures and Results}",
    eprint = "2002.07253",
    archivePrefix = "arXiv",
    primaryClass = "physics.ins-det",
    doi = "10.22323/1.373.0024",
    journal = "PoS",
    volume = "Vertex2019",
    pages = "024",
    year = "2020"
}

@article{Karshenboim:2008zj,
    author = "Karshenboim, Savely G.",
    title = "{A Constraint on antigravity of antimatter from precision spectroscopy of simple atoms}",
    eprint = "0811.1008",
    archivePrefix = "arXiv",
    primaryClass = "gr-qc",
    doi = "10.1134/S1063773709100028",
    journal = "Astron. Lett.",
    volume = "35",
    pages = "663",
    year = "2009"}

@article{2014-Vargas,
  title = {Laboratory tests of Lorentz and $CPT$ symmetry with muons},
  author = {Gomes, Andr\'e H. and Kosteleck\'y, V. Alan and Vargas, Arnaldo J.},
  journal = {Phys. Rev. D},
  volume = {90},
  issue = {7},
  pages = {076009},
  numpages = {24},
  year = {2014},
  month = {10},
  publisher = {American Physical Society},
  doi = {10.1103/PhysRevD.90.076009},
}

@article{2016-Jung,
author = {Jungmann ,Klaus P.},
title = {Precision Muonium Spectroscopy},
journal = {Journal of the Physical Society of Japan},
volume = {85},
number = {9},
pages = {091004},
year = {2016},
doi = {10.7566/JPSJ.85.091004},
}

@article{Frugiuele:2019drl,
    author = "Frugiuele, Claudia and P\'erez-R\'\i{}os, Jes\'us and Peset, Clara",
    title = "{Current and future perspectives of positronium and muonium spectroscopy as dark sectors probe}",
    eprint = "1902.08585",
    archivePrefix = "arXiv",
    primaryClass = "hep-ph",
    reportNumber = "TUM-1188/19",
    doi = "10.1103/PhysRevD.100.015010",
    journal = "Phys. Rev. D",
    volume = "100",
    number = "1",
    pages = "015010",
    year = "2019"
}

@article{2020-MUSEUM,
    title="{New precise spectroscopy of the hyperfine structure in muonium with a high-intensity pulsed muon beam}",
    author={S. Kanda and Y. Fukao and Y. Ikedo and K. Ishida and M. Iwasaki and D. Kawall and N. Kawamura and K. M. Kojima and N. Kurosawa and Y. Matsuda and T. Mibe and Y. Miyake and S. Nishimura and N. Saito and Y. Sato and S. Seo and K. Shimomura and P. Strasser and K. S. Tanaka and T. Tanaka and H. A. Torii and A. Toyoda and Y. Ueno},
    year={2020},
    eprint={2004.05862},
    archivePrefix={arXiv},
    primaryClass={hep-ex}
}

@article{Delaunay:2021uph,
    author = "Delaunay, C\'edric and Ohayon, Ben and Soreq, Yotam",
    title = "{Towards an independent determination of muon g-2 from muonium spectroscopy}",
    eprint = "2106.11998",
    archivePrefix = "arXiv",
    primaryClass = "hep-ph",
    month = "6",
    year = "2021"
}

@article{Eides:2018rph,
    author = "Eides, Michael I.",
    title = "{Hyperfine Splitting in Muonium: Accuracy of the Theoretical Prediction}",
    eprint = "1812.10881",
    archivePrefix = "arXiv",
    primaryClass = "hep-ph",
    doi = "10.1016/j.physletb.2019.06.011",
    journal = "Phys. Lett. B",
    volume = "795",
    pages = "113--116",
    year = "2019"
}

@book{Yaouanc:2011Book,
        address = {Oxford, {UK}},
        title = {Muon Spin Rotation, Relaxation, and Resonance: Applications to Condensed Matter},
        publisher = {Oxford University Press},
        author = {Yaouanc, {A.} and Dalmas de R\'{e}otier, P.},
        year = {2011},
}

@article {Fleming:2011Science,
	author = {Fleming, Donald G. and Arseneau, Donald J. and Sukhorukov, Oleksandr and Brewer, Jess H. and Mielke, Steven L. and Schatz, George C. and Garrett, Bruce C. and Peterson, Kirk A. and Truhlar, Donald G.},
	title = {Kinetic Isotope Effects for the Reactions of Muonic Helium and Muonium with H$_2$},
	volume = {331},
	number = {6016},
	pages = {448--450},
	year = {2011},
	doi = {10.1126/science.1199421},
	publisher = {American Association for the Advancement of Science},
	journal = {Science}
}

@article{Shafir:2005prb,
  title = {Demonstrating multibit magnetic memory in the ${\mathrm{Fe}}_{8}$ high-spin molecule by muon spin rotation},
  author = {Shafir, Oren and Keren, Amit and Maegawa, Satoru and Ueda, Miki and Amato, Alex and Baines, Chris},
  journal = {Phys. Rev. B},
  volume = {72},
  pages = {092410},
  year = {2005},
  publisher = {American Physical Society},
  doi = {10.1103/PhysRevB.72.092410}
}

@article{Tesi:2018ChemComm,
author ="Tesi, Lorenzo and Salman, Zaher and Cimatti, Irene and Pointillart, Fabrice and Bernot, Kevin and Mannini, Matteo and Sessoli, Roberta",
title  ="Isotope effects on the spin dynamics of single-molecule magnets probed using muon spin spectroscopy",
journal  ="Chem. Commun.",
year  ="2018",
volume  ="54",
pages  ="7826-7829",
publisher  ="The Royal Society of Chemistry",
doi  ="10.1039/C8CC04703K"
}

@article{Kiefl:2016acsnano,
author = {Kiefl, Evan and Mannini, Matteo and Bernot, Kevin and Yi, Xiaohui and Amato, Alex and Leviant, Tom and Magnani, Agnese and Prokscha, Thomas and Suter, Andreas and Sessoli, Roberta and Salman, Zaher},
title = {Robust Magnetic Properties of a Sublimable Single-Molecule Magnet},
journal = {ACS Nano},
volume = {10},
pages = {5663-5669},
year = {2016},
doi = {10.1021/acsnano.6b01817}
}

@article{Forslund:2019SciRep,
author = {Forslund, Ola Kenji and Andreica, Daniel and Sassa, Yasmine and Nozaki, Hiroshi and
Umegaki, Izumi and Nocerino, Elisabetta and Jonsson, Viktor and Tjernberg, Oscar and Guguchia, Zurab and Shermadini, Zurab and Khasanov, Rustem and Isobe, Masahiko and Takagi, Hidenori and Ueda, Yutaka and Sugiyama, Jun and Mansson, Martin},
title = {Magnetic phase diagram of K$_2$Cr$_8$O$_{16}$ clarified by high-pressure muon spin spectroscopy},
journal = {Scientific Reports},
volume = {9},
pages = {1141},
year = {2019},
doi = {10.1038/s41598-018-37844-5}
}

@article{Lancaster:2013PhysScrip,
	doi = {10.1088/0031-8949/88/06/068506},
	year = 2013,
	publisher = {{IOP} Publishing},
	volume = {88},
	number = {6},
	pages = {068506},
	author = {Tom Lancaster and Stephen J Blundell and Francis L Pratt},
	title = {Another dimension: investigations of molecular magnetism using muon{\textendash}spin relaxation},
	journal = {Physica Scripta}
}

@article{Tustain:2020NPJQM,
author = {Tustain, Katherine and Ward-O’Brien, Brendan and Bert, Fabrice and Han, Tianheng and Luetkens, Hubertus and Lancaster, Tom and Huddart, Benjamin M. and Baker, Peter J. and Clark, Lucy},
title = {From magnetic order to quantum disorder in the Zn-barlowite series of S = 1/2 kagomé antiferromagnets},
journal = {npj Quantum Materials},
volume = {5},
pages = {74},
year = {2020},
doi = {10.1038/s41535-020-00276-4}
}

@article{Gao:2019NatPhys,
author = {Gao, Bin and Chen, Tong and Tam, David W. and Huang, Chien-Lung and Sasmal, Kalyan and Adroja, Devashibhai T. and Ye, Feng and Cao, Huibo and Sala, Gabriele and Stone, Matthew B. and Baines, Christopher and Verezhak, Joel A. T. and Hu, Haoyu and Chung, Jae-Ho and Xu, Xianghan and Cheong, Sang-Wook and Nallaiyan, Manivannan and Spagna, Stefano and Maple, M. Brian and Nevidomskyy, Andriy H. and Morosan, Emilia and Chen, Gang and Dai, Pengchen},
title = {Experimental signatures of a three-dimensional quantum spin liquid in effective spin-1/2 Ce$_2$Zr$_2$O$_7$ pyrochlore},
journal = {Nature Physics},
volume = {15},
pages = {1052-1057},
year = {2019},
doi = {10.1038/s41567-019-0577-6}
}

@article{Pratt:2016JSPS,
author = {Pratt ,Francis},
title = {Superconductivity and Magnetism in Organic Materials Studied with $\mu$SR},
journal = {Journal of the Physical Society of Japan},
volume = {85},
pages = {091008},
year = {2016},
doi = {10.7566/JPSJ.85.091008}
}

@article{Wang:2016JPSJ,
author = {Wang ,Ke and Schulz ,Leander and Willis ,Maureen and Zhang ,Sijie and Misquitta ,Alston J. and Drew ,Alan J.},
title = {Spintronic and Electronic Phenomena in Organic Molecules Measured with $\mu$SR},
journal = {Journal of the Physical Society of Japan},
volume = {85},
pages = {091011},
year = {2016},
doi = {10.7566/JPSJ.85.091011}
}

@article{Pratt:JPhysCondMatt2004,
	doi = {10.1088/0953-8984/16/40/019},
	year = 2004,
	volume = {16},
	pages = {S4779--S4796},
	author = {F L Pratt},
	title = {Muon spin relaxation as a probe of electron motion in conducting polymers},
	journal = {Journal of Physics: Condensed Matter}
}

@article{McKenzie:2013PRE,
  title = {Muoniated spin probes in the discotic liquid crystal HHTT: Rapid electron spin relaxation in the hexagonal columnar and isotropic phases},
  author = {McKenzie, Iain and Cammidge, Andrew N. and Gopee, Hemant and Dilger, Herbert and Scheuermann, Robert and Stoykov, Alexey and Jayasooriya, Upali A.},
  journal = {Phys. Rev. E},
  volume = {87},
  pages = {012504},
  numpages = {8},
  year = {2013},
  publisher = {American Physical Society},
  doi = {10.1103/PhysRevE.87.012504}
}

@article{Guguchia:2020NatComm,
author = {Guguchia, Z. and Verezhak, J. A. T. and Gawryluk, D. J. and Tsirkin, S. S. and Yin, J.-X. and Belopolski, I. and Zhou, H. and Simutis, G. and Zhang, S.-S. and Cochran, T. A. and Chang, G. and Pomjakushina, E. and Keller, L. and Skrzeczkowska, Z. and Wang, Q. and Lei, H. C. and Khasanov, R. and Amato, A. and Jia, S. and Neupert, T. and Luetkens, H. and Hasan, M. Z.},
title = {Tunable anomalous Hall conductivity through volume-wise magnetic competition in a topological kagomé magnet},
journal = {Nature Communications},
volume = {11},
pages = {559},
year = {2020},
doi = {0.1038/s41467-020-14325-w}
}

@article{Guguchia:2019NPJQM,
author = {Guguchia, Zurab and Gawryluk, Dariusz J. and Brzezinska, Marta and Tsirkin, Stepan S. and Khasanov, Rustem and Pomjakushina, Ekaterina and von Rohr, Fabian O. and Verezhak, Joel A. T. and Hasan, M. Zahid and Neupert, Titus and Luetkens, Hubertus and Amato, Alex},
title = {Nodeless superconductivity and its evolution with pressure in the layered dirac semimetal 2M-WS$_2$},
journal = {npj Quantum Materials},
volume = {4},
pages = {50},
year = {2019},
doi = {10.1038/s41535-019-0189-5}
}

@article{Krieger:2020PRL,
  title = {Proximity-Induced Odd-Frequency Superconductivity in a Topological Insulator},
  author = {Krieger, Jonas A. and Pertsova, Anna and Giblin, Sean R. and D\"obeli, Max and Prokscha, Thomas and Schneider, Christof W. and Suter, Andreas and Hesjedal, Thorsten and Balatsky, Alexander V. and Salman, Zaher},
  journal = {Phys. Rev. Lett.},
  volume = {125},
  pages = {026802},
  numpages = {6},
  year = {2020},
  publisher = {American Physical Society},
  doi = {10.1103/PhysRevLett.125.026802}
}

@article{Krieger:2019PRB,
  title = {Do topology and ferromagnetism cooperate at the $\mathrm{EuS}/{\mathrm{Bi}}_{2}{\mathrm{Se}}_{3}$ interface?},
  author = {Krieger, J. A. and Ou, Y. and Caputo, M. and Chikina, A. and D\"obeli, M. and Husanu, M.-A. and Keren, I. and Prokscha, T. and Suter, A. and Chang, Cui-Zu and Moodera, J. S. and Strocov, V. N. and Salman, Z.},
  journal = {Phys. Rev. B},
  volume = {99},
  issue = {6},
  pages = {064423},
  numpages = {10},
  year = {2019},
  publisher = {American Physical Society},
  doi = {10.1103/PhysRevB.99.064423}
}

@article{Alberto:2018PhysRevMat,
  title = {Slow-muon study of quaternary solar-cell materials: Single layers and $p\ensuremath{-}n$ junctions},
  author = {Alberto, H. V. and Vil\~ao, R. C. and Vieira, R. B. L. and Gil, J. M. and Weidinger, A. and Sousa, M. G. and Teixeira, J. P. and da Cunha, A. F. and Leit\~ao, J. P. and Salom\'e, P. M. P. and Fernandes, P. A. and T\"orndahl, T. and Prokscha, T. and Suter, A. and Salman, Z.},
  journal = {Phys. Rev. Materials},
  volume = {2},
  pages = {025402},
  numpages = {11},
  year = {2018},
  doi = {10.1103/PhysRevMaterials.2.025402}
}

@article{Curado:2020ApplMatTod,
title = {Front passivation of Cu(In,Ga)Se2 solar cells using Al2O3: Culprits and benefits},
journal = {Applied Materials Today},
volume = {21},
pages = {100867},
year = {2020},
doi = {10.1016/j.apmt.2020.100867},
author = {M.A. Curado and J.P. Teixeira and M. Monteiro and E.F.M. Ribeiro and R.C. Vilão and H.V. Alberto and J.M.V. Cunha and T.S. Lopes and K. Oliveira and O. Donzel-Gargand and A. Hultqvist and S. Calderon and M.A. Barreiros and W. Chiappim and J.P. Leitão and A.G. Silva and T. Prokscha and C. Vinhais and P.A. Fernandes and P.M.P. Salomé}
}

@article{Woerle:2019PRB,
  title = {Interaction of low-energy muons with defect profiles in proton-irradiated Si and $4H$-SiC},
  author = {Woerle, Judith and Prokscha, Thomas and Hall\'en, Anders and Grossner, Ulrike},
  journal = {Phys. Rev. B},
  volume = {100},
  pages = {115202},
  year = {2019},
  doi = {10.1103/PhysRevB.100.115202}
}

@article{Woerle:2020PRAppl,
  title = {Muon Interaction with Negative-$U$ and High-Spin-State Defects: Differentiating Between $\mathrm{C}$ and $\mathrm{Si}$ Vacancies in $4H$-$\mathrm{Si}\mathrm{C}$},
  author = {Woerle, J. and Bathen, M.E. and Prokscha, T. and Galeckas, A. and Ayedh, H.M. and Vines, L. and Grossner, U.},
  journal = {Phys. Rev. Applied},
  volume = {14},
  pages = {054053},
  year = {2020},
  publisher = {American Physical Society},
  doi = {10.1103/PhysRevApplied.14.054053}
}

@article{Prokscha:2020PRAppl,
  title = {Direct Observation of Hole Carrier-Density Profiles and Their Light-Induced Manipulation at the Surface of $\mathrm{Ge}$},
  author = {Prokscha, T. and Chow, K.-H. and Salman, Z. and Stilp, E. and Suter, A.},
  journal = {Phys. Rev. Applied},
  volume = {14},
  pages = {014098},
  year = {2020},
  doi = {10.1103/PhysRevApplied.14.014098}
}

@ARTICLE{Bakule:2004ContPhys,
   author = "P. Bakule and E. Morenzoni",
   title = "Generation and applications of slow polarized muons",
   year = "2004",
   doi = "10.1080/00107510410001676803",
   journal = "Contemp. \ Phys.",
   volume = "45",
   pages = "203"
}

@ARTICLE{Morenzoni:2004JPCondMatt,
   author = "E. Morenzoni and T. Prokscha and A. Suter and H. Luetkens and R. Khasanov",
   title = "Nano-scale thin film investigations with slow polarized muons",
   year = "2004",
   doi = "10.1088/0953-8984/16/40/010",
   journal = "J. \ Phys.: \ Cond. \ Matt.",
   volume = "16",
   pages = "S4583"
}

@article{Prokscha:2014DepthDependence,
	title = {Depth dependence of the ionization energy of shallow hydrogen states in {ZnO} and {CdS}},
	volume = {90},
	doi = {10.1103/PhysRevB.90.235303},
	journal = {Physical Review B},
	author = {Prokscha, T. and Luetkens, H. and Morenzoni, E. and Nieuwenhuys, G. J. and Suter, A. and D\"obeli, M. and Horisberger, M. and Pomjakushina, E.},
	year = {2014},
	pages = {235303}
}

@article{Zhou:2021simulation,
      title="{Simulation studies for upgrading the surface muon beamline $\mu$E4 at PSI}", 
      author={Luping Zhou and Xiaojie Ni and Zaher Salman and Andreas Suter and Jingyu Tang and Vjeran Vrankovic and Thomas Prokscha},
      year={2021},
      eprint={2108.04986},
      archivePrefix={arXiv},
      primaryClass={physics.acc-ph}
}

@article{Yamazaki:2020muontransmission,
	title = {Muon {Cyclotron} for {Transmission} {Muon} {Microscope}},
	volume = {Cyclotrons2019},
	doi = {10.18429/JACOW-CYCLOTRONS2019-TUP024},
	journal = {Proceedings of the 22nd International Conference on Cyclotrons and their Applications},
	author = {Yamazaki, Takayuki and Adachi, Toshikazu and Goto, Akira and Kumata, Yukio and Kusuoka, Shinya and Miyake, Yasuhiro and Nagatani, Yukinori and Ohnishi, Jun-Ichi and Onda, Takashi and Tsutsui, Hiroshi},
	collaborator = {Lowry (Ed.), Conradie and Garrett (Ed.), John, De Villiers and RW (Ed.), Volker, Schaa},
	year = {2020}
}

@article{aqeel:2017probing,
	title = {Probing current-induced magnetic fields in {Au}{\textbar}{YIG} heterostructures with low-energy muon spin spectroscopy},
	volume = {110},
	doi = {10.1063/1.4975487},
	journal = {Applied Physics Letters},
	author = {Aqeel, A. and Vera-Marun, I. J. and Salman, Z. and Prokscha, T. and Suter, A. and van Wees, B. J. and Palstra, T. T. M.},
	year = {2017},
	pages = {062409}
}

@article{moorsom:reversible2020,
	title = {Reversible spin storage in metal oxide—fullerene heterojunctions},
	volume = {6},
	doi = {10.1126/sciadv.aax1085},
	urldate = {2020-03-21},
	journal = {Science Advances},
	author = {Moorsom, T. and Rogers, M. and Scivetti, I. and Bandaru, S. and Teobaldi, G. and Valvidares, M. and Flokstra, M. and Lee, S. and Stewart, R. and Prokscha, T. and Gargiani, P. and Alosaimi, N. and Stefanou, G. and Ali, M. and Ma’Mari, F. Al and Burnell, G. and Hickey, B. J. and Cespedes, O.},
	year = {2020},
	pages = {eaax1085}
}

@misc{woerle:low-energy2020,
	title = {Low-{Energy} {Muons} as a {Tool} for a {Depth}-{Resolved} {Analysis} of the {SiO2}/{4H}-{SiC} {Interface}},
	journal = {Materials Science Forum},
	author = {Woerle, Judith and Prokscha, Thomas and Grossner, Ulrike},
	year = {2020},
	doi = {10.4028/www.scientific.net/MSF.1004.581},
	note = {Conference Name: Silicon Carbide and Related Materials 2019
   Pages: 581-586
   Volume: 1004}
}

@article{miyake:ultra2013,
	title = {Ultra slow muon microscopy by laser resonant ionization at {J}-{PARC}, {MUSE}},
	volume = {216},
	doi = {10.1007/s10751-012-0759-4},
	journal = {Hyperfine Interactions},
	author = {Miyake, Y. and Ikedo, Y. and Shimomura, K. and Strasser, P. and Kawamura, N. and Nishiyama, K. and Koda, A. and Fujimori, H. and Makimura, S. and Nakamura, J. and Nagatomo, T. and Kadono, R. and Torikai, E. and Iwasaki, M. and Wada, S. and Saito, N. and Okamura, K. and Yokoyama, K. and Ito, T. and Higemoto, W.},
	year = {2013},
	pages = {79--83}
}

@article{janka:intense2020,
	title = {Intense beam of metastable {Muonium}},
	volume = {80},
	doi = {10.1140/epjc/s10052-020-8400-1},
	journal = {The European Physical Journal C},
	author = {Janka, G. and Ohayon, B. and Burkley, Z. and Gerchow, L. and Kuroda, N. and Ni, X. and Nishi, R. and Salman, Z. and Suter, A. and Tuzi, M. and Vigo, C. and Prokscha, T. and Crivelli, P.},
	year = {2020},
	pages = {804}
}

@article{li:coexistence2011,
	title = {Coexistence of magnetic order and two-dimensional superconductivity at {LaAlO3}/{SrTiO3} interfaces},
	volume = {7},
	doi = {10.1038/nphys2080},
	journal = {Nature Physics},
	author = {Li, Lu and Richter, C. and Mannhart, J. and Ashoori, R. C.},
	year = {2011},
	pages = {762--766}
}

@article{bert:direct2011,
	title = {Direct imaging of the coexistence of ferromagnetism and superconductivity at the {LaAlO3}/{SrTiO3} interface},
	volume = {7},
	doi = {10.1038/nphys2079},
	journal = {Nature Physics},
	author = {Bert, Julie A. and Kalisky, Beena and Bell, Christopher and Kim, Minu and Hikita, Yasuyuki and Hwang, Harold Y. and Moler, Kathryn A.},
	year = {2011},
	pages = {767--771}
}

@article{luke:time-reversal1998,
	title = {Time-reversal symmetry-breaking superconductivity in {Sr2RuO4}},
	volume = {394},
	doi = {10.1038/29038},
	journal = {Nature},
	author = {Luke, G. M. and Fudamoto, Y. and Kojima, K. M. and Larkin, M. I. and Merrin, J. and Nachumi, B. and Uemura, Y. J. and Maeno, Y. and Mao, Z. Q. and Mori, Y. and Nakamura, H. and Sigrist, M.},
	year = {1998},
	pages = {558--561}
}

@article{Khaw:2016ofi,
    author = "Khaw, K. S. and Antognini, A. and Prokscha, T. and Kirch, K. and Liszkay, L. and Salman, Z. and Crivelli, P.",
    title = "{Spatial confinement of muonium atoms}",
    eprint = "1606.05840",
    archivePrefix = "arXiv",
    primaryClass = "physics.atom-ph",
    doi = "10.1103/PhysRevA.94.022716",
    journal = "Phys. Rev. A",
    volume = "94",
    number = "2",
    pages = "022716",
    year = "2016"
}

@article{sugiyama:li-ion2015,
	title = {Li-ion diffusion in Li$_4$Ti$_5$O$_{12}$ and LiTi$_2$O$_4$ battery materials detected by muon spin spectroscopy},
	volume = {92},
	doi = {10.1103/PhysRevB.92.014417},
	journal = {Physical Review B},
	author = {Sugiyama, Jun and Nozaki, Hiroshi and Umegaki, Izumi and Mukai, Kazuhiko and Miwa, Kazutoshi and Shiraki, Susumu and Hitosugi, Taro and Suter, Andreas and Prokscha, Thomas and Salman, Zaher and Lord, James S. and Månsson, Martin},
	year = {2015},
	pages = {014417}
}

@article{sugiyama:spin2019,
	title = {Spin polarized beam for battery materials research: $\mu_{\pm SR}$ and $\beta_{-NMR}$ },
	volume = {240},
	doi = {10.1007/s10751-019-1560-4},
	journal = {Hyperfine Interactions},
	author = {Sugiyama, Jun},
	year = {2019},
	pages = {17}
}

@article{jiao:chiral2020,
	title = {Chiral superconductivity in heavy-fermion metal {UTe2}},
	volume = {579},
	doi = {10.1038/s41586-020-2122-2},
    journal = {Nature},
	author = {Jiao, Lin and Howard, Sean and Ran, Sheng and Wang, Zhenyu and Rodriguez, Jorge Olivares and Sigrist, Manfred and Wang, Ziqiang and Butch, Nicholas P. and Madhavan, Vidya},
	year = {2020},
	pages = {523--527}
}

@article{yokoyama:photoexcited2017,
	title = {Photoexcited {Muon} {Spin} {Spectroscopy}: {A} {New} {Method} for {Measuring} {Excess} {Carrier} {Lifetime} in {Bulk} {Silicon}},
	volume = {119},
	doi = {10.1103/PhysRevLett.119.226601},
	journal = {Physical Review Letters},
	author = {Yokoyama, K. and Lord, J. S. and Miao, J. and Murahari, P. and Drew, A. J.},
	year = {2017},
	pages = {226601}
}

@article{kohler:application1981,
	title = {Application of muonic {X}-ray techniques to the elemental analysis of archeological objects},
	volume = {187},
	doi = {10.1016/0029-554X(81)90389-X},
	journal = {Nuclear Instruments and Methods in Physics Research},
	author = {Köhler, E. and Bergmann, R. and Daniel, H. and Ehrhart, P. and Hartmann, F.J.},
	year = {1981},
	pages = {563--568}
}

@article{daniel:application1984,
	title = {Application of {X} rays from negative muons},
	volume = {3},
	doi = {10.1016/0168-583X(84)90338-0},
	journal = {Nuclear Instruments and Methods in Physics Research Section B: Beam Interactions with Materials and Atoms},
	author = {Daniel, H.},
	year = {1984},
	pages = {65--70}
}

@article{ninomiya:development2010,
	title = {Development of elemental analysis by muonic {X}-ray measurement in {J}-{PARC}},
	volume = {225},
	doi = {10.1088/1742-6596/225/1/012040},
	journal = {Journal of Physics: Conference Series},
	author = {Ninomiya, K and Nagatomo, T and Kubo, K M and Strasser, P and Kawamura, N and Shimomura, K and Miyake, Y and Saito, T and Higemoto, W},
	year = {2010},
	pages = {012040}
}

@article{hillier:probing2016,
	title = {Probing beneath the surface without a scratch — {Bulk} non-destructive elemental analysis using negative muons},
	volume = {125},
	doi = {10.1016/j.microc.2015.11.031},
	journal = {Microchemical Journal},
	author = {Hillier, A. D. and Paul, D. McK. and Ishida, K.},
	year = {2016},
	pages = {203--207}
}

@article{Ohayon:2021qof,
    author = "Ohayon, B. and others",
    title = "{Precision measurement of the Lamb shift in Muonium}",
    eprint = "2108.12891",
    archivePrefix = "arXiv",
    primaryClass = "physics.atom-ph",
    month = "8",
    year = "2021"
}

@article{Abela1999,
author = {Abela, R. and Amato, A. and Baines, C. and Donath, X. and Erne, R. and George, D. C. and Herlach, D. and Irminger, G. and Reid, I. D. and Renker, D. and Solt, G. and Suhi, D. and Werner, M. and Zimmermann, U.},
doi = {10.1023/A:1017046817431},
issn = {03043843},
journal = {Hyperfine Interactions},
number = {1-8},
pages = {575--578},
title = {{Muons on request (MORE): Combining advantages of continuous and pulsed muon beams}},
volume = {120-121},
year = {1999}
}

@article{Blondel:2013ia,
    author = "Blondel, A. and others",
    title = "{Research Proposal for an Experiment to Search for the Decay $\mu \to eee$}",
    eprint = "1301.6113",
    archivePrefix = "arXiv",
    primaryClass = "physics.ins-det",
    month = "1",
    year = "2013"
}

@article{Arndt:2020obb,
    author = "Arndt, K. and others",
    collaboration = "Mu3e",
    title = "{Technical design of the phase I Mu3e experiment}",
    eprint = "2009.11690",
    archivePrefix = "arXiv",
    primaryClass = "physics.ins-det",
    doi = "10.1016/j.nima.2021.165679",
    journal = "Nucl. Instrum. Meth. A",
    volume = "1014",
    pages = "165679",
    year = "2021"
}

@article{Pruna:2016spf,
    author = "Pruna, G.M. and Signer, A. and Ulrich, Y.",
    title = "{Fully differential NLO predictions for the rare muon decay}",
    eprint = "1611.03617",
    archivePrefix = "arXiv",
    primaryClass = "hep-ph",
    reportNumber = "PSI-PR-16-14, ZU-TH-41-16",
    doi = "10.1016/j.physletb.2016.12.039",
    journal = "Phys. Lett. B",
    volume = "765",
    pages = "280--284",
    year = "2017"
}

@article{SINDRUM:1985vbg,
    author = "Bertl, Wilhelm H. and others",
    collaboration = "SINDRUM",
    title = "{Search for the Decay $\mu^+ \to e^+ e^+ e^-$}",
    reportNumber = "SIN-PR-85-06",
    doi = "10.1016/0550-3213(85)90308-6",
    journal = "Nucl. Phys. B",
    volume = "260",
    pages = "1--31",
    year = "1985"
}

@article{Perrevoort:2018ttp,
    author = "Perrevoort, Ann-Kathrin",
    collaboration = "Mu3e",
    title = "{The Rare and Forbidden: Testing Physics Beyond the Standard Model with Mu3e}",
    eprint = "1812.00741",
    archivePrefix = "arXiv",
    primaryClass = "hep-ex",
    doi = "10.21468/SciPostPhysProc.1.052",
    journal = "SciPost Phys. Proc.",
    volume = "1",
    pages = "052",
    year = "2019"
}

@phdthesis{Perrevoort:2018okj,
    author = "Perrevoort, Ann-Kathrin",
    title = "{Sensitivity Studies on New Physics in the Mu3e Experiment and Development of Firmware for the Front-End of the Mu3e Pixel Detector}",
    doi = "10.11588/heidok.00024585",
    school = "Heidelberg University",
    year = "2018"
}

@phdthesis{HughesPhD,
    author = "Hughes, Sean",
    title = "{Background studies for the Mu3e experiment}",
    school = "Liverpool University",
    pubstate = "In preparation"
}

@article{vomBruch:2017fqw,
    author = "vom Bruch, Dorothea",
    editor = "Germain, C. and Grasland, H. and Lounis, A. and Rousseau, D. and Varouchas, D.",
    collaboration = "Mu3e",
    title = "{Online Data Reduction using Track and Vertex Reconstruction on GPUs for the Mu3e Experiment}",
    doi = "10.1051/epjconf/201715000013",
    journal = "EPJ Web Conf.",
    volume = "150",
    pages = "00013",
    year = "2017"
}

@inproceedings{vomBruch:2015kla,
    author = "vom Bruch, Dorothea",
    collaboration = "Mu3e",
    title = "{Track and Vertex Reconstruction on GPUs for the Mu3e Experiment}",
    booktitle = "{GPU Computing in High-Energy Physics}",
    doi = "10.3204/DESY-PROC-2014-05/13",
    month = "6",
    year = "2015"
}

@PHDTHESIS{DissVomBruch2017,
  author = {D. vom Bruch},
  title = {Pixel Sensor Evaluation and Online Event Selection for the Mu3e
	Experiment},
  school = {Heidelberg University},
  year = {2017}
}

@article{Kozlinskiy:2017wyl,
    author = "Kozlinskiy, Alexandr",
    editor = "Germain, C. and Grasland, H. and Lounis, A. and Rousseau, D. and Varouchas, D.",
    collaboration = "Mu3e",
    title = "{Track reconstruction for the Mu3e experiment based on a novel Multiple Scattering fit}",
    doi = "10.1051/epjconf/201715000005",
    journal = "EPJ Web Conf.",
    volume = "150",
    pages = "00005",
    year = "2017"
}

@article{Berger:2016vak,
    author = {Berger, Niklaus and Kiehn, Moritz and Kozlinskiy, Alexandr and Sch\"oning, Andre},
    title = "{A New Three-Dimensional Track Fit with Multiple Scattering}",
    eprint = "1606.04990",
    archivePrefix = "arXiv",
    primaryClass = "physics.ins-det",
    doi = "10.1016/j.nima.2016.11.012",
    journal = "Nucl. Instrum. Meth. A",
    volume = "844",
    pages = "135",
    year = "2017"
}

@MASTERSTHESIS{Grefkes2021,
  author = {Grefkes, N.},
  title = {{Mu3e Final Focus Study}},
  school = {Heidelberg University},
  year = {2021},
  type = {Bachelor Thesis}
}

@MASTERSTHESIS{Leuschner2019,
  author = {Leuschner, H.},
  title = {{Mu3e-$\gamma$ upgrade simulations for the Mu3e experiment}},
  school = {Heidelberg University},
  year = {2019},
  type = {Master Thesis}
}

@article{Kozlinskiy:2014rwa,
    author = {Kozlinskiy, A. and Sch\"oning, A. and Kiehn, M. and Berger, N. and Schenk, S.},
    title = "{A new track reconstruction algorithm for the Mu3e experiment based on a fast multiple scattering fit}",
    doi = "10.1088/1748-0221/9/12/C12012",
    journal = "JINST",
    volume = "9",
    number = "12",
    pages = "C12012",
    year = "2014"
}

@MASTERSTHESIS{Schenk2013,
  author = {Schenk, S.},
  title = {{A fast vertex fit for the Mu3e experiment}},
  school = {Heidelberg University},
  year = {2013},
  type = {Bachelor Thesis}
}

@article{Baldini2021symmetry,
    author = "A. M. Baldini et al., MEG~II collaboration",
    collaboration = "MEG~II",
    title = "{The Search for $\mu^+\to e^+ \gamma$ with $10^{–14}$ Sensitivity: The Upgrade of the MEG Experiment}",
    doi = "10.3390/sym13091591",
    journal = "Symmetry",
    volume = "13",
    pages = "19",
    year = "2021"
}

@article{Tsuji:2019zuj,
    author = "Tsuji, Naoki and Liu, Linghui and Torimaru, Tatsuro and Mori, Toshinori and Ootani, Wataru",
    collaboration = "CALICE",
    title = "{Study on Granularity Optimization for ILD Hadron Calorimeter}",
    doi = "10.7566/JPSCP.27.012015",
    journal = "JPS Conf. Proc.",
    volume = "27",
    pages = "012015",
    year = "2019"
}

@article{ParticleDataGroup:2020ssz,
    author = "Zyla, P. A. and others",
    collaboration = "Particle Data Group",
    title = "{Review of Particle Physics}",
    doi = "10.1093/ptep/ptaa104",
    journal = "PTEP",
    volume = "2020",
    number = "8",
    pages = "083C01",
    year = "2020"
}

@article{MEGA:2001cwl,
    author = "Ahmed, M. and others",
    collaboration = "MEGA",
    title = "{Search for the lepton family number nonconserving decay $\mu^+\to e^+ \gamma$}",
    eprint = "hep-ex/0111030",
    archivePrefix = "arXiv",
    doi = "10.1103/PhysRevD.65.112002",
    journal = "Phys. Rev. D",
    volume = "65",
    pages = "112002",
    year = "2002"
}

@inproceedings{Oya:2021iqx,
    author = "Oya, Atsushi and Ieki, Kei and Ochi, Atsuhiko and Onda, Rina and Ootani, Wataru and Yamamoto, Kensuke",
    title = "{Development of high-rate capable and ultra-low mass Resistive Plate Chamber with Diamond-Like Carbon}",
    eprint = "2109.13525",
    archivePrefix = "arXiv",
    primaryClass = "physics.ins-det",
    month = "9",
    year = "2021"
}

@article{CREMA:2016idx,
    author = "Pohl, Randolf and others",
    collaboration = "CREMA",
    title = "{Laser spectroscopy of muonic deuterium}",
    doi = "10.1126/science.aaf2468",
    journal = "Science",
    volume = "353",
    number = "6300",
    pages = "669--673",
    year = "2016"
}

@article{Antognini:1900ns,
    author = {Antognini, A. and Nez, F. and Schuhmann, K. and Amaro, F.~D. and Biraben, F. and Cardoso, J.~M.~R. and Covita, D.~S. and Dax, A. and Dhawan, S. and Diepold, M. and Fernandes, L.~M.~P. and Giesen, A. and Gouvea, A.~L. and Graf, T. and H{\"a}nsch, T.~W. and Indelicato, P. and Julien, L. and Kao, C. and Knowles, P. and Kottmann, F. and Le Bigot, E. and Liu, Y. and Lopes, J~.A.~M. and Ludhova, L. and Monteiro, C.~M.~B. and Mulhauser, F. and Nebel, T. and Rabinowitz, P. and dos Santos, J.~M.~F. and Schaller, L.~A. and Schwob, C. and Taqqu, D. and Veloso, J.~F.~C.~A. and Vogelsang, J. and Pohl, R.},
    title = "{Proton Structure from the Measurement of $2S-2P$ Transition Frequencies of Muonic Hydrogen}",
    doi = "10.1126/science.1230016",
    journal = "Science",
    volume = "339",
    pages = "417--420",
    year = "2013"
}

@article{Krauth:2021foz,
    author = "Krauth, Julian J. and others",
    title = "{Measuring the \ensuremath{\alpha}-particle charge radius with muonic helium-4 ions}",
    doi = "10.1038/s41586-021-03183-1",
    journal = "Nature",
    volume = "589",
    number = "7843",
    pages = "527--531",
    year = "2021"
}

@article{Wauters:2021cze,
    author = "Wauters, Frederik and Knecht, Andreas",
    collaboration = "muX",
    title = "{The $muX$ project}",
    eprint = "2108.10765",
    archivePrefix = "arXiv",
    primaryClass = "physics.ins-det",
    month = "8",
    year = "2021"
}

@article{Missimer:1984hx,
      author         = "Missimer, John H. and Simons, Leopold M.",
      title          = "{The Neutral Weak Current of the Muon}",
      journal        = "Phys. Rept.",
      volume         = "118",
      pages          = "179",
      year           = "1985",
      SLACcitation   = "%%CITATION = PRPLC,118,179;%%",
}

@article{PhysRevLett.78.4363,
  title = {Metastability of the Muonic Boron $2\mathit{S}$ State},
  author = {Kirch, K. and Abbott, D. and Bach, B. and DeCecco, P. and Hauser, P. and Horv\'ath, D. and Kottmann, F. and Missimer, J. and Siegel, R. T. and Simons, L. M. and Viel, D.},
  journal = {Phys. Rev. Lett.},
  volume = {78},
  issue = {23},
  pages = {4363--4366},
  numpages = {0},
  year = {1997},
  publisher = {American Physical Society},
  doi = {10.1103/PhysRevLett.78.4363},
}

@article{Karshenboim:2014tka,
    author = "Karshenboim, Savely G. and McKeen, David and Pospelov, Maxim",
    title = "{Constraints on muon-specific dark forces}",
    eprint = "1401.6154",
    archivePrefix = "arXiv",
    primaryClass = "hep-ph",
    doi = "10.1103/PhysRevD.90.073004",
    journal = "Phys. Rev. D",
    volume = "90",
    number = "7",
    pages = "073004",
    year = "2014",
    note = "[Addendum: Phys.Rev.D 90, 079905 (2014)]"
}

@article{Crivellin:2021bkd,
    author = "Crivellin, Andreas and Hoferichter, Martin and Kirk, Matthew and Manzari, Claudio Andrea and Schnell, Luc",
    title = "{First-generation new physics in simplified models: from low-energy parity violation  to the LHC}",
    eprint = "2107.13569",
    archivePrefix = "arXiv",
    primaryClass = "hep-ph",
    reportNumber = "CERN-TH-2021-112, PSI-21-16, ZU-TH 32/21",
    doi = "10.1007/JHEP10(2021)221",
    journal = "JHEP",
    volume = "10",
    pages = "221",
    year = "2021"
}

@article{KESSLER2001187,
title = {The GAMS4 flat crystal facility},
journal = {Nuclear Instruments and Methods in Physics Research Section A: Accelerators, Spectrometers, Detectors and Associated Equipment},
volume = {457},
number = {1},
pages = {187-202},
year = {2001},
issn = {0168-9002},
doi = {https://doi.org/10.1016/S0168-9002(00)00753-1},
author = {E.G. Kessler and M.S. Dewey and R.D. Deslattes and A. Henins and H.G. Börner and M. Jentschel and H. Lehmann}
}

@article{Paul:2020cnx,
    author = "Paul, Nancy and Bian, Guojie and Azuma, Toshiyuki and Okada, Shinji and Indelicato, Paul",
    title = "{Testing Quantum Electrodynamics with Exotic Atoms}",
    eprint = "2011.09715",
    archivePrefix = "arXiv",
    primaryClass = "physics.atom-ph",
    doi = "10.1103/PhysRevLett.126.173001",
    journal = "Phys. Rev. Lett.",
    volume = "126",
    number = "17",
    pages = "173001",
    year = "2021"
}

@article{Kraft-Bermuth:2018mgz,
    author = {Kraft-Bermuth, Saskia and Hengstler, Daniel and Egelhof, Peter and Enss, Christian and Fleischmann, Andreas and Keller, Michael and St\"ohlker, Thomas},
    title = "{Microcalorimeters for X-Ray Spectroscopy of Highly Charged Ions at Storage Rings}",
    doi = "10.3390/atoms6040059",
    journal = "Atoms",
    volume = "6",
    number = "4",
    pages = "59",
    year = "2018"
}

@article{WINKLER2015203,
title = {256-pixel microcalorimeter array for high-resolution $\gamma$-ray spectroscopy of mixed-actinide materials},
journal = {Nuclear Instruments and Methods in Physics Research Section A: Accelerators, Spectrometers, Detectors and Associated Equipment},
volume = {770},
pages = {203-210},
year = {2015},
issn = {0168-9002},
doi = {https://doi.org/10.1016/j.nima.2014.09.049},
author = {R. Winkler and A.S. Hoover and M.W. Rabin and D.A. Bennett and W.B. Doriese and J.W. Fowler and J. Hays-Wehle and R.D. Horansky and C.D. Reintsema and D.R. Schmidt and L.R. Vale and J.N. Ullom}
}

@article{Paolozzi_2019,
	doi = {10.1088/1748-0221/14/02/p02009},
	year = 2019,
	publisher = {{IOP} Publishing},
	volume = {14},
	number = {02},
	pages = {P02009--P02009},
	author = {L. Paolozzi and Y. Bandi and R. Cardarelli and S. D{\'{e}}bieux and Y. Favre and D. Ferr{\`{e}}re and D. Forshaw and D. Hayakawa and G. Iacobucci and M. Kaynak and A. Miucci and M. Nessi and E. Ripiccini and H. Rücker and P. Valerio and M. Weber},
	title = {Characterization of the demonstrator of the fast silicon monolithic {ASIC} for the {TT}-{PET} project},
	journal = {Journal of Instrumentation}%,
}

@article{SINDRUMII:2006dvw,
    author = "Bertl, Wilhelm H. and others",
    collaboration = "SINDRUM II",
    title = "{A Search for muon to electron conversion in muonic gold}",
    doi = "10.1140/epjc/s2006-02582-x",
    journal = "Eur. Phys. J. C",
    volume = "47",
    pages = "337--346",
    year = "2006"
}

@article{Kitano:2002mt,
    author = "Kitano, Ryuichiro and Koike, Masafumi and Okada, Yasuhiro",
    title = "{Detailed calculation of lepton flavor violating muon electron conversion rate for various nuclei}",
    eprint = "hep-ph/0203110",
    archivePrefix = "arXiv",
    reportNumber = "KEK-TH-808",
    doi = "10.1103/PhysRevD.76.059902",
    journal = "Phys. Rev. D",
    volume = "66",
    pages = "096002",
    year = "2002",
    note = "[Erratum: Phys.Rev.D 76, 059902 (2007)]"
}

@article{Tassielli:2021D1,
  author = "Tassielli, Giovanni F.",
  title = "{A proposal of a drift chamber for the IDEA experiment for a future e+e- collider}",
  doi = "10.22323/1.390.0877",
  journal = "PoS",
  year = 2021,
  volume = "ICHEP2020",
  pages = "877"
}

@article{Tassielli:2020wap,
    author = "Tassielli, G. F. and others",
    title = "{The Drift Chamber of the MEG II experiment}",
    eprint = "2006.02378",
    archivePrefix = "arXiv",
    primaryClass = "physics.ins-det",
    doi = "10.1088/1748-0221/15/09/C09051",
    journal = "JINST",
    volume = "15",
    number = "09",
    pages = "C09051",
    year = "2020"
}

@article{Chiarello:2017yft,
    author = "Chiarello, G. and Chiri, C. and Cocciolo, G. and Corvaglia, A. and Grancagnolo, F. and Panareo, M. and Pepino, A. and Renga, F. and Tassielli, G. F. and Voena, C.",
    editor = "Shekhtman, Lev",
    title = "{Application of the Cluster Counting/Timing techniques to improve the performances of high transparency Drift Chamber for modern HEP experiments}",
    doi = "10.1088/1748-0221/12/07/C07021",
    journal = "JINST",
    volume = "12",
    number = "07",
    pages = "C07021",
    year = "2017"
}

@article{Bencivenni:2007zz,
    author = "Bencivenni, G. and Domenici, D.",
    editor = "Hrubec, Josef and Jeitler, Manfred and Krammer, Manfred and Regler, Meinhard and Badurek, Gerald",
    title = "{An ultra-light cylindrical GEM detector as inner tracker at KLOE-2}",
    doi = "10.1016/j.nima.2007.07.082",
    journal = "Nucl. Instrum. Meth. A",
    volume = "581",
    pages = "221--224",
    year = "2007"
}

@article{Bencivenni:2014exa,
    author = "Bencivenni, G. and De Oliveira, R. and Morello, G. and Lener, M. Poli",
    title = "{The micro-Resistive WELL detector: a compact spark-protected single amplification-stage MPGD}",
    eprint = "1411.2466",
    archivePrefix = "arXiv",
    primaryClass = "physics.ins-det",
    doi = "10.1088/1748-0221/10/02/P02008",
    journal = "JINST",
    volume = "10",
    number = "02",
    pages = "P02008",
    year = "2015"
}

@article{Ketzer:2013laa,
    author = "Ketzer, Bernhard",
    editor = "Bergauer, T. and Badurek, G. and Dragicevic, M. and Friedl, M. and Hrubec, J. and Jeitler, M. and Krammer, M.",
    collaboration = "GEM-TPC, ALICE TPC",
    title = "{A Time Projection Chamber for High-Rate Experiments: Towards an Upgrade of the ALICE TPC}",
    eprint = "1303.6694",
    archivePrefix = "arXiv",
    primaryClass = "physics.ins-det",
    doi = "10.1016/j.nima.2013.08.027",
    journal = "Nucl. Instrum. Meth. A",
    volume = "732",
    pages = "237--240",
    year = "2013"
}

@article{Capdevilla:2021rwo,
    author = "Capdevilla, Rodolfo and Curtin, David and Kahn, Yonatan and Krnjaic, Gordan",
    title = "{A No-Lose Theorem for Discovering the New Physics of $(g-2)_\mu$ at Muon Colliders}",
    eprint = "2101.10334",
    archivePrefix = "arXiv",
    primaryClass = "hep-ph",
    reportNumber = "FERMILAB-PUB-21-012-T",
    month = "1",
    year = "2021"
}

@article{Danilkin:2021icn,
    author = "Danilkin, Igor and Hoferichter, Martin and Stoffer, Peter",
    title = "{A dispersive estimate of scalar contributions to hadronic light-by-light scattering}",
    eprint = "2105.01666",
    archivePrefix = "arXiv",
    primaryClass = "hep-ph",
    reportNumber = "INT-PUB-21-11, UWThPh 2021-3",
    doi = "10.1016/j.physletb.2021.136502",
    journal = "Phys. Lett. B",
    volume = "820",
    pages = "136502",
    year = "2021"
}

@article{Zanke:2021wiq,
    author = "Zanke, Marvin and Hoferichter, Martin and Kubis, Bastian",
    title = "{On the transition form factors of the axial-vector resonance $f_1(1285)$ and its decay into $e^+e^-$}",
    eprint = "2103.09829",
    archivePrefix = "arXiv",
    primaryClass = "hep-ph",
    doi = "10.1007/JHEP07(2021)106",
    journal = "JHEP",
    volume = "07",
    pages = "106",
    year = "2021"
}

@article{Bijnens:2021jqo,
    author = "Bijnens, Johan and Hermansson-Truedsson, Nils and Laub, Laetitia and Rodr\'iguez-S\'anchez, Antonio",
    title = "{The two-loop perturbative correction to the $(g - 2)_{\mu}$ HLbL at short distances}",
    eprint = "2101.09169",
    archivePrefix = "arXiv",
    primaryClass = "hep-ph",
    reportNumber = "LU TP 21-03",
    doi = "10.1007/JHEP04(2021)240",
    journal = "JHEP",
    volume = "04",
    pages = "240",
    year = "2021"
}

@article{Bijnens:2020xnl,
    author = "Bijnens, Johan and Hermansson-Truedsson, Nils and Laub, Laetitia and Rodr\'iguez-S\'anchez, Antonio",
    title = "{Short-distance HLbL contributions to the muon anomalous magnetic moment beyond perturbation theory}",
    eprint = "2008.13487",
    archivePrefix = "arXiv",
    primaryClass = "hep-ph",
    reportNumber = "LU TP 20-47",
    doi = "10.1007/JHEP10(2020)203",
    journal = "JHEP",
    volume = "10",
    pages = "203",
    year = "2020"
}

@article{Hoferichter:2020lap,
    author = "Hoferichter, Martin and Stoffer, Peter",
    title = "{Asymptotic behavior of meson transition form factors}",
    eprint = "2004.06127",
    archivePrefix = "arXiv",
    primaryClass = "hep-ph",
    reportNumber = "INT-PUB-20-015",
    doi = "10.1007/JHEP05(2020)159",
    journal = "JHEP",
    volume = "05",
    pages = "159",
    year = "2020"
}

@article{Achasov:2020iys,
    author = "Achasov, M. N. and others",
    collaboration = "SND",
    title = "{Measurement of the $e^+e^- \to\pi^+\pi^- $ process cross section with the SND detector at the VEPP-2000 collider in the energy region $0.525<\sqrt{s}<0.883$ GeV}",
    eprint = "2004.00263",
    archivePrefix = "arXiv",
    primaryClass = "hep-ex",
    doi = "10.1007/JHEP01(2021)113",
    journal = "JHEP",
    volume = "01",
    pages = "113",
    year = "2021"
}

@article{Campanario:2019mjh,
    author = "Campanario, Francisco and Czy\.z, Henryk and Gluza, Janusz and Jeli\'nski, Tomasz and Rodrigo, Germ\'an and Tracz, Szymon and Zhuridov, Dmitry",
    title = "{Standard model radiative corrections in the pion form factor measurements do not explain the $a_\mu$ anomaly}",
    eprint = "1903.10197",
    archivePrefix = "arXiv",
    primaryClass = "hep-ph",
    doi = "10.1103/PhysRevD.100.076004",
    journal = "Phys. Rev. D",
    volume = "100",
    number = "7",
    pages = "076004",
    year = "2019"
}

@article{Chao:2021tvp,
    author = "Chao, En-Hung and Hudspith, Renwick J. and G\'erardin, Antoine and Green, Jeremy R. and Meyer, Harvey B. and Ottnad, Konstantin",
    title = "{Hadronic light-by-light contribution to $(g-2)_\mu $ from lattice QCD: a complete calculation}",
    eprint = "2104.02632",
    archivePrefix = "arXiv",
    primaryClass = "hep-lat",
    doi = "10.1140/epjc/s10052-021-09455-4",
    journal = "Eur. Phys. J. C",
    volume = "81",
    number = "7",
    pages = "651",
    year = "2021"
}

@article{Borsanyi:2020mff,
    author = "Borsanyi, Sz. and others",
    title = "{Leading hadronic contribution to the muon magnetic moment from lattice QCD}",
    eprint = "2002.12347",
    archivePrefix = "arXiv",
    primaryClass = "hep-lat",
    doi = "10.1038/s41586-021-03418-1",
    journal = "Nature",
    volume = "593",
    number = "7857",
    pages = "51--55",
    year = "2021"
}

@article{Lehner:2020crt,
    author = "Lehner, Christoph and Meyer, Aaron S.",
    title = "{Consistency of hadronic vacuum polarization between lattice QCD and the R-ratio}",
    eprint = "2003.04177",
    archivePrefix = "arXiv",
    primaryClass = "hep-lat",
    doi = "10.1103/PhysRevD.101.074515",
    journal = "Phys. Rev. D",
    volume = "101",
    pages = "074515",
    year = "2020"
}

@article{Colangelo:2020lcg,
    author = "Colangelo, Gilberto and Hoferichter, Martin and Stoffer, Peter",
    title = "{Constraints on the two-pion contribution to hadronic vacuum polarization}",
    eprint = "2010.07943",
    archivePrefix = "arXiv",
    primaryClass = "hep-ph",
    reportNumber = "UWThPh 2020-25",
    doi = "10.1016/j.physletb.2021.136073",
    journal = "Phys. Lett. B",
    volume = "814",
    pages = "136073",
    year = "2021"
}

@article{Colangelo:2021nkr,
    author = "Colangelo, Gilberto and Hagelstein, Franziska and Hoferichter, Martin and Laub, Laetitia and Stoffer, Peter",
    title = "{Short-distance constraints for the longitudinal component of the hadronic light-by-light amplitude: an update}",
    eprint = "2106.13222",
    archivePrefix = "arXiv",
    primaryClass = "hep-ph",
    reportNumber = "PSI-PR-21-13, UWThPh 2021-7",
    doi = "10.1140/epjc/s10052-021-09513-x",
    journal = "Eur. Phys. J. C",
    volume = "81",
    number = "8",
    pages = "702",
    year = "2021"
}

@article{Malaescu:2020zuc,
    author = "Malaescu, Bogdan and Schott, Matthias",
    title = "{Impact of correlations between $a_{\mu }$ and $\alpha _\text {QED}$ on the EW fit}",
    eprint = "2008.08107",
    archivePrefix = "arXiv",
    primaryClass = "hep-ph",
    doi = "10.1140/epjc/s10052-021-08848-9",
    journal = "Eur. Phys. J. C",
    volume = "81",
    number = "1",
    pages = "46",
    year = "2021"
}

@article{Keshavarzi:2020bfy,
    author = "Keshavarzi, Alexander and Marciano, William J. and Passera, Massimo and Sirlin, Alberto",
    title = "{Muon $g-2$ and $\Delta \alpha$ connection}",
    eprint = "2006.12666",
    archivePrefix = "arXiv",
    primaryClass = "hep-ph",
    reportNumber = "MAN/HEP/2020/006",
    doi = "10.1103/PhysRevD.102.033002",
    journal = "Phys. Rev. D",
    volume = "102",
    number = "3",
    pages = "033002",
    year = "2020"
}

@article{Crivellin:2020zul,
    author = "Crivellin, Andreas and Hoferichter, Martin and Manzari, Claudio Andrea and Montull, Marc",
    title = "{Hadronic Vacuum Polarization: $(g-2)_\mu$ versus Global Electroweak Fits}",
    eprint = "2003.04886",
    archivePrefix = "arXiv",
    primaryClass = "hep-ph",
    reportNumber = "PSI-PR-20-04, UZ-TH 06/20",
    doi = "10.1103/PhysRevLett.125.091801",
    journal = "Phys. Rev. Lett.",
    volume = "125",
    number = "9",
    pages = "091801",
    year = "2020"
}

@article{Parker:2018vye,
    author = {Parker, Richard H. and Yu, Chenghui and Zhong, Weicheng and Estey, Brian and M\"uller, Holger},
    title = "{Measurement of the fine-structure constant as a test of the Standard Model}",
    eprint = "1812.04130",
    archivePrefix = "arXiv",
    primaryClass = "physics.atom-ph",
    doi = "10.1126/science.aap7706",
    journal = "Science",
    volume = "360",
    pages = "191",
    year = "2018"
}

@article{Morel:2020dww,
    author = {Morel, L\'eo and Yao, Zhibin and Clad\'e, Pierre and Guellati-Kh\'elifa, Sa\"\i{}da},
    title = "{Determination of the fine-structure constant with an accuracy of 81 parts per trillion}",
    doi = "10.1038/s41586-020-2964-7",
    journal = "Nature",
    volume = "588",
    number = "7836",
    pages = "61--65",
    year = "2020"
}

@article{Volkov:2019phy,
    author = "Volkov, Sergey",
    title = "{Calculating the five-loop QED contribution to the electron anomalous magnetic moment: Graphs without lepton loops}",
    eprint = "1909.08015",
    archivePrefix = "arXiv",
    primaryClass = "hep-ph",
    doi = "10.1103/PhysRevD.100.096004",
    journal = "Phys. Rev. D",
    volume = "100",
    number = "9",
    pages = "096004",
    year = "2019"
}

@article{Laporta:2017okg,
    author = "Laporta, Stefano",
    title = "{High-precision calculation of the 4-loop contribution to the electron g-2 in QED}",
    eprint = "1704.06996",
    archivePrefix = "arXiv",
    primaryClass = "hep-ph",
    doi = "10.1016/j.physletb.2017.06.056",
    journal = "Phys. Lett. B",
    volume = "772",
    pages = "232--238",
    year = "2017"
}

@article{Hanneke:2008tm,
    author = "Hanneke, D. and Fogwell, S. and Gabrielse, G.",
    title = "{New Measurement of the Electron Magnetic Moment and the Fine Structure Constant}",
    eprint = "0801.1134",
    archivePrefix = "arXiv",
    primaryClass = "physics.atom-ph",
    doi = "10.1103/PhysRevLett.100.120801",
    journal = "Phys. Rev. Lett.",
    volume = "100",
    pages = "120801",
    year = "2008"
}

@article{Albahri:2021mtf,
    author = "Albahri, T. and others",
    collaboration = "Muon $g-2$",
    title = "{Beam dynamics corrections to the Run-1 measurement of the muon anomalous magnetic moment at Fermilab}",
    eprint = "2104.03240",
    archivePrefix = "arXiv",
    primaryClass = "physics.acc-ph",
    reportNumber = "FERMILAB-PUB-21-133-E",
    doi = "10.1103/PhysRevAccelBeams.24.044002",
    journal = "Phys. Rev. Accel. Beams",
    volume = "24",
    pages = "044002",
    year = "2021"
}

@article{Abi:2021gix,
	archiveprefix = {arXiv},
	author = {Abi, B. and others},
	collaboration = "Muon $g-2$",
	doi = {10.1103/PhysRevLett.126.141801},
	eprint = {2104.03281},
	journal = {Phys. Rev. Lett.},
	month = {4},
	pages = {141801},
	primaryclass = {hep-ex},
	reportnumber = {FERMILAB-PUB-21-132-E},
	title = {{Measurement of the Positive Muon Anomalous Magnetic Moment to 0.46 ppm}},
	volume = {126},
	year = {2021}
}

@article{Albahri:2021kmg,
	archiveprefix = {arXiv},
	author = {Albahri, T. and others},
	collaboration = "Muon $g-2$",
	doi = {10.1103/PhysRevA.103.042208},
	eprint = {2104.03201},
	journal = {Phys. Rev. A},
	pages = {042208},
	primaryclass = {hep-ex},
	reportnumber = {FERMILAB-PUB-21-109-E},
	title = {{Magnetic Field Measurement and Analysis for the Muon $g-2$ Experiment at Fermilab}},
	volume = {103},
	year = {2021}
}

@article{Albahri:2021ixb,
	archiveprefix = {arXiv},
	author = {Albahri, T. and others},
	collaboration = "Muon $g-2$",
	doi = {10.1103/PhysRevD.103.072002},
	eprint = {2104.03247},
	journal = {Phys. Rev. D},
	pages = {072002},
	primaryclass = {hep-ex},
	reportnumber = {FERMILAB-PUB-21-183-E},
	title = {{Measurement of the anomalous precession frequency of the muon in the Fermilab Muon $g-2$ experiment}},
	volume = {103},
	year = {2021}
}

@article{Carena:1994bv,
    author = "Carena, Marcela and Olechowski, M. and Pokorski, S. and Wagner, C. E. M.",
    title = "{Electroweak symmetry breaking and bottom - top Yukawa unification}",
    eprint = "hep-ph/9402253",
    archivePrefix = "arXiv",
    reportNumber = "MPI-PH-93-103, CERN-TH-7163-94",
    doi = "10.1016/0550-3213(94)90313-1",
    journal = "Nucl. Phys. B",
    volume = "426",
    pages = "269--300",
    year = "1994"
}

@article{Ananthanarayan:1991xp,
    author = "Ananthanarayan, B. and Lazarides, George and Shafi, Q.",
    title = "{Top mass prediction from supersymmetric guts}",
    reportNumber = "BA-91-25, CERN-TH-6057-91",
    doi = "10.1103/PhysRevD.44.1613",
    journal = "Phys. Rev. D",
    volume = "44",
    pages = "1613--1615",
    year = "1991"
}

@article{Dedes:2001fv,
    author = "Dedes, Athanasios and Dreiner, Herbert K. and Nierste, Ulrich",
    title = "{Correlation of $B_s \to \mu^{+} \mu^{-}$ and $(g-2)_\mu$ in minimal supergravity}",
    eprint = "hep-ph/0108037",
    archivePrefix = "arXiv",
    reportNumber = "CERN-TH-2001-211",
    doi = "10.1103/PhysRevLett.87.251804",
    journal = "Phys. Rev. Lett.",
    volume = "87",
    pages = "251804",
    year = "2001"
}

@article{Chattopadhyay:1995ae,
    author = "Chattopadhyay, U. and Nath, Pran",
    title = "{Probing supergravity grand unification in the Brookhaven g-2 experiment}",
    eprint = "hep-ph/9507386",
    archivePrefix = "arXiv",
    reportNumber = "NSF-ITP-95-64, NUB-TH-3125-95",
    doi = "10.1103/PhysRevD.53.1648",
    journal = "Phys. Rev. D",
    volume = "53",
    pages = "1648--1657",
    year = "1996"
}

@article{Lopez:1993vi,
    author = "Lopez, Jorge L. and Nanopoulos, Dimitri V. and Wang, Xu",
    title = "{Large $(g-2)_\mu$ in $SU(5) \times U(1)$ supergravity models}",
    eprint = "hep-ph/9308336",
    archivePrefix = "arXiv",
    reportNumber = "CERN-TH-6986-93, CTP-TAMU-44-93, ACT-17-93",
    doi = "10.1103/PhysRevD.49.366",
    journal = "Phys. Rev. D",
    volume = "49",
    pages = "366--372",
    year = "1994"
}

@article{Bauer:2015knc,
    author = "Bauer, Martin and Neubert, Matthias",
    title = "{Minimal Leptoquark Explanation for the R$_{D^{(*)}}$ , R$_K$ , and $(g-2)_\mu$ Anomalies}",
    eprint = "1511.01900",
    archivePrefix = "arXiv",
    primaryClass = "hep-ph",
    reportNumber = "MITP-15-100",
    doi = "10.1103/PhysRevLett.116.141802",
    journal = "Phys. Rev. Lett.",
    volume = "116",
    number = "14",
    pages = "141802",
    year = "2016"
}

@article{Cheung:2001ip,
    author = "Cheung, Kingman",
    title = "{Muon anomalous magnetic moment and leptoquark solutions}",
    eprint = "hep-ph/0102238",
    archivePrefix = "arXiv",
    reportNumber = "NSC-NCTS-010219",
    doi = "10.1103/PhysRevD.64.033001",
    journal = "Phys. Rev. D",
    volume = "64",
    pages = "033001",
    year = "2001"
}

@article{Chakraverty:2001yg,
    author = "Chakraverty, Debrupa and Choudhury, Debajyoti and Datta, Anindya",
    title = "{A Nonsupersymmetric resolution of the anomalous muon magnetic moment}",
    eprint = "hep-ph/0102180",
    archivePrefix = "arXiv",
    doi = "10.1016/S0370-2693(01)00419-1",
    journal = "Phys. Lett. B",
    volume = "506",
    pages = "103--108",
    year = "2001"
}

@article{Park:2001uc,
    author = "Park, Seong Chan and Song, H. S.",
    title = "{Muon anomalous magnetic moment and the stabilized Randall-Sundrum scenario}",
    eprint = "hep-ph/0103072",
    archivePrefix = "arXiv",
    reportNumber = "SNUTP-01-010",
    doi = "10.1016/S0370-2693(01)00417-8",
    journal = "Phys. Lett. B",
    volume = "506",
    pages = "99--102",
    year = "2001"
}

@article{Xiong:2001rt,
    author = "Xiong, Zhao-Hua and Yang, Jin Min",
    title = "{Muon anomalous magnetic moment in technicolor models}",
    eprint = "hep-ph/0102259",
    archivePrefix = "arXiv",
    doi = "10.1016/S0370-2693(01)00521-4",
    journal = "Phys. Lett. B",
    volume = "508",
    pages = "295--300",
    year = "2001"
}

@article{Das:2001it,
    author = "Das, Prasanta and Kumar Rai, Santosh and Raychaudhuri, Sreerup",
    editor = "Caparthy, J.",
    title = "{Anomalous magnetic moment of the muon in a composite model}",
    eprint = "hep-ph/0102242",
    archivePrefix = "arXiv",
    month = "2",
    year = "2001"
}

@article{Crivellin:2015hha,
    author = "Crivellin, Andreas and Heeck, Julian and Stoffer, Peter",
    title = "{A perturbed lepton-specific two-Higgs-doublet model facing experimental hints for physics beyond the Standard Model}",
    eprint = "1507.07567",
    archivePrefix = "arXiv",
    primaryClass = "hep-ph",
    reportNumber = "CERN-PH-TH-2015-168, ULB-15-13",
    doi = "10.1103/PhysRevLett.116.081801",
    journal = "Phys. Rev. Lett.",
    volume = "116",
    number = "8",
    pages = "081801",
    year = "2016"
}

@article{Broggio:2014mna,
    author = "Broggio, Alessandro and Chun, Eung Jin and Passera, Massimo and Patel, Ketan M. and Vempati, Sudhir K.",
    title = "{Limiting two-Higgs-doublet models}",
    eprint = "1409.3199",
    archivePrefix = "arXiv",
    primaryClass = "hep-ph",
    doi = "10.1007/JHEP11(2014)058",
    journal = "JHEP",
    volume = "11",
    pages = "058",
    year = "2014"
}

@article{Farley:2003wt,
    author = "Farley, F. J. M. and Jungmann, K. and Miller, J. P. and Morse, W. M. and Orlov, Y. F. and Roberts, B. L. and Semertzidis, Y. K. and Silenko, A. and Stephenson, E. J.",
    title = "{A New method of measuring electric dipole moments in storage rings}",
    eprint = "hep-ex/0307006",
    archivePrefix = "arXiv",
    doi = "10.1103/PhysRevLett.93.052001",
    journal = "Phys. Rev. Lett.",
    volume = "93",
    pages = "052001",
    year = "2004"
}

@article{Iinuma:2016zfu,
    author = "Iinuma, Hiromi and Nakayama, Hisayoshi and Oide, Katsunobu and Sasaki, Ken-ichi and Saito, Naohito and Mibe, Tsutomu and Abe, Mitsushi",
    title = "{Three-dimensional spiral injection scheme for the g-2/EDM experiment at J-PARC}",
    doi = "10.1016/j.nima.2016.05.126",
    journal = "Nucl. Instrum. Meth. A",
    volume = "832",
    pages = "51--62",
    year = "2016"
}

@article{Gorringe:2015cma,
    author = "Gorringe, T. P. and Hertzog, D. W.",
    title = "{Precision Muon Physics}",
    eprint = "1506.01465",
    archivePrefix = "arXiv",
    primaryClass = "hep-ex",
    doi = "10.1016/j.ppnp.2015.06.001",
    journal = "Prog. Part. Nucl. Phys.",
    volume = "84",
    pages = "73--123",
    year = "2015"
}

@article{Bennett:2006fi,
    author = "Bennett, G. W. and others",
    collaboration = "Muon $g-2$",
    title = "{Final Report of the Muon E821 Anomalous Magnetic Moment Measurement at BNL}",
    eprint = "hep-ex/0602035",
    archivePrefix = "arXiv",
    doi = "10.1103/PhysRevD.73.072003",
    journal = "Phys. Rev. D",
    volume = "73",
    pages = "072003",
    year = "2006"
}

@article{Bennett:2008dy,
    author = "Bennett, G. W. and others",
    collaboration = "Muon $g-2$",
    title = "{An Improved Limit on the Muon Electric Dipole Moment}",
    eprint = "0811.1207",
    archivePrefix = "arXiv",
    primaryClass = "hep-ex",
    doi = "10.1103/PhysRevD.80.052008",
    journal = "Phys. Rev. D",
    volume = "80",
    pages = "052008",
    year = "2009"
}

@article{Pruna:2017tif,
    author = "Pruna, Giovanni Marco",
    title = "{Leptonic CP violation in the charged sector and effective field theory approach}",
    eprint = "1710.08311",
    archivePrefix = "arXiv",
    primaryClass = "hep-ph",
    reportNumber = "FPCP2017-PRUNA, PSI-PR-17-17",
    doi = "10.22323/1.304.0016",
    journal = "PoS",
    volume = "FPCP2017",
    pages = "016",
    year = "2017"
}

@article{Andreev:2018ayy,
    author = "Andreev, V. and others",
    collaboration = "ACME",
    title = "{Improved limit on the electric dipole moment of the electron}",
    doi = "10.1038/s41586-018-0599-8",
    journal = "Nature",
    volume = "562",
    number = "7727",
    pages = "355--360",
    year = "2018"
}

@article{Crivellin:2018qmi,
    author = "Crivellin, Andreas and Hoferichter, Martin and Schmidt-Wellenburg, Philipp",
    title = "{Combined explanations of $(g-2)_{\mu,e}$ and implications for a large muon EDM}",
    eprint = "1807.11484",
    archivePrefix = "arXiv",
    primaryClass = "hep-ph",
    reportNumber = "INT-PUB-18-039, PSI-PR-18-09",
    doi = "10.1103/PhysRevD.98.113002",
    journal = "Phys. Rev. D",
    volume = "98",
    number = "11",
    pages = "113002",
    year = "2018"
}

@article{Grange:2015fou,
    author = "Grange, J. and others",
    collaboration = "Muon $g-2$",
    title = "{Muon $(g-2)$ Technical Design Report}",
    eprint = "1501.06858",
    archivePrefix = "arXiv",
    primaryClass = "physics.ins-det",
    reportNumber = "FERMILAB-FN-0992-E, FERMILAB-DESIGN-2014-02",
    month = "1",
    year = "2015"
}

@article{Abe:2019thb,
    author = "Abe, M. and others",
    title = "{A New Approach for Measuring the Muon Anomalous Magnetic Moment and Electric Dipole Moment}",
    eprint = "1901.03047",
    archivePrefix = "arXiv",
    primaryClass = "physics.ins-det",
    doi = "10.1093/ptep/ptz030",
    journal = "PTEP",
    volume = "2019",
    number = "5",
    pages = "053C02",
    year = "2019"
}

@article{Holdom:1985ag,
    author = "Holdom, Bob",
    title = "{Two U(1)'s and Epsilon Charge Shifts}",
    reportNumber = "UTPT-85-30",
    doi = "10.1016/0370-2693(86)91377-8",
    journal = "Phys. Lett. B",
    volume = "166",
    pages = "196--198",
    year = "1986"
}

@article{Pospelov:2008zw,
    author = "Pospelov, Maxim",
    title = "{Secluded U(1) below the weak scale}",
    eprint = "0811.1030",
    archivePrefix = "arXiv",
    primaryClass = "hep-ph",
    doi = "10.1103/PhysRevD.80.095002",
    journal = "Phys. Rev. D",
    volume = "80",
    pages = "095002",
    year = "2009"
}

@article{Chen:2015vqy,
    author = "Chen, Chien-Yi and Davoudiasl, Hooman and Marciano, William J. and Zhang, Cen",
    title = "{Implications of a light \textquotedblleft{}dark Higgs\textquotedblright{} solution to the $g_{\mu}$-2 discrepancy}",
    eprint = "1511.04715",
    archivePrefix = "arXiv",
    primaryClass = "hep-ph",
    doi = "10.1103/PhysRevD.93.035006",
    journal = "Phys. Rev. D",
    volume = "93",
    number = "3",
    pages = "035006",
    year = "2016"
}

@article{Davoudiasl:2018fbb,
    author = "Davoudiasl, Hooman and Marciano, William J.",
    title = "{Tale of two anomalies}",
    eprint = "1806.10252",
    archivePrefix = "arXiv",
    primaryClass = "hep-ph",
    doi = "10.1103/PhysRevD.98.075011",
    journal = "Phys. Rev. D",
    volume = "98",
    number = "7",
    pages = "075011",
    year = "2018"
}

@article{Czarnecki:2001pv,
    author = "Czarnecki, Andrzej and Marciano, William J.",
    title = "{The Muon anomalous magnetic moment: A Harbinger for 'new physics'}",
    eprint = "hep-ph/0102122",
    archivePrefix = "arXiv",
    reportNumber = "ALBERTA-THY-03-01, BNL-HET-01-4",
    doi = "10.1103/PhysRevD.64.013014",
    journal = "Phys. Rev. D",
    volume = "64",
    pages = "013014",
    year = "2001"
}

@article{Stockinger:1900zz,
    author = {St\"ockinger, Dominik},
    title = "{Muon (g \ensuremath{-} 2) and Physics Beyond the Standard Model}",
    doi = "10.1142/9789814271844_0012",
    journal = "Adv. Ser. Direct. High Energy Phys.",
    volume = "20",
    pages = "393--438",
    year = "2009"
}

@article{Giudice:2012ms,
    author = "Giudice, G. F. and Paradisi, P. and Passera, M.",
    title = "{Testing new physics with the electron g-2}",
    eprint = "1208.6583",
    archivePrefix = "arXiv",
    primaryClass = "hep-ph",
    reportNumber = "CERN-PH-TH-2012-017-a",
    doi = "10.1007/JHEP11(2012)113",
    journal = "JHEP",
    volume = "11",
    pages = "113",
    year = "2012"
}

@article{Crivellin:2019mvj,
    author = "Crivellin, Andreas and Hoferichter, Martin",
    title = "{Combined Explanations of  $(g-2)_\mu$,  $(g-2)_e$ and Implications for a Large Muon EDM\ensuremath{*}}",
    eprint = "1905.03789",
    archivePrefix = "arXiv",
    primaryClass = "hep-ph",
    reportNumber = "PSI-PR-19-07",
    doi = "10.22323/1.360.0009",
    journal = "PoS",
    volume = "ALPS2019",
    pages = "009",
    year = "2020"
}

@article{Chupp:2017rkp,
    author = "Chupp, Timothy and Fierlinger, Peter and Ramsey-Musolf, Michael and Singh, Jaideep",
    title = "{Electric dipole moments of atoms, molecules, nuclei, and particles}",
    eprint = "1710.02504",
    archivePrefix = "arXiv",
    primaryClass = "physics.atom-ph",
    reportNumber = "ACFI-T17-18",
    doi = "10.1103/RevModPhys.91.015001",
    journal = "Rev. Mod. Phys.",
    volume = "91",
    number = "1",
    pages = "015001",
    year = "2019"
}

@article{Aoyama:2020ynm,
    author = "Aoyama, T. and others",
    title = "{The anomalous magnetic moment of the muon in the Standard Model}",
    eprint = "2006.04822",
    archivePrefix = "arXiv",
    primaryClass = "hep-ph",
    reportNumber = "FERMILAB-PUB-20-207-T, INT-PUB-20-021, KEK Preprint 2020-5,
  MITP/20-028, KEK Preprint 2020-5, MITP/20-028, CERN-TH-2020-075, IFT-UAM/CSIC-20-74, LMU-ASC 18/20, LTH 1234,
  LU TP 20-20, LTH 1234, LU TP 20-20, MAN/HEP/2020/003, PSI-PR-20-06, UWThPh 2020-14, ZU-TH 18/20",
    doi = "10.1016/j.physrep.2020.07.006",
    journal = "Phys. Rept.",
    volume = "887",
    pages = "1--166",
    year = "2020"
}

@Article{	  aoyama:2012wk,
  author	= "Aoyama, Tatsumi and Hayakawa, Masashi and Kinoshita,
		  Toichiro and Nio, Makiko",
  title		= "{Complete Tenth-Order QED Contribution to the Muon $g-2$}",
  journal	= "Phys. Rev. Lett.",
  volume	= "109",
  year		= "2012",
  pages		= "111808",
  doi		= "10.1103/PhysRevLett.109.111808",
  eprint	= "1205.5370",
  archiveprefix	= "arXiv",
  primaryclass	= "hep-ph",
  reportnumber	= "RIKEN-QHP-26",
  slaccitation	= "%%CITATION = ARXIV:1205.5370;%%"
}

@article{Aoyama:2019ryr,
    author = "Aoyama, Tatsumi and Kinoshita, Toichiro and Nio, Makiko",
    title = "{Theory of the Anomalous Magnetic Moment of the Electron}",
    doi = "10.3390/atoms7010028",
    journal = "Atoms",
    volume = "7",
    number = "1",
    pages = "28",
    year = "2019"
}

@Article{	  czarnecki:2002nt,
  author	= "Czarnecki, Andrzej and Marciano, William J. and
		  Vainshtein, Arkady",
  title		= "{Refinements in electroweak contributions to the muon
		  anomalous magnetic moment}",
  journal	= "Phys. Rev.",
  volume	= "D67",
  year		= "2003",
  pages		= "073006",
  doi		= "10.1103/PhysRevD.67.073006",
 note		= "[Erratum: Phys. Rev. {D73}, 119901 (2006)]",
  eprint	= "hep-ph/0212229",
  archiveprefix	= "arXiv",
  primaryclass	= "hep-ph",
  reportnumber	= "ALBERTA-THY-18-02, BNL-HET-02-25, TPI-MINN-02-46,
		  UMN-TH-2119-02"
}

@Article{	  gnendiger:2013pva,
  author	= "Gnendiger, C. and St{\"o}ckinger, D. and
		  St{\"o}ckinger-Kim, H.",
  title		= "{The electroweak contributions to $(g-2)_{\mu}$ after the
		  Higgs boson mass measurement}",
  journal	= "Phys. Rev.",
  volume	= "D88",
  year		= "2013",
  pages		= "053005",
  doi		= "10.1103/PhysRevD.88.053005",
  eprint	= "1306.5546",
  archiveprefix	= "arXiv",
  primaryclass	= "hep-ph"
}

@Article{	  davier:2017zfy,
  author	= "Davier, Michel and Hoecker, Andreas and Malaescu, Bogdan
		  and Zhang, Zhiqing",
  title		= "{Reevaluation of the hadronic vacuum polarisation
		  contributions to the Standard Model predictions of the muon
		  $g-2$ and ${\alpha (m_Z^2)}$ using newest hadronic
		  cross-section data}",
  journal	= "Eur. Phys. J.",
  volume	= "C77",
  year		= "2017",
  number	= "12",
  pages		= "827",
  doi		= "10.1140/epjc/s10052-017-5161-6",
  eprint	= "1706.09436",
  archiveprefix	= "arXiv",
  primaryclass	= "hep-ph"
}

@Article{	  keshavarzi:2018mgv,
  author	= "Keshavarzi, Alexander and Nomura, Daisuke and Teubner,
		  Thomas",
  title		= "{Muon $g-2$ and $\alpha(M_Z^2)$: a new data-based
		  analysis}",
  journal	= "Phys. Rev.",
  volume	= "D97",
  year		= "2018",
  number	= "11",
  pages		= "114025",
  doi		= "10.1103/PhysRevD.97.114025",
  eprint	= "1802.02995",
  archiveprefix	= "arXiv",
  primaryclass	= "hep-ph",
  reportnumber	= "LTH 1153, KEK-TH-2035, LTH-1153, YITP-18-09, LTH 1153;
		  KEK-TH-2035; YITP-18-09"
}

@Article{	  colangelo:2018mtw,
  author	= "Colangelo, Gilberto and Hoferichter, Martin and Stoffer,
		  Peter",
  title		= "{Two-pion contribution to hadronic vacuum polarization}",
  journal	= "JHEP",
  volume	= "02",
  year		= "2019",
  pages		= "006",
  doi		= "10.1007/JHEP02(2019)006",
  eprint	= "1810.00007",
  archiveprefix	= "arXiv",
  primaryclass	= "hep-ph",
  reportnumber	= "INT-PUB-18-048"
}

@Article{	  hoferichter:2019gzf,
  author	= "Hoferichter, Martin and Hoid, Bai-Long and Kubis, Bastian",
  title		= "{Three-pion contribution to hadronic vacuum polarization}",
  journal	= "JHEP",
  volume	= "08",
  year		= "2019",
  pages		= "137",
  doi		= "10.1007/JHEP08(2019)137",
  eprint	= "1907.01556",
  archiveprefix	= "arXiv",
  primaryclass	= "hep-ph",
  reportnumber	= "INT-PUB-19-030"
}

@Article{	  davier:2019can,
  author	= "Davier, M. and Hoecker, A. and Malaescu, B. and Zhang, Z.",
  title		= "{A new evaluation of the hadronic vacuum polarisation
		  contributions to the muon anomalous magnetic moment and to
		  $\alpha(m_Z^2)$}",
  journal	= "Eur. Phys. J.",
  volume	= "C80",
  year		= "2020",
  number	= "3",
  pages		= "241",
  doi		= "10.1140/epjc/s10052-020-7792-2",
  note          = "[Erratum: Eur. Phys. J. {C80}, 410 (2020)]", 
  eprint	= "1908.00921",
  archiveprefix	= "arXiv",
  primaryclass	= "hep-ph"
}

@Article{	  keshavarzi:2019abf,
  author	= "Keshavarzi, Alexander and Nomura, Daisuke and Teubner,
		  Thomas",
  title		= "{The $g-2$ of charged leptons, $\alpha(M_Z^2)$ and the
		  hyperfine splitting of muonium}",
  journal	= "Phys. Rev.",
  volume	= "D101",
  year		= "2020",
  pages		= "014029",
  doi		= "10.1103/PhysRevD.101.014029",
  eprint	= "1911.00367",
  archiveprefix	= "arXiv",
  primaryclass	= "hep-ph"
}

@Article{	  kurz:2014wya,
  author	= "Kurz, Alexander and Liu, Tao and Marquard, Peter and
		  Steinhauser, Matthias",
  title		= "{Hadronic contribution to the muon anomalous magnetic
		  moment to next-to-next-to-leading order}",
  journal	= "Phys. Lett.",
  volume	= "B734",
  year		= "2014",
  pages		= "144-147",
  doi		= "10.1016/j.physletb.2014.05.043",
  eprint	= "1403.6400",
  archiveprefix	= "arXiv",
  primaryclass	= "hep-ph",
  reportnumber	= "SFB-CPP-14-19, TTP14-009, DESY-14-038, LPN14-056",
  slaccitation	= "%%CITATION = ARXIV:1403.6400;%%"
}

@Article{	  melnikov:2003xd,
  author	= "Melnikov, Kirill and Vainshtein, Arkady",
  title		= "{Hadronic light-by-light scattering contribution to the
		  muon anomalous magnetic moment revisited}",
  journal	= "Phys. Rev.",
  volume	= "D70",
  year		= "2004",
  pages		= "113006",
  doi		= "10.1103/PhysRevD.70.113006",
  eprint	= "hep-ph/0312226",
  archiveprefix	= "arXiv",
  primaryclass	= "hep-ph",
  reportnumber	= "UH-511-1041-03, FTPI-MINN-03-36, UMN-TH-2224-03",
  slaccitation	= "%%CITATION = HEP-PH/0312226;%%"
}

@Article{	  masjuan:2017tvw,
  archiveprefix	= {arXiv},
  author	= {Masjuan, Pere and S{\'a}nchez-Puertas, Pablo},
  doi		= {10.1103/PhysRevD.95.054026},
  eprint	= {1701.05829},
  journal	= {Phys. Rev.},
  number	= {5},
  pages		= {054026},
  primaryclass	= {hep-ph},
  slaccitation	= {%%CITATION = ARXIV:1701.05829;%%},
  title		= {{Pseudoscalar-pole contribution to the $(g_{\mu}-2)$: a
		  rational approach}},
  volume	= {D95},
  year		= {2017}
}

@article{Colangelo:2017fiz,
      author         = "Colangelo, Gilberto and Hoferichter, Martin and Procura,
                        Massimiliano and Stoffer, Peter",
      title          = "{Dispersion relation for hadronic light-by-light
                        scattering: two-pion contributions}",
      journal        = "JHEP",
      volume         = "04",
      year           = "2017",
      pages          = "161",
      doi            = "10.1007/JHEP04(2017)161",
      eprint         = "1702.07347",
      archivePrefix  = "arXiv",
      primaryClass   = "hep-ph",
      reportNumber   = "INT-PUB-17-009, CERN-TH-2017-041, NSF-KITP-17-036",
      SLACcitation   = "%%CITATION = ARXIV:1702.07347;%%"
}

@Article{	  hoferichter:2018kwz,
  archiveprefix	= {arXiv},
  author	= {Hoferichter, Martin and Hoid, Bai-Long and Kubis, Bastian
		  and Leupold, Stefan and Schneider, Sebastian P.},
  doi		= {10.1007/JHEP10(2018)141},
  eprint	= {1808.04823},
  journal	= {JHEP},
  pages		= {141},
  primaryclass	= {hep-ph},
  reportnumber	= {INT-PUB-18-042},
  slaccitation	= {%%CITATION = ARXIV:1808.04823;%%},
  title		= {{Dispersion relation for hadronic light-by-light
		  scattering: pion pole}},
  volume	= {10},
  year		= {2018}
}

@Article{	  gerardin:2019vio,
  archiveprefix	= {arXiv},
  author	= {G{\'e}rardin, Antoine and Meyer, Harvey B. and Nyffeler,
		  Andreas},
  doi		= {10.1103/PhysRevD.100.034520},
  eprint	= {1903.09471},
  journal	= {Phys. Rev.},
  number	= {3},
  pages		= {034520},
  primaryclass	= {hep-lat},
  reportnumber	= {MITP-19-014},
  slaccitation	= {%%CITATION = ARXIV:1903.09471;%%},
  title		= {{Lattice calculation of the pion transition form factor
		  with $N_f=2+1$ Wilson quarks}},
  volume	= {D100},
  year		= {2019}
}

@Article{	  bijnens:2019ghy,
  archiveprefix	= {arXiv},
  author	= {Bijnens, Johan and Hermansson-Truedsson, Nils and
		  Rodr{\'i}guez-S{\'a}nchez, Antonio},
  doi		= {10.1016/j.physletb.2019.134994},
  eprint	= {1908.03331},
  journal	= {Phys. Lett.},
  pages		= {134994},
  primaryclass	= {hep-ph},
  reportnumber	= {LU TP 19-38},
  slaccitation	= {%%CITATION = ARXIV:1908.03331;%%},
  title		= {{Short-distance constraints for the HLbL contribution to
		  the muon anomalous magnetic moment}},
  volume	= {B798},
  year		= {2019}
}

@Article{	  colangelo:2019uex,
  archiveprefix	= {arXiv},
  author	= {Colangelo, Gilberto and Hagelstein, Franziska and
		  Hoferichter, Martin and Laub, Laetitia and Stoffer, Peter},
  doi		= {10.1007/JHEP03(2020)101},
  journal	= {JHEP},
  pages		= {101},
  volume	= {03},
  year		= {2020},
  eprint	= {1910.13432},
  primaryclass	= {hep-ph},
  reportnumber	= {INT-PUB-19-051},
  slaccitation	= {%%CITATION = ARXIV:1910.13432;%%},
  title		= {{Longitudinal short-distance constraints for the hadronic
		  light-by-light contribution to $(g-2)_\mu$ with large-$N_c$
		  Regge models}}
}

@article{Blum:2019ugy,
      author         = "Blum, Thomas and Christ, Norman and Hayakawa, Masashi and
                        Izubuchi, Taku and Jin, Luchang and Jung, Chulwoo and
                        Lehner, Christoph",
      title          = "{The hadronic light-by-light scattering contribution to
                        the muon anomalous magnetic moment from lattice QCD}",
      journal        = "Phys. Rev. Lett.",
      volume         = "124",
      year           = "2020",
      number         = "13",
      pages          = "132002",
      doi            = "10.1103/PhysRevLett.124.132002",
      eprint         = "1911.08123",
      archivePrefix  = "arXiv",
      primaryClass   = "hep-lat"
}

@Article{	  colangelo:2014qya,
  author	= "Colangelo, Gilberto and Hoferichter, Martin and Nyffeler,
		  Andreas and Passera, Massimo and Stoffer, Peter",
  title		= "{Remarks on higher-order hadronic corrections to the muon
		  $g-2$}",
  journal	= "Phys. Lett.",
  volume	= "B735",
  year		= "2014",
  pages		= "90-91",
  doi		= "10.1016/j.physletb.2014.06.012",
  eprint	= "1403.7512",
  archiveprefix	= "arXiv",
  primaryclass	= "hep-ph"
}

@TechReport{MUonE:LoI,
  author	= "Abbiendi, G and others",
  collaboration = "{MUonE}",
  title		= "{Letter of Intent: the MUonE project}",
  number	= "CERN-SPSC-2019-026, SPSC-I-252",
  month		= "6",
  year		= "2019",
  reportnumber	= "CERN-SPSC-2019-026",
  url		= "https://cds.cern.ch/record/2677471",
}

@article{Banerjee:2020tdt,
    author = "Banerjee, Pulak and others",
    title = "{Theory for muon-electron scattering @ 10 ppm: A report of the MUonE theory initiative}",
    eprint = "2004.13663",
    archivePrefix = "arXiv",
    primaryClass = "hep-ph",
    reportNumber = "IFIC/20-16, LAPTH-017/20, PSI-PR-20-05, TTP20-018, ZU-TH 11/20",
    doi = "10.1140/epjc/s10052-020-8138-9",
    journal = "Eur. Phys. J. C",
    volume = "80",
    number = "6",
    pages = "591",
    year = "2020"
}

@article{Schwinger:1948iu,
    author = "Schwinger, Julian S.",
    title = "{On Quantum electrodynamics and the magnetic moment of the electron}",
    doi = "10.1103/PhysRev.73.416",
    journal = "Phys. Rev.",
    volume = "73",
    pages = "416--417",
    year = "1948"
}

@article{Kusch:1948mvb,
    author = "Kusch, P. and Foley, H. M.",
    title = "{The Magnetic Moment of the Electron}",
    doi = "10.1103/PhysRev.74.250",
    journal = "Phys. Rev.",
    volume = "74",
    number = "3",
    pages = "250",
    year = "1948"
}

@article{He:1991qd,
    author = "He, Xiao-Gang and Joshi, Girish C. and Lew, H. and Volkas, R. R.",
    title = "{Simplest Z-prime model}",
    reportNumber = "CERN-TH-6084-91, UM-P-91-32, OZ-91-07",
    doi = "10.1103/PhysRevD.44.2118",
    journal = "Phys. Rev. D",
    volume = "44",
    pages = "2118--2132",
    year = "1991"
}

@article{Foot:1990mn,
    author = "Foot, Robert",
    title = "{New Physics From Electric Charge Quantization?}",
    reportNumber = "MAD/TH/90-14",
    doi = "10.1142/S0217732391000543",
    journal = "Mod. Phys. Lett. A",
    volume = "6",
    pages = "527--530",
    year = "1991"
}

@article{He:1990pn,
    author = "He, X. G. and Joshi, Girish C. and Lew, H. and Volkas, R. R.",
    title = "{New Z-prime Phenomenology}",
    reportNumber = "UM-P-90/42, OZ-P-90/16",
    doi = "10.1103/PhysRevD.43.R22",
    journal = "Phys. Rev. D",
    volume = "43",
    pages = "22--24",
    year = "1991"
}

@article{Altmannshofer:2016oaq,
    author = "Altmannshofer, Wolfgang and Carena, Marcela and Crivellin, Andreas",
    title = "{$L_\mu - L_\tau$ theory of Higgs flavor violation and $(g-2)_\mu$}",
    eprint = "1604.08221",
    archivePrefix = "arXiv",
    primaryClass = "hep-ph",
    reportNumber = "PSI-PR-16-04, FERMILAB-PUB-16-131-T",
    doi = "10.1103/PhysRevD.94.095026",
    journal = "Phys. Rev. D",
    volume = "94",
    number = "9",
    pages = "095026",
    year = "2016"
}

@article{Chivukula:1987fw,
    author = "Chivukula, R.Sekhar and Georgi, H. and Randall, Lisa",
    title = "{A Composite Technicolor Standard Model of Quarks}",
    doi = "10.1016/0550-3213(87)90638-9",
    journal = "Nucl. Phys. B",
    volume = "292",
    pages = "93--108",
    year = "1987"
}

@article{Hall:1990ac,
    author = "Hall, L.J. and Randall, Lisa",
    title = "{Weak scale effective supersymmetry}",
    reportNumber = "UCB-PTH-90/13, LBL-28879",
    doi = "10.1103/PhysRevLett.65.2939",
    journal = "Phys. Rev. Lett.",
    volume = "65",
    pages = "2939--2942",
    year = "1990"
}

@article{Buras:2000dm,
    author = "Buras, A.J. and Gambino, P. and Gorbahn, M. and Jager, S. and Silvestrini, L.",
    title = "{Universal unitarity triangle and physics beyond the standard model}",
    eprint = "hep-ph/0007085",
    archivePrefix = "arXiv",
    reportNumber = "TUM-HEP-379-00, CERN-TH-2000-190",
    doi = "10.1016/S0370-2693(01)00061-2",
    journal = "Phys. Lett. B",
    volume = "500",
    pages = "161--167",
    year = "2001"
}

@article{DAmbrosio:2002vsn,
    author = "D'Ambrosio, G. and Giudice, G.F. and Isidori, G. and Strumia, A.",
    title = "{Minimal flavor violation: An Effective field theory approach}",
    eprint = "hep-ph/0207036",
    archivePrefix = "arXiv",
    reportNumber = "CERN-TH-2002-147, IFUP-TH-2002-17",
    doi = "10.1016/S0550-3213(02)00836-2",
    journal = "Nucl. Phys. B",
    volume = "645",
    pages = "155--187",
    year = "2002"
}

@inproceedings{Butler:2017afk,
    author = "Butler, Joel Nathan",
    collaboration = "CMS",
    title = "{Highlights and Perspectives from the CMS Experiment}",
    booktitle = "{5th Large Hadron Collider Physics Conference}",
    eprint = "1709.03006",
    archivePrefix = "arXiv",
    primaryClass = "hep-ex",
    reportNumber = "CMS-CR-2017-226, FERMILAB-CONF-17-366-CMS",
    month = "9",
    year = "2017"
}

@article{Masetti:2018btj,
    author = "Masetti, Lucia",
    editor = "Ricciardi, Giulia and De Nardo, Guglielmo and Merola, Mario and Sciacca, Crisostomo",
    collaboration = "ATLAS",
    title = "{ATLAS results and prospects with focus on beyond the Standard Model}",
    reportNumber = "ATL-PHYS-PROC-2018-160",
    doi = "10.1016/j.nuclphysbps.2019.03.009",
    journal = "Nucl. Part. Phys. Proc.",
    volume = "303-305",
    pages = "43--48",
    year = "2018"
}

@article{Aaij:2014pli,
    author = "Aaij, R. and others",
    collaboration = "LHCb",
    title = "{Differential branching fractions and isospin asymmetries of $B \to K^{(*)} \mu^+ \mu^-$ decays}",
    eprint = "1403.8044",
    archivePrefix = "arXiv",
    primaryClass = "hep-ex",
    reportNumber = "LHCB-PAPER-2014-006, CERN-PH-EP-2014-055",
    doi = "10.1007/JHEP06(2014)133",
    journal = "JHEP",
    volume = "06",
    pages = "133",
    year = "2014"
}

@article{Aaij:2014ora,
    author = "Aaij, Roel and others",
    collaboration = "LHCb",
    title = "{Test of lepton universality using $B^{+}\rightarrow K^{+}\ell^{+}\ell^{-}$ decays}",
    eprint = "1406.6482",
    archivePrefix = "arXiv",
    primaryClass = "hep-ex",
    reportNumber = "CERN-PH-EP-2014-140, LHCB-PAPER-2014-024",
    doi = "10.1103/PhysRevLett.113.151601",
    journal = "Phys. Rev. Lett.",
    volume = "113",
    pages = "151601",
    year = "2014"
}

@article{Aaij:2015esa,
    author = "Aaij, Roel and others",
    collaboration = "LHCb",
    title = "{Angular analysis and differential branching fraction of the decay $B^0_s\to\phi\mu^+\mu^-$}",
    eprint = "1506.08777",
    archivePrefix = "arXiv",
    primaryClass = "hep-ex",
    reportNumber = "CERN-PH-EP-2015-145, LHCB-PAPER-2015-023",
    doi = "10.1007/JHEP09(2015)179",
    journal = "JHEP",
    volume = "09",
    pages = "179",
    year = "2015"
}

@article{Aaij:2015oid,
    author = "Aaij, Roel and others",
    collaboration = "LHCb",
    title = "{Angular analysis of the $B^{0} \to K^{*0} \mu^{+} \mu^{-}$ decay using 3 fb$^{-1}$ of integrated luminosity}",
    eprint = "1512.04442",
    archivePrefix = "arXiv",
    primaryClass = "hep-ex",
    reportNumber = "CERN-PH-EP-2015-314, LHCB-PAPER-2015-051",
    doi = "10.1007/JHEP02(2016)104",
    journal = "JHEP",
    volume = "02",
    pages = "104",
    year = "2016"
}

@article{Khachatryan:2015isa,
    author = "Khachatryan, Vardan and others",
    collaboration = "CMS",
    title = "{Angular analysis of the decay $B^0 \to K^{*0} \mu^+ \mu^-$ from pp collisions at $\sqrt  s = 8$ TeV}",
    eprint = "1507.08126",
    archivePrefix = "arXiv",
    primaryClass = "hep-ex",
    reportNumber = "CMS-BPH-13-010, CERN-PH-EP-2015-178",
    doi = "10.1016/j.physletb.2015.12.020",
    journal = "Phys. Lett. B",
    volume = "753",
    pages = "424--448",
    year = "2016"
}

@article{ATLAS:2017dlm,
    author = "Carli, Ina",
    collaboration = "ATLAS",
    title = "{Angular analysis of $B^0_d \to K^* \mu^+ \mu^-$ decay with the ATLAS detector}",
    reportNumber = "ATL-PHYS-PROC-2017-138",
    doi = "10.22323/1.304.0043",
    journal = "PoS",
    volume = "FPCP2017",
    pages = "043",
    year = "2017"
}

@article{Sirunyan:2017dhj,
    author = "Sirunyan, Albert M and others",
    collaboration = "CMS",
    title = "{Measurement of angular parameters from the decay $\mathrm{B}^0 \to \mathrm{K}^{*0} \mu^+ \mu^-$ in proton-proton collisions at $\sqrt{s} = $ 8 TeV}",
    eprint = "1710.02846",
    archivePrefix = "arXiv",
    primaryClass = "hep-ex",
    reportNumber = "CMS-BPH-15-008, CERN-EP-2017-240",
    doi = "10.1016/j.physletb.2018.04.030",
    journal = "Phys. Lett. B",
    volume = "781",
    pages = "517--541",
    year = "2018"
}

@article{Aaij:2017vbb,
    author = "Aaij, R. and others",
    collaboration = "LHCb",
    title = "{Test of lepton universality with $B^{0} \rightarrow K^{*0}\ell^{+}\ell^{-}$ decays}",
    eprint = "1705.05802",
    archivePrefix = "arXiv",
    primaryClass = "hep-ex",
    reportNumber = "LHCB-PAPER-2017-013, CERN-EP-2017-100",
    doi = "10.1007/JHEP08(2017)055",
    journal = "JHEP",
    volume = "08",
    pages = "055",
    year = "2017"
}

@article{Cirigliano:2005ck,
    author = "Cirigliano, Vincenzo and Grinstein, Benjamin and Isidori, Gino and Wise, Mark B.",
    title = "{Minimal flavor violation in the lepton sector}",
    eprint = "hep-ph/0507001",
    archivePrefix = "arXiv",
    reportNumber = "UCSD-PTH-05-11, CALT-68-2566",
    doi = "10.1016/j.nuclphysb.2005.08.037",
    journal = "Nucl. Phys. B",
    volume = "728",
    pages = "121--134",
    year = "2005"
}

@article{Djouadi:1989md,
    author = "Djouadi, A. and Kohler, T. and Spira, M. and Tutas, J.",
    title = "{$(e b)$, $(e t)$ type leptoquarks at $e$ $p$ colliners}",
    reportNumber = "PITHA 89/12",
    doi = "10.1007/BF01560270",
    journal = "Z. Phys. C",
    volume = "46",
    pages = "679--686",
    year = "1990"
}

@article{ColuccioLeskow:2016dox,
    author = {Coluccio Leskow, Estefania and D'Ambrosio, Giancarlo and Crivellin, Andreas and M\"uller, Dario},
    title = "{$(g-2)\mu$, lepton flavor violation, and $Z$ decays with leptoquarks: Correlations and future prospects}",
    eprint = "1612.06858",
    archivePrefix = "arXiv",
    primaryClass = "hep-ph",
    reportNumber = "PSI-PR-16-17, ZU-TH-44-16",
    doi = "10.1103/PhysRevD.95.055018",
    journal = "Phys. Rev. D",
    volume = "95",
    number = "5",
    pages = "055018",
    year = "2017"
}

@article{Crivellin:2020tsz,
    author = "Crivellin, Andreas and Mueller, Dario and Saturnino, Francesco",
    title = "{Correlating $h\to \mu^+\mu^-$  to the Anomalous Magnetic Moment of the Muon via Leptoquarks}",
    eprint = "2008.02643",
    archivePrefix = "arXiv",
    primaryClass = "hep-ph",
    reportNumber = "CERN-TH-2020-134, PSI-PR-20-12, ZU-TH 28/20",
    doi = "10.1103/PhysRevLett.127.021801",
    journal = "Phys. Rev. Lett.",
    volume = "127",
    number = "2",
    pages = "021801",
    year = "2021"
}

@article{Crivellin:2020mjs,
    author = {Crivellin, Andreas and Greub, Christoph and M\"uller, Dario and Saturnino, Francesco},
    title = "{Scalar Leptoquarks in Leptonic Processes}",
    eprint = "2010.06593",
    archivePrefix = "arXiv",
    primaryClass = "hep-ph",
    reportNumber = "CERN-TH-2020-167, PSI-20-17, ZU-TH 38/20",
    doi = "10.1007/JHEP02(2021)182",
    journal = "JHEP",
    volume = "02",
    pages = "182",
    year = "2021"
}

@article{Falkowski:2013jya,
    author = "Falkowski, Adam and Straub, David M. and Vicente, Avelino",
    title = "{Vector-like leptons: Higgs decays and collider phenomenology}",
    eprint = "1312.5329",
    archivePrefix = "arXiv",
    primaryClass = "hep-ph",
    reportNumber = "LPT-ORSAY-13-141",
    doi = "10.1007/JHEP05(2014)092",
    journal = "JHEP",
    volume = "05",
    pages = "092",
    year = "2014"
}

@article{Dermisek:2013gta,
    author = "Dermisek, Radovan and Raval, Aditi",
    title = "{Explanation of the Muon g-2 Anomaly with Vectorlike Leptons and its Implications for Higgs Decays}",
    eprint = "1305.3522",
    archivePrefix = "arXiv",
    primaryClass = "hep-ph",
    doi = "10.1103/PhysRevD.88.013017",
    journal = "Phys. Rev. D",
    volume = "88",
    pages = "013017",
    year = "2013"
}

@article{Crivellin:2021rbq,
    author = "Crivellin, Andreas and Hoferichter, Martin",
    title = "{Consequences of chirally enhanced explanations of $(g-2)_\mu$ for $h\to \mu\mu$ and $Z\to \mu\mu$}",
    eprint = "2104.03202",
    archivePrefix = "arXiv",
    primaryClass = "hep-ph",
    reportNumber = "CERN-TH-2021-050, PSI-PR-21-04, ZU-TH 14/21",
    doi = "10.1007/JHEP07(2021)135",
    journal = "JHEP",
    volume = "07",
    pages = "135",
    year = "2021"
}

@article{Arnan:2019uhr,
    author = "Arnan, Pere and Crivellin, Andreas and Fedele, Marco and Mescia, Federico",
    title = "{Generic Loop Effects of New Scalars and Fermions in $b\to s\ell^+\ell^-$, $(g-2)_\mu$ and a Vector-like $4^\text{th}$ Generation}",
    eprint = "1904.05890",
    archivePrefix = "arXiv",
    primaryClass = "hep-ph",
    reportNumber = "PSI-PR-19-05, ZU-TH-18/19",
    doi = "10.1007/JHEP06(2019)118",
    journal = "JHEP",
    volume = "06",
    pages = "118",
    year = "2019"
}

@article{Kowalska:2017iqv,
    author = "Kowalska, Kamila and Sessolo, Enrico Maria",
    title = "{Expectations for the muon g-2 in simplified models with dark matter}",
    eprint = "1707.00753",
    archivePrefix = "arXiv",
    primaryClass = "hep-ph",
    reportNumber = "DO-TH-17-13, QFET-2017-11",
    doi = "10.1007/JHEP09(2017)112",
    journal = "JHEP",
    volume = "09",
    pages = "112",
    year = "2017"
}

@article{Frank:2020smf,
    author = "Frank, Mariana and Saha, Ipsita",
    title = "{Muon anomalous magnetic moment in two-Higgs-doublet models with vectorlike leptons}",
    eprint = "2008.11909",
    archivePrefix = "arXiv",
    primaryClass = "hep-ph",
    doi = "10.1103/PhysRevD.102.115034",
    journal = "Phys. Rev. D",
    volume = "102",
    number = "11",
    pages = "115034",
    year = "2020"
}

@article{Chun:2020uzw,
    author = "Chun, Eung Jin and Mondal, Tanmoy",
    title = "{Explaining $g-2$ anomalies in two Higgs doublet model with vector-like leptons}",
    eprint = "2009.08314",
    archivePrefix = "arXiv",
    primaryClass = "hep-ph",
    reportNumber = "KIAS - P20050",
    doi = "10.1007/JHEP11(2020)077",
    journal = "JHEP",
    volume = "11",
    pages = "077",
    year = "2020"
}

@article{Ferreira:2021gke,
    author = "Ferreira, P. M. and Gon\c{c}alves, B. L. and Joaquim, F. R. and Sher, Marc",
    title = "{$(g-2)_\mu$ in the 2HDM and slightly beyond -- an updated view}",
    eprint = "2104.03367",
    archivePrefix = "arXiv",
    primaryClass = "hep-ph",
    month = "4",
    year = "2021"
}

@article{Abe:2015oca,
    author = "Abe, Tomohiro and Sato, Ryosuke and Yagyu, Kei",
    title = "{Lepton-specific two Higgs doublet model as a solution of muon $g-2$ anomaly}",
    eprint = "1504.07059",
    archivePrefix = "arXiv",
    primaryClass = "hep-ph",
    reportNumber = "KEK-TH-1803",
    doi = "10.1007/JHEP07(2015)064",
    journal = "JHEP",
    volume = "07",
    pages = "064",
    year = "2015"
}

@article{Ilisie:2015tra,
    author = "Ilisie, Victor",
    title = "{New Barr-Zee contributions to $(g-2)_\mu$ in two-Higgs-doublet models}",
    eprint = "1502.04199",
    archivePrefix = "arXiv",
    primaryClass = "hep-ph",
    reportNumber = "FTUV-15-0213, IFIC-15-09",
    doi = "10.1007/JHEP04(2015)077",
    journal = "JHEP",
    volume = "04",
    pages = "077",
    year = "2015"
}

@article{Wang:2014sda,
    author = "Wang, Lei and Han, Xiao-Fang",
    title = "{A light pseudoscalar of 2HDM confronted with muon g-2 and experimental constraints}",
    eprint = "1412.4874",
    archivePrefix = "arXiv",
    primaryClass = "hep-ph",
    doi = "10.1007/JHEP05(2015)039",
    journal = "JHEP",
    volume = "05",
    pages = "039",
    year = "2015"
}

@article{Hou:2021qmf,
    author = "Hou, Wei-Shu and Kumar, Girish",
    title = "{Charged lepton flavor violation in light of Muon $g-2$}",
    eprint = "2107.14114",
    archivePrefix = "arXiv",
    primaryClass = "hep-ph",
    month = "7",
    year = "2021"
}

@article{Wang:2021fkn,
    author = "Wang, Hong-Xin and Wang, Lei and Zhang, Yang",
    title = "{muon $g-2$ anomaly and $\mu$-$\tau$-philic Higgs doublet with a light CP-even component}",
    eprint = "2104.03242",
    archivePrefix = "arXiv",
    primaryClass = "hep-ph",
    month = "4",
    year = "2021"
}

@article{Iguro:2020rby,
    author = "Iguro, Syuhei and Omura, Yuji and Takeuchi, Michihisa",
    title = "{Probing $\mu\tau$ flavor-violating solutions for the muon $g-2$ anomaly at Belle II}",
    eprint = "2002.12728",
    archivePrefix = "arXiv",
    primaryClass = "hep-ph",
    doi = "10.1007/JHEP09(2020)144",
    journal = "JHEP",
    volume = "09",
    pages = "144",
    year = "2020"
}

@article{Crivellin:2019dun,
    author = {Crivellin, Andreas and M\"uller, Dario and Wiegand, Christoph},
    title = "{$b\to s\ell^+\ell^-$ transitions in two-Higgs-doublet models}",
    eprint = "1903.10440",
    archivePrefix = "arXiv",
    primaryClass = "hep-ph",
    reportNumber = "PSI-PR-19-02, ZU-TH 10/19",
    doi = "10.1007/JHEP06(2019)119",
    journal = "JHEP",
    volume = "06",
    pages = "119",
    year = "2019"
}

@article{Cao:2009as,
    author = "Cao, Junjie and Wan, Peihua and Wu, Lei and Yang, Jin Min",
    title = "{Lepton-Specific Two-Higgs Doublet Model: Experimental Constraints and Implication on Higgs Phenomenology}",
    eprint = "0909.5148",
    archivePrefix = "arXiv",
    primaryClass = "hep-ph",
    doi = "10.1103/PhysRevD.80.071701",
    journal = "Phys. Rev. D",
    volume = "80",
    pages = "071701",
    year = "2009"
}

@article{Darme:2021qzw,
    author = "Darm\'e, Luc and Fedele, Marco and Kowalska, Kamila and Sessolo, Enrico Maria",
    title = "{Flavour anomalies and the muon $g-2$ from feebly interacting particles}",
    eprint = "2106.12582",
    archivePrefix = "arXiv",
    primaryClass = "hep-ph",
    reportNumber = "TTP21-020, P3H-21-047",
    month = "6",
    year = "2021"
}

@article{Buen-Abad:2021fwq,
    author = "Buen-Abad, Manuel A. and Fan, JiJi and Reece, Matthew and Sun, Chen",
    title = "{Challenges for an axion explanation of the muon $g-2$ measurement}",
    eprint = "2104.03267",
    archivePrefix = "arXiv",
    primaryClass = "hep-ph",
    month = "4",
    year = "2021"
}

@article{Cadeddu:2021dqx,
    author = "Cadeddu, M. and Cargioli, N. and Dordei, F. and Giunti, C. and Picciau, E.",
    title = "{Muon and electron g-2 and proton and cesium weak charges implications on dark Zd models}",
    eprint = "2104.03280",
    archivePrefix = "arXiv",
    primaryClass = "hep-ph",
    doi = "10.1103/PhysRevD.104.L011701",
    journal = "Phys. Rev. D",
    volume = "104",
    number = "1",
    pages = "011701",
    year = "2021"
}

@article{Darme:2020sjf,
    author = "Darm\'e, L. and Giacchino, Federica and Nardi, Enrico and Raggi, Mauro",
    title = "{Invisible decays of axion-like particles: constraints and prospects}",
    eprint = "2012.07894",
    archivePrefix = "arXiv",
    primaryClass = "hep-ph",
    doi = "10.1007/JHEP06(2021)009",
    journal = "JHEP",
    volume = "06",
    pages = "009",
    year = "2021"
}

@article{LHCb:2021zwz,
    author = "Aaij, Roel and others",
    collaboration = "LHCb",
    title = "{Branching fraction measurements of the rare $B^0_s\rightarrow\phi\mu^+\mu^-$ and $B^0_s\rightarrow f_2^\prime(1525)\mu^+\mu^-$ decays}",
    eprint = "2105.14007",
    archivePrefix = "arXiv",
    primaryClass = "hep-ex",
    reportNumber = "LHCb-PAPER-2021-014, CERN-EP-2021-092",
    month = "5",
    year = "2021"
}

@article{LHCb:2020zud,
    collaboration = "LHCb",
    title = "{Combination of the ATLAS, CMS and LHCb results on the $B^0_{(s)} \to \mu^+ \mu^-$ decays}",
    reportNumber = "LHCb-CONF-2020-002, CERN-LHCb-CONF-2020-002",
    month = "8",
    year = "2020"
}

@article{Ciuchini:2020gvn,
    author = "Ciuchini, Marco and Fedele, Marco and Franco, Enrico and Paul, Ayan and Silvestrini, Luca and Valli, Mauro",
    title = "{Lessons from the $B^{0,+}\to K^{*0,+}\mu^+\mu^-$ angular analyses}",
    eprint = "2011.01212",
    archivePrefix = "arXiv",
    primaryClass = "hep-ph",
    reportNumber = "DESY 20-190, DESY-20-190, HU-EP-20/30, HU 20/30, TTP20-037, P3H-20-064, UCI-TR 2020-18",
    doi = "10.1103/PhysRevD.103.015030",
    journal = "Phys. Rev. D",
    volume = "103",
    number = "1",
    pages = "015030",
    year = "2021"
}

@article{Hurth:2020ehu,
    author = "Hurth, T. and Mahmoudi, F. and Neshatpour, S.",
    title = "{Model independent analysis of the angular observables in $B^{0} \to K^{*0} \mu^+ \mu^-$ and $B^{+} \to K^{*+} \mu^+ \mu^-$}",
    eprint = "2012.12207",
    archivePrefix = "arXiv",
    primaryClass = "hep-ph",
    reportNumber = "CERN-TH-2020-222, MITP-20-085",
    doi = "10.1103/PhysRevD.103.095020",
    journal = "Phys. Rev. D",
    volume = "103",
    pages = "095020",
    year = "2021"
}

@article{Alok:2020bia,
    author = "Alok, Ashutosh Kumar and Kumbhakar, Suman and Uma Sankar, S.",
    title = "{A unique discrimination between new physics scenarios in $b\rightarrow s\mu^+\mu^-$ anomalies}",
    eprint = "2001.04395",
    archivePrefix = "arXiv",
    primaryClass = "hep-ph",
    month = "1",
    year = "2020"
}

@article{Altmannshofer:2021qrr,
    author = "Altmannshofer, Wolfgang and Stangl, Peter",
    title = "{New Physics in Rare B Decays after Moriond 2021}",
    eprint = "2103.13370",
    archivePrefix = "arXiv",
    primaryClass = "hep-ph",
    month = "3",
    year = "2021"
}

@inproceedings{Alguero:2021anc,
    author = "Alguer\'o, Marcel and Capdevila, Bernat and Descotes-Genon, S\'ebastien and Matias, Joaquim and Novoa-Brunet, Mart\'\i{}n",
    title = "{${b\to s\ell\ell}$ global fits after Moriond 2021 results}",
    booktitle = "{55th Rencontres de Moriond on QCD and High Energy Interactions}",
    eprint = "2104.08921",
    archivePrefix = "arXiv",
    primaryClass = "hep-ph",
    month = "4",
    year = "2021"
}

@article{Ohayon:2021dec,
    author = "Ohayon, Ben and Burkley, Zakary and Crivelli, Paolo",
    title = "{Current status and prospects of muonium spectroscopy at PSI}",
    doi = "10.21468/SciPostPhysProc.5.029",
    journal = "SciPost Phys. Proc.",
    volume = "5",
    pages = "029",
    year = "2021"
}

@article{2000-1S2S,
  title = {Measurement of the $1\mathit{s}\ensuremath{-}2\mathit{s}$ Energy Interval in Muonium},
  author = {Meyer, V. and Bagayev, S. N. and Baird, P. E. G. and Bakule, P. and Boshier, M. G. and Breitr\"uck, A. and Cornish, S. L. and Dychkov, S. and Eaton, G. H. and Grossmann, A. and H\"ubl, D. and Hughes, V. W. and Jungmann, K. and Lane, I. C. and Liu, Yi-Wei and Lucas, D. and Matyugin, Y. and Merkel, J. and zu Putlitz, G. and Reinhard, I. and Sandars, P. G. H. and Santra, R. and Schmidt, P. V. and Scott, C. A. and Toner, W. T. and Towrie, M. and Tr\"ager, K. and Willmann, L. and Yakhontov, V.},
  journal = {Phys. Rev. Lett.},
  volume = {84},
  issue = {6},
  pages = {1136--1139},
  numpages = {0},
  year = {2000},
  month = {2},
  publisher = {American Physical Society},
  doi = {10.1103/PhysRevLett.84.1136},
}

@article{Savely,
    author = "Karshenboim, Savely G.",
    title = "{Precision physics of simple atoms: QED tests, nuclear structure and fundamental constants}",
    eprint = "hep-ph/0509010",
    archivePrefix = "arXiv",
    doi = "10.1016/j.physrep.2005.08.008",
    journal = "Phys. Rept.",
    volume = "422",
    pages = "1--63",
    year = "2005"
}

@article{NA64:2021xzo,
    author = "Andreev, Yu. M. and others",
    collaboration = "NA64",
    title = "{Constraints on New Physics in Electron $g-2$ from a Search for Invisible Decays of a Scalar, Pseudoscalar, Vector, and Axial Vector}",
    eprint = "2102.01885",
    archivePrefix = "arXiv",
    primaryClass = "hep-ex",
    reportNumber = "CERN-EP-2021-017",
    doi = "10.1103/PhysRevLett.126.211802",
    journal = "Phys. Rev. Lett.",
    volume = "126",
    number = "21",
    pages = "211802",
    year = "2021"
}

@article{Fadeev:2018rfl,
    author = "Fadeev, Pavel and Stadnik, Yevgeny V. and Ficek, Filip and Kozlov, Mikhail G. and Flambaum, Victor V. and Budker, Dmitry",
    title = "{Revisiting spin-dependent forces mediated by new bosons: Potentials in the coordinate-space representation for macroscopic- and atomic-scale experiments}",
    eprint = "1810.10364",
    archivePrefix = "arXiv",
    primaryClass = "hep-ph",
    doi = "10.1103/PhysRevA.99.022113",
    journal = "Phys. Rev. A",
    volume = "99",
    number = "2",
    pages = "022113",
    year = "2019"
}

@article{ATLAS:2021tfo,
    author = "Aad, G., and others",
    collaboration = "ATLAS",
    title = "{The ATLAS Fast TracKer system}",
    eprint = "2101.05078",
    archivePrefix = "arXiv",
    primaryClass = "physics.ins-det",
    reportNumber = "CERN-EP-2020-232",
    doi = "10.1088/1748-0221/16/07/P07006",
    journal = "JINST",
    volume = "16",
    pages = "P07006",
    year = "2021"
}

@article{Dittmeier:2020cct,
    author = "Dittmeier, Sebastian",
    collaboration = "ATLAS TDAQ",
    title = "{The ATLAS Hardware Track Trigger design towards first prototypes}",
    reportNumber = "ATL-DAQ-PROC-2019-032",
    doi = "10.22323/1.373.0049",
    journal = "PoS",
    volume = "Vertex2019",
    pages = "049",
    year = "2020"
}

@article{beer2014enhancement,
  title={Enhancement of muonium emission rate from silica aerogel with a laser-ablated surface},
  author={Beer, GA and Fujiwara, Y and Hirota, S and Ishida, K and Iwasaki, M and Kanda, S and Kawai, H and Kawamura, N and Kitamura, R and Lee, S and others},
  journal={Progress of Theoretical and Experimental Physics},
  volume={2014},
  number={9},
  year={2014},
  publisher={Oxford University Press},
  doi={ 10.1093/ptep/ptu116}
}

@phdthesis{kim2015,
    title    = {{Towards next generation fundamental precision measurements with muons}},
    school   = {{ETH Z\"urich Research Collection}},
    author   = {Khaw,~Kim~S.},
    year     = {2015},
    url = {https://www.research-collection.ethz.ch/handle/20.500.11850/102926}
}

@article{taqqu2011ultraslow,
  title={Ultraslow Muonium for a Muon beam of ultra high quality},
  author={Taqqu, D},
  journal={Physics Procedia},
  volume={17},
  pages={216--223},
  year={2011},
  publisher={Elsevier},
  doi={ 10.1016/j.phpro.2011.06.039}
}

@inproceedings{kirch2014testing,
  title={Testing antimatter gravity with muonium},
  author={Kirch, Klaus and Khaw, Kim Siang},
  booktitle={International Journal of Modern Physics: Conference Series},
  volume={30},
  pages={1460258},
  year={2014},
  organization={World Scientific},
  doi={ 10.1142/S2010194514602580}
}

@article{soter2021development,
title={{Development of a cold atomic muonium beam for next generation atomic physics and gravity experiments}},
author={Anna Soter and  Andreas Knecht},
journal={SciPost Phys. Proc.},
issue={5},
pages={31},
year={2021},
publisher={SciPost},
doi={10.21468/SciPostPhysProc.5.031},
}

@article{saarela1993hydrogen,
  title={Hydrogen isotope and {$^3$He} impurities in liquid {$^4$He}},
  author={Saarela, M and Krotscheck, E},
  journal={Journal of Low Temperature Physics},
  volume={90},
  number={5},
  pages={415--449},
  year={1993},
  publisher={Springer},
  doi={ 10.1007/BF00681890}
}

@article{marin1998atomic,
  title={Atomic and Molecular Hydrogen Impurities in Liquid 4 He},
  author={Marin, JM and Boronat, J and Casulleras, J},
  journal={Journal of Low Temperature Physics},
  volume={110},
  number={1},
  pages={205--211},
  year={1998},
  doi={ 10.1023/A:1022599725388},
  publisher={Springer}
}

@article{oberthaler1996inertial,
  title={Inertial sensing with classical atomic beams},
  author={Oberthaler, Markus K and Bernet, Stefan and Rasel, Ernst M and Schmiedmayer, J{\"o}rg and Zeilinger, Anton},
  journal={Physical Review A},
  volume={54},
  number={4},
  pages={3165},
  year={1996},
  publisher={APS},
  doi={ 10.1103/PhysRevA.54.3165}
}

@article{mcmorran2008model,
  title={Model for partial coherence and wavefront curvature in grating interferometers},
  author={McMorran, Ben and Cronin, Alexander D},
  journal={Physical Review A},
  volume={78},
  number={1},
  pages={013601},
  year={2008},
  publisher={APS},
  doi={ 10.1103/PhysRevA.78.013601}
}

@book{goodman2015statistical,
  title={Statistical optics},
  author={Goodman, Joseph W},
  year={2015},
  publisher={John Wiley \& Sons}
}

@article{TWIST:2011jfx,
    author = "Bayes, R. and others",
    collaboration = "TWIST",
    title = "{Experimental Constraints on Left-Right Symmetric Models from Muon Decay}",
    doi = "10.1103/PhysRevLett.106.041804",
    journal = "Phys. Rev. Lett.",
    volume = "106",
    pages = "041804",
    year = "2011"
}

@article{TWIST:2011egd,
    author = "Bueno, J. F. and others",
    collaboration = "TWIST",
    title = "{Precise measurement of parity violation in polarized muon decay}",
    eprint = "1104.3632",
    archivePrefix = "arXiv",
    primaryClass = "hep-ex",
    doi = "10.1103/PhysRevD.84.032005",
    journal = "Phys. Rev. D",
    volume = "84",
    pages = "032005",
    year = "2011"
}

@article{Webber:2010zf,
    author = "Webber, D. M. and others",
    collaboration = "MuLan",
    title = "{Measurement of the Positive Muon Lifetime and
                  Determination of the Fermi Constant to
                  Part-per-Million Precision}",
    eprint = "1010.0991",
    archivePrefix = "arXiv",
    primaryClass = "hep-ex",
    doi = "10.1103/PhysRevLett.106.079901",
    journal = "Phys. Rev. Lett.",
    volume = "106",
    pages = "041803",
    year = "2011"
}

@article{Fetscher:1986uj,
    author = "Fetscher, W. and Gerber, H. J. and Johnson, K. F.",
    title = "{Muon Decay: Complete Determination of the Interaction
                  and Comparison with the Standard Model}",
    reportNumber = "ETHZ-IMP-P86/3",
    doi = "10.1016/0370-2693(86)91239-6",
    journal = "Phys. Lett. B",
    volume = "173",
    pages = "102--106",
    year = "1986"
}

\newpage 
\printacronyms

\end{document}